%% file: BASS_Arxiv.tex
\documentclass[apj]{emulateapj}

\usepackage[backref,breaklinks,colorlinks,citecolor=blue]{hyperref}
\usepackage{hyperref}
\usepackage[all]{hypcap}
\renewcommand*{\backref}[1]{[#1]}
\usepackage{graphicx, natbib, subfigure}

\usepackage{apjfonts}
\usepackage{amsmath}
\usepackage{textcomp}
\usepackage{listings}

\newcommand{\kms}{\hbox{km~s$^{-1}$}}
\newcommand{\Mjup}{$M_{\mathrm{Jup}}$}

\newcommand{\masyr}{$\mathrm{mas}\ \mathrm{yr}^{-1}$}
\newcommand{\asyr}{$\arcsec\ \mathrm{yr}^{-1}$}

\slugcomment{Accepted for publication in ApJ; Oct 16 2014.}

\shorttitle{BANYAN. V. AN ALL-SKY SURVEY FOR LMSs and BDs in YMGs}
\shortauthors{Gagn\'e et al.}

\begin{document}

\title{BANYAN. V. A SYSTEMATIC ALL-SKY SURVEY FOR NEW VERY LATE-TYPE LOW-MASS STARS AND BROWN DWARFS IN NEARBY YOUNG MOVING GROUPS}

\author{Jonathan Gagn\'e\altaffilmark{1},\, David Lafreni\`ere\altaffilmark{1},\, Ren\'e Doyon\altaffilmark{1},\, Lison Malo\altaffilmark{2,1}, \'Etienne Artigau\altaffilmark{1}}
\affil{\altaffilmark{1} D\'epartement de Physique, Universit\'e de Montr\'eal, C.P. 6128 Succ. Centre-ville, Montr\'eal, Qc H3C 3J7, Canada; \href{mailto:jonathan.gagne@astro.umontreal.ca}{jonathan.gagne@astro.umontreal.ca}}
\affil{\altaffilmark{2} Canada-France-Hawaii Telescope, 65-1238 Mamalahoa Hwy, Kamuela, HI 96743, USA}

\begin{abstract}

We present the BANYAN All-Sky Survey (\href{http://www.astro.umontreal.ca/\textasciitilde gagne/BASS.php}{\emph{BASS}}) catalog, consisting of 228 new late-type (M4--L6) candidate members of nearby young moving groups (YMGs) with an expected false-positive rate of $\sim$ 13\%. This sample includes 79 new candidate young brown dwarfs and 22 planetary-mass objects. These candidates were identified through the first systematic all-sky survey for late-type low-mass stars and brown dwarfs in YMGs. We cross-matched the \href{http://www.ipac.caltech.edu/2mass/}{\emph{2MASS}} and \href{http://wise2.ipac.caltech.edu/docs/release/allwise/}{\emph{AllWISE}} catalogs outside of the galactic plane to build a sample of 98\,970 potential $\geq$~M5 dwarfs in the solar neighborhood and calculated their proper motions with typical precisions of 5--15 \masyr. We selected highly probable candidate members of several YMGs from this sample using the Bayesian Analysis for Nearby Young AssociatioNs~II tool (\href{http://www.astro.umontreal.ca/\textasciitilde gagne/banyanII.php}{BANYAN~II}). We used the most probable statistical distances inferred from \href{http://www.astro.umontreal.ca/\textasciitilde gagne/banyanII.php}{BANYAN~II} to estimate the spectral type and mass of these candidate YMG members. We used this unique sample to show tentative signs of mass segregation in the AB~Doradus moving group and the Tucana-Horologium and Columba associations. The \href{http://www.astro.umontreal.ca/\textasciitilde gagne/BASS.php}{\emph{BASS}} sample has already been successful in identifying several new young brown dwarfs in earlier publications, and will be of great interest in studying the initial mass function of YMGs and for the search of exoplanets by direct imaging; the input sample of potential close-by $\geq$~M5 dwarfs will be useful to study the kinematics of low-mass stars and brown dwarfs and search for new proper motion pairs.

\end{abstract}


\section{INTRODUCTION}

A few decades ago, several groups of stars sharing similar galactic space velocities have been identified in the solar neighborhood. These similar kinematics are a consequence of the young age (typically 10--200~Myr) of these groups (i.e. young moving groups; YMGs), which formed from a common origin. The closest and youngest YMGs include the TW~Hydrae association (TWA; \citealp{1989ApJ...343L..61D}, \citealp{2004ARAA..42..685Z}; 5 -- 15~Myr; \citealp{2013ApJ...762..118W}), $\beta$~Pictoris ($\beta$PMG; \citealp{2001ApJ...562L..87Z}; 20 -- 26~Myr; \citealp{2014arXiv1409.2737M}, \citealp{2014ApJ...792...37M}, \citealp{2014MNRAS.438L..11B}), Tucana-Horologium (THA; \citealp{2000AJ....120.1410T}, \citealp{2000ApJ...535..959Z}; 20 -- 40~Myr; \citealp{2014AJ....147..146K}), Carina (CAR; 20 -- 40~Myr; \citealp{2008hsf2.book..757T}), Columba (COL; 20 -- 40~Myr; \citealp{2008hsf2.book..757T}),  Argus (ARG; 30 -- 50~Myr; \citealp{2000MNRAS.317..289M}) and AB~Doradus (ABDMG;  \citealp{2004ApJ...613L..65Z}; 110 -- 130~Myr ; \citealp{2005ApJ...628L..69L}, \citealp{2013ApJ...766....6B}). Identifying these YMGs was made possible with the advent of the \href{http://www.rssd.esa.int/index.php?project=HIPPARCOS}{\emph{HIPPARCOS}} survey \citep{1997AA...323L..49P}, which provided parallax measurements for $\sim$ 120,000 bright stars. Because of its limited sensitivity and the fact that it operated at visible wavelengths, this survey mainly studied stars with spectral types earlier than $\sim$~K0. Identifying the missing later-type, low-mass members of YMGs is of great interest for multiple reasons: it would provide constraints on the low-mass end of their initial mass function (IMF) and accessible benchmarks for cool, low-pressure atmospheres, similar to those of directly imaged giant planets (e.g. \citealp{2012AA...548A..26D}; \citealp{2013AJ....145....2F}; \citealp{2013ApJ...777L..20L}). Furthermore, direct imaging of exoplanets around these low-mass members would be facilitated by their proximity and the fact that younger planets are hotter, and thus brighter
(e.g. see \citealp{2012ApJ...753..142B}; \citealp{2012ApJ...756...69B}; \citealp{2013AA...553L...5D}; \citealp{2013ApJ...774...55B}; \citealp{2014ApJ...787....5N}). For these reasons, a large number of studies were aimed at finding these missing low-mass members and refine our understanding of YMGs (see \citealp{2003AJ....125..825T}; \citealp{2004AJ....127.2246W}; \citealp{2006AA...460..695T}; \citealp{2007ApJ...669L..97L}; \citealp{2009ApJ...699..649S}; \citealp{2009AA...506..799B}; \citealp{2009AJ....137.3632L}; \citealp{2010AJ....140..119S}; \citealp{2010AJ....140.1486L}; \citealp{2010ApJ...714...45L}; \citealp{2010ApJ...715L.165R}; \citealp{2011ApJ...727...62R}; \citealp{2011MNRAS.411..117K}; \citealp{2012AJ....143...80S}; \citealp{2012ApJ...754...39S}; \citealp{2012ApJ...752...56F}; \citealp{2012ApJ...758...56S}; \citealp{2012AA...548A..26D}; \citealp{2012AJ....144..109S}; \citealp{2013ApJ...762...88M}; \citealp{2013AJ....145....2F}; \citealp{2013ApJ...762..118W}; \citealp{2013MNRAS.435.1376M}; \citealp{2013ApJ...774..101R}; \citealp{2013ApJ...777L..20L}; \citealp{2013ApJ...779..153H}; \citealp{2014AJ....147...34S}; \citealp{2014AJ....147..146K}; \citealp{2014ApJ...783..121G}; \citealp{2014AA...562A.127B}; \citealp{2014ApJ...785L..14G}; \citealp{2014ApJ...788...81M}; \citealp{2014AJ....147...85R}; \citealp{2014ApJ...792...37M}; \citealp{2014AA...564A..55M}; \citealp{2014ApJ...792L..17G}; \citealp{2014AA...568A...6Z}; \citealp{2014arXiv1409.2737M}).

The identification of later-type members of nearby YMGs is a challenging task in the absence of reliable parallax and radial velocity (RV) measurements since their members are spread on large regions of the celestial sphere. Furthermore, obtaining parallax and RV measurements for such faint targets is time-consuming. Careful pre-selection of candidates is thus essential to keep the follow-up effort to a manageable size. Efforts have already been made in identifying late-type members in YMGs, notably by selecting X-ray or UV-bright stars (\citealp{2008hsf2.book..757T}, \citealp{2011ApJ...727...62R}, \citealp{2012ApJ...758...56S}) and by comparing their proper motions to those of known members with the convergent point proper motion analysis (CPA; \citealp{2001MNRAS.328...45M}, \citealp{2013ApJ...774..101R}). However, this method does not use all available measurements (e.g. photometry, magnitude of proper motion, RV and parallax), therefore it generally suffers from a large contamination of field stars that have proper motions similar to those of YMG members by pure chance, as well as cross-contamination between different YMG candidates. In particular, some YMGs such as COL, $\beta$PMG and TWA happen to share similar proper motion distributions as viewed from the Earth, which makes it difficult to differentiate their members using only sky position and the direction of proper motion without radial velocity measurements.

To address these problems, \cite{2013ApJ...762...88M} developed the Bayesian Analysis for Nearby Young AssociatioNs (\href{http://www.astro.umontreal.ca/\textasciitilde malo/banyan.php}{BANYAN}\footnote{Publicly available at \url{http://www.astro.umontreal.ca/\textasciitilde malo/banyan.php}.}), a statistical tool based on Bayesian inference, to identify strong K5--M5 candidate members of YMGs primarily from a sample of X-ray bright sources. In addition to proper motion and sky position, this tool takes advantage of $I_C$ and \emph{J} photometry measurements to ensure that candidate members fall in a region of the color-magnitude diagram (CMD) consistent with other YMG members; younger low-mass stars (LMSs) and brown dwarfs (BDs) are inflated and thus brighter than field stars as they are still undergoing gravitational contraction. This approach provides a more robust set of candidates, as well as most probable distance and RV predictions. However, this study is still limited to detecting candidates with spectral types earlier than $\sim$ M5, and photometric measurements in the $I_C$ band are required to take CMD information into account. In parallel, \cite{2014ApJ...783..121G} presented \href{http://www.astro.umontreal.ca/\textasciitilde gagne/banyanII.php}{BANYAN~II}\footnote{Publicly available at \url{http://www.astro.umontreal.ca/\textasciitilde gagne/banyanII.php}.}, a new selection tool based on BANYAN that includes several improvements (e.g. a better modeling of YMGs spatial and kinematic properties and an extensive treatment of contamination and completeness), and is specifically designed to identify ~> M5 YMG candidates, by relying on two different CMDs constructed with photometry from the Two Micron All-Sky Survey (\href{http://www.ipac.caltech.edu/2mass/}{\emph{2MASS}};  \citealp{2006AJ....131.1163S}) and the \href{http://irsa.ipac.caltech.edu/Missions/wise.html}{\emph{WISE}} survey \citep{2010AJ....140.1868W}. This tool was used in \cite{2014ApJ...783..121G} to identify 39 new M5--L4 candidate members among known young field LMSs and BDs. Recently, \cite{2014AJ....147..146K} identified 129 new K3--M6 strong candidate members of THA by carrying extensive RV measurements of targets selected for having proper motion and CMD positions similar to those of other THA members. Their results indicate that samples based on \href{http://www.galex.caltech.edu/}{\emph{GALEX}} (\href{http://vizier.u-strasbg.fr/viz-bin/VizieR?-meta.foot&-source=I/252}{\emph{USNO--A2.0}} (\href{http://vizier.u-strasbg.fr/viz-bin/VizieR}{VizieR} catalog \href{http://vizier.u-strasbg.fr/viz-bin/VizieR?-meta.foot&-source=II/312}{\emph{II/312}} and \citealp{2005ApJ...619L...1M}) or \href{http://heasarc.gsfc.nasa.gov/docs/rosat/}{\emph{ROSAT}} (\href{http://vizier.u-strasbg.fr/viz-bin/VizieR?-meta.foot&-source=I/252}{\emph{USNO--A2.0}} (\href{http://vizier.u-strasbg.fr/viz-bin/VizieR}{VizieR} catalog \href{http://vizier.u-strasbg.fr/viz-bin/VizieR?-meta.foot&-source=IX/29}{\emph{IX/29}} and \citealp{1999AA...349..389V}) miss candidates later than $\sim$ M2 at distances beyond $\gtrsim$ 40 pc.

We present here the BANYAN All-sky Survey (\href{http://www.astro.umontreal.ca/\textasciitilde gagne/BASS.php}{\emph{BASS}}), which is the first all-sky, systematic survey for $\geq$ M5 LMSs and BDs in YMGs. The whole \href{http://www.ipac.caltech.edu/2mass/}{\emph{2MASS}} and \href{http://wise2.ipac.caltech.edu/docs/release/allwise/}{\emph{AllWISE}} \citep{2014ApJ...783..122K} catalogs outside of the galactic plane ($|b| > 15$\textdegree) were cross-matched, yielding proper motions with typical precisions of a few \masyr. Color-quality cuts as well as the \href{http://www.astro.umontreal.ca/\textasciitilde gagne/banyanII.php}{BANYAN~II} tool were used to select 153 high- and 21 modest-probability candidate members of YMGs, for which near-infrared (NIR) colors are consistent with $\geq$ M5 spectral types. The \href{http://www.astro.umontreal.ca/\textasciitilde gagne/BASS.php}{\emph{BASS}} survey has already generated a wealth of new discoveries, including a triple M5 + M5 + planetary-mass companion in THA (\citealp{2013AA...553L...5D}; \href{http://www.astro.umontreal.ca/\textasciitilde gagne/J0103inprep.php}{J.~Gagn\'e et al., in preparation}), an M5 + L4 host--planet system candidate member of THA (\href{http://www.astro.umontreal.ca/\textasciitilde gagne/banyanVI.php}{\'E.~Artigau et al., in preparation}), a new L-type candidate member of TWA \citep{2014ApJ...785L..14G} and a new low-gravity L4$\beta$ BD candidate member of ARG \citep{2014ApJ...792L..17G}. A NIR and optical spectroscopic follow-up of all candidates that will be presented here is undergoing; first results were presented in \cite{2013MmSAI..84..916G} and more will be presented in a subsequent paper (\href{http://www.astro.umontreal.ca/\textasciitilde gagne/banyanVII.php}{J.~Gagn\'e et al., in preparation}).

In \hyperref[sec:crossmatch]{Section~\ref*{sec:crossmatch}}, we detail our method for cross-matching the \href{http://www.ipac.caltech.edu/2mass/}{\emph{2MASS}} and \href{http://wise2.ipac.caltech.edu/docs/release/allwise/}{\emph{AllWISE}} catalogs, which we follow by a description of the various color-quality cuts applied, and how we use the \href{http://www.astro.umontreal.ca/\textasciitilde gagne/banyanII.php}{BANYAN~II} tool to select candidates members of YMGs (\hyperref[sec:filters]{Section~\ref*{sec:filters}}). In 
\hyperref[sec:lit]{Section~\ref*{sec:lit}}, we present all information available in the literature for the \href{http://www.astro.umontreal.ca/\textasciitilde gagne/BASS.php}{\emph{BASS}} catalog, which we used to update the membership probability when relevant. In \hyperref[sec:recov]{Section~\ref*{sec:recov}}, we evaluate the recovery rate of the \href{http://www.astro.umontreal.ca/\textasciitilde gagne/BASS.php}{\emph{BASS}} sample for known $\geq$ M5 candidate members and bona fide members of YMGs. We then present various characteristics of the updated \href{http://www.astro.umontreal.ca/\textasciitilde gagne/BASS.php}{\emph{BASS}} catalog in \hyperref[sec:updated]{Section~\ref*{sec:updated}}. In \hyperref[sec:common_pm]{Section~\ref*{sec:common_pm}}, we search for new common proper motion pairs among our sample, and we tentatively investigate mass segregation in \hyperref[sec:mseg]{Section~\ref*{sec:mseg}}. Conclusion are presented in Section~\hyperref[sec:conclusion]{Section~\ref*{sec:conclusion}}. The \emph{Low-Priority} \emph{BASS} (\href{http://www.astro.umontreal.ca/\textasciitilde gagne/LP-BASS.php}{\emph{LP-BASS}}) sample, consisting of objects only marginally redder than field dwarfs, is presented in Appendix, along with our full input sample of 98\,970 potential close-by $\geq$~M5 dwarfs.

\section{CROSS-MATCHING THE \href{http://www.ipac.caltech.edu/2mass/}{\emph{2MASS}} AND \href{http://wise2.ipac.caltech.edu/docs/release/allwise/}{\emph{AllWISE}} CATALOGS}\label{sec:crossmatch}

Cross-matching the \href{http://www.ipac.caltech.edu/2mass/}{\emph{2MASS}} and \href{http://wise2.ipac.caltech.edu/docs/release/allwise/}{\emph{AllWISE}} catalogs ($\sim$ 470 million and $\sim$ 750 million entries respectively) without the use of significant computational resources is a challenge that must be tackled in a strategic way. Fortunately, the NASA Infrared Science Archive (\href{http://irsa.ipac.caltech.edu/}{IRSA}\footnote{Available at \url{http://irsa.ipac.caltech.edu/}}; \citealp{2010AAS...21543805G}) provides useful tools to achieve this. In a first step, we have built two distinct queries for the \href{http://www.ipac.caltech.edu/2mass/}{\emph{2MASS}} and \href{http://wise2.ipac.caltech.edu/docs/release/allwise/}{\emph{AllWISE}} catalogs to target only potential nearby $\geq$ M5 dwarfs. We start from spectral type-color relations described in \cite{2013ApJS..208....9P}, \cite{2011ApJS..197...19K} and \cite{2012ApJS..201...19D} to select only targets that have NIR colors consistent with $\geq$~M5 spectral types, which we subsequently relax to include all currently known young dwarfs in the same range of spectral types (see \citealp{2014ApJ...783..121G} for an extensive list of known young LMSs and BDs in the field). We target only regions of the sky located more than 15 degrees away from the galactic plane, require that measurements of $J$, $H$, $K_S$, \emph{W1} and \emph{W2} photometry have a reasonable quality, and that no contamination or saturation flags are problematic. We also reject sources spatially resolved in \href{http://www.ipac.caltech.edu/2mass/}{\emph{2MASS}} but not in \href{http://wise2.ipac.caltech.edu/docs/release/allwise/}{\emph{AllWISE}}. In the Appendix, we list the requirements in the form of two Structured Query Language (SQL) statements that were used to perform all-sky \href{http://irsa.ipac.caltech.edu/}{IRSA} queries, which correspond to\footnote{See the column descriptions of the \href{http://www.ipac.caltech.edu/2mass/}{\emph{2MASS}} User's Guide \url{http://www.ipac.caltech.edu/2mass/releases/allsky/doc/sec2\_2a.html} and the \href{http://wise2.ipac.caltech.edu/docs/release/allwise/}{\emph{AllWISE}} User's Guide \url{http://wise2.ipac.caltech.edu/docs/release/allwise/expsup/sec2\_1a.html} for additional information on the keywords.} :

\begin{itemize}
	\item The absolute galactic latitude $|b|$ of both \href{http://www.ipac.caltech.edu/2mass/}{\emph{2MASS}} and \href{http://wise2.ipac.caltech.edu/docs/release/allwise/}{\emph{AllWISE}} counterparts respect $|b| > 15$ \textdegree.
	\item $J > 2$, $H > 2$, $K_S > 2$, $W1 > 2$ and $W2 > 2$.
	\item $0.506 < J - H < 2$, $0.269 < H - K_S < 1.6$ and $0.168 < W1 - W2 < 2.5$.
	\item $W1 - W2 < ( 0.96\cdot(W2 - W3) - 0.96 )$ if $W3$ is detected with SNR $> 5$ and not saturated \citep{2011ApJS..197...19K}.
	\item If a \href{http://www.ipac.caltech.edu/2mass/}{\emph{2MASS}} counterpart is identified in the \href{http://wise2.ipac.caltech.edu/docs/release/allwise/}{\emph{AllWISE}} catalog, it must be at least at an angular distance 0\farcs3 from the \href{http://wise2.ipac.caltech.edu/docs/release/allwise/}{\emph{AllWISE}} coordinates (i.e., to reject low proper motion objects) and respect $0.153 < K_S - W1 < 2$ in addition to the \href{http://www.ipac.caltech.edu/2mass/}{\emph{2MASS}} color cuts described above.	
	\item The blue magnitude $B$, which is either the Johnson $B_J$ magnitude of a \emph{Tycho 2} \citep{2000AA...355L..27H} counterpart, or the photographic blue magnitude of a \emph{USNO--A2.0} \citep{1998usno.book.....M} counterpart of the \href{http://www.ipac.caltech.edu/2mass/}{\emph{2MASS}} object (\href{http://www.ipac.caltech.edu/2mass/releases/allsky/doc/sec2_2a.html#b_m_opt}{\emph{B\_M\_OPT}} keyword) is either undetected or has $B - J \geq 4.048$.
	\item The red or visible $VR$ magnitude, which is either the Johnson $V_J$ magnitude of a \emph{Tycho 2} counterpart, or the photographic red magnitude of a \href{http://vizier.u-strasbg.fr/viz-bin/VizieR?-meta.foot&-source=I/252}{\emph{USNO--A2.0}} (\href{http://vizier.u-strasbg.fr/viz-bin/VizieR}{VizieR} catalog \href{http://vizier.u-strasbg.fr/viz-bin/VizieR?-meta.foot&-source=I/252}{\emph{I/252}}) counterpart of the \href{http://www.ipac.caltech.edu/2mass/}{\emph{2MASS}} object (\href{http://www.ipac.caltech.edu/2mass/releases/allsky/doc/sec2_2a.html#vr_m_opt}{\emph{VR\_M\_OPT}} keyword) is either undetected or has $VR - J \geq 2.63$ and $B - VR \geq 1.3$.
	\item At least two \href{http://www.ipac.caltech.edu/2mass/}{\emph{2MASS}} bands have excellent (A) or good (B) photometric quality flags.
	\item No \href{http://www.ipac.caltech.edu/2mass/}{\emph{2MASS}} band has a poor (D, E or F) or undetected (X or U) quality flags.
	\item The \href{http://wise2.ipac.caltech.edu/docs/release/allwise/}{\emph{AllWISE}} photometric quality flags of the $W1$ and $W2$ bands are either excellent (A) or good (B).
	\item The angular distance between the object and its closest neighbor is at least 6\farcs4 in \href{http://www.ipac.caltech.edu/2mass/}{\emph{2MASS}}, to ensure that they are resolved in \href{http://wise2.ipac.caltech.edu/docs/release/allwise/}{\emph{AllWISE}}.
	\item There are less than 0.2\% of saturated pixels in the profile fitting regions of both the $W1$ and $W2$ bands in \href{http://wise2.ipac.caltech.edu/docs/release/allwise/}{\emph{AllWISE}}.
	\item  The source is detected in the $W1$ and $W2$ \href{http://wise2.ipac.caltech.edu/docs/release/allwise/}{\emph{AllWISE}} bands with a statistical significance larger than $5\sigma$.
	\item  The reduced $\chi^2$ of the profile fits for the $W1$ and $W2$ \href{http://wise2.ipac.caltech.edu/docs/release/allwise/}{\emph{AllWISE}} bands both respect $\chi^2 < 5$.
	\item The \href{http://www.ipac.caltech.edu/2mass/}{\emph{2MASS}} read flags do not contain 0 (no detection in any band), 6 (not detected in one band) or 9 (nominally detected in one band because of confused regions) for any band.
	\item The \href{http://www.ipac.caltech.edu/2mass/}{\emph{2MASS}} blend flag is 1 (only one component was fit simultaneously for photometry) for all bands.
	\item The \href{http://www.ipac.caltech.edu/2mass/}{\emph{2MASS}} contamination flag is 0 (not contaminated) for all bands.
	\item The \href{http://www.ipac.caltech.edu/2mass/}{\emph{2MASS}} extragalactic contamination flag is 0 (resolved and not extended).
	\item The \href{http://www.ipac.caltech.edu/2mass/}{\emph{2MASS}} minor planet flag is 0 (not associated with a known solar system object).
	\item The \href{http://wise2.ipac.caltech.edu/docs/release/allwise/}{\emph{AllWISE}} contamination flags of the $W1$ and $W2$ bands do not correspond to potentially spurious detections (D, due to a diffraction spike; P, due to detector persistence; H, due to the scattered light of a bright nearby source; or O, due to an optical ghost caused by a nearby bright source).
	\item  The \href{http://wise2.ipac.caltech.edu/docs/release/allwise/}{\emph{AllWISE}} extended flag is either 0 (consistent with a point source) or 1 (goodness-of-fit of the profile fitting is larger than 3 in at least one band).
\end{itemize}

These queries generated two lists: 2\,762\,191 objects from \href{http://www.ipac.caltech.edu/2mass/}{\emph{2MASS}} and 76\,883\,849 objects from \href{http://wise2.ipac.caltech.edu/docs/release/allwise/}{\emph{AllWISE}}. To avoid obtaining very large output file sizes, we downloaded only designations, RA and DEC positions, as well as \href{http://www.ipac.caltech.edu/2mass/}{\emph{2MASS}} unique identifiers at this stage (keyword \href{http://www.ipac.caltech.edu/2mass/releases/allsky/doc/sec2_2a.html#pts_key}{\emph{CNTR}} in the \href{http://www.ipac.caltech.edu/2mass/}{\emph{2MASS}} catalog, and \href{http://wise2.ipac.caltech.edu/docs/release/allwise/expsup/sec2_1a.html#tmass_key}{\emph{TMASS\_KEY}} in the \href{http://wise2.ipac.caltech.edu/docs/release/allwise/}{\emph{AllWISE}} catalog; the \href{http://irsa.ipac.caltech.edu/}{IRSA} team already identified \href{http://www.ipac.caltech.edu/2mass/}{\emph{2MASS}}--\href{http://wise2.ipac.caltech.edu/docs/release/allwise/}{\emph{AllWISE}} cross-matches within 3\textquotedbl). We then locally rejected all objects located in the following star-forming regions to avoid heavily reddened contaminants : Orion (5h29m < RA < 5h41m and -06\textdegree37\textquotesingle\ < DEC < -02\textdegree25\textquotesingle; \citealp{1999ApJ...521..671B}), Taurus (3h50m < RA < 5h15m and 15\textdegree < DEC < 32\textdegree; \citealp{2004ApJ...617.1216L}), Chamaeleon (10h45m < RA < 11h30m and -78\textdegree30\textquotesingle\ < DEC < -76\textdegree; \citealp{2007ApJS..173..104L}; \citealp{2012AA...539A.151A}) and Upper Scorpius (15h35m < RA < 16h45m and -30\textdegree\ < DEC < -21\textdegree; \citealp{2011MNRAS.418.1231D}). We subsequently counted the number of \href{http://www.ipac.caltech.edu/2mass/}{\emph{2MASS}} neighbors in a 3\textquotesingle\ radius around each target in the \href{http://www.ipac.caltech.edu/2mass/}{\emph{2MASS}} subset, and rejected all those with more than 71 neighbors to avoid densely populated regions. This number was chosen so that none of the known young brown dwarfs in the field and outside of the galactic plane were rejected. This cut down the number of \href{http://www.ipac.caltech.edu/2mass/}{\emph{2MASS}} targets to 2\,178\,389. We then locally cross-matched the unique \href{http://www.ipac.caltech.edu/2mass/}{\emph{2MASS}} identifiers of both catalogs to construct list A, consisting of 169\,934 \href{http://www.ipac.caltech.edu/2mass/}{\emph{2MASS}} sources which already had an \href{http://wise2.ipac.caltech.edu/docs/release/allwise/}{\emph{AllWISE}} counterpart identified in the latter catalog. The remaining unmatched 2\,008\,455 \href{http://www.ipac.caltech.edu/2mass/}{\emph{2MASS}} sources, as well as the 75\,478\,161 \href{http://wise2.ipac.caltech.edu/docs/release/allwise/}{\emph{AllWISE}} sources with null \href{http://www.ipac.caltech.edu/2mass/}{\emph{2MASS}} keys, were saved as lists B and C, respectively. \href{http://wise2.ipac.caltech.edu/docs/release/allwise/}{\emph{AllWISE}} sources with non-null \href{http://www.ipac.caltech.edu/2mass/}{\emph{2MASS}} entries that were not cross-matched this way were rejected, since they must have failed at least one of the \href{http://www.ipac.caltech.edu/2mass/}{\emph{2MASS}} constraints described above.

\begin{figure}
	\centering
	\subfigure{\includegraphics[width=0.495\textwidth]{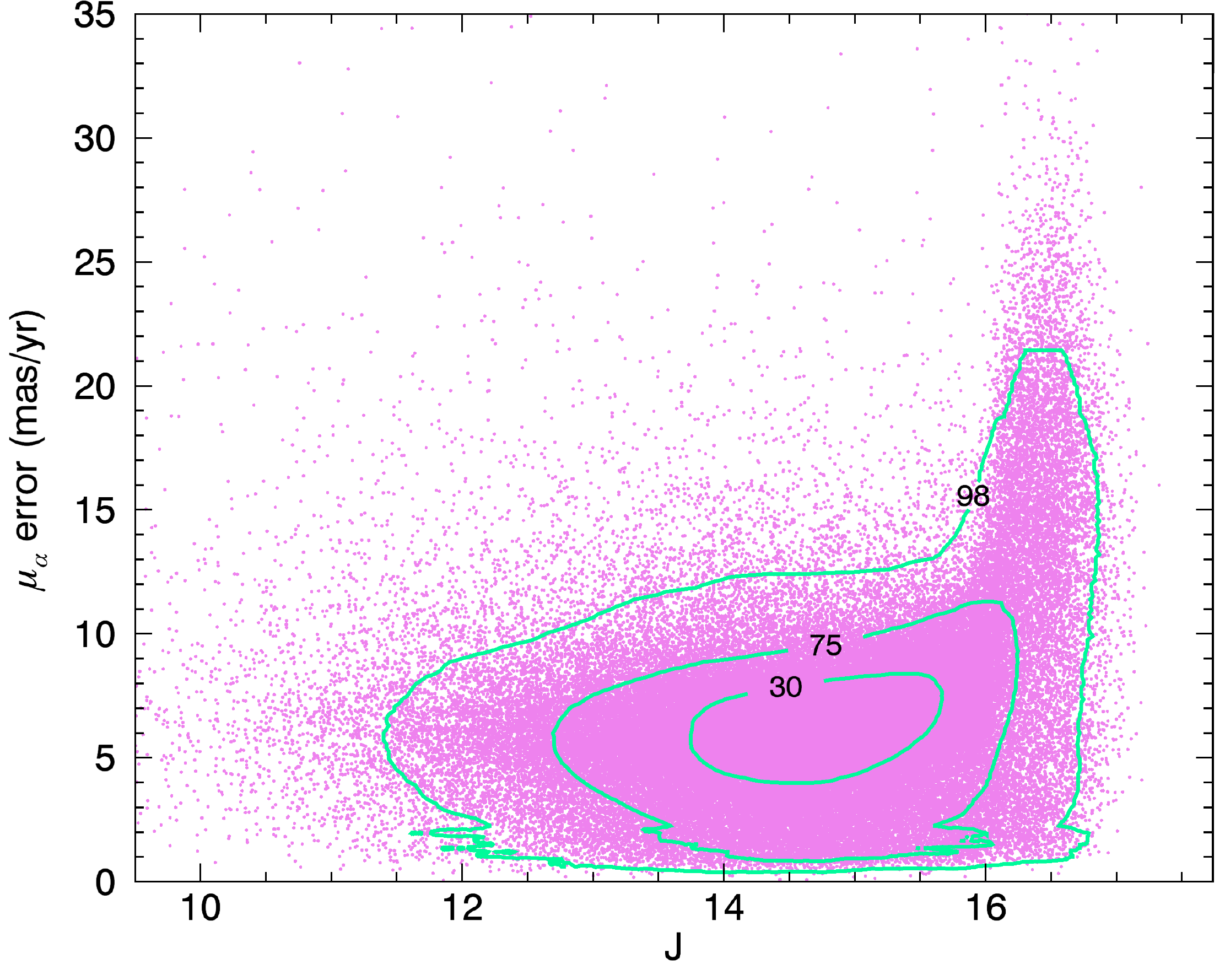}}
	\subfigure{\includegraphics[width=0.495\textwidth]{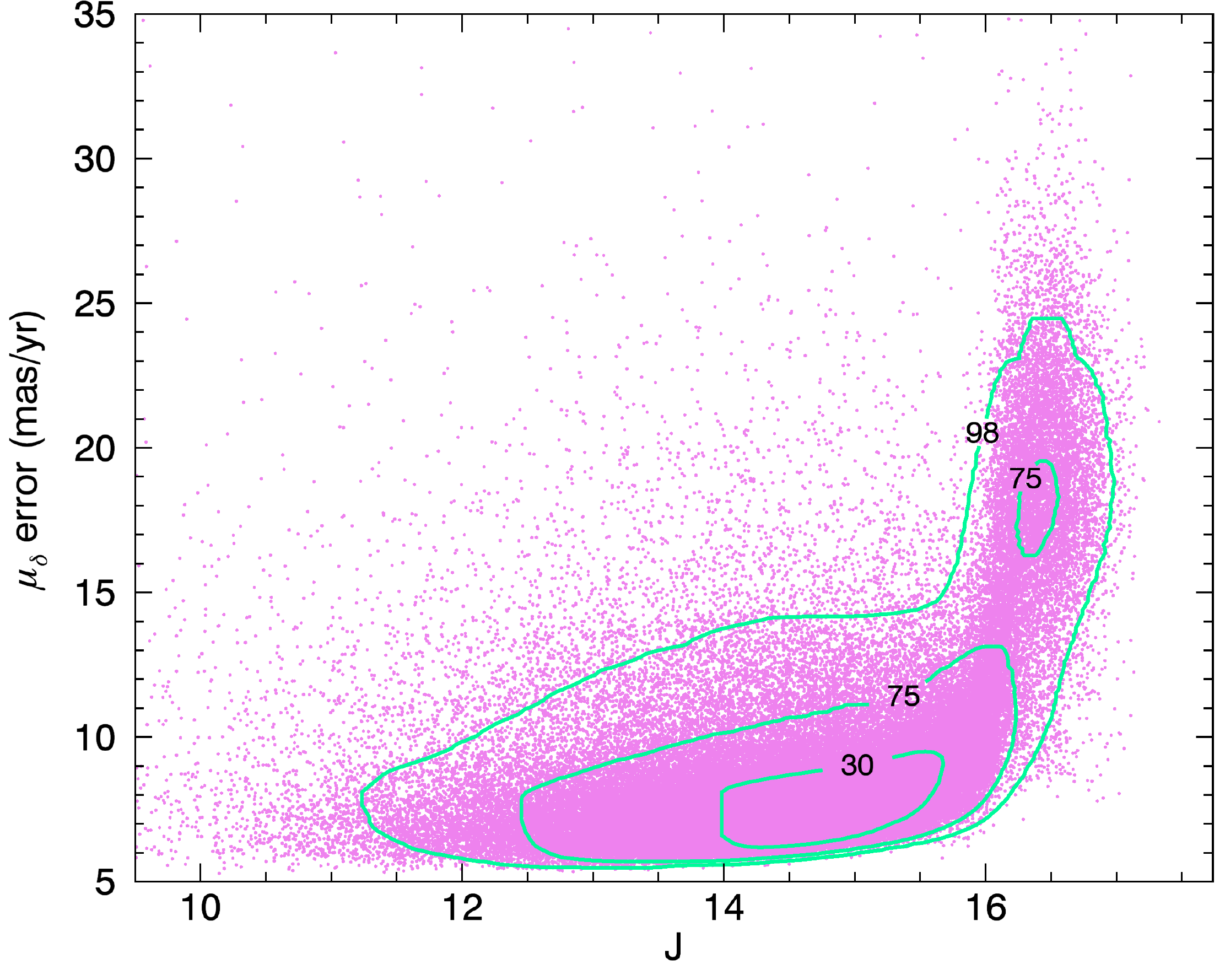}}
	\caption{Proper motion precision as a function of \href{http://www.ipac.caltech.edu/2mass/}{\emph{2MASS}} J magnitude in List~A (pink points; see \hyperref[sec:crossmatch]{Section~\ref*{sec:crossmatch}}). Green contour lines respectively include 10\%, 75\% and 98\% of all data points. In the case of bright objects ($J < 16$), typical precisions are 3--10 \masyr\ ($\mu_\alpha\cos\delta$) and 5--10 \masyr\ ($\mu_\delta$), whereas they can go down to $\sim$ 25\masyr\ for fainter objects.}
	\label{fig:pmprec}
\end{figure}

We created preliminary cross-matches by identifying the closest \href{http://wise2.ipac.caltech.edu/docs/release/allwise/}{\emph{AllWISE}} entry in List C to each \href{http://www.ipac.caltech.edu/2mass/}{\emph{2MASS}} entry in list B. A total of 2\,001\,246 of those preliminary matches were separated by distances larger than 25\textquotedbl\ (equivalent to a proper motion $> 2.2$ \asyr) or had $K_S - W1 < 0.153$ or $K_S - W1 > 2$, and were rejected. For each \href{http://www.ipac.caltech.edu/2mass/}{\emph{2MASS}} component of the remaining 7\,209 pairs (separated by angular distances of $\delta$), we subsequently downloaded all \href{http://wise2.ipac.caltech.edu/docs/release/allwise/}{\emph{AllWISE}} entries within $\delta$, and verified that the closest entry with a null \emph{2MASS\_KEY} corresponded to our preliminary match. We also verified that the \emph{2MASS\_KEY} was not assigned to any other nearby \href{http://wise2.ipac.caltech.edu/docs/release/allwise/}{\emph{AllWISE}} source. This step has rejected 767 objects. In a final step, we downloaded all \href{http://www.ipac.caltech.edu/2mass/}{\emph{2MASS}} and \href{http://wise2.ipac.caltech.edu/docs/release/allwise/}{\emph{AllWISE}} entries in a radius $\delta$+3\textquotedbl\ around every \href{http://wise2.ipac.caltech.edu/docs/release/allwise/}{\emph{AllWISE}} component of the 5\,876 remaining pairs, and removed all \href{http://irsa.ipac.caltech.edu/}{IRSA}-identified cross-matches. We use a search radius of $\delta$+3\textquotedbl\ in this step to ensure that we retrieve all \href{http://www.ipac.caltech.edu/2mass/}{\emph{2MASS}}--\href{http://wise2.ipac.caltech.edu/docs/release/allwise/}{\emph{AllWISE}} matches in the \href{http://wise2.ipac.caltech.edu/docs/release/allwise/}{\emph{AllWISE}} catalog in a radius $\delta$, since those matches can be separated by up to 3\textquotedbl. We then verified that the closest \href{http://www.ipac.caltech.edu/2mass/}{\emph{2MASS}} entry among those objects not already cross-matched by \href{http://irsa.ipac.caltech.edu/}{IRSA} corresponded to the \href{http://www.ipac.caltech.edu/2mass/}{\emph{2MASS}} component of the preliminary pairs: this filter rejected 2\,367 objects. The 3\,509 pairs that survived all these selection criteria were added to List A. We then used \href{http://www.ipac.caltech.edu/2mass/}{\emph{2MASS}} and \href{http://wise2.ipac.caltech.edu/docs/release/allwise/}{\emph{AllWISE}} astrometry to determine proper motions for all 173\,443 objects in this supplemented List A, and rejected the 74\,473 sources with a total proper motion lower than 30~\masyr, or with a total proper motion measurement at $< 5\sigma$, to reject extragalactic contaminants and red giants.

Proper motions were calculated directly from entries in both the \href{http://www.ipac.caltech.edu/2mass/}{\emph{2MASS}} and \href{http://wise2.ipac.caltech.edu/docs/release/allwise/}{\emph{AllWISE}} catalogs. The right ascension (\emph{RA}) and declination (\emph{DEC}) entries were used for the astrometric position of both catalogs; the \emph{SIGRA} and \emph{SIGDEC} entries of \href{http://wise2.ipac.caltech.edu/docs/release/allwise/}{\emph{AllWISE}} were used as a measurement error, and the \emph{ERR\_MAJ} ($\sigma_{\mathrm{MAJ}}$), \emph{ERR\_MIN} ($\sigma_{\mathrm{MIN}}$) and \emph{ERR\_ANG} ($\sigma_\theta$) entries of the \href{http://www.ipac.caltech.edu/2mass/}{\emph{2MASS}} catalog were projected back to errors on right ascension ($\sigma_\alpha$) and declination ($\sigma_\delta$) with :

\begin{align}
	\sigma_{\alpha} &= \sqrt{(\sigma_{\mathrm{MAJ}}\sin\sigma_\theta)^2 + (\sigma_{\mathrm{MIN}}\cos\sigma_\theta)^2}\cdot\cos\delta \\
	\sigma_{\delta} &= \sqrt{(\sigma_{\mathrm{MAJ}}\cos\sigma_\theta)^2 + (\sigma_{\mathrm{MIN}}\sin\sigma_\theta)^2}
\end{align}

where $\delta$ is the \href{http://www.ipac.caltech.edu/2mass/}{\emph{2MASS}} declination. The epochs corresponding to these astrometric measurements were taken from the \emph{JDATE} and \emph{W1MJDMEAN} entries in the respective catalogs. \emph{W1MJDMEAN} corresponds to the mean epoch of all \href{http://wise2.ipac.caltech.edu/docs/release/allwise/}{\emph{AllWISE}} exposures taken in the $W1$ band. The uncertainty on the \href{http://www.ipac.caltech.edu/2mass/}{\emph{2MASS}} epoch is taken to be 30 s, as described in the \href{http://www.ipac.caltech.edu/2mass/releases/allsky/doc/sec2_2b.html}{\emph{2MASS} User's Guide}, and the uncertainty on the \href{http://wise2.ipac.caltech.edu/docs/release/allwise/}{\emph{AllWISE}} epoch is taken in a conservative way as half of the maximal distance between all exposures (from the \emph{W1MJDMAX} and \emph{W1MJDMIN} entries). We analytically propagated all measurement errors (astrometric and temporal) of both catalogs, assuming they were all independent, to obtain the measurement errors on our \href{http://www.ipac.caltech.edu/2mass/}{\emph{2MASS}}--\href{http://wise2.ipac.caltech.edu/docs/release/allwise/}{\emph{AllWISE}} proper motions. The positional accuracy of the \href{http://www.ipac.caltech.edu/2mass/}{\emph{2MASS}} and \href{http://wise2.ipac.caltech.edu/docs/release/allwise/}{\emph{AllWISE}} catalogs vary from $\sim$ 0\farcs05 for bright sources ($J \lesssim 14$), to 0\farcs1--0\farcs4 (\href{http://www.ipac.caltech.edu/2mass/}{\emph{2MASS}}) and 0\farcs06--0\farcs15 (\href{http://wise2.ipac.caltech.edu/docs/release/allwise/}{\emph{AllWISE}}) for fainter sources. The final set of 98\,970 objects contains probable nearby $>$~M5 dwarfs with measurements of proper motion above 30 \masyr. We list this sample in the Appendix, since this sample provides a great opportunity to study the kinematics of LMSs and BDs in the solar neighborhood. In \hyperref[fig:pmprec]{Figure~\ref*{fig:pmprec}}, we show that typical measurement errors on proper motions are 5--10 \masyr\ for bright objects ($J < 16$), or 5--25 \masyr\ for fainter objects.

\begin{figure*}
	\centering
	\subfigure{\includegraphics[width=0.495\textwidth]{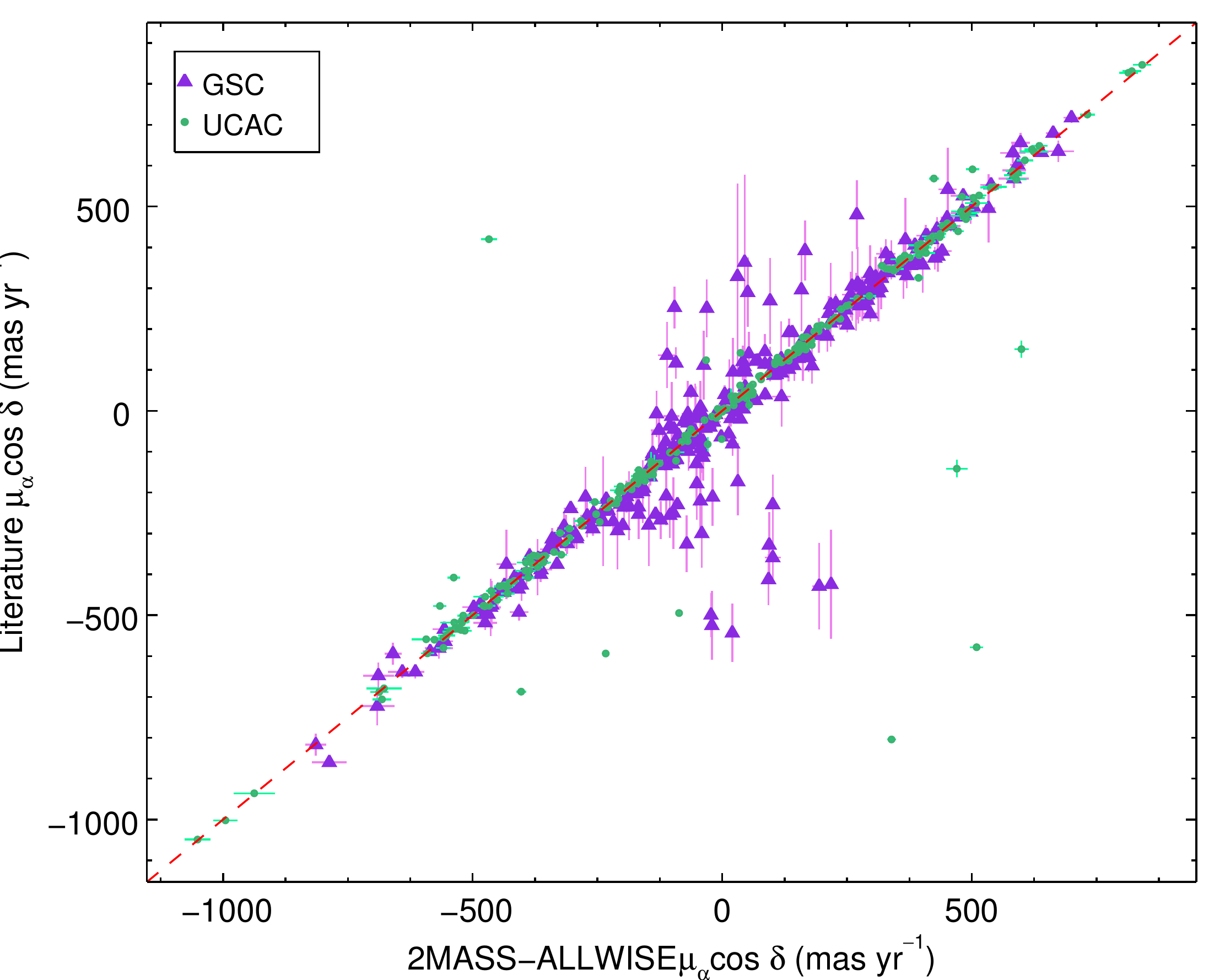}}
	\subfigure{\includegraphics[width=0.495\textwidth]{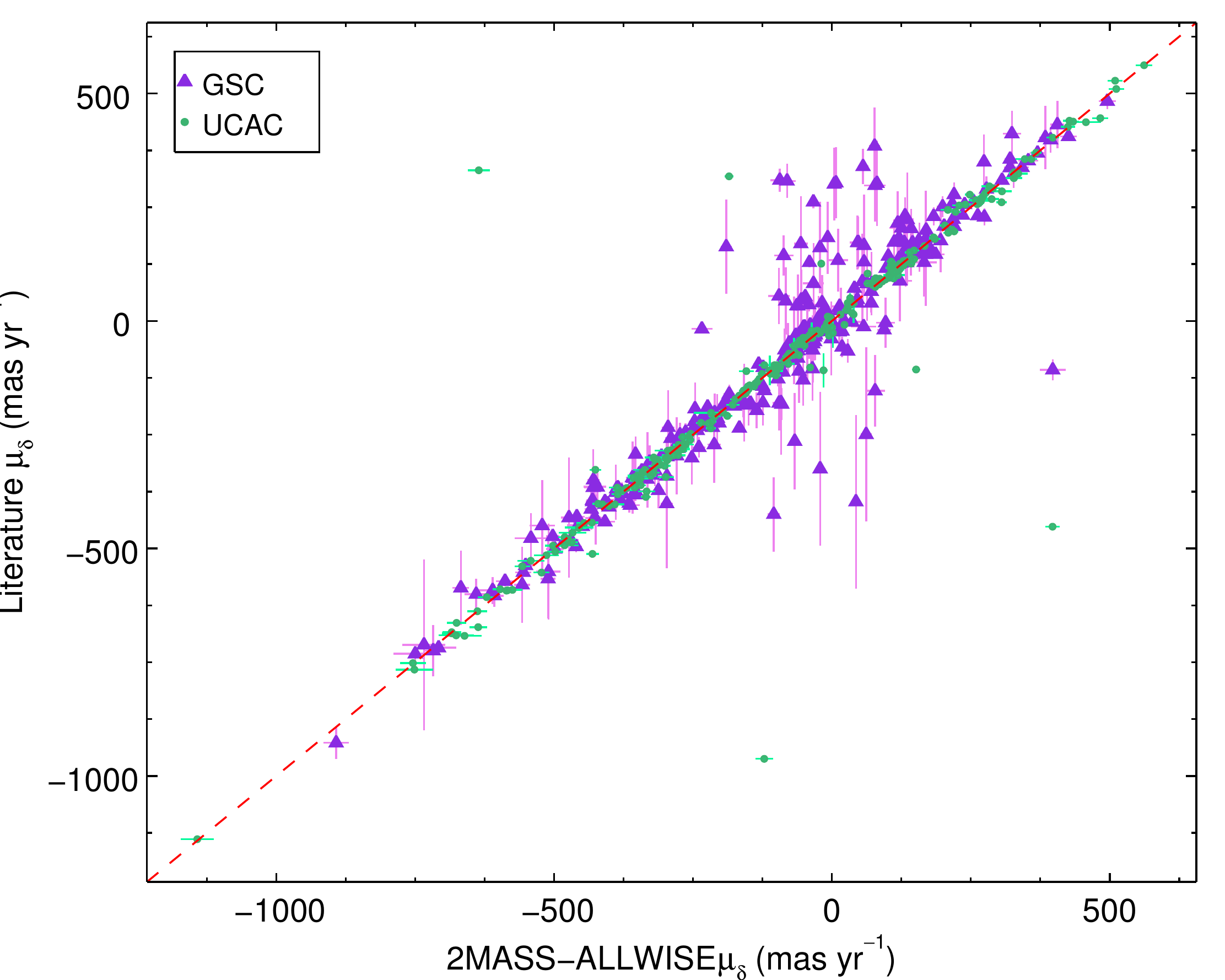}}
	\caption{Comparison between proper motions determined from the \href{http://www.ipac.caltech.edu/2mass/}{\emph{2MASS}} and \href{http://wise2.ipac.caltech.edu/docs/release/allwise/}{\emph{AllWISE}} datasets and measurements in the literature, for a random subset of the the input sample of 98\,970 objects. We only display 500 random objets per bin of $\sim$ 200 \masyr, to improve visibility. Measurements from the literature were obtained from the Initial Gaia Source List (\href{http://vizier.u-strasbg.fr/viz-bin/VizieR}{VizieR} catalog \href{http://vizier.u-strasbg.fr/viz-bin/VizieR?-meta.foot&-source=I/324}{\emph{I/324/igsl3}}) which cross-matches \href{http://vizier.u-strasbg.fr/viz-bin/VizieR?-meta.foot&-source=I/315}{\emph{UCAC3}} (\citealp{2009yCat.1315....0Z}; green circles) and the \href{http://vizier.u-strasbg.fr/viz-bin/VizieR?-meta.foot&-source=I/305}{Guide Star Catalog} (\citealp{2008AJ....136..735L}; purple triangles). The reduced chi-square values for $\mu_\alpha\cos\delta$ and $\mu_\delta$ are 1.27 and 1.03, respectively.}
	\label{fig:LIT_PM}
\end{figure*}

We cross-matched our input sample with the Initial Gaia Source List (\href{http://vizier.u-strasbg.fr/viz-bin/VizieR}{VizieR} catalog \href{http://vizier.u-strasbg.fr/viz-bin/VizieR?-meta.foot&-source=I/324}{\emph{I/324/igsl3}}) to obtain proper motions from the \href{http://vizier.u-strasbg.fr/viz-bin/VizieR?-meta.foot&-source=I/315}{\emph{UCAC3}} (\href{http://vizier.u-strasbg.fr/viz-bin/VizieR}{VizieR} catalog \href{http://vizier.u-strasbg.fr/viz-bin/VizieR?-meta.foot&-source=I/315}{\emph{I/315}}; \citealp{2009yCat.1315....0Z}) and the \href{http://vizier.u-strasbg.fr/viz-bin/VizieR?-meta.foot&-source=I/305}{Guide Star Catalog} (\href{http://vizier.u-strasbg.fr/viz-bin/VizieR?-meta.foot&-source=I/305}{GSC}; \href{http://vizier.u-strasbg.fr/viz-bin/VizieR}{VizieR} catalog \href{http://vizier.u-strasbg.fr/viz-bin/VizieR?-meta.foot&-source=I/305}{\emph{I/305}} and \citealp{2008AJ....136..735L}), and present in \hyperref[fig:LIT_PM]{Figure~\ref*{fig:LIT_PM}} a comparison to the proper motions we derived from \href{http://www.ipac.caltech.edu/2mass/}{\emph{2MASS}}--\href{http://wise2.ipac.caltech.edu/docs/release/allwise/}{\emph{AllWISE}}. We find reduced $\chi^2$ values of 1.27 and 1.03 for $\mu_\alpha\cos\delta$ and $\mu_\delta$, respectively, which indicates that our measurement errors are representative of the differences between our proper motions and those in the catalogs mentioned above. However, there are a few cases where the literature proper motions are significantly discrepant from the \href{http://www.ipac.caltech.edu/2mass/}{\emph{2MASS}}--\href{http://wise2.ipac.caltech.edu/docs/release/allwise/}{\emph{AllWISE}} measurements. We investigated the 25/3\,873 worst cases in \href{http://vizier.u-strasbg.fr/viz-bin/VizieR?-meta.foot&-source=I/315}{\emph{UCAC3}} where either $\mu_\alpha\cos\delta$ or $\mu_\delta$ were discrepant by more than 300~\masyr. In 24/25 cases, we found other measurements in the literature that matched the \href{http://www.ipac.caltech.edu/2mass/}{\emph{2MASS}}--\href{http://wise2.ipac.caltech.edu/docs/release/allwise/}{\emph{AllWISE}} measurement within a few $\sigma$ (typically less than $1\sigma$), indicating that the \href{http://vizier.u-strasbg.fr/viz-bin/VizieR?-meta.foot&-source=I/315}{\emph{UCAC3}} measurement might be at fault. The other case (\href{http://simbad.u-strasbg.fr/simbad/sim-id?submit=display&bibdisplay=refsum&bibyear1=1850&bibyear2=$currentYear&Ident=2MASS%20J17274680%2B5200079}{2MASS~J17274680+5200079}) corresponds to a 6\farcs5 binary which is barely above the angular resolution of \href{http://wise2.ipac.caltech.edu/docs/release/allwise/}{\emph{AllWISE}} (6\farcs1 in the $W1$ band and 6\farcs4 in the $W2$ band). \cite{2013ApJ...774..101R} indicate that they observe a small systematic distortion ($<$ 15~\masyr) for their $\mu_\alpha\cos\delta$ measurements from \href{http://www.ipac.caltech.edu/2mass/}{\emph{2MASS}}--\href{http://irsa.ipac.caltech.edu/Missions/wise.html}{\emph{WISE}} as a function  of galactic latitude. They propose a correction factor, which would increase our reduced $\chi^2$ value to 1.27 to 1.82. This indicates that such a distortion is not clearly seen in our sample, and we thus choose not to include it in the present work. We conclude that the proper motions derived from \href{http://www.ipac.caltech.edu/2mass/}{\emph{2MASS}}--\href{http://wise2.ipac.caltech.edu/docs/release/allwise/}{\emph{AllWISE}} are reliable and will use only those measurements of proper motion for the remainder of this work. This will ensure that our selection criteria are more homogeneous, which will be helpful in an eventual characterization of the young population in the \href{http://www.astro.umontreal.ca/\textasciitilde gagne/BASS.php}{\emph{BASS}} survey.

\section{IDENTIFICATION OF CANDIDATE YOUNG MOVING GROUP MEMBERS}\label{sec:filters}

\begin{figure*}
	\centering
	\subfigure{\includegraphics[width=0.495\textwidth]{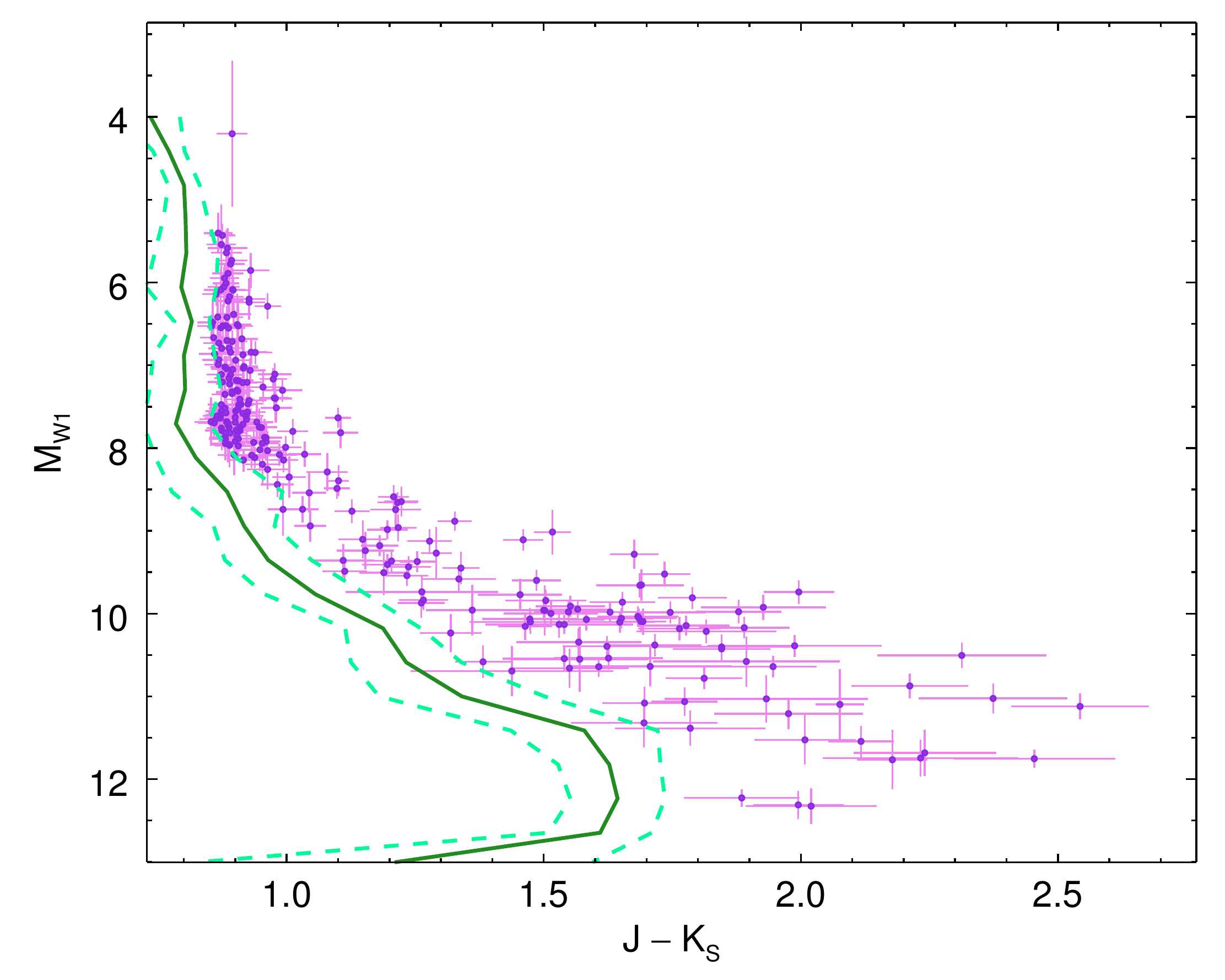}}
	\subfigure{\includegraphics[width=0.495\textwidth]{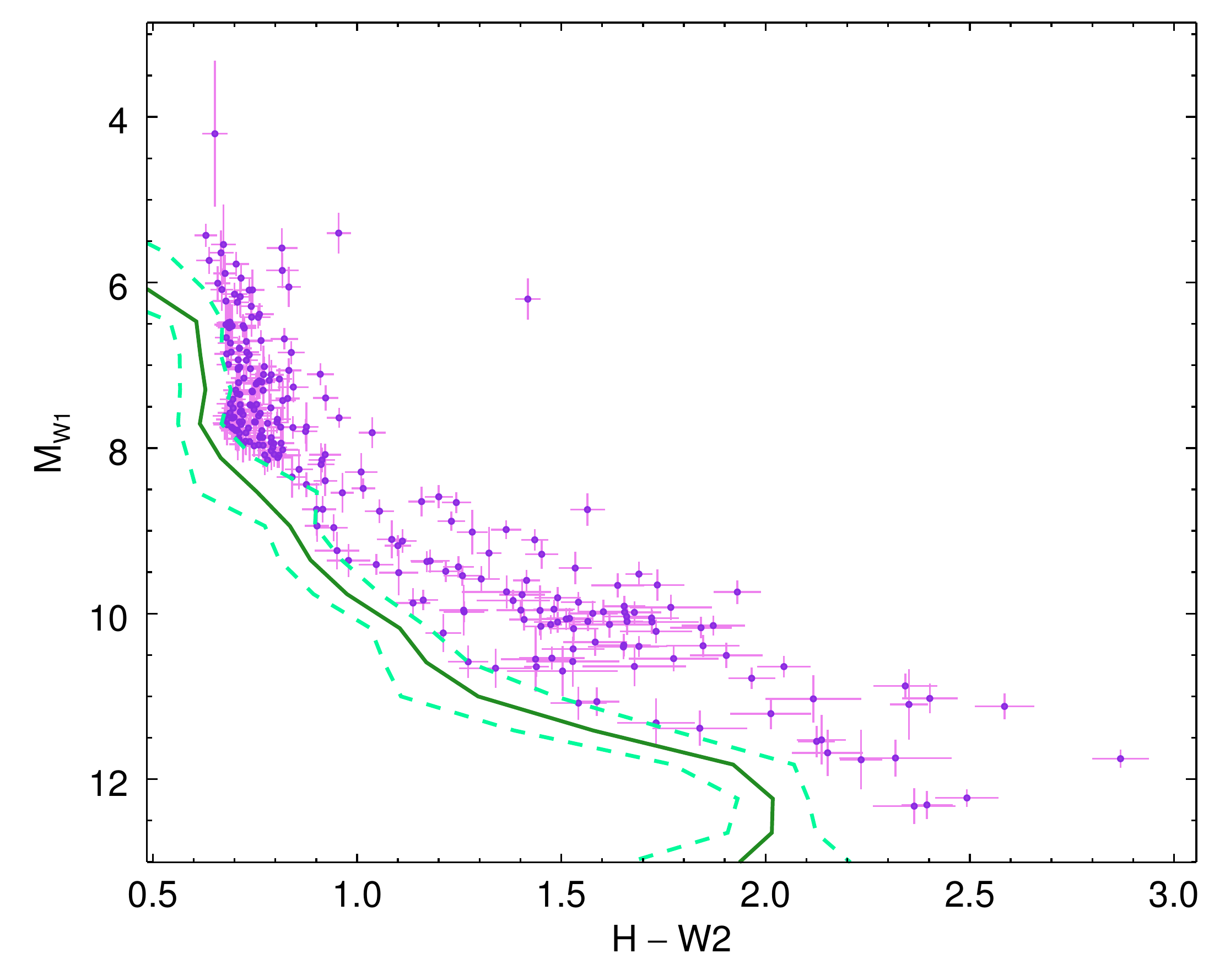}}
	\caption{Positions of all objects in the \href{http://www.astro.umontreal.ca/\textasciitilde gagne/BASS.php}{\emph{BASS}} sample in two different CMDs (purple points), compared with the field sequence (thick green line) and its scatter (dashed green lines). We used the statistical distances of the most probable hypothesis from the \href{http://www.astro.umontreal.ca/\textasciitilde gagne/banyanII.php}{BANYAN~II} tool to compute absolute magnitudes. The positions of all \href{http://www.astro.umontreal.ca/\textasciitilde gagne/BASS.php}{\emph{BASS}} candidates are consistent with them being young objects brighter and/or redder than the field sequence.}
	\label{fig:CMD}
\end{figure*}

We used \href{http://www.astro.umontreal.ca/\textasciitilde gagne/banyanII.php}{BANYAN~II} \citep{2014ApJ...783..121G} to compute the membership probability of all 98\,970 potential close-by $\geq$~M5 dwarfs identified in the previous section (List A). The \href{http://www.astro.umontreal.ca/\textasciitilde gagne/banyanII.php}{BANYAN~II} tool takes sky position, proper motion and \href{http://www.ipac.caltech.edu/2mass/}{\emph{2MASS}} and \href{http://wise2.ipac.caltech.edu/docs/release/allwise/}{\emph{AllWISE}} photometry as inputs and determines, using a naive Bayesian classifier, the membership probability that an object belongs to seven YMGs (TWA, $\beta$PMG, THA, COL, CAR, ARG, ABDMG) and the field population, which constitutes our eight hypotheses. Probability Density Functions (PDFs) are computed for every hypothesis and on each point of a regular $500 \times 500$ grid of distances and RVs spanning 0.1 to 200 pc and -35 to 35 \kms\ respectively, by comparing galactic positions (\emph{XYZ}) and space velocities (\emph{UVW}) to the spatial and kinematic model (SKM) of the respective hypotheses, as well as comparing \href{http://www.ipac.caltech.edu/2mass/}{\emph{2MASS}} and \href{http://wise2.ipac.caltech.edu/docs/release/allwise/}{\emph{AllWISE}} magnitudes to a photometric model. All measurement errors are propagated and considered in this comparison. SKMs of YMGs were built by fitting 3-dimensional ellipsoids, with unconstrained axes orientations, over the population of bona fide members with signs of youth as well as parallax and RV measurements (see \citealp{2013ApJ...762...88M} and \citealp{2014ApJ...783..121G} for a complete list). For the field hypothesis, similar ellipsoids were fitted to synthetic objects drawn from the Besan\c{c}on galactic model (\href{http://www.astro.umontreal.ca/\textasciitilde gagne/ACRobin2014inprep.php}{A.~C.~Robin et al. in preparation}, \citealp{2012AA...538A.106R}) at distances of $<$ 200 pc. The photometric model consists of an old and a young field sequence in two CMD diagrams: absolute $W1$ as a function of $H-W2$ and absolute $W1$ as a function of $J-K_S$. The positions of maxima and characteristic widths of the resulting posterior PDFs yield a statistical distance and RV prediction, assuming the object fulfills the  respective hypothesis. The same PDFs are marginalized to a final probability by numerically integrating them along the whole grid. Optionally, parallax and RV measurements can be included to derive a more robust probability. In these cases, the corresponding dimension of the marginalization grid is eliminated. The Prior probabilities in the Bayesian classifier are set to the respective population estimates of each hypotheses, considering the magnitude of proper motion and galactic latitude of the object. Additionally, equal-luminosity binary hypotheses for the field and all YMGs are supplemented to our set of hypotheses, where the CMDs are shifted up by 0.75 magnitudes. Objects for which the binary hypothesis has a higher probability will be flagged as potential binaries, and only the binary hypotheses will be used when we analyze known binary systems. A naive Bayesian classifier implicitly considers that all input parameters are independent, which is generally not the case here. Using such an analysis with dependent input parameters will generally provide a good classification, however the Bayesian probability will be biased and thus not interpretable in an absolute way (e.g. a set of candidates with a Bayesian probabilities of 90\% will not necessarily include a fraction of contaminants equal to 10\%; \citealp{Hand:2001tr}, \citealp{Russek:1983ed}). To address this, \citealp{2014ApJ...783..121G} performed a Monte Carlo analysis using all SKM and photometric models described above to estimate the field contamination probability as a function of Bayesian probability for different hypotheses. They find that Bayesian probabilities are generally pessimistic, except for YMGs which are most subject to contamination (ARG, ABDMG, $\beta$PMG and COL) when no parallax measurement is included. When a parallax measurement is included, the contamination probability are becomes very low  ($\lesssim$ 20\% when the Bayesian probability is larger than $\sim$ 10--40\% depending on the YMG). These results provide a translation for the Bayesian probability output by \href{http://www.astro.umontreal.ca/\textasciitilde gagne/banyanII.php}{BANYAN~II} to an expected contamination rate. \cite{2014ApJ...783..121G} showed that bona fide members within $<$ 1$\sigma$ of their YMG's SKM all have a Bayesian probability $>$ 95\% associated with a membership to their respective YMG, whereas peripheral (1--2.5$\sigma$) bona fide members have a Bayesian probability between 10--95\%. For more details about the \href{http://www.astro.umontreal.ca/\textasciitilde gagne/banyanII.php}{BANYAN~II} tool, the reader is referred to \cite{2014ApJ...783..121G}.

After applying \href{http://www.astro.umontreal.ca/\textasciitilde gagne/banyanII.php}{BANYAN~II} to our input sample (list A), we rejected all objects with a Bayesian probability $<$ 10\% of being a member to a YMG, or with an estimated contamination rate $>$ 50\%. At this point we are left with 983 candidates. We used statistical distances of the most probable hypotheses to place all candidates in the two CMDs described above, and rejected all candidates that did not have NIR colors at least 1$\sigma$ redder than the field sequence. These filters cut down the candidate list to 273 objects. Another set of 275 candidates located to the right of the field sequence by an amount less than 1$\sigma$ were used to build the low-priority \href{http://www.astro.umontreal.ca/\textasciitilde gagne/BASS.php}{\emph{BASS}} catalog (\href{http://www.astro.umontreal.ca/\textasciitilde gagne/LP-BASS.php}{\emph{LP-BASS}}) which is discussed in the Appendix of this paper. The \href{http://wise2.ipac.caltech.edu/docs/release/allwise/}{\emph{AllWISE}} catalog includes \href{http://irsa.ipac.caltech.edu/Missions/wise.html}{\emph{WISE}} observations that were performed in its warm phase, hence in some cases, the measurement of \emph{W1} or \emph{W2} can be saturated. To avoid overlooking such saturated targets, we repeated all steps described above using the \href{http://irsa.ipac.caltech.edu/Missions/wise.html}{\emph{WISE}} catalog instead of \href{http://wise2.ipac.caltech.edu/docs/release/allwise/}{\emph{AllWISE}}, and supplemented our sample with the additional 26 objects uncovered this way (96 in the case of \href{http://www.astro.umontreal.ca/\textasciitilde gagne/LP-BASS.php}{\emph{LP-BASS}}). We subsequently used the \href{http://irsa.ipac.caltech.edu/applications/DUST/}{IRSA dust extinction tool}\footnote{Available at \url{http://irsa.ipac.caltech.edu/applications/DUST/}} to remove 9 objects displaying extinction larger than 0.4 mag, potentially corresponding to distant contaminants reddened by interstellar matter in our line of sight. Another 3 objects listed in the the \emph{2MASS extended sources} catalog (\href{http://vizier.u-strasbg.fr/viz-bin/VizieR}{VizieR} catalog \href{http://cdsarc.u-strasbg.fr/viz-bin/VizieR-3?-source=VII/233/xsc}{\emph{VII/233/xsc}}) were rejected. In a final step, we visually inspected all \href{http://www.sdss.org/}{\emph{SDSS}}, \href{http://archive.stsci.edu/cgi-bin/dss_form}{\emph{DSS}}, \href{http://www.ipac.caltech.edu/2mass/}{\emph{2MASS}} and \href{http://wise2.ipac.caltech.edu/docs/release/allwise/}{\emph{AllWISE}} acquisition images to flag any object with a suspicious shape or evidence of interstellar absorption in the surrounding 5$'$. No such occurrence was found, which indicates the filters described above were efficient in preventing such contaminating objects. The resulting \href{http://www.astro.umontreal.ca/\textasciitilde gagne/BASS.php}{\emph{BASS}} catalog is presented in \hyperref[tab:bass1]{Table~\ref*{tab:bass1}}. We divide the sample in two sections: those with a contamination probability lower than 15\% are grouped in a \emph{High Probability} section, whereas those with a contamination probability between 15--50\% are grouped in the \emph{Modest Probability} section.

\capstartfalse
\input{bass_table_1.tex}
\capstarttrue

In \hyperref[tab:incompleteness]{Table~\ref*{tab:incompleteness}}, we present the fraction of members in each moving group that would fail our galactic plane and proper motion filters, assuming that our SKM models are accurate. We obtained these quantities by drawing a million synthetic objects from a gaussian random distribution represented by each SKM and assessing what fraction fails each filter. We used the estimated recovery rate of the \href{http://www.astro.umontreal.ca/\textasciitilde gagne/banyanII.php}{BANYAN~II} tool for each YMG (see \citealp{2014ApJ...783..121G}) corresponding to our tolerated field contamination of $< 50$\% and combined all these sources of incompleteness to estimate that the \href{http://www.astro.umontreal.ca/\textasciitilde gagne/BASS.php}{\emph{BASS}} sample is complete at the 6--90\% level in the range of spectral types considered here, depending on the YMG in question. The YMGs that would benefit the most from a search within the galactic plane are ARG and CAR, and to a lesser extent $\beta$PMG, ABDMG and TWA. However, such a survey would present a significant challenge for two reasons ; (1) a cross-match between the \href{http://www.ipac.caltech.edu/2mass/}{\emph{2MASS}} and \href{http://wise2.ipac.caltech.edu/docs/release/allwise/}{\emph{AllWISE}} catalogs would require the use of powerful algorithms because of crowded regions; and (2) a new free parameter would have to be added to the analysis, describing the effect of reddening by interstellar medium on the CMD sequence of field stars (e.g. this effect could be represented by a reddening vector of unknown amplitude in both CMDs that are used in the \href{http://www.astro.umontreal.ca/\textasciitilde gagne/banyanII.php}{BANYAN~II} tool). We note that even if those two hurdles would be overcome, we expect the field contamination to remain very high within the galactic plane, unless the survey benefits from RV and parallax measurements for a large number of objects. The only YMG which is significantly affected by our low proper motion cut is COL. Since this filter serves the main purpose of rejecting distant extragalactic and red giant contaminants, starting from a sample of targets with distance measurements would allow relaxing this filter and accessing to a larger number of COL candidates. The final major obstacle to identify efficiently a large number of candidate members of ARG, COL, $\beta$PMG and ABDMG is the low recovery rate intrinsic to a naive Bayesian classifier in the situation where no information is known on the RV and distance of the input sample. It could be expected that adopting a more complex method, which could for example take account of the dependency of input parameters, would help to draw the most possible information from a sample without RV and distance measurements. However, \cite{Hand:2001tr} suggest otherwise by demonstrating that a naive Bayesian classifier performs much better than could be expect in these conditions. This would leave only three foreseeable options to attack this aspect of our survey completeness; (1) allow for significantly more contaminants in our sample and perform an extensive spectroscopic follow-up; (2) start from a sample that includes RV and parallax measurements; or (3) identify new readily-accessible observables, such as new filters in color-color diagrams, that could distinguish YMG members from field interlopers.

\capstartfalse
\input{bass_table_2_short.tex}
\capstarttrue

\begin{figure*}
	\begin{center}
 	\includegraphics[width=0.995\textwidth]{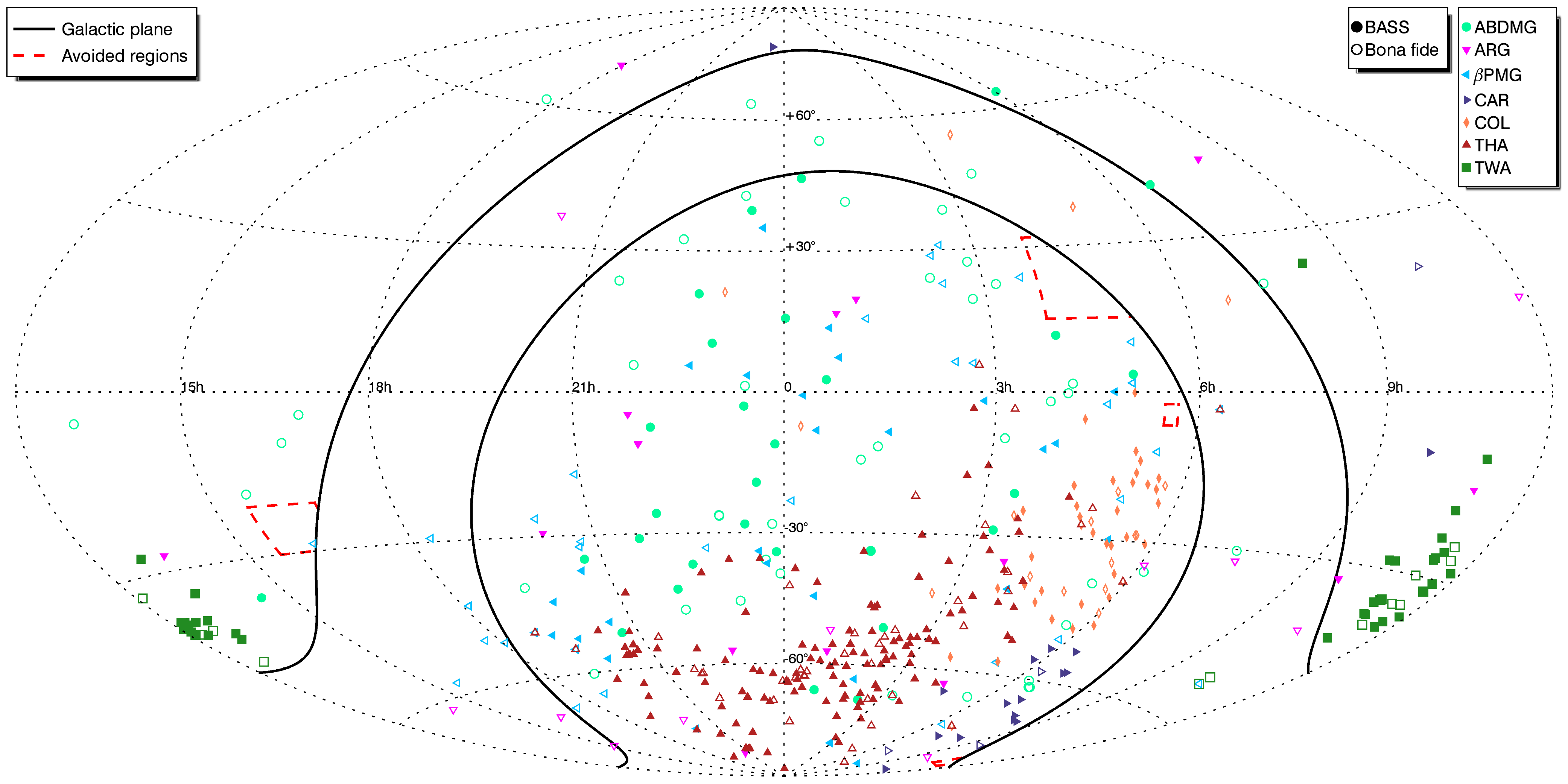}
	\end{center}
	\caption{Sky position of all \href{http://www.astro.umontreal.ca/\textasciitilde gagne/BASS.php}{\emph{BASS}} candidates (filled symbols), compared with currently known bona fide members (open symbols) of each YMG considered here. Thick black lines delimit the galactic plane within $\pm 15$\textdegree\ of galactic latitude, and the dahsed red lines delimit regions that were avoided in our search for YMG candidates (see 
\hyperref[sec:crossmatch]{Section~\ref*{sec:crossmatch}}).}
	\label{fig:radec}
\end{figure*}

\section{A LITERATURE SEARCH FOR ADDITIONAL INFORMATION}\label{sec:lit}

\begin{figure*}
	\centering
	\subfigure{\includegraphics[width=0.495\textwidth]{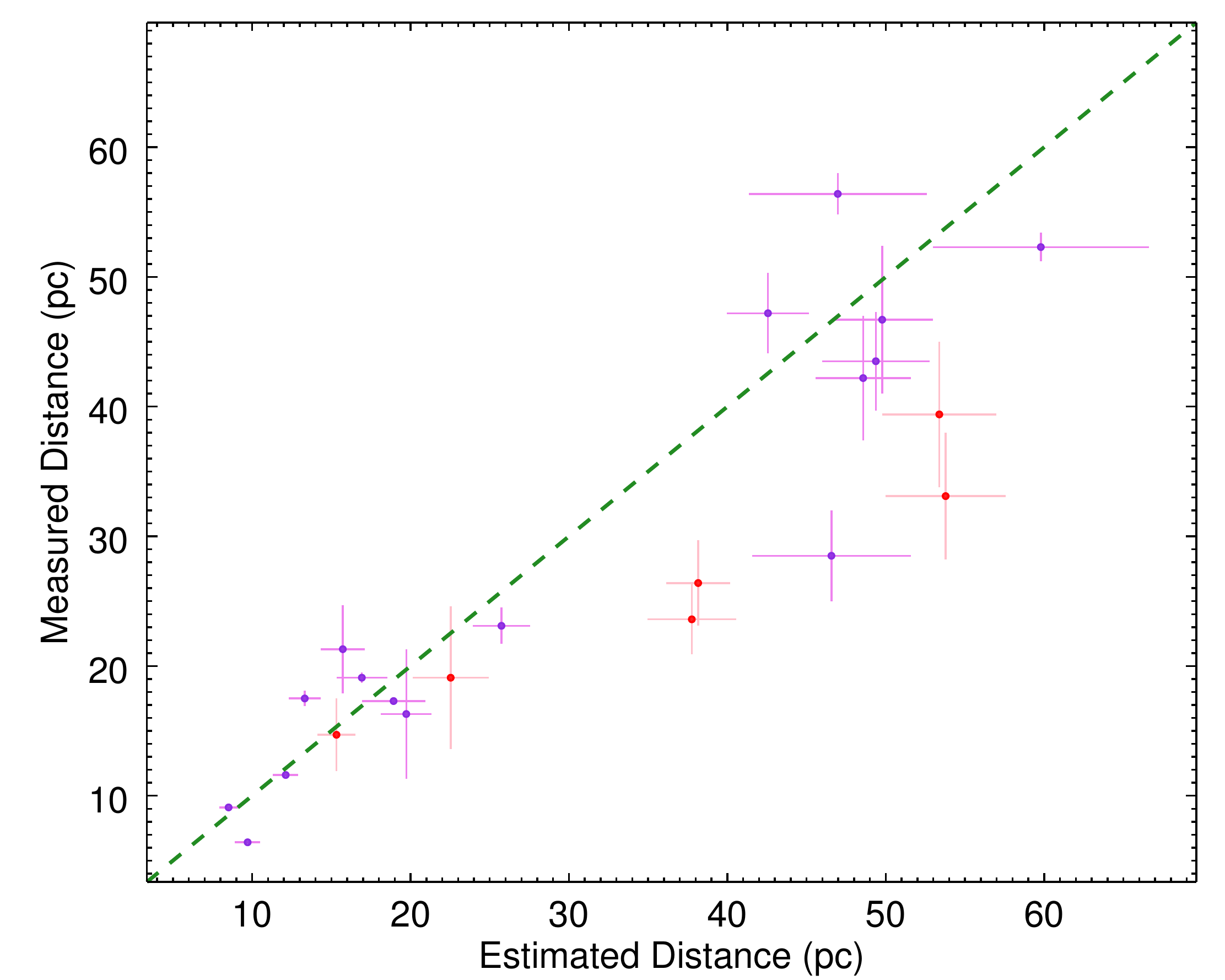}}
	\subfigure{\includegraphics[width=0.495\textwidth]{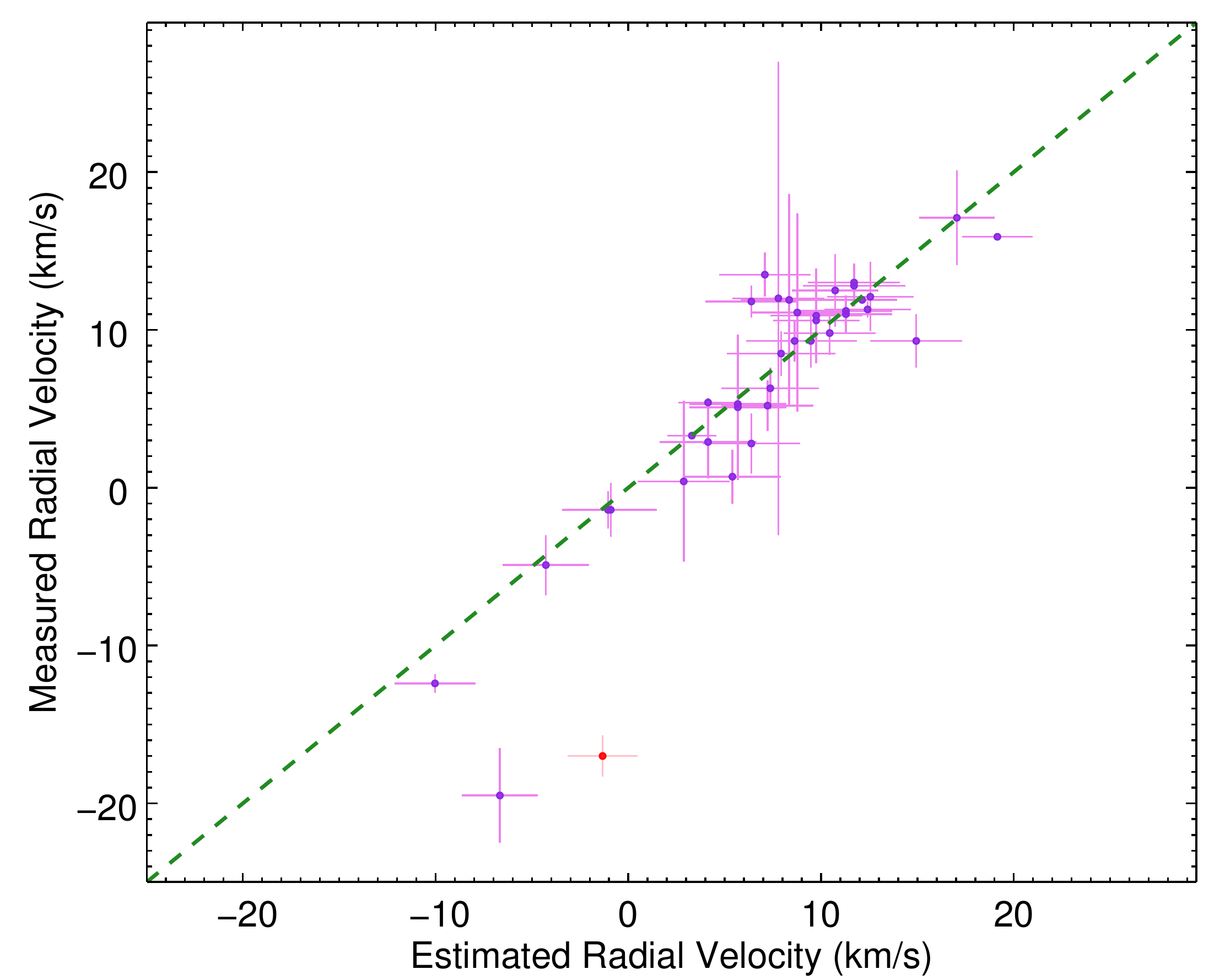}}
	\caption{Comparison of statistical RV and distance predictions from \href{http://www.astro.umontreal.ca/\textasciitilde gagne/banyanII.php}{BANYAN~II} to measurements found in the literature. The dashed green line has a unit slope and intersects with the origin. Measurements which corroborated the most probable hypothesis are displayed in purple, whereas those favoring a different YMG are displayed in red. Measurements which are significantly discrepant and thus rejecting possible YMG memberships are not displayed here.}
	\label{fig:drv_comp}
\end{figure*}

We searched for any additional information in the literature for all candidates in \hyperref[tab:bass1]{Table~\ref*{tab:bass1}} using the \href{http://simbad.u-strasbg.fr/}{SIMBAD} and \href{http://vizier.u-strasbg.fr/viz-bin/VizieR}{VizieR} web tools. We found 122 objects for which at least one of RV, parallax, spectral type, signs of youth or any other relevant information was available, including 60 known candidates or bona fide members of the YMGs considered here. There are only 4 known bona fide members included in those: \href{http://simbad.u-strasbg.fr/simbad/sim-id?submit=display&bibdisplay=refsum&bibyear1=1850&bibyear2=$currentYear&Ident=2MASS%20J00452143%2B1634446}{2MASS~J00452143+1634446} (ARG; \citealp{2014AA...568A...6Z} and \hyperref[sec:indiv]{Section~\ref*{sec:indiv}}) \href{http://simbad.u-strasbg.fr/simbad/sim-id?submit=display&bibdisplay=refsum&bibyear1=1850&bibyear2=$currentYear&Ident=2MASS%20J01231125-6921379}{2MASS~J01231125--6921379} (THA; \citealp{2014ApJ...783..121G}), \href{http://simbad.u-strasbg.fr/simbad/sim-id?submit=display&bibdisplay=refsum&bibyear1=1850&bibyear2=$currentYear&Ident=GJ%202022}{GJ 2022} (ABDMG; \citealp{2012PhDT.......100R}, \citealp{2012ApJ...758...56S} and \citealp{2014AJ....147...85R}), \href{http://simbad.u-strasbg.fr/simbad/sim-id?submit=display&bibdisplay=refsum&bibyear1=1850&bibyear2=$currentYear&Ident=2MASS%20J03552337%2B1133437}{2MASS~J03552337+1133437} (ABDMG; \citealp{2013AJ....145....2F},  \citealp{2013AN....334...85L}). We list these 59 objects in \hyperref[tab:lit]{Table~\ref*{tab:lit}}, with an updated Bayesian probability in light of these additional measurements. In \hyperref[fig:drv_comp]{Figure~\ref*{fig:drv_comp}}, we compare the \href{http://www.astro.umontreal.ca/\textasciitilde gagne/banyanII.php}{BANYAN~II} statistical predictions for the RV and distance to measurements found in the literature, and show that the reduced $\chi^2$ values are 1.32 and 0.84, respectively. This indicates that errors on statistical predictions are representative of the scatter observed here.

\subsection{Estimates of Spectral Types}\label{sec:spt}

\begin{figure}
	\begin{center}
		\subfigure[Before Correction]{\includegraphics[width=0.495\textwidth]{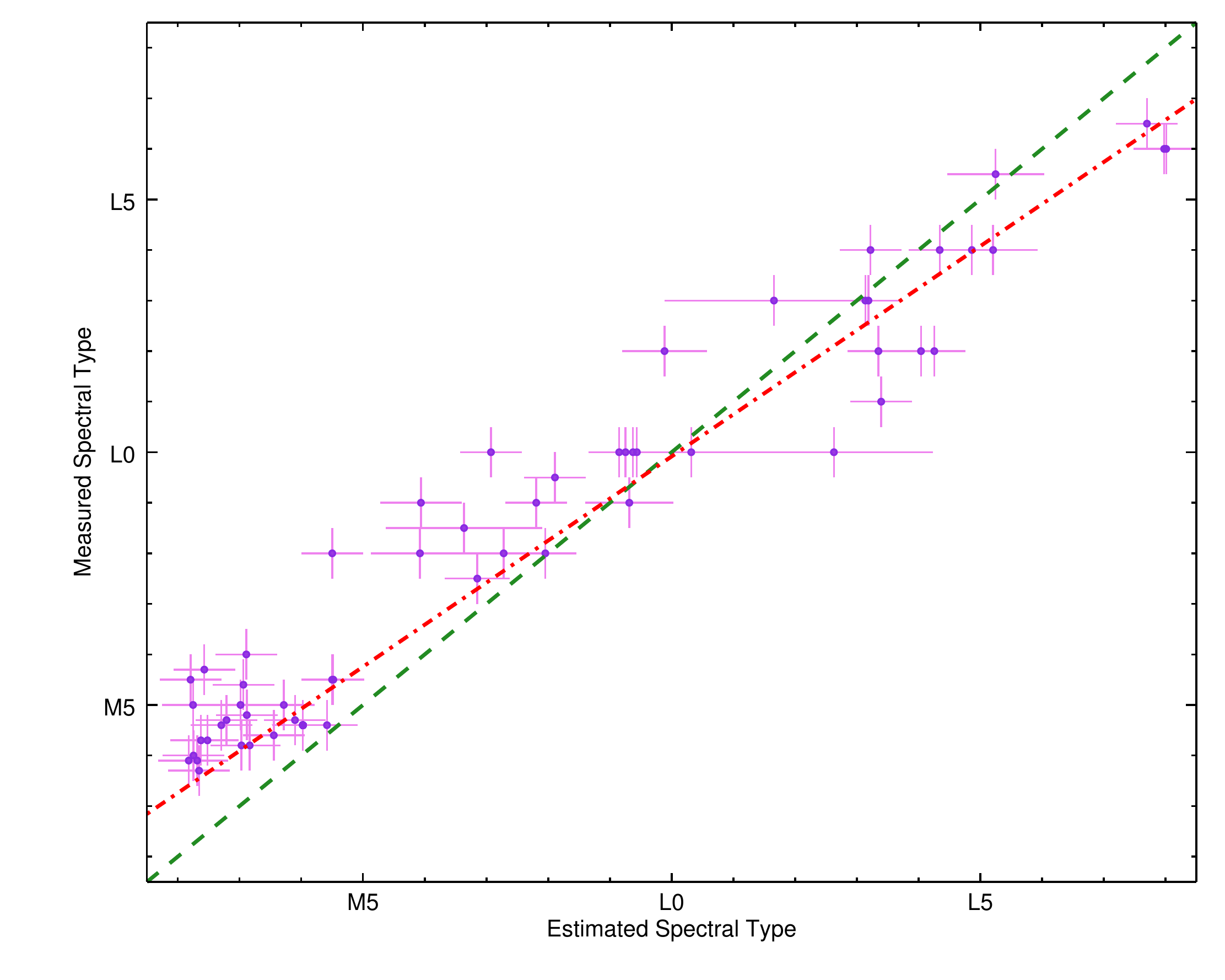}}
		\subfigure[After Correction]{\includegraphics[width=0.495\textwidth]{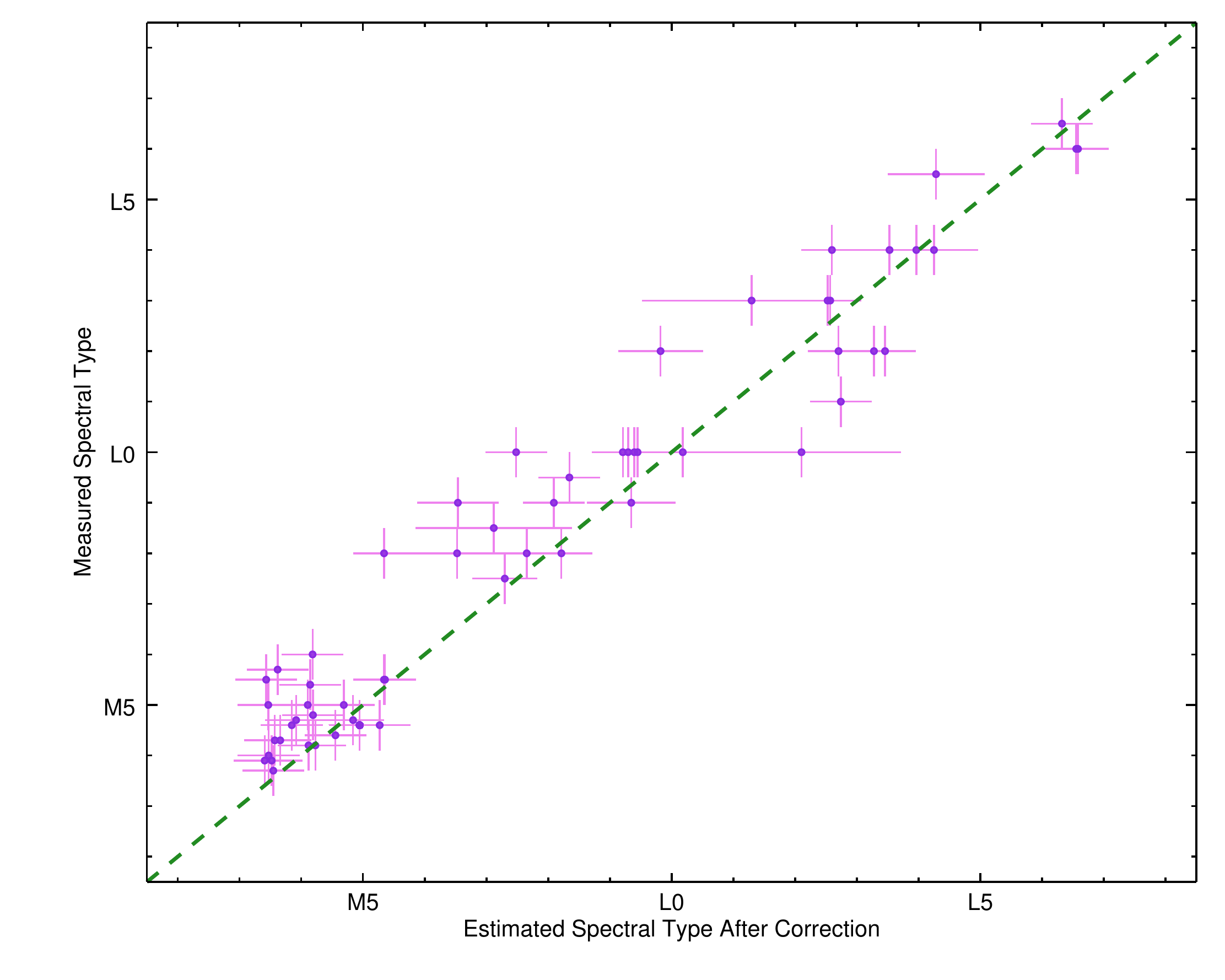}} 	
	\end{center}
	\caption{Estimated spectral types obtained from \href{http://www.ipac.caltech.edu/2mass/}{\emph{2MASS}} and \href{http://wise2.ipac.caltech.edu/docs/release/allwise/}{\emph{AllWISE}} photometry as well as statistical distances from \href{http://www.astro.umontreal.ca/\textasciitilde gagne/banyanII.php}{BANYAN~II}, compared with measurements available in the literature from optical or NIR spectroscopy. The dashed green line has a unit slope and intersects with the origin. Our estimates are reliable within $\sim$~1.5 subtype in the M5--L6 range, but tend to overestimate (underestimate) later (earlier) spectral types. To account for this effect, we adjusted a linear correction to the estimated spectral types (red dashed line; top panel). Corrected estimations of spectral types are displayed in the bottom panel.}
	\label{fig:spt_estim}
\end{figure}

We used \href{http://www.ipac.caltech.edu/2mass/}{\emph{2MASS}} and \href{http://wise2.ipac.caltech.edu/docs/release/allwise/}{\emph{AllWISE}} $J$, $H$, $K_S$, \emph{W1} and \emph{W2} magnitudes with the statistical distance associated to the most probable hypothesis from \href{http://www.astro.umontreal.ca/\textasciitilde gagne/banyanII.php}{BANYAN~II} to assign a tentative spectral type to all candidates identified here. We used the \emph{Database of Ultracool Parallaxes}\footnote{Available at\\ \url{http://www.cfa.harvard.edu/\textasciitilde tdupuy/plx/Database\_of\_Ultracool\_Parallaxes.html}} \citep{2012ApJS..201...19D} to compare the position of each candidate with the corresponding spectral type -- magnitude sequence (spanning the M5--T9 range) and derived a PDF in each case as a function of spectral type. We then combined these PDFs in a likelihood analysis, and used the maximal position of the final PDF to assign a most probable spectral type to each object. In \hyperref[fig:spt_estim]{Figure~\ref*{fig:spt_estim}}, we compare our spectral type estimates to measurements available in the literature and show that these estimates are reliable to within $\sim$~2.5 subtypes.


We note a clear trend where we tend to underestimate spectral types for $<$ M5 objects and overestimate those of $>$ L5 objects. We used a linear fit to characterize this systematic trend and obtain a correction for our estimated spectral types:

\begin{align}
	\mathrm{SpT}_{\mathrm{corr}} = 1.64 + 0.81 \cdot \mathrm{SpT}_{\mathrm{estim}},
\end{align}

where 0 corresponds to the M0 spectral type. We used this equation to correct all estimated spectral types listed in Tables~\ref{tab:bass1} and \ref{tab:allbass}. Before the correction, the reduced $\chi^2$ value for our estimated spectral types is 2.51, and the estimated--measured spectral type differences display a standard deviation of 1.1 subtypes. After the correction, the reduced $\chi^2$ and standard deviation become 1.0 and 0.8 subtypes, respectively.

\begin{figure}
	\begin{center}
 	\includegraphics[width=0.495\textwidth]{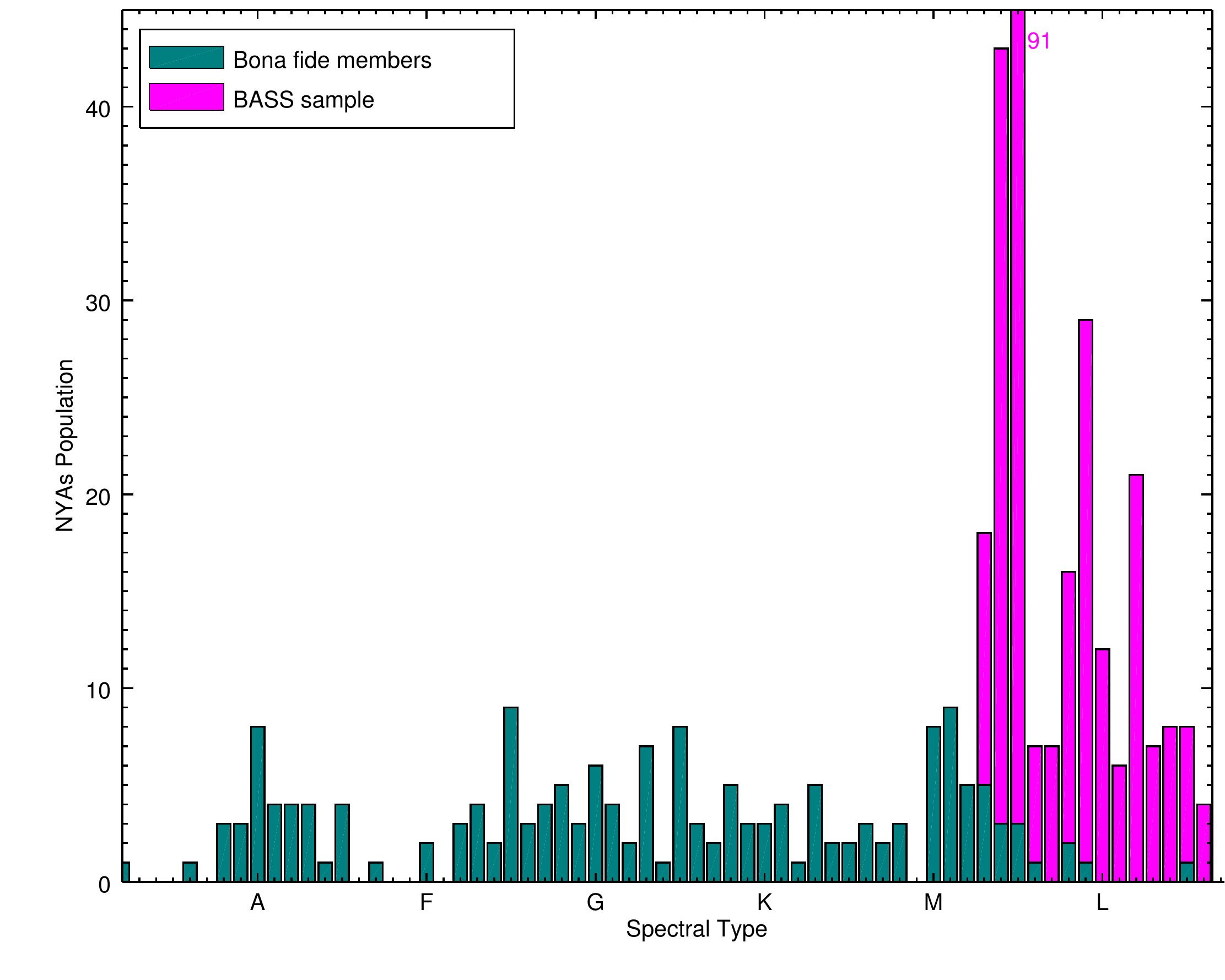}
	\end{center}
	\caption{Estimated spectral types (violet bars) for the \href{http://www.astro.umontreal.ca/\textasciitilde gagne/BASS.php}{\emph{BASS}} sample, compared with the current bona fide population of all YMGs considered here (green bars). The M5 spectral bin has a value of 91: the vertical range has been shortened for clarity. The \href{http://www.astro.umontreal.ca/\textasciitilde gagne/BASS.php}{\emph{BASS}} sample targets YMG candidates in a range of spectral types which is yet largely unexplored.}
	\label{fig:spt_hist}
\end{figure}

In \hyperref[fig:spt_hist]{Figure~\ref*{fig:spt_hist}}, we use spectral type measurements when available or estimates of spectral types otherwise to compare the \href{http://www.astro.umontreal.ca/\textasciitilde gagne/BASS.php}{\emph{BASS}} sample with current bona fide members in YMGs. This Figure clearly demonstrates that a significant fraction of the \href{http://www.astro.umontreal.ca/\textasciitilde gagne/BASS.php}{\emph{BASS}} candidates have a later spectral type than most known members of YMGs, which outlines that we are entering a yet poorly explored mass regime of the YMG population.

\subsection{Comments on Individual Objects}\label{sec:indiv}

In this Section, we present comments on individual objects which deserve further discussion. All those already discussed in \citeauthor{2014ApJ...783..121G} (\citeyear{2014ApJ...783..121G}; see the Reference column in \hyperref[tab:lit]{Table~\ref*{tab:lit}}) will not be discussed here, unless new information is available.

\href{http://simbad.u-strasbg.fr/simbad/sim-id?submit=display&bibdisplay=refsum&bibyear1=1850&bibyear2=$currentYear&Ident=1RXS%20J003902.4%2B133009}{2MASS~J00390342+1330170} has been identified by \cite{2012AJ....143...80S} as a candidate member of ABDMG with X-ray and near-UV emission indicative of a young, early-M dwarf, however they do not estimate a spectral type. We find that this object has a Bayesian probability of 84.3\% and 7.5\% for $\beta$PMG and ABDMG, respectively. We thus assign it as a candidate member of $\beta$PMG, but we note that there is an expected $\sim$~10\% contamination rate from ABDMG to $\beta$PMG for such a result (see \citealp{2014ApJ...783..121G}).

\href{http://simbad.u-strasbg.fr/simbad/sim-id?submit=display&bibdisplay=refsum&bibyear1=1850&bibyear2=$currentYear&Ident=2MASS%20J00452143%2B1634446}{2MASS~J00452143+1634446} was reported by \cite{2014ApJ...783..121G} as a candidate member of ARG with unusually red NIR colors for its L2 spectral type. \cite{2010ApJ...723..684B} measured a RV of $3.4 \pm 0.2$ \kms, and \cite{2014AA...568A...6Z} measured a trigonometric distance of $17.5 \pm 0.6$, pc, which bring the Bayesian probability of the ARG membership hypothesis to 98.0\%. \cite{2014AA...568A...6Z} also derived an isochronal age of 10--100 Myr and detected lithium in its atmosphere. As noted by \cite{2014AA...568A...6Z}, all evidence points towards a membership to ARG, hence we propose that this $\sim$ 15~\Mjup\ object is a bona fide member of this association.

\href{http://simbad.u-strasbg.fr/simbad/sim-id?submit=display&bibdisplay=refsum&bibyear1=1850&bibyear2=$currentYear&Ident=2MASS%20J01033563-5515561}{2MASS~J01033563--5515561} was first identified as a highly probable candidate to THA in early versions of the \href{http://www.astro.umontreal.ca/\textasciitilde gagne/BASS.php}{\emph{BASS}} sample. \cite{2013AA...553L...5D} used high contrast imaging to search for low-mass companions around \href{http://www.astro.umontreal.ca/\textasciitilde gagne/BASS.php}{\emph{BASS}} candidates and demonstrated that this object is in fact an M5+M5, 0\farcs26 tight binary harboring a 12--14~\Mjup\ substellar companion at a separation of 1\farcs78. They note that the NIR colors of the companion are indicative of a young L-type object, which is consistent with the THA membership. Subsequently, \cite{2014AJ....147..146K} and \cite{2014ApJ...788...81M} independently measured RVs of $4.0 \pm 2.0$ \kms\ and $7.3 \pm 2.6$ \kms\ respectively, whereas the latter independently identifies it as a candidate member of THA. We combined both RV measurements to obtain RV = $5.2 \pm 1.6$ \kms. \cite{2014AJ....147...85R} measured a trigonometric distance of $47.2 \pm 3.1$ pc, in good agreement with our statistical distance of $42.3 \pm 3$ pc (which is at 1.1$\sigma$ from the measurement). Without using any RV measurement, they argue that its kinematics are more consistent with CAR rather than THA. They also use empirical isochrones for YMGs to show that the system is over-luminous for THA or CAR even when binarity is taken into account, which could mean that it is possibly younger, or an even higher-order multiple system. When not using the RV measurement in \href{http://www.astro.umontreal.ca/\textasciitilde gagne/banyanII.php}{BANYAN~II}, we obtain a Bayesian probability of 98.9\%, 0.7\% and $2\cdot 10^{-7}$ for THA, ABDMG and CAR, respectively. The statistical RVs associated to these hypotheses are respectively $7.2 \pm 2.5$ \kms, $10.8 \pm 1.8$ \kms\ and $14.0 \pm 2.0$ \kms. Both the measured RVs are consistent with the THA hypothesis (at 0.7$\sigma$) and not consistent with CAR (at 3.0$\sigma$), which strengthens the THA hypothesis even more. Once we include the RV measurement, the THA hypothesis clearly dominates with a Bayesian probability of 99.9\% for THA and $2\cdot 10^{-10}$ for CAR. We thus suggest that this system is a bona fide member of THA, since it has all measurements needed to be considered as such (i.e. complete $XYZUVW$ kinematics and signs of youth). This system will be discussed in more details in a subsequent paper (\href{http://www.astro.umontreal.ca/\textasciitilde gagne/J0103inprep.php}{J.~Gagn\'e et al., in preparation}).

\href{http://simbad.u-strasbg.fr/simbad/sim-id?submit=display&bibdisplay=refsum&bibyear1=1850&bibyear2=$currentYear&Ident=2MASS%20J01243060-3355014}{2MASS~J01243060--3355014} (\href{http://simbad.u-strasbg.fr/simbad/sim-id?submit=display&bibdisplay=refsum&bibyear1=1850&bibyear2=$currentYear&Ident=2MASS%20J01243060-3355014}{GJ~2022~B}) was identified by \cite{2003AJ....125..332J} as a co-moving companion to the tight 1\farcs8 M4+M4 binary \href{http://simbad.u-strasbg.fr/simbad/sim-id?Ident=%401212443&Name=GJ%20%202022}{GJ~2202~AC}. \cite{2009ApJ...699..649S} used the X-ray emission and low \ion{K}{1} EW of the latter to constrain its age between 40--300~Myr, and \cite{2012ApJ...758...56S} measured a trigonometric distance of $25.1~\pm~1.0$ pc and a RV of $18.3~\pm~1.5$ \kms\ for \href{http://simbad.u-strasbg.fr/simbad/sim-id?submit=display&bibdisplay=refsum&bibyear1=1850&bibyear2=$currentYear&Ident=2MASS%20J01243060-3355014}{GJ~2022~B}. They use this information to identify this object as a new bona fide member of ABDMG. \cite{2014AJ....147...85R} subsequently measured a trigonometric distance of $25.8~\pm~1.4$ pc; we combined both distance measurements in an error-weighted average to obtain $25.3~\pm~0.8$ pc. We find that the ABDMG membership, distance and RV measurements are all consistent with our results from \href{http://www.astro.umontreal.ca/\textasciitilde gagne/banyanII.php}{BANYAN~II}; the predicted RV of $18.3~\pm~2.0$ \kms\ is consistent with the measurement, and the statistical distance of $26.1~\pm~1.6$ pc is at $< 1\sigma$ of the combined distance measurements. Including youth, RV, distance and spectral types in our analysis yields a membership probability of 99.98\% for the ABDMG hypothesis, associated with a field contamination probability of $<$ 0.1\%. This is consistent with the conclusions of \cite{2012ApJ...758...56S} and \cite{2014AJ....147...85R} that this system is a bona fide member of ABDMG. We note that \cite{2012ApJ...758...56S} refer to the wide companion as GJ~2022~C, whereas \cite{2003AJ....125..332J} and \cite{2014AJ....147...85R} refer to it as GJ~2022~B. We adopt the latter to preserve historical nomenclature, as proposed by \cite{2014AJ....147...85R}.

\href{http://simbad.u-strasbg.fr/simbad/sim-id?submit=display&bibdisplay=refsum&bibyear1=1850&bibyear2=$currentYear&Ident=2MASS%20J01303563-4445411}{2MASS~J01303563--4445411} was identified as an M9 dwarf by \cite{2008AJ....136.1290R} and \cite{2009AJ....137....1F}. Subsequently, \cite{2011AJ....141....7D} resolved this system as an M9+L6 pair with a 3\farcs2 separation. They note that the companion displays red colors for its spectral type, at 1.7$\sigma$ of the field L6 BDs, but the primary has normal NIR colors for its spectral type, which could be an indication that the companion has an unusually dusty atmosphere. They show that the optical spectrum of the primary does not display H$\alpha$ or Li, which indicates a minimal age of 250~Myr. Furthermore, a resolved NIR spectrum of the L6 companion does not display typical signs of youth such as a triangular $H$-band continuum. We thus conclude that this system must be a false positive in our analysis, despite its 90.6\% Bayesian probability of being a member of THA, since its age is not consistent with any YMG in the solar neighborhood.

\href{http://simbad.u-strasbg.fr/simbad/sim-id?submit=display&bibdisplay=refsum&bibyear1=1850&bibyear2=$currentYear&Ident=2MASS%20J02212859-6831400}{2MASS~J02212859--6831400} has been identified as an M8 dwarf by \cite{2008AJ....136.1290R}, and \cite{2009AJ....137....1F} indicate that it is unusually red and for its spectral type and displays signs of low-gravity. \cite{2012ApJ...752...56F} measured a trigonometric distance of $39.4 \pm 5.6$ pc. This object was not considered as a strong candidate member of any YMG in \citealp{2014ApJ...783..121G}, but here we find it as a candidate member of ABDMG with a Bayesian probability of 40.8\% and a contamination probability of $<$ 0.1\%. This discrepancy is due to the \href{http://www.ipac.caltech.edu/2mass/}{\emph{2MASS}}--\href{http://wise2.ipac.caltech.edu/docs/release/allwise/}{\emph{AllWISE}} proper motion, which is at 2.2$\sigma$ or 5.0~\masyr\ ($\mu_\alpha\cos\delta$) and 1.7$\sigma$ ($\mu_\delta$) or 9.1~\masyr\ of the proper motion used in the analysis of \cite{2014ApJ...783..121G} (which was measured by \citealp{2012ApJ...752...56F}). We visually inspected the \href{http://www.ipac.caltech.edu/2mass/}{\emph{2MASS}} and \href{http://wise2.ipac.caltech.edu/docs/release/allwise/}{\emph{AllWISE}} Atlas images and found that our cross-match between both catalogs is unambiguous, however it is possible that this candidate is a false positive in our analysis. A measurement of RV will be necessary to better constrain the membership of this object.

\href{http://simbad.u-strasbg.fr/simbad/sim-id?submit=display&bibdisplay=refsum&bibyear1=1850&bibyear2=$currentYear&Ident=2MASS%20J02401209-5305527}{2MASS~J02401209--5305527} was reported as an M9.5 BD by \cite{2010AA...517A..53M}. They measured the equivalent width (EW) of the \ion{Na}{1} doublet at 8170--8200~\AA\ to be EW = $5.5~\pm~0.8$ \AA. It is well known that low-gravity objects have a lower-than-normal \ion{Na}{1} EW, however no classification scheme using this measurement extends to such a late spectral type. We note that this EW is low compared with other M9.5 BDs in their sample, for which \ion{Na}{1} EWs range from 5.9 to 9.7~\AA\ with an average and standard deviation of 7.3 and 1.3~\AA\ respectively. However, it is higher than the \ion{Na}{1} EW of low-gravity field BDs in their sample (\href{http://simbad.u-strasbg.fr/simbad/sim-id?submit=display&bibdisplay=refsum&bibyear1=1850&bibyear2=$currentYear&Ident=2MASS%20J04433761%2B0002051}{2MASS~J04433761+0002051} with $3.6~\pm~0.8$ \AA\ and \href{http://simbad.u-strasbg.fr/simbad/sim-id?submit=display&bibdisplay=refsum&bibyear1=1850&bibyear2=$currentYear&Ident=2MASS%20J06085283-2753583}{2MASS~J06085283--2753583} with $5.0~\pm~0.7$ \AA). NIR spectroscopy would be useful to clarify the age of \href{http://simbad.u-strasbg.fr/simbad/sim-id?submit=display&bibdisplay=refsum&bibyear1=1850&bibyear2=$currentYear&Ident=2MASS%20J02401209-5305527}{2MASS~J02401209--5305527}.

\href{http://simbad.u-strasbg.fr/simbad/sim-id?submit=display&bibdisplay=refsum&bibyear1=1850&bibyear2=$currentYear&Ident=2MASS%20J03014892-5903021}{2MASS~J03014892--5903021} and \href{http://simbad.u-strasbg.fr/simbad/sim-id?submit=display&bibdisplay=refsum&bibyear1=1850&bibyear2=$currentYear&Ident=2MASS%20J03252938-4312299}{2MASS~J03252938--4312299} have both been identified as M9 dwarfs by \cite{2008AJ....136.1290R}. \cite{2010AA...517A..53M} measured the equivalent width of their 8170--8200~\AA\ \ion{Na}{1} doublets and find $4.5~\pm~0.8$ \AA\ and $5.1~\pm~0.8$ \AA, respectively. They also revised the spectral type of \href{http://simbad.u-strasbg.fr/simbad/sim-id?submit=display&bibdisplay=refsum&bibyear1=1850&bibyear2=$currentYear&Ident=2MASS%20J03252938-4312299}{2MASS~J03252938--4312299} to M8.5. In a similar way to \href{http://simbad.u-strasbg.fr/simbad/sim-id?submit=display&bibdisplay=refsum&bibyear1=1850&bibyear2=$currentYear&Ident=2MASS%20J02401209-5305527}{2MASS~J02401209--5305527}, they have not flagged either objects as low-gravity, but both display the lowest \ion{Na}{1} EW of all objects of their respective spectral types, except for Upper Scorpius candidates. NIR spectroscopy would be useful in clearly identifying potential signs of low-gravity in these objects.

\href{http://simbad.u-strasbg.fr/simbad/sim-id?submit=display&bibdisplay=refsum&bibyear1=1850&bibyear2=$currentYear&Ident=LP%20944-20}{2MASS~J03393521--3525440} (\href{http://simbad.u-strasbg.fr/simbad/sim-id?submit=display&bibdisplay=refsum&bibyear1=1850&bibyear2=$currentYear&Ident=LP%20944-20}{LP~944--20}) was identified as an M9 dwarf by \cite{2001ApJ...548..908L}. They used their lithium detection to constrain its age below 1~Gyr. \cite{2013ApJ...772...79A} updated its spectral classification to an intermediate gravity L0$\beta$; \cite{2002AJ....124..519R} and \cite{2009ApJ...705.1416R} measured a RV which \cite{2014ApJ...783..121G} combined to obtain $9.3 \pm 1.7$~\kms; \cite{2014AJ....147...94D} measured a trigonometric distance of $6.41 \pm 0.04$~pc. \cite{2014ApJ...783..121G} used a previous parallax measurement from \citeauthor{1996MNRAS.281..644T} (\citeyear{1996MNRAS.281..644T}); $5.0 \pm 0.1$~pc) with the \href{http://www.astro.umontreal.ca/\textasciitilde gagne/banyanII.php}{BANYAN~II} tool to derive a Bayesian probability of 17.5\% that this is a member of ARG. However, \cite{2003AA...400..297R} indicated that it is a candidate member to the purported $\sim~200$~Myr old Castor moving group (CAS; \citealp{1998AA...339..831B}). They thus use an alternate Bayesian analysis similar to BANYAN~I \citep{2013ApJ...762...88M} but including a SKM of CAS built from members reported by \cite{1998AA...339..831B} and find a significantly larger Bayesian probability for CAS (99.7\%). More recently, \cite{2013AJ....146..154M} used updated distance and RV measurements of the original CAS members to demonstrate that they are too far apart in velocity space to be a part of a moving group of common origin. They thus argue that CAS likely a dynamical stream rather than a moving group, which is in line with the results of \cite{2012ApJ...754L..20M}, \cite{2012ApJ...761L...3M} and \cite{2013ApJ...778....5Z}. The difference in UVW space between \href{http://simbad.u-strasbg.fr/simbad/sim-id?submit=display&bibdisplay=refsum&bibyear1=1850&bibyear2=$currentYear&Ident=LP%20944-20}{LP~944--20} and ARG is considerable (9.7~\kms) and comparable to its distance to Fomalhaut (13.5~\kms). We conclude that \href{http://simbad.u-strasbg.fr/simbad/sim-id?submit=display&bibdisplay=refsum&bibyear1=1850&bibyear2=$currentYear&Ident=LP%20944-20}{LP~944--20} is likely a contaminant in our analysis, which could possibly be explained by the fact that our SKM model of field stars, derived from the Besan\c{c}on galactic model (\citealp{2014AA...569A..13R}; \citealp{2012AA...538A.106R}), does not explicitly include such dynamical streams that could act as an additional source of contamination.

\href{http://simbad.u-strasbg.fr/simbad/sim-id?submit=display&bibdisplay=refsum&bibyear1=1850&bibyear2=$currentYear&Ident=2MASS%20J05002100%2B0330501}{2MASS~J05002100+0330501} was identified as an L4 dwarf by \cite{2008AJ....136.1290R} and \cite{2010ApJ...723..684B} measured a RV of $15.9 \pm 0.2$~\kms, from which we obtain a 62.8\% membership probability associated with ABDMG. However, \cite{2008AJ....136.1290R} specified that this object displays no notable peculiarities and would be a good spectral standard. While NIR spectroscopy could unambiguously rule out low-gravity, it is likely that this object is a field contaminant in our analysis.

\href{http://simbad.u-strasbg.fr/simbad/sim-id?submit=display&bibdisplay=refsum&bibyear1=1850&bibyear2=$currentYear&Ident=2MASS%20J05012406-0010452}{2MASS~J05012406--0010452} has been identified by \cite{2008AJ....136.1290R} as an L4 BD with signs of low-gravity in its optical spectrum. \cite{2009AJ....137.3345C} updated its classification to L4$\gamma$ using its optical spectrum, and \cite{2013ApJ...772...79A} classified it as L3$\gamma$ using NIR spectroscopy. \cite{2012ApJ...752...56F} measured a trigonometric distance of $13.1 \pm 0.8$ pc. \citealp{2014ApJ...783..121G} considered this object and found no obvious candidacy to any YMG considered here. However, we find that it has a 64.7\% Bayesian probability of being a member of COL, associated with a 2.3\% contamination probability. The discrepancy between this result and that of \citealp{2014ApJ...783..121G} is due to the $\mu_\delta$ proper motion measurement from \href{http://www.ipac.caltech.edu/2mass/}{\emph{2MASS}}--\href{http://wise2.ipac.caltech.edu/docs/release/allwise/}{\emph{AllWISE}}, which is at 2.8$\sigma$ of the value they used (which was measured by \citealp{2012ApJ...752...56F}). We visually inspected the \href{http://www.ipac.caltech.edu/2mass/}{\emph{2MASS}} and \href{http://wise2.ipac.caltech.edu/docs/release/allwise/}{\emph{AllWISE}} Atlas images and found that our cross-match between both catalogs is unambiguous. Much like the case of \href{http://simbad.u-strasbg.fr/simbad/sim-id?submit=display&bibdisplay=refsum&bibyear1=1850&bibyear2=$currentYear&Ident=2MASS%20J02212859-6831400}{2MASS~J02212859--6831400}, a RV measurement will be needed to better constrain the membership of this object, but it is plausible that this object is a false-positive in our analysis. 

\href{http://simbad.u-strasbg.fr/simbad/sim-id?submit=display&bibdisplay=refsum&bibyear1=1850&bibyear2=$currentYear&Ident=2MASS%20J10584787-1548172}{2MASS~J10584787--1548172} (\href{http://simbad.u-strasbg.fr/simbad/sim-id?submit=display&bibdisplay=refsum&bibyear1=1850&bibyear2=$currentYear&Ident=2MASS%20J10584787-1548172}{DENIS--P J1058.7--1548}) has been identified as an L3 dwarf by \cite{2002ApJ...564..466G} and \cite{2002AJ....124.1170D} measured a trigonometric distance of $17.3 \pm 0.3$~pc, from which we obtain a 93.1\% membership probability to ARG. \cite{2008AJ....136.1290R} measured H$\alpha$ emission in its optical spectrum, but reported no further peculiarities. \cite{2014AJ....147...34S} subsequently measured the gravity-sensitive H$_2$($K$) in its NIR spectrum and obtain a value of $1.021$, which is consistent with a field L3 dwarf. It is thus likely that this object is a field contaminant in our analysis.

\href{http://simbad.u-strasbg.fr/simbad/sim-id?submit=display&bibdisplay=refsum&bibyear1=1850&bibyear2=$currentYear&Ident=2MASS%20J12474428-3816464}{2MASS~J12474428--3816464} has been identified by \cite{2014ApJ...785L..14G} as a low-gravity M9$\gamma$ candidate member of TWA, as part of the initial follow-up of the \href{http://www.astro.umontreal.ca/\textasciitilde gagne/BASS.php}{\emph{BASS}} survey. They note that its kinematics are discrepant with TWA albeit its low probability of being a field contaminant: its kinematics would match with TWA if it was placed further away, however this would make it over-luminous compared to young BDs of the same spectral type and age. It could be expected that this is a contaminant from the Lower-Centaurus-Crux region (LCC; $\sim$10--20 Myr; \citealp{1999AJ....117..354D}) of the Scorpius-Centaurus complex, but its distance ($\sim$~120~pc) would also make it over-luminous. It is possible that this object could be an unresolved binary and located further away, between TWA and LCC: this is reminiscent of \href{http://simbad.u-strasbg.fr/simbad/sim-id?submit=display&bibdisplay=refsum&bibyear1=1850&bibyear2=$currentYear&Ident=TWA%2029}{TWA~29} and \href{http://simbad.u-strasbg.fr/simbad/sim-id?submit=display&bibdisplay=refsum&bibyear1=1850&bibyear2=$currentYear&Ident=TWA%2031}{TWA~31}, and might strengthen the proposition of \citeauthor{2003ApJ...599..342S} (\citeyear{2003ApJ...599..342S}; see also \citealp{2012ApJ...754...39S}) that TWA could actually be part of the LCC.

\href{http://simbad.u-strasbg.fr/simbad/sim-id?submit=display&bibdisplay=refsum&bibyear1=1850&bibyear2=$currentYear&Ident=2MASS%20J14252798-3650229}{2MASS~J14252798--3650229} has been identified as an L5 BD by \cite{2009AJ....137....1F}. Including RV and trigonometric distance measurements from \cite{2010ApJ...723..684B} and \cite{2014AJ....147...85R} respectively, we find a 99.6\% probability that this object is a member of ABDMG, with 0.1\% contamination probability. Only signs of youth need to be confirmed before we can consider this object as a new bona fide member of ABDMG, however we note that its has NIR colors $J-K_S = 1.94$, hence 1$\sigma$ redder than field L5 dwarfs, which could be an indication of youth.

\href{http://simbad.u-strasbg.fr/simbad/sim-id?submit=display&bibdisplay=refsum&bibyear1=1850&bibyear2=$currentYear&Ident=2MASS%20J17571539%2B7042011}{2MASS~J17571539+7042011} (\href{http://simbad.u-strasbg.fr/simbad/sim-id?submit=display&bibdisplay=refsum&bibyear1=1850&bibyear2=$currentYear&Ident=2MASS%20J17571539%2B7042011}{LP~44--162}) has been identified as an M7.5 dwarf by \cite{2000AJ....120.1085G}. \citealp{2010PASP..122.1195T} and \citealp{2012ApJ...747L..38T} measured its radial velocity, which we combine in an error-weighted mean to obtain $-12.4 \pm 0.6$~\kms. \cite{2009AJ....137.4109L} measured a trigonometric distance of $19.1 \pm 0.4$~pc and report that it is significantly over-luminous compared to dwarfs of the same colors, and propose that it might be an unresolved multiple. We find a Bayesian probability of 91.0\% that this is a member of ARG. However, \cite{2012AJ....144...99D} obtained high-resolution NIR spectroscopy and report pseudo-equivalent widths of \ion{K}{1} lines in the $J$ band which are consistent with M7.5 field dwarfs (e.g. see \citealp{2013ApJ...772...79A}). It is thus plausible that this object is a false positive in our analysis, despite its high probability.

\href{http://simbad.u-strasbg.fr/simbad/sim-id?submit=display&bibdisplay=refsum&bibyear1=1850&bibyear2=$currentYear&Ident=SIMP%20J215434.5-105530.8}{SIMP~J21543454--1055308} has been independently discovered in the \emph{SIMP} survey for field BDs (\citealp{2009AIPC.1094..493A}; \href{http://www.astro.umontreal.ca/\textasciitilde gagne/simp.php}{J.~Robert et al., in preparation}). A NIR spectroscopic follow-up revealed that this object is a low-gravity L4$\beta$ BD with an estimated mass of $10 \pm 0.5$~\Mjup, well into the planetary regime, if it is a member of ARG as suspected \citep{2014ApJ...792L..17G}.

\href{http://simbad.u-strasbg.fr/simbad/sim-id?submit=display&bibdisplay=refsum&bibyear1=1850&bibyear2=$currentYear&Ident=2MASS%20J23225384%2B7847386}{2MASS~J23225384+7847386} has been identified as an M5 proper motion companion to \href{http://simbad.u-strasbg.fr/simbad/sim-id?Ident=V*+V368+Cep}{V~368~Cep} and \href{http://simbad.u-strasbg.fr/simbad/sim-id?submit=display&bibdisplay=refsum&bibyear1=1850&bibyear2=$currentYear&Ident=LSPM%20J2322%2B7847}{LSPM~J2322+7847} by \cite{2007ApJ...668L.155M}. Using the X-ray luminosity of \href{http://simbad.u-strasbg.fr/simbad/sim-id?Ident=V*+V368+Cep}{V~368~Cep} as well as an isochrone analysis on all three components, they estimated an age of $\sim$~50~Myr for the system. Using the RV measurement from \cite{2007yCat.3254....0K}, and combined trigonometric distances measurements from \cite{2007yCat.3254....0K} and \cite{2014ApJ...784..156D}, we find that this object has a 29.7\% probability of being a member of CAR, with a contamination probability of 1.0\%. The estimated age of this system is consistent with that of CAR, which makes it a compelling candidate member, even if its Bayesian probability is somewhat low. This low probability is a consequence of its galactic position $XYZ$ = ($-8.7 \pm 2.5$,$16.1 \pm 4.6$,$5.5 \pm 1.6$) pc, at 2.5$\sigma$ of our spatial model for CAR. We note however that its kinematics are a very good match to CAR with $UVW$ = ($-10.1\pm 5.2$,$-23.5\pm 2.9$,$-6.3\pm 1.0$), at only 0.5$\sigma$ of our kinematic model. This could be an indication that CAR is in fact spatially larger than our present model, which would not be surprising since it was built from the only 7 currently known bona fide members. We thus suggest that \href{http://simbad.u-strasbg.fr/simbad/sim-id?submit=display&bibdisplay=refsum&bibyear1=1850&bibyear2=$currentYear&Ident=2MASS%20J23225384%2B7847386}{2MASS~J23225384+7847386} is probably a member in CAR, and that we might be currently missing more objects like this one as a result of our spatial and kinematic model for this association being too narrowly confined. Finding additional objects like this one will be needed to better constrain the SKM of CAR. \cite{2001MNRAS.328...45M} suggested that \href{http://simbad.u-strasbg.fr/simbad/sim-id?Ident=V*+V368+Cep}{V~368~Cep} is a member of the Pleiades moving group (PMG; also called the Local Association), however we find that its kinematics are much more consistent with those of CAR, at only 1.5~\kms\ of our dynamical model, compared to a difference of 5.5~\kms\ with the kinematics of the PMG \citep{2001MNRAS.328...45M}. \cite{2005AA...430..165F} demonstrated that the PMG is likely a dynamical stream with a large spread in age rather than a coeval moving group, hence the age constraint acts as a further indication that a membership to CAR is more likely.

\section{RECOVERY OF KNOWN CANDIDATES AND MEMBERS OF YOUNG MOVING GROUPS}\label{sec:recov}

In this Section, we assess the fraction of known $\geq$~M5 candidate members of YMGs that are recovered in the \href{http://www.astro.umontreal.ca/\textasciitilde gagne/BASS.php}{\emph{BASS}} and \href{http://www.astro.umontreal.ca/\textasciitilde gagne/LP-BASS.php}{\emph{LP-BASS}} catalogs. We identified a total of 98 candidate members of the YMGs considered here in the literature (\citealp{2012AJ....144..109S}; \citealp{2012ApJ...758...56S}; \citealp{2013ApJ...762...88M}; \citealp{2013ApJ...774..101R}; \citealp{2014ApJ...783..121G}; \citealp{2014AJ....147..146K} and references therein). We do not include low-probability candidates from \cite{2014ApJ...783..121G} here, since they have a contamination probability of $>$ 50\% by definition, which ensures that they are not listed in the \href{http://www.astro.umontreal.ca/\textasciitilde gagne/BASS.php}{\emph{BASS}} catalog. We find that a total of 55/98 of all these candidates are recovered in \href{http://www.astro.umontreal.ca/\textasciitilde gagne/BASS.php}{\emph{BASS}} (see \hyperref[tab:lit]{Table~\ref*{tab:lit}}), whereas 8 others are recovered in \href{http://www.astro.umontreal.ca/\textasciitilde gagne/LP-BASS.php}{\emph{LP-BASS}} (see the Appendix), hence making up for 64\% of currently known candidate members. All 35 candidates not recovered here are listed in \hyperref[tab:miss]{Table~\ref*{tab:miss}}, along with a list of the filters which caused them to be rejected. We note that 17 of those 36 candidates were missed only because they were cut from our input sample because of quality filters (i.e. low galactic latitude, low proper motion, large number of \href{http://www.ipac.caltech.edu/2mass/}{\emph{2MASS}} neighbors, poor \href{http://www.ipac.caltech.edu/2mass/}{\emph{2MASS}} or \href{http://wise2.ipac.caltech.edu/docs/release/allwise/}{\emph{AllWISE}} photometric quality or NIR colors too blue), whereas 18 were missed at least because of a low Bayesian probability, high contamination probability or position in a CMD diagram derived from its statistical distance. Considering only the known candidate members that were part of our input search sample, the \href{http://www.astro.umontreal.ca/\textasciitilde gagne/BASS.php}{\emph{BASS}} and \href{http://www.astro.umontreal.ca/\textasciitilde gagne/LP-BASS.php}{\emph{LP-BASS}} catalogs thus recover 68\% of them.

\capstartfalse
\input{bass_table_3.tex}
\capstarttrue

\capstartfalse
\input{bass_table_4.tex}
\capstarttrue

\section{THE UPDATED \href{http://www.astro.umontreal.ca/\textasciitilde gagne/BASS.php}{\emph{BASS}} SAMPLE}\label{sec:updated}

We present in \hyperref[tab:allbass]{Table~\ref*{tab:allbass}} a complete list of the \href{http://www.astro.umontreal.ca/\textasciitilde gagne/BASS.php}{\emph{BASS}} sample, which contains only objects respecting all criteria mentioned in Sections~\ref{sec:crossmatch}--\ref{sec:filters} after taking account of all information available in the literature. We list in this table all the contamination probability of all objects, obtained from the Monte Carlo analysis described in \hyperref[sec:filters]{Section~\ref*{sec:filters}}, as well as statistical estimates for their distance and RV. We refer to this list as the \href{http://www.astro.umontreal.ca/\textasciitilde gagne/BASS.php}{\emph{BASS}} sample for the remainder of this work. We used the individual contamination probability of all candidate members to estimate an average contamination fraction from field stars of 2.4\% and 29.5\% for the high probability and modest probability samples, respectively. These estimates of contamination do not take account of possible cross-contamination between the YMGs considered here, or other, older nearby associations not considered, e.g. Carina-Near ($\sim$~200~Myr; \citealp{2006ApJ...649L.115Z}), the Ursa Major moving ($\sim$~500~Myr; \citealp{2003AJ....125.1980K}) and the Hercules-Lyra association ($\sim$~250~Myr; \citealp{2013AA...556A..53E}). Another way to assess a minimal contamination rate is to count the fraction of candidates with RV, distance or spectra in the literature which were rejected from these measurements. This estimate yields a larger contamination rate of 12.6\% (11/87) for the high probability candidates. Small number statistics prevent an accurate estimation for the low-probability candidates: only 37 had such measurements in the literature, from which 4 were rejected. We rather choose to scale the observed 12.6\% contamination fraction of the high-probability sample with the ratio of predicted contamination fractions of both samples to estimate a more reliable expected contamination fraction of $\sim$ 71\% for the modest probability \href{http://www.astro.umontreal.ca/\textasciitilde gagne/BASS.php}{\emph{BASS}} sample.

\begin{figure*}[p]
	\begin{center}
	\includegraphics[width=0.995\textwidth]{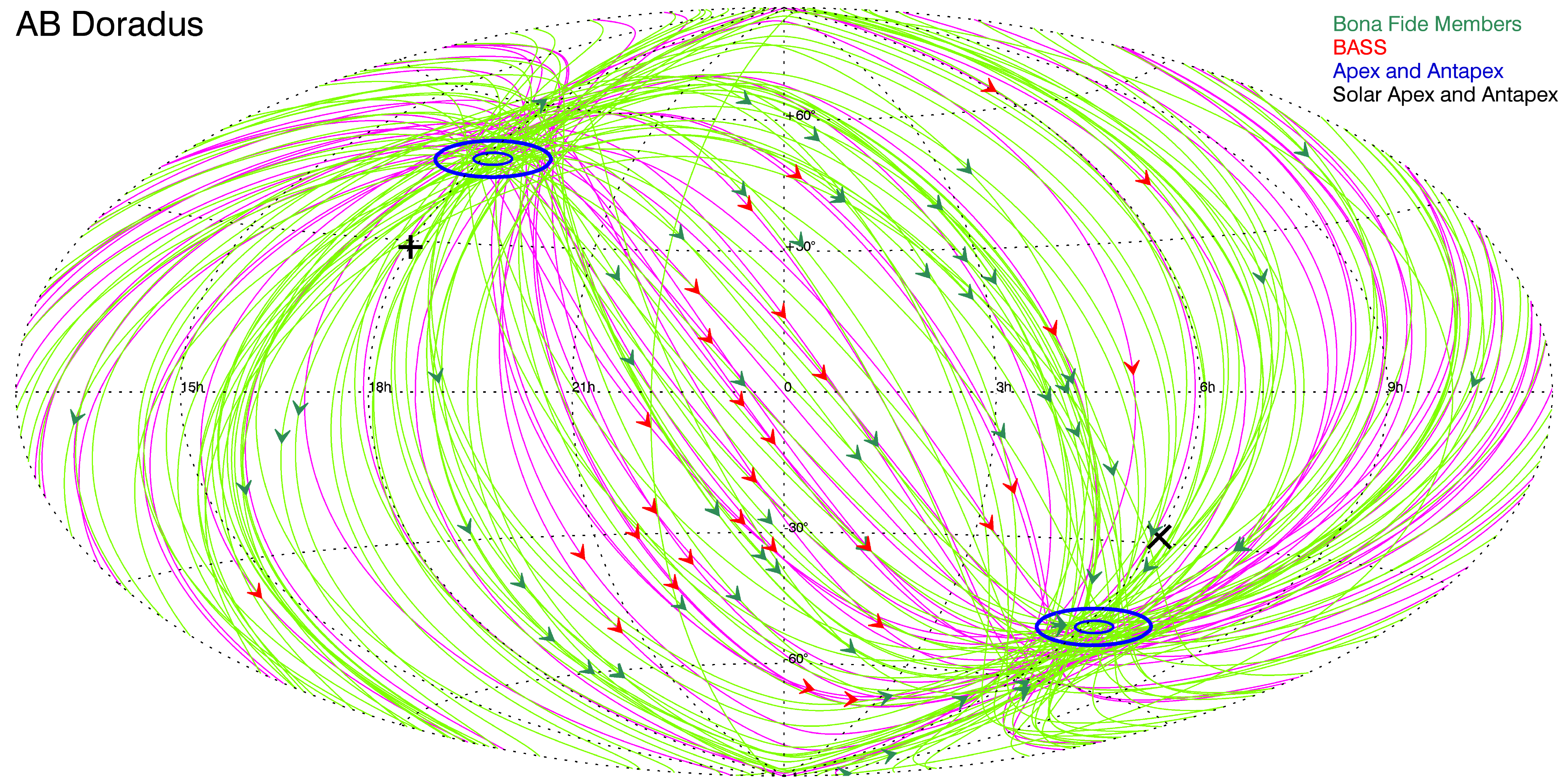}
	\end{center}
	\caption{Proper motion as a function of sky position for candidate members of AB Doradus in the \href{http://www.astro.umontreal.ca/\textasciitilde gagne/BASS.php}{\emph{BASS}} Catalog (red arrows and lines), compared with currently known bona fide members (light green; see \citealp{2014ApJ...783..121G}). The proper motions of candidate members and bona fide members all converge to the apex and antapex of ABDMG (blue circles), which is a well known property of YMGs.}
	\label{fig:pm_maps_abdmg}
\end{figure*}

\begin{figure*}[p]
	\begin{center}
 	\includegraphics[width=0.995\textwidth]{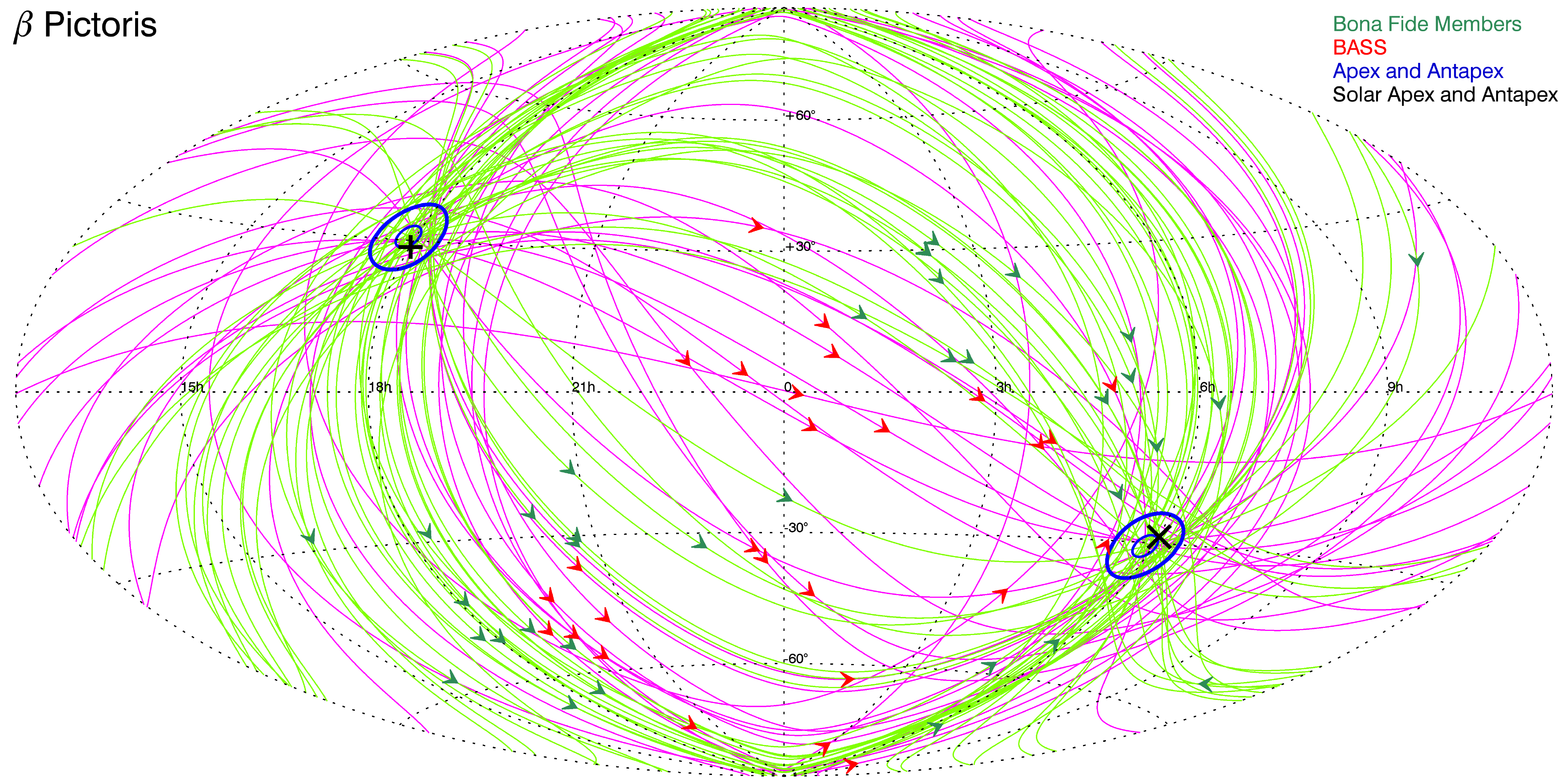}
	\end{center}
	\caption{Proper motion as a function of sky position for \href{http://www.astro.umontreal.ca/\textasciitilde gagne/BASS.php}{\emph{BASS}} candidate members and bona fide members of $\beta$PMG. Colors and symbols are defined in the same way as in \hyperref[fig:pm_maps_abdmg]{Figure~\ref*{fig:pm_maps_abdmg}}}
	\label{fig:pm_maps_bpmg}
\end{figure*}

\begin{figure*}[p]
	\begin{center}
 	\includegraphics[width=0.995\textwidth]{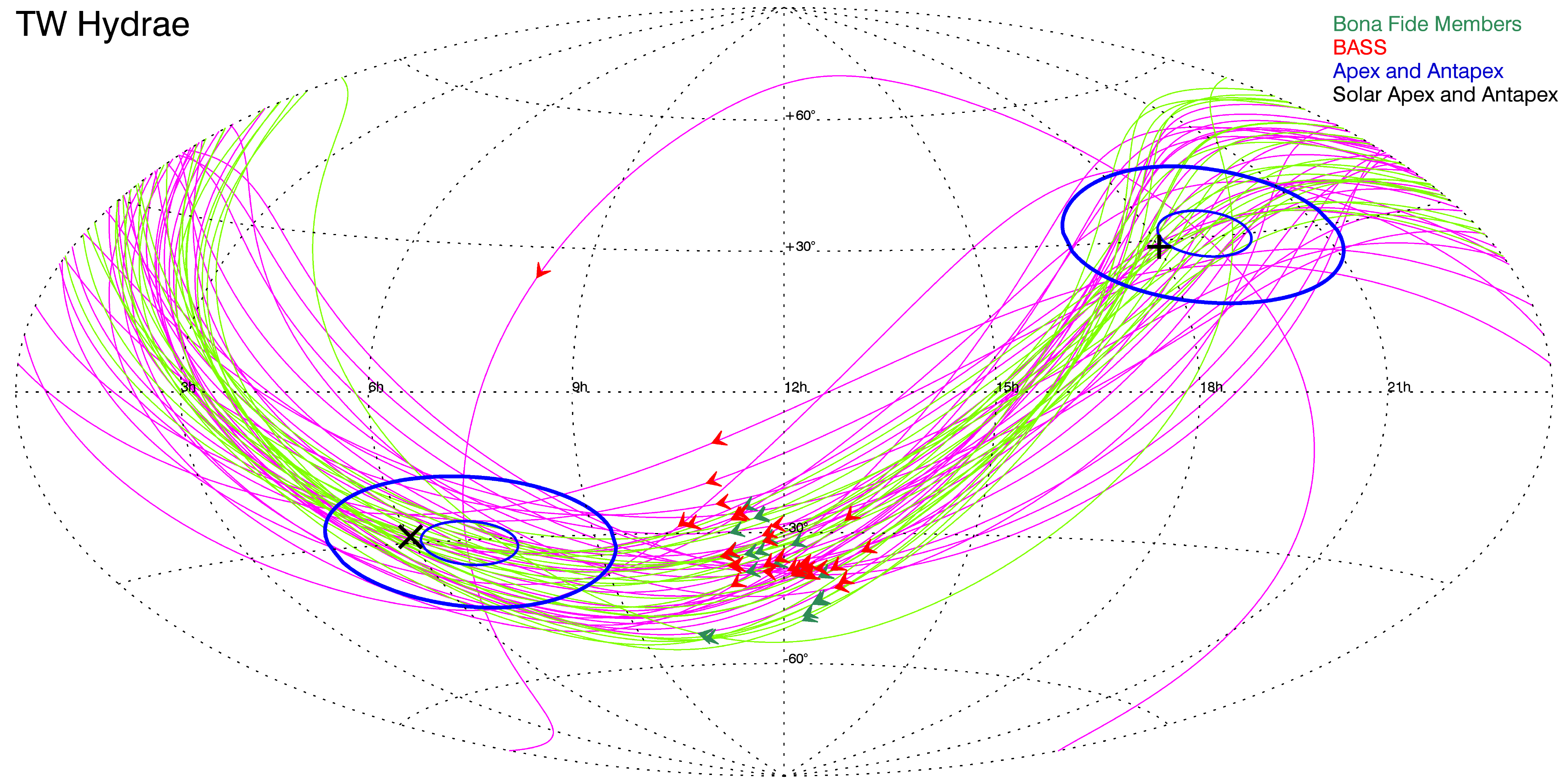}
	\end{center}
	\caption{Proper motion as a function of sky position for \href{http://www.astro.umontreal.ca/\textasciitilde gagne/BASS.php}{\emph{BASS}} candidate members and bona fide members of TWA. Colors and symbols are defined in the same way as in \hyperref[fig:pm_maps_abdmg]{Figure~\ref*{fig:pm_maps_abdmg}}}
	\label{fig:pm_maps_twa}
\end{figure*}

\begin{figure*}[p]
	\begin{center}
 	\includegraphics[width=0.995\textwidth]{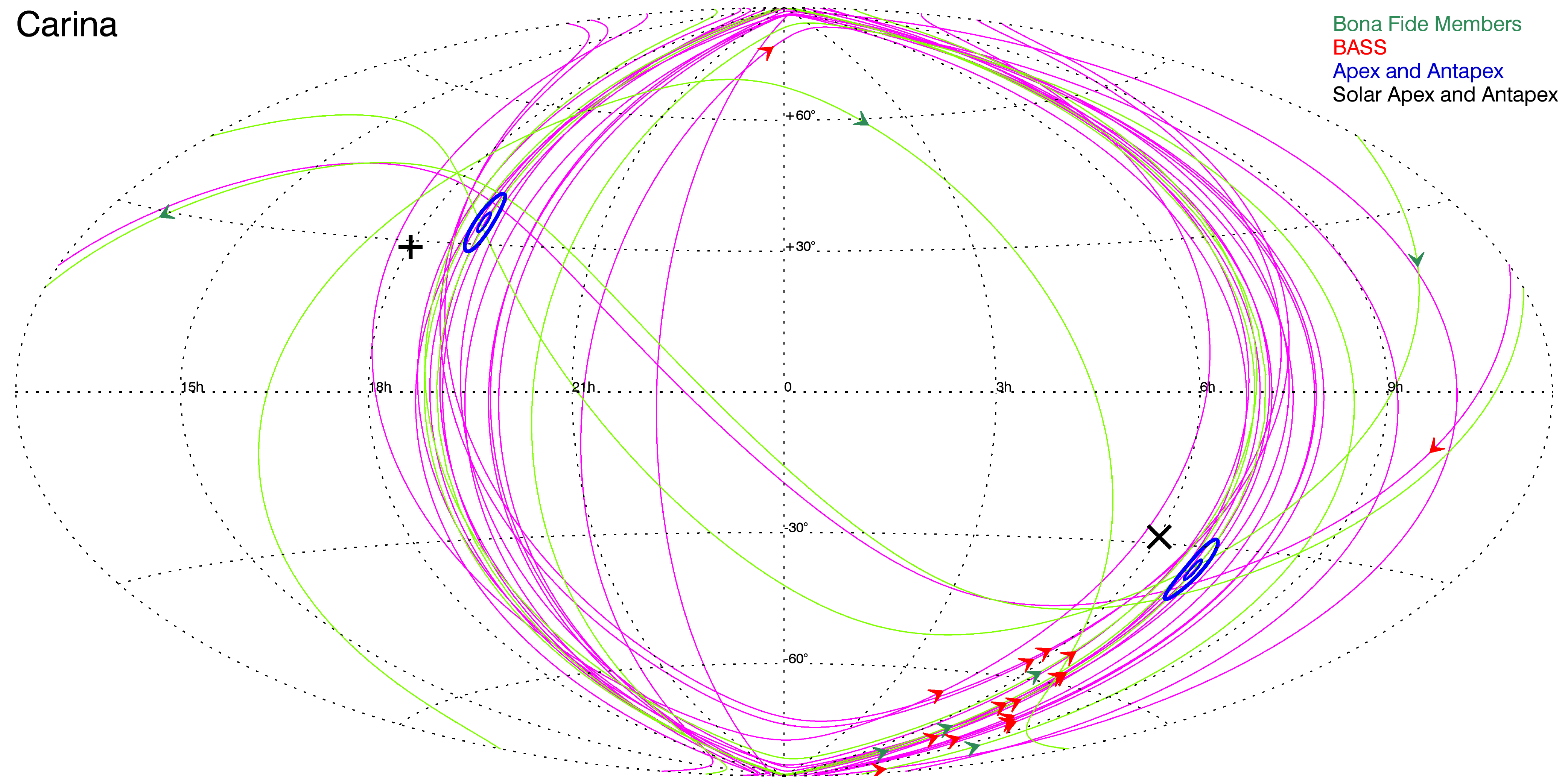}
	\end{center}
	\caption{Proper motion as a function of sky position for \href{http://www.astro.umontreal.ca/\textasciitilde gagne/BASS.php}{\emph{BASS}} candidate members and bona fide members of CAR. Colors and symbols are defined in the same way as in \hyperref[fig:pm_maps_abdmg]{Figure~\ref*{fig:pm_maps_abdmg}}}
	\label{fig:pm_maps_car}
\end{figure*}

\begin{figure*}[p]
	\begin{center}
 	\includegraphics[width=0.995\textwidth]{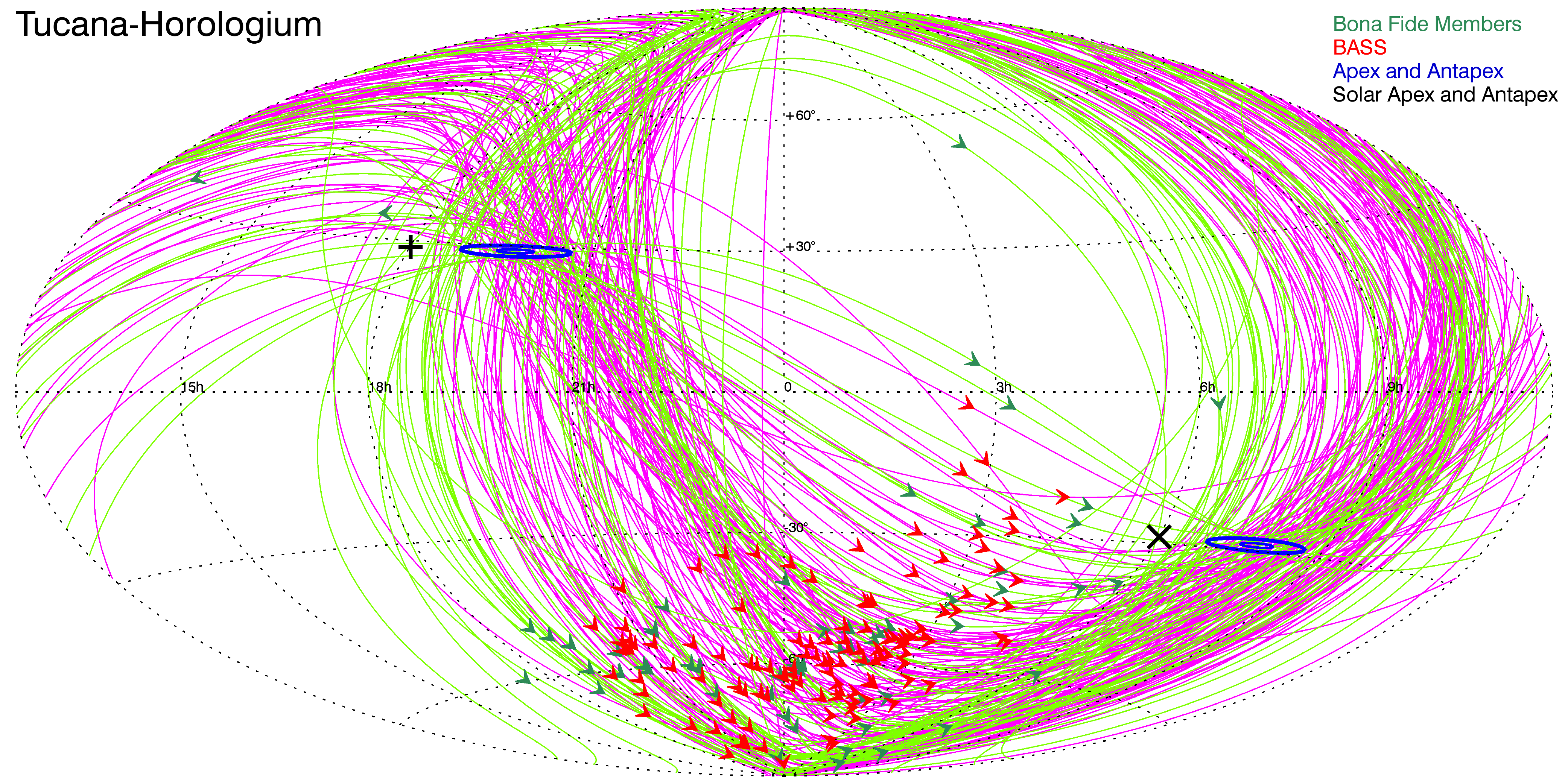}
	\end{center}
	\caption{Proper motion as a function of sky position for \href{http://www.astro.umontreal.ca/\textasciitilde gagne/BASS.php}{\emph{BASS}} candidate members and bona fide members of THA. Colors and symbols are defined in the same way as in \hyperref[fig:pm_maps_abdmg]{Figure~\ref*{fig:pm_maps_abdmg}}}
	\label{fig:pm_maps_tucana}
\end{figure*}

\begin{figure*}[p]
	\begin{center}
 	\includegraphics[width=0.995\textwidth]{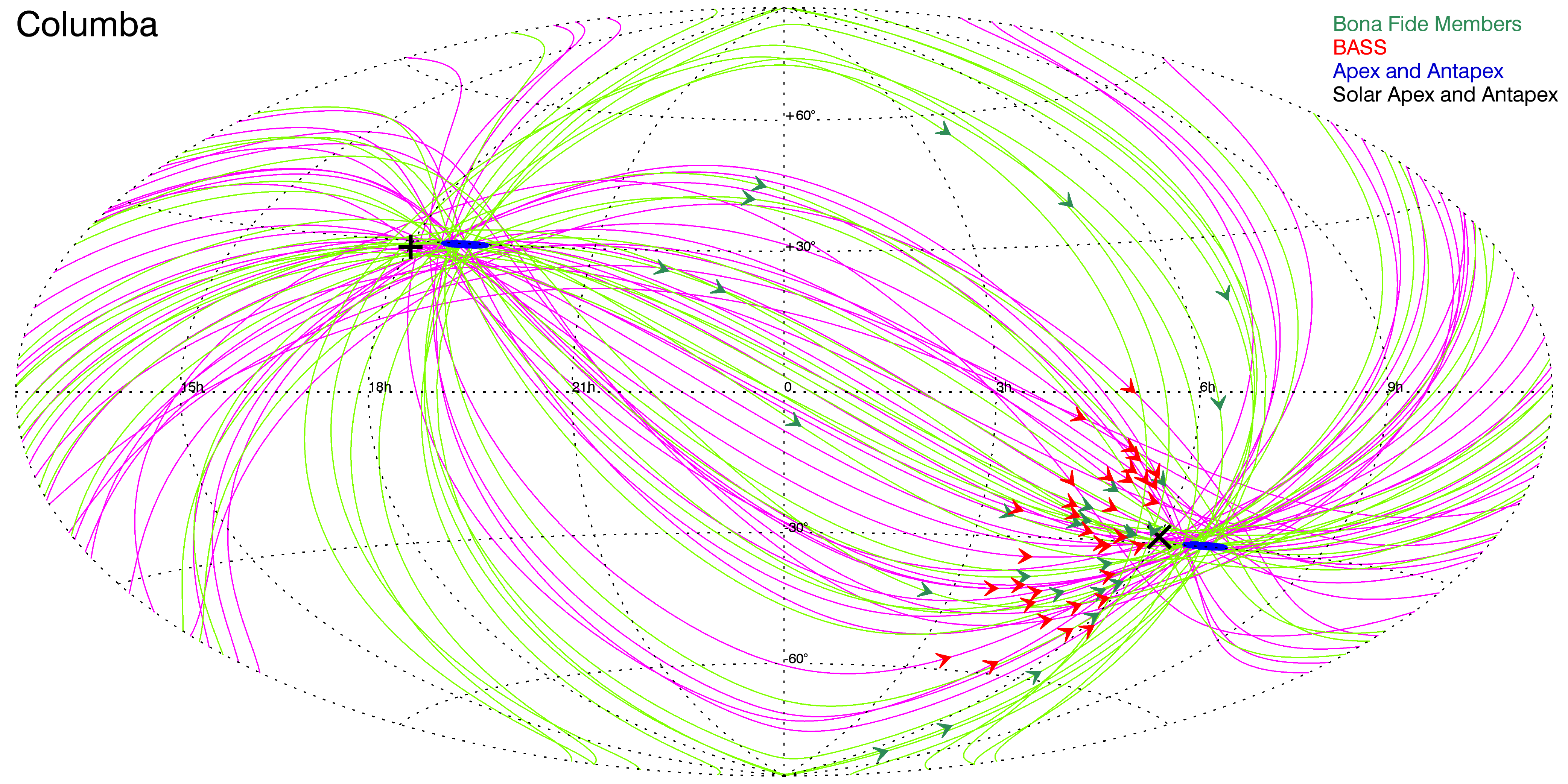}
	\end{center}
	\caption{Proper motion as a function of sky position for \href{http://www.astro.umontreal.ca/\textasciitilde gagne/BASS.php}{\emph{BASS}} candidate members and bona fide members of COL. Colors and symbols are defined in the same way as in \hyperref[fig:pm_maps_abdmg]{Figure~\ref*{fig:pm_maps_abdmg}}}
	\label{fig:pm_maps_columba}
\end{figure*}

\begin{figure*}[p]
	\begin{center}
 	\includegraphics[width=0.995\textwidth]{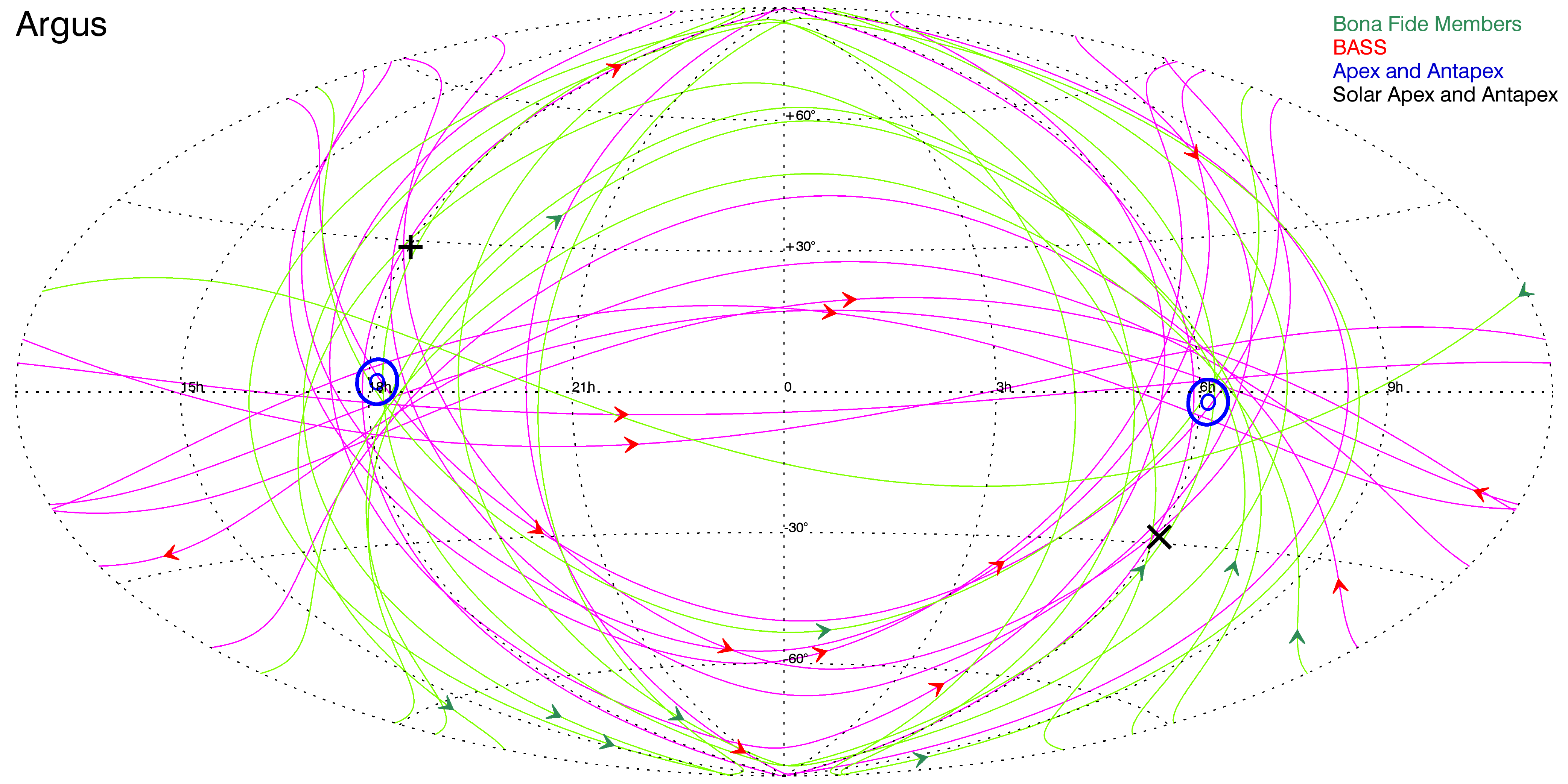}
	\end{center}
	\caption{Proper motion as a function of sky position for \href{http://www.astro.umontreal.ca/\textasciitilde gagne/BASS.php}{\emph{BASS}} candidate members and bona fide members of ARG. Colors and symbols are defined in the same way as in \hyperref[fig:pm_maps_abdmg]{Figure~\ref*{fig:pm_maps_abdmg}}}
	\label{fig:pm_maps_arg}
\end{figure*}

\begin{figure*}[p]
	\center
	\subfigure[\emph{XYZ}]{\includegraphics[width=0.495\textwidth]{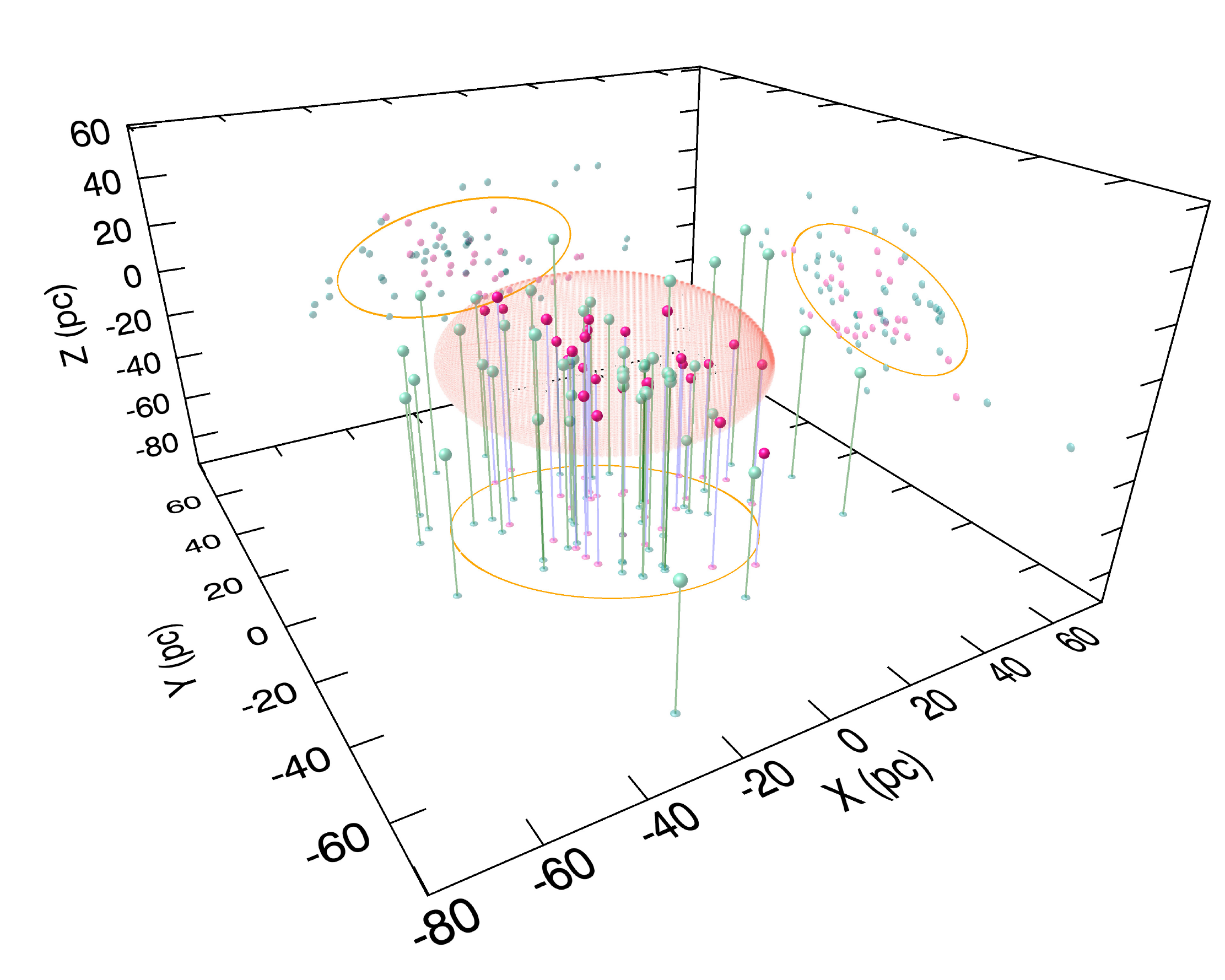}}
	\subfigure[\emph{UVW}]{\includegraphics[width=0.495\textwidth]{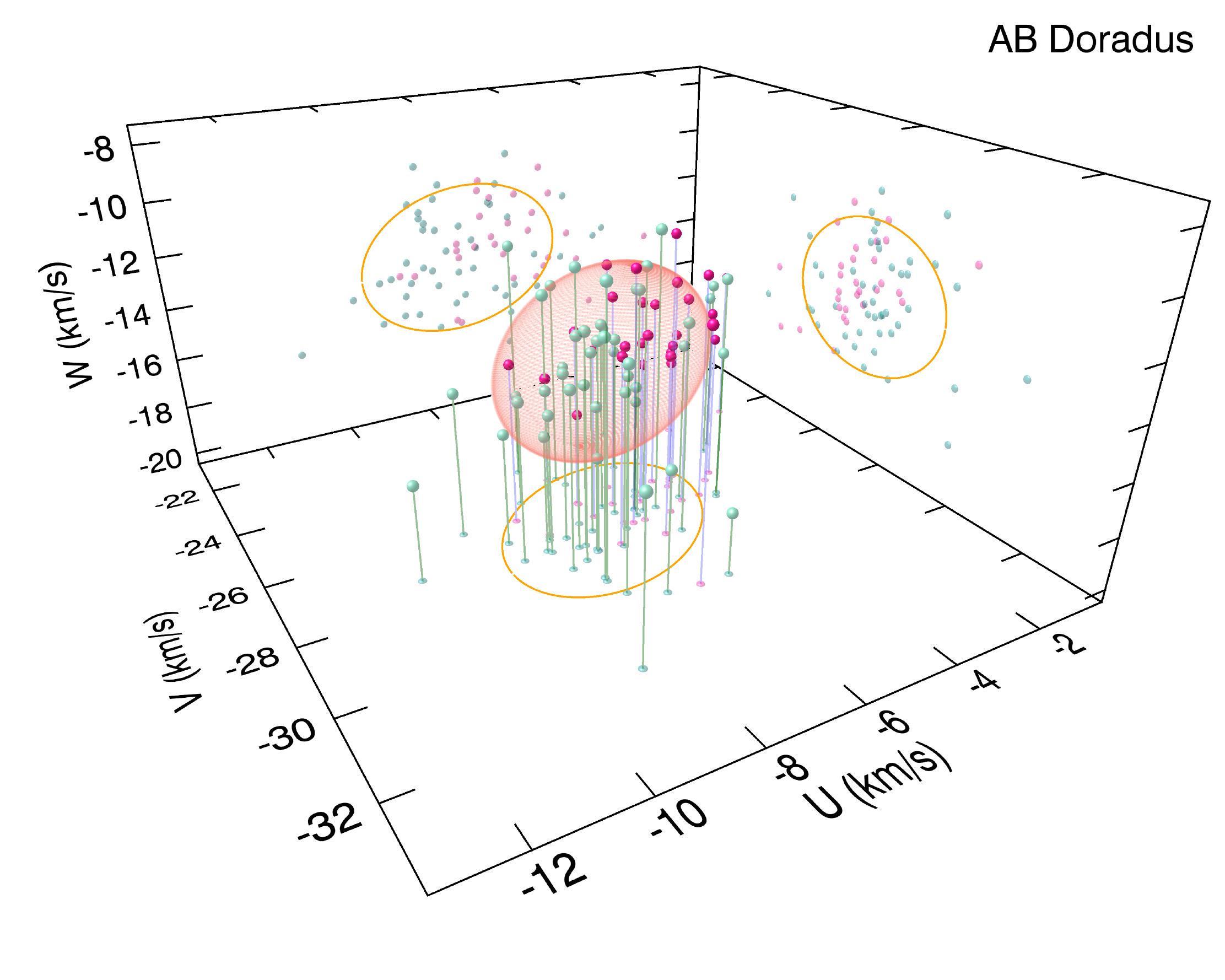}}
	\caption{Most probable galactic positions \emph{XYZ} and space velocities \emph{UVW} based on \href{http://www.astro.umontreal.ca/\textasciitilde gagne/banyanII.php}{BANYAN~II} statistical distances and RVs for all \href{http://www.astro.umontreal.ca/\textasciitilde gagne/BASS.php}{\emph{BASS}} candidate members in ABDMG (red points), compared with bona fide members (green points), as well as the spatial and kinematic ellipsoid models used in \href{http://www.astro.umontreal.ca/\textasciitilde gagne/banyanII.php}{BANYAN~II} (orange ellipsoids; see \citealp{2014ApJ...783..121G} for more details). All points and models are projected on the three normal planes for a better clarity.}
	\label{fig:xyzuvw_abdmg}
\end{figure*}

\begin{figure*}[p]
	\center
	\subfigure[\emph{XYZ}]{\includegraphics[width=0.495\textwidth]{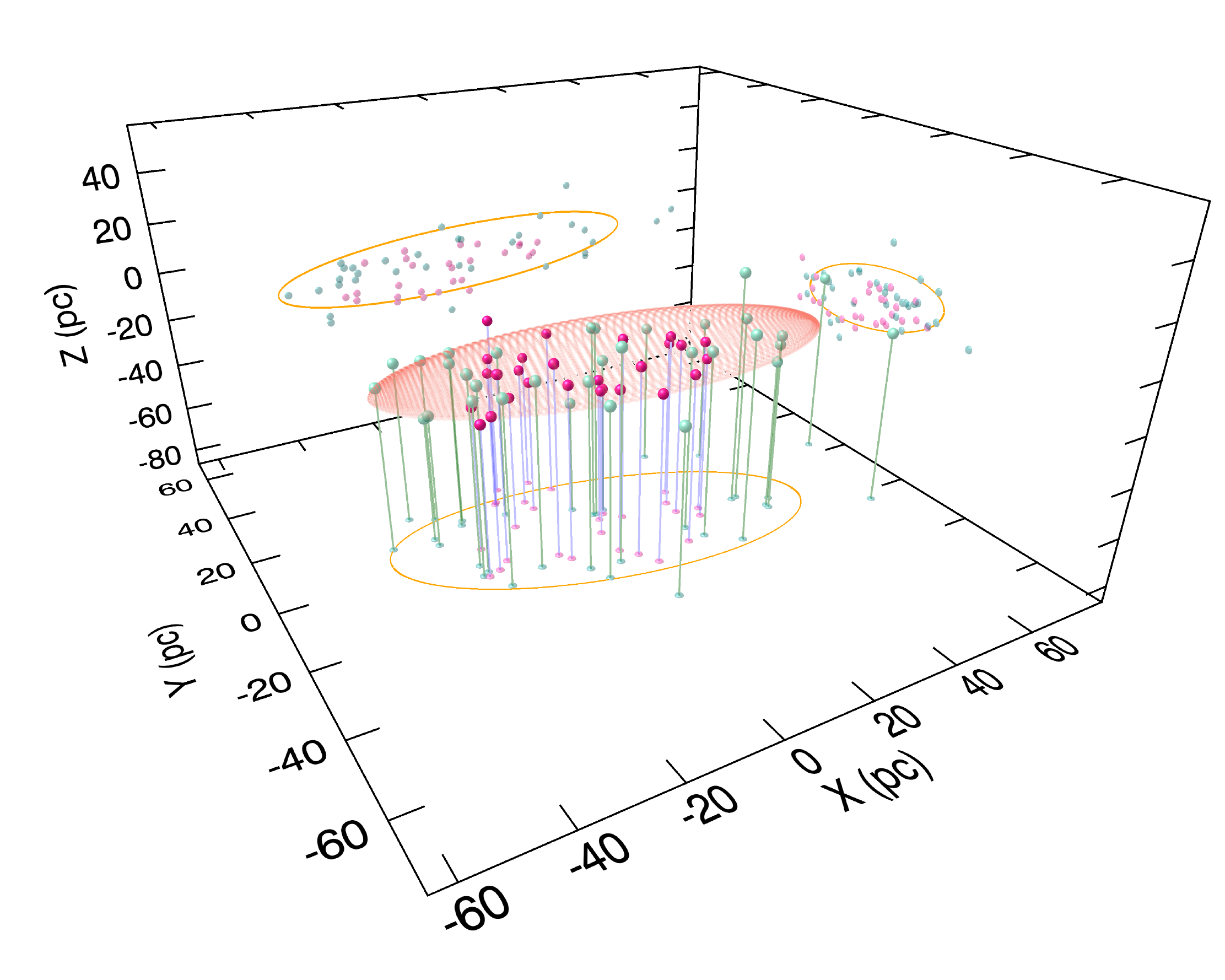}}
	\subfigure[\emph{UVW}]{\includegraphics[width=0.495\textwidth]{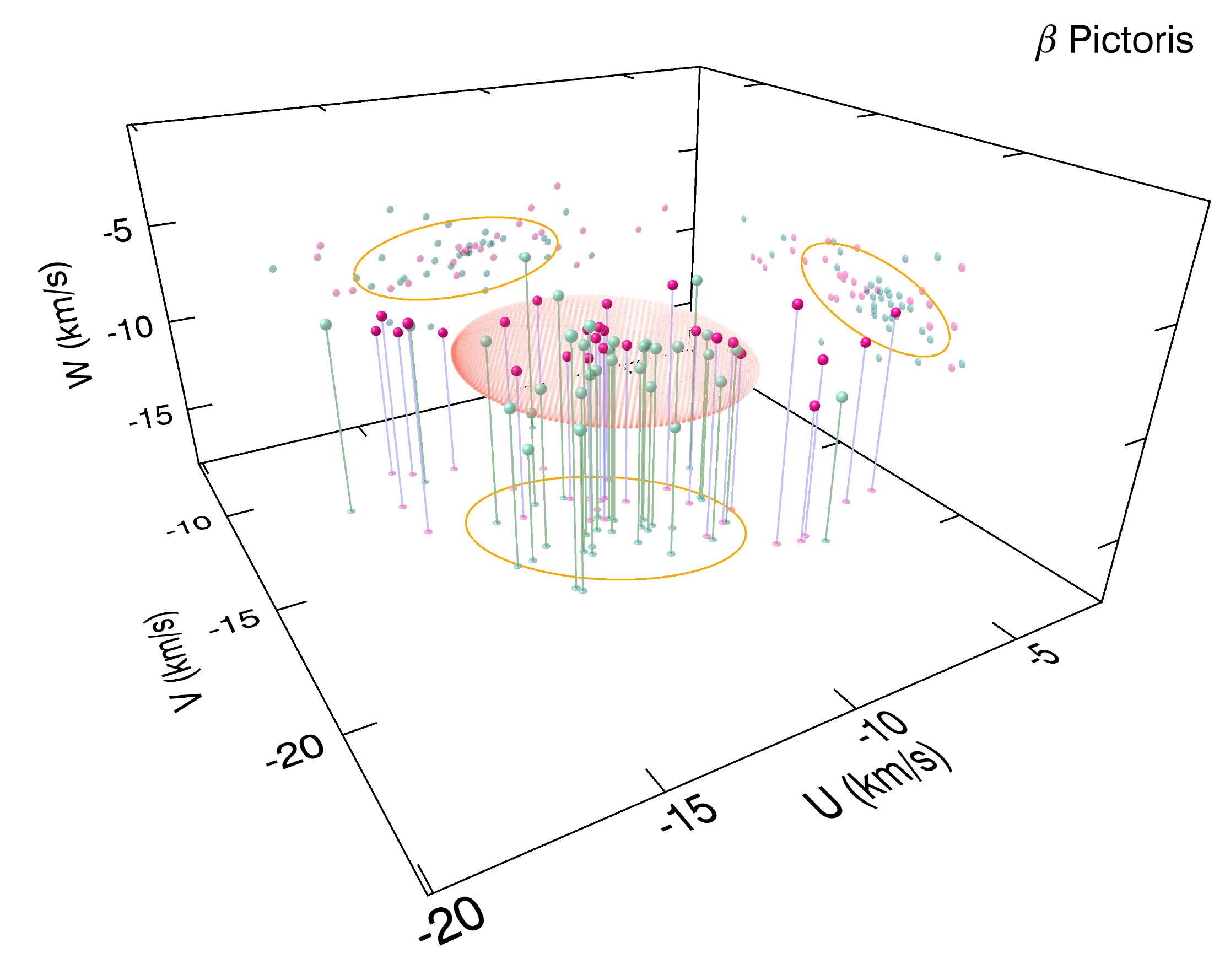}}
	\caption{Most probable galactic positions \emph{XYZ} and space velocities \emph{UVW} based on \href{http://www.astro.umontreal.ca/\textasciitilde gagne/banyanII.php}{BANYAN~II} statistical distances and RVs for all \href{http://www.astro.umontreal.ca/\textasciitilde gagne/BASS.php}{\emph{BASS}} candidate members in $\beta$PMG compared with bona fide members. Colors and symbols are defined in the same way as in \hyperref[fig:xyzuvw_abdmg]{Figure~\ref*{fig:xyzuvw_abdmg}}.}
	\label{fig:xyzuvw_bpmg}
\end{figure*}

\begin{figure*}[p]
	\center
	\subfigure[\emph{XYZ}]{\includegraphics[width=0.495\textwidth]{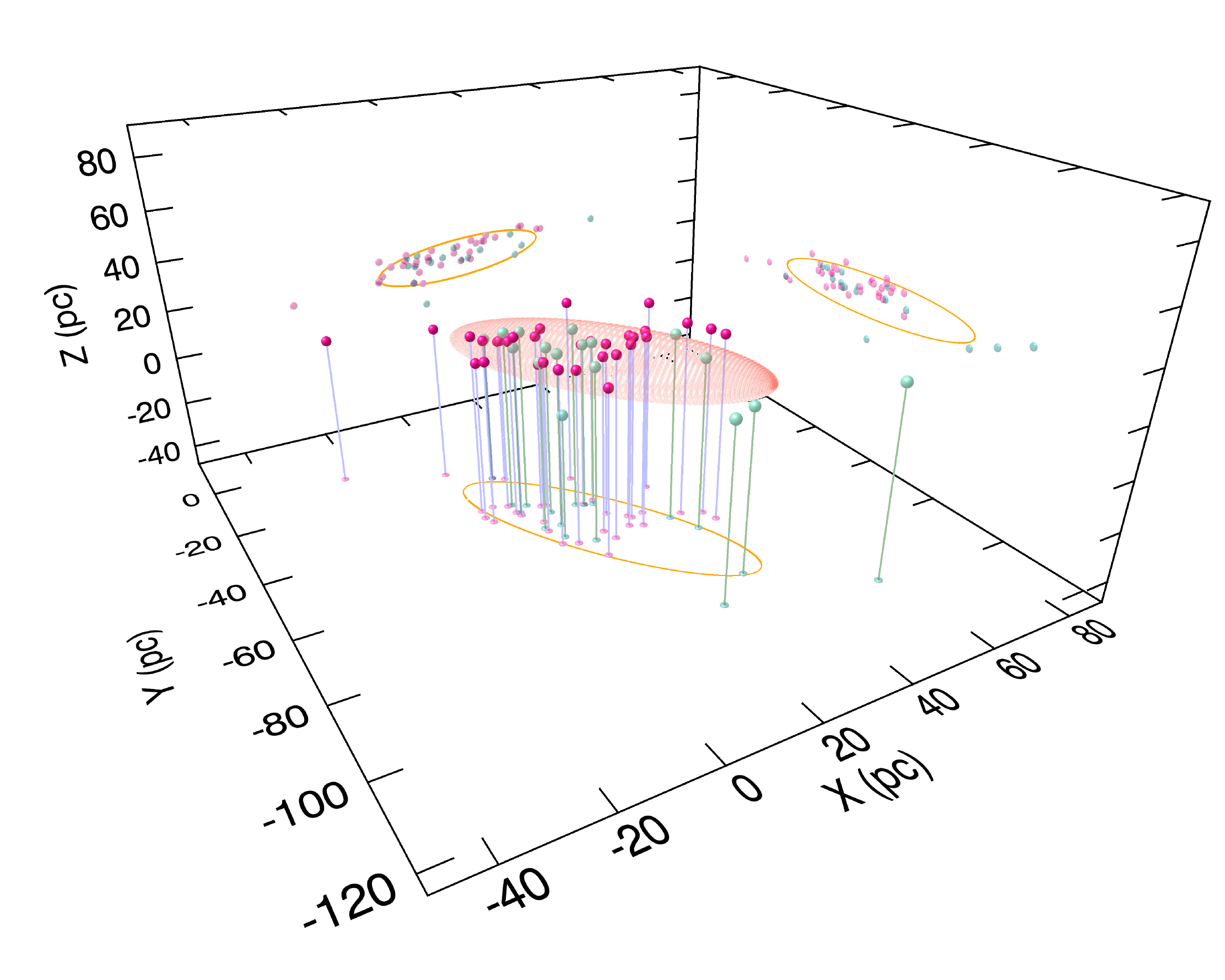}}
	\subfigure[\emph{UVW}]{\includegraphics[width=0.495\textwidth]{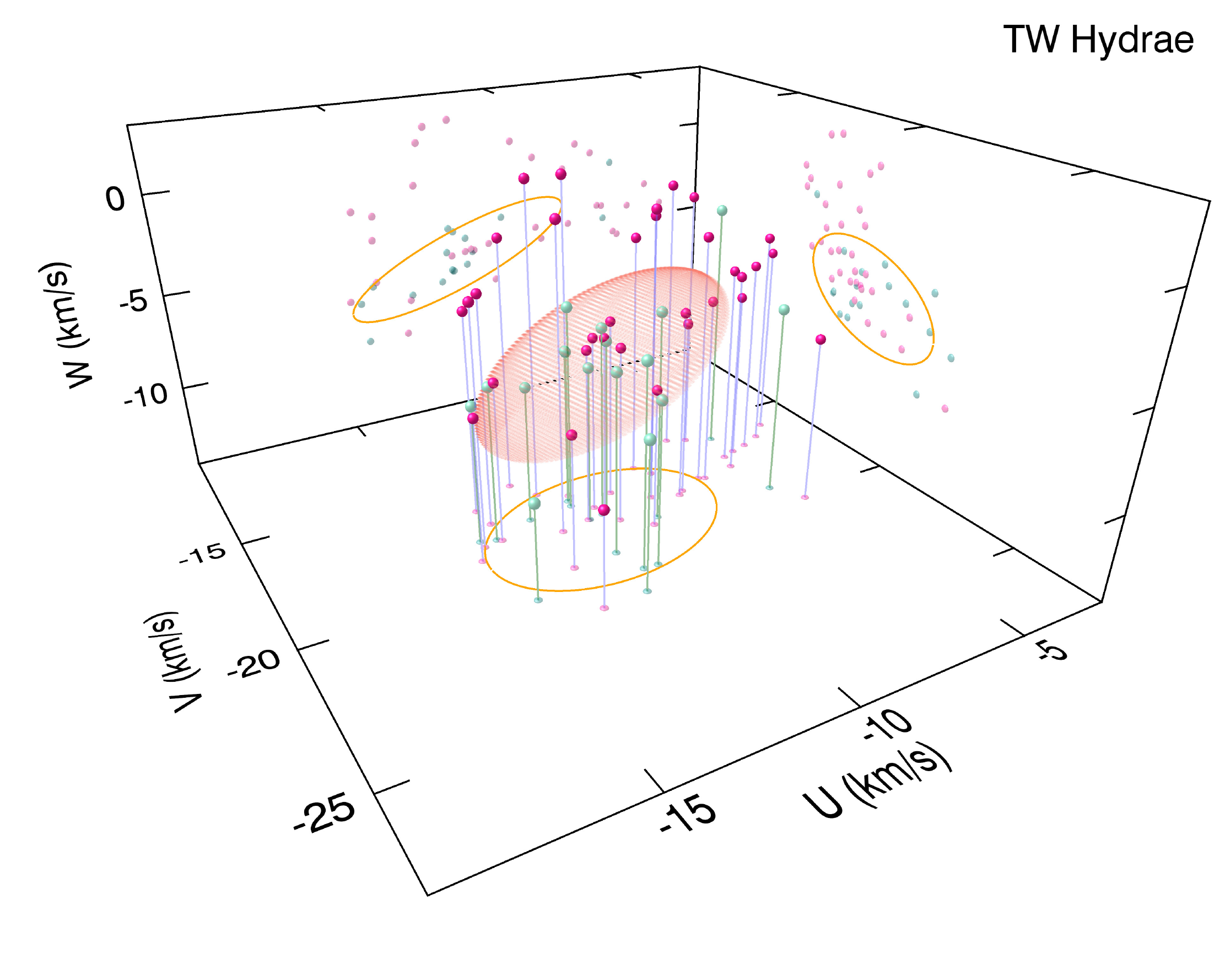}}
	\caption{Most probable galactic positions \emph{XYZ} and space velocities \emph{UVW} based on \href{http://www.astro.umontreal.ca/\textasciitilde gagne/banyanII.php}{BANYAN~II} statistical distances and RVs for all \href{http://www.astro.umontreal.ca/\textasciitilde gagne/BASS.php}{\emph{BASS}} candidate members in TWA compared with bona fide members. Colors and symbols are defined in the same way as in \hyperref[fig:xyzuvw_abdmg]{Figure~\ref*{fig:xyzuvw_abdmg}}. We note that a fraction of \href{http://www.astro.umontreal.ca/\textasciitilde gagne/BASS.php}{\emph{BASS}} candidates have kinematics slightly discrepant with those of TWA. It is possible that contamination from the Lower-Centaurus-Crux causes this (i.e. \citealp{2012ApJ...754...39S}), however a follow-up of these candidates will be needed to confirm this.}
	\label{fig:xyzuvw_twa}
\end{figure*}

\begin{figure*}[p]
	\center
	\subfigure[\emph{XYZ}]{\includegraphics[width=0.495\textwidth]{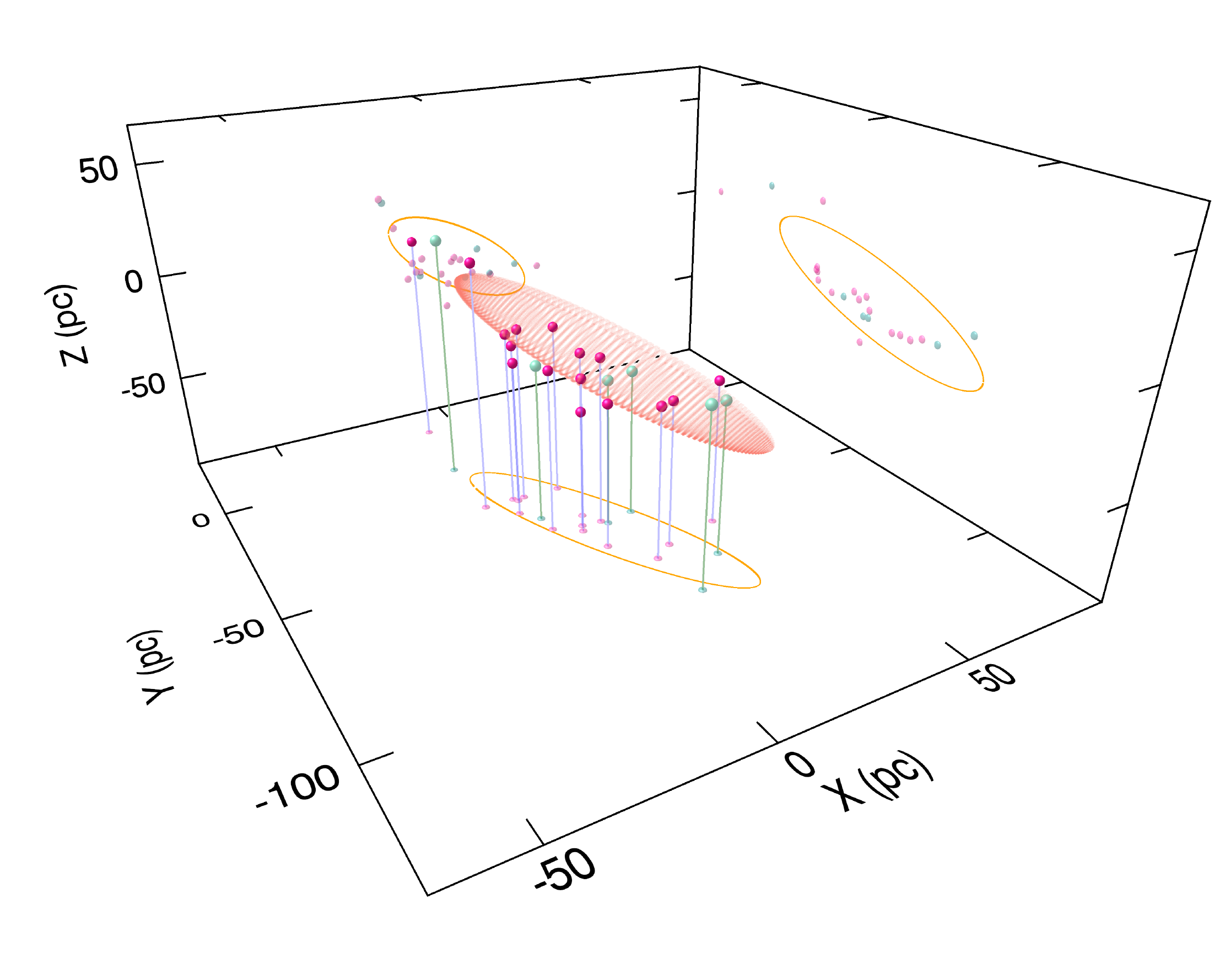}}
	\subfigure[\emph{UVW}]{\includegraphics[width=0.495\textwidth]{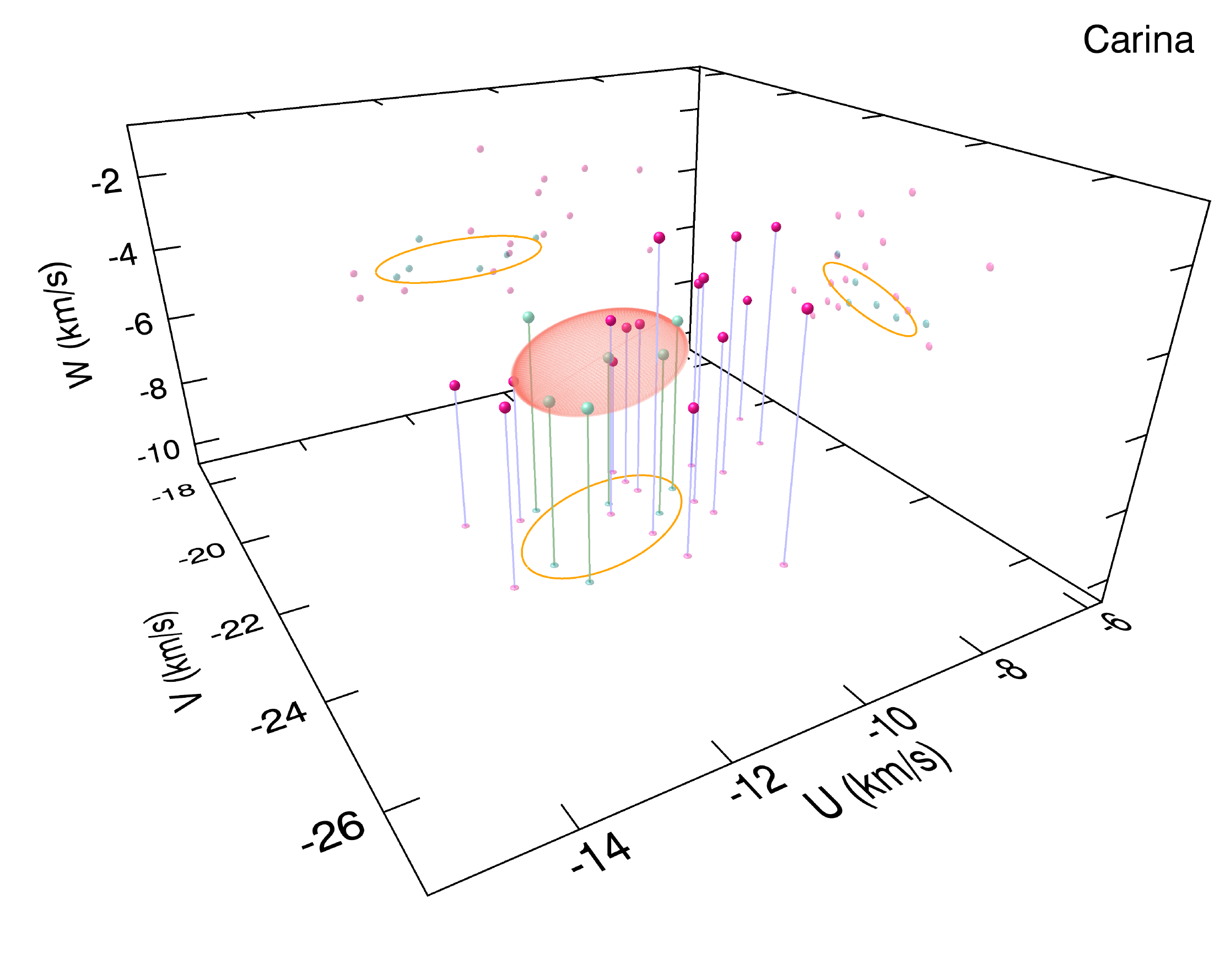}}
	\caption{Most probable galactic positions \emph{XYZ} and space velocities \emph{UVW} based on \href{http://www.astro.umontreal.ca/\textasciitilde gagne/banyanII.php}{BANYAN~II} statistical distances and RVs for all \href{http://www.astro.umontreal.ca/\textasciitilde gagne/BASS.php}{\emph{BASS}} candidate members in CAR compared with bona fide members. Colors and symbols are defined in the same way as in \hyperref[fig:xyzuvw_abdmg]{Figure~\ref*{fig:xyzuvw_abdmg}}. We note that the SKMs presented here (orange ellipsoids) are based on only 7 bona fide members, and they are thus most probably incomplete (see \citealp{2014ApJ...783..121G} for a discussion). It can be seen that \href{http://www.astro.umontreal.ca/\textasciitilde gagne/BASS.php}{\emph{BASS}} candidates preferentially fall in a region slightly outside of the kinematic model, which potentially points out to an overlooked region of CAR members in the kinematic space.}
	\label{fig:xyzuvw_car}
\end{figure*}

\begin{figure*}[p]
	\center
	\subfigure[\emph{XYZ}]{\includegraphics[width=0.495\textwidth]{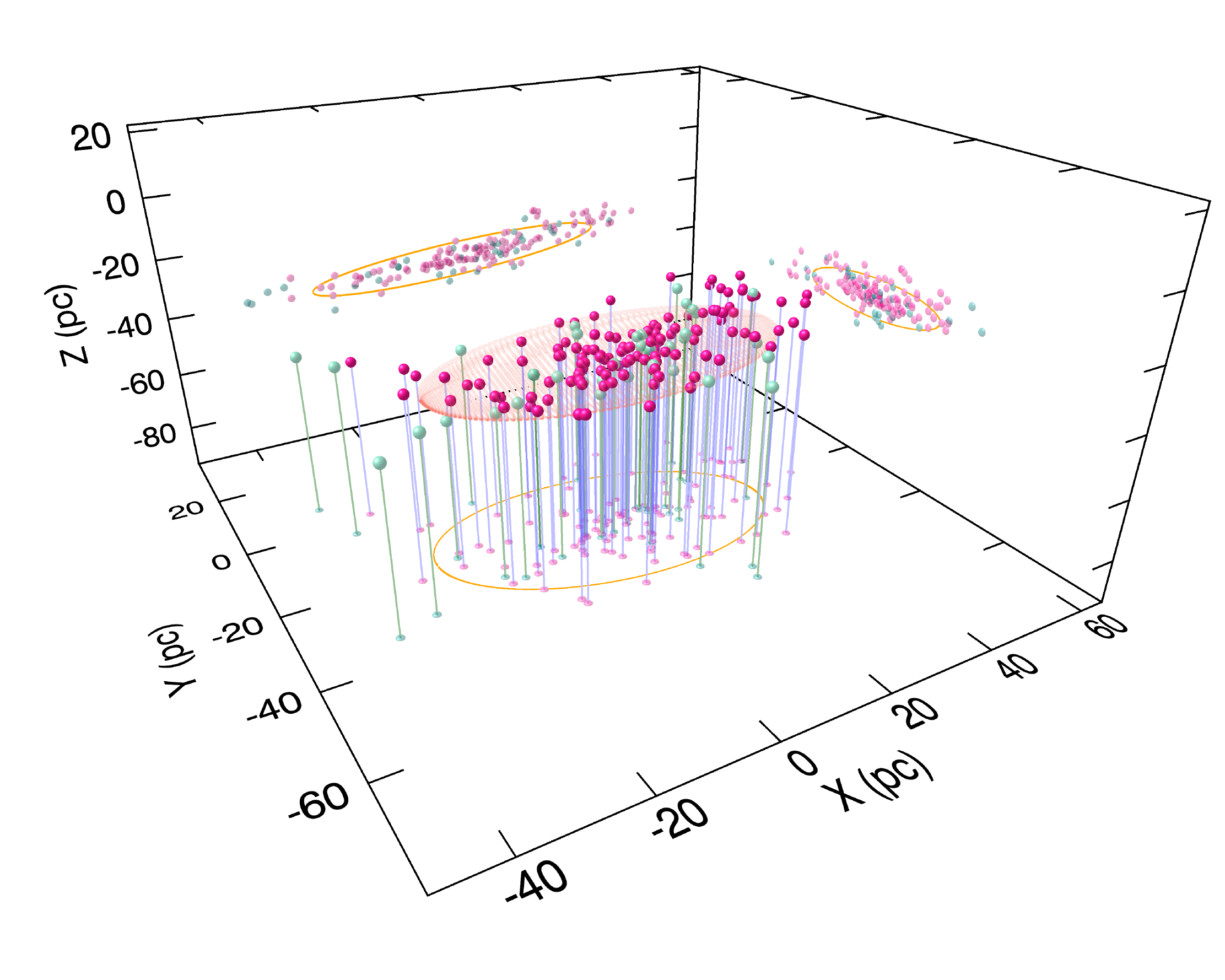}}
	\subfigure[\emph{UVW}]{\includegraphics[width=0.495\textwidth]{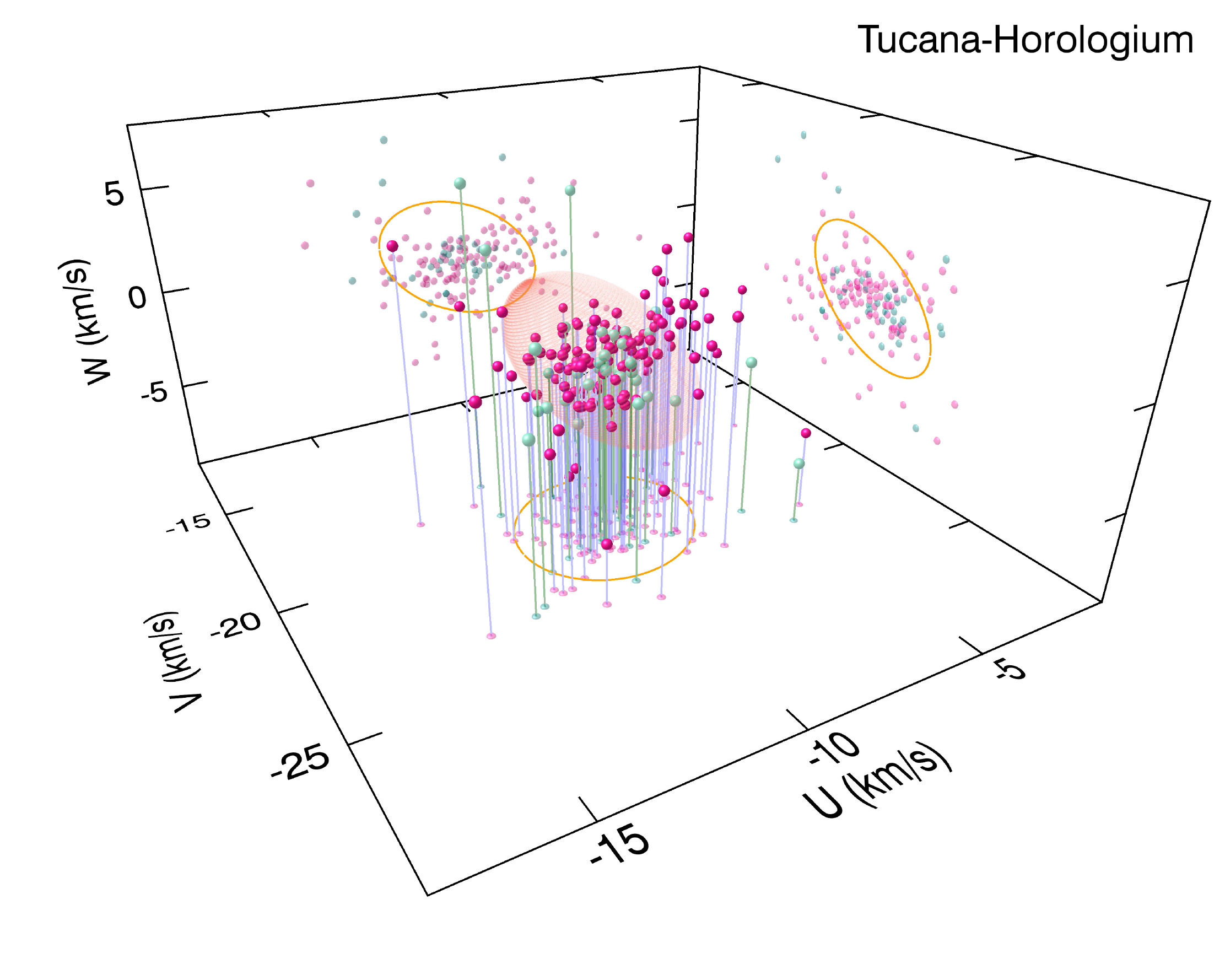}}
	\caption{Most probable galactic positions \emph{XYZ} and space velocities \emph{UVW} based on \href{http://www.astro.umontreal.ca/\textasciitilde gagne/banyanII.php}{BANYAN~II} statistical distances and RVs for all \href{http://www.astro.umontreal.ca/\textasciitilde gagne/BASS.php}{\emph{BASS}} candidate members in THA compared with bona fide members. Colors and symbols are defined in the same way as in \hyperref[fig:xyzuvw_abdmg]{Figure~\ref*{fig:xyzuvw_abdmg}}. As noted by \cite{2014AJ....147..146K}, the spatial distribution of THA is significantly thinner in the \emph{Z} direction and thus forms a plane in the \emph{XYZ} space.}
	\label{fig:xyzuvw_tha}
\end{figure*}

\begin{figure*}[p]
	\center
	\subfigure[\emph{XYZ}]{\includegraphics[width=0.495\textwidth]{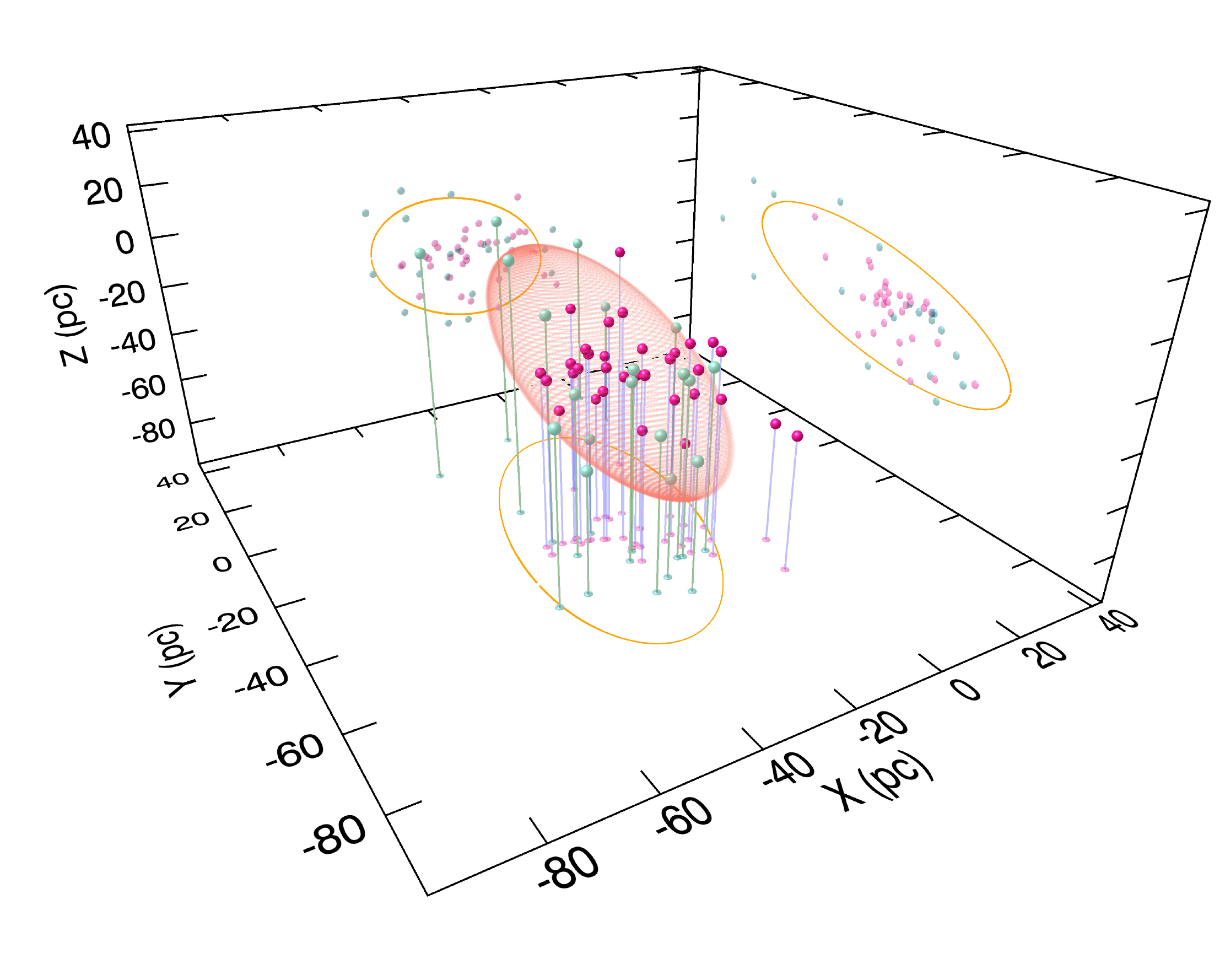}}
	\subfigure[\emph{UVW}]{\includegraphics[width=0.495\textwidth]{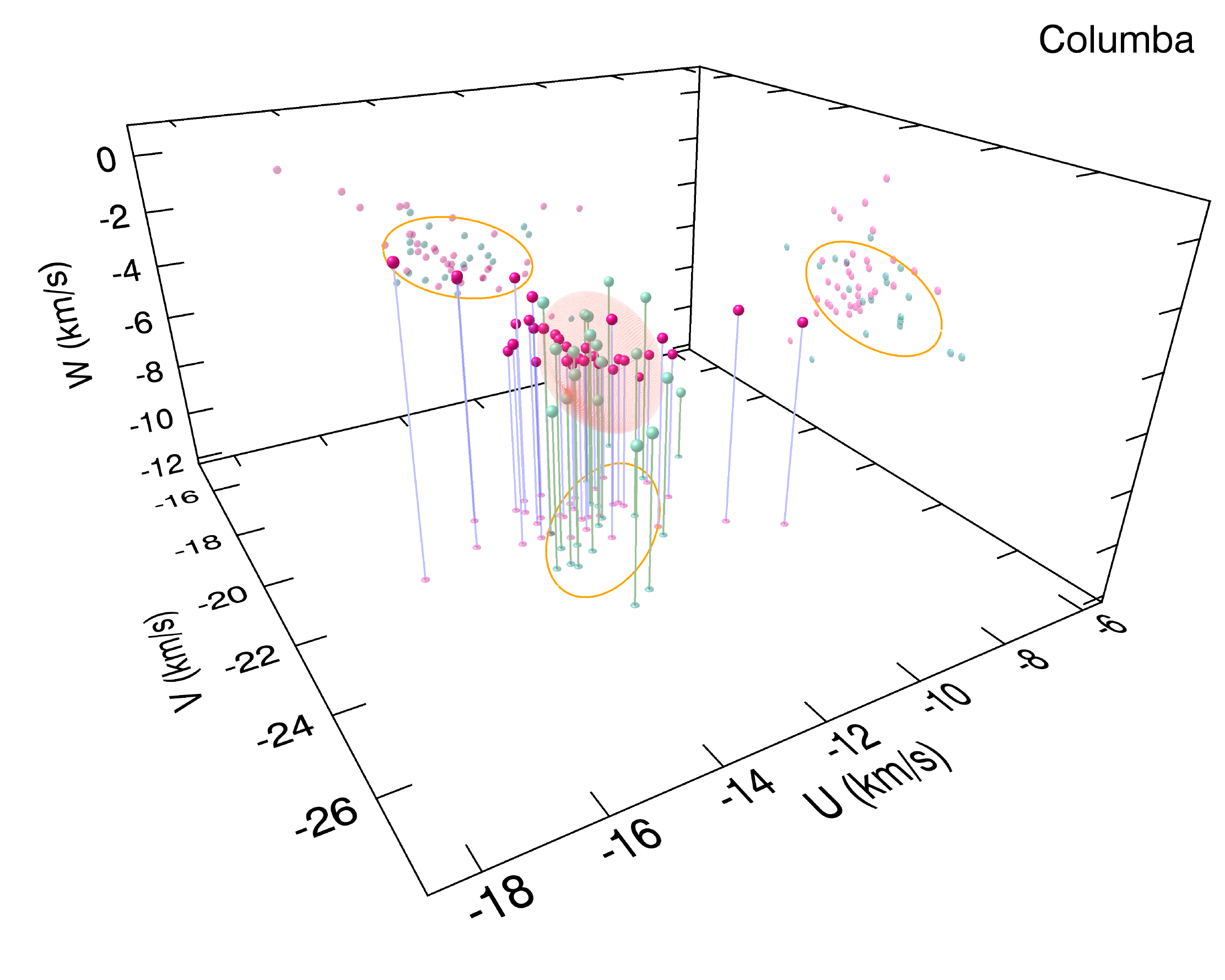}}
	\caption{Most probable galactic positions \emph{XYZ} and space velocities \emph{UVW} based on \href{http://www.astro.umontreal.ca/\textasciitilde gagne/banyanII.php}{BANYAN~II} statistical distances and RVs for all \href{http://www.astro.umontreal.ca/\textasciitilde gagne/BASS.php}{\emph{BASS}} candidate members in COL compared with bona fide members. Colors and symbols are defined in the same way as in \hyperref[fig:xyzuvw_abdmg]{Figure~\ref*{fig:xyzuvw_abdmg}}.}
	\label{fig:xyzuvw_col}
\end{figure*}

\begin{figure*}[p]
	\center
	\subfigure[\emph{XYZ}]{\includegraphics[width=0.495\textwidth]{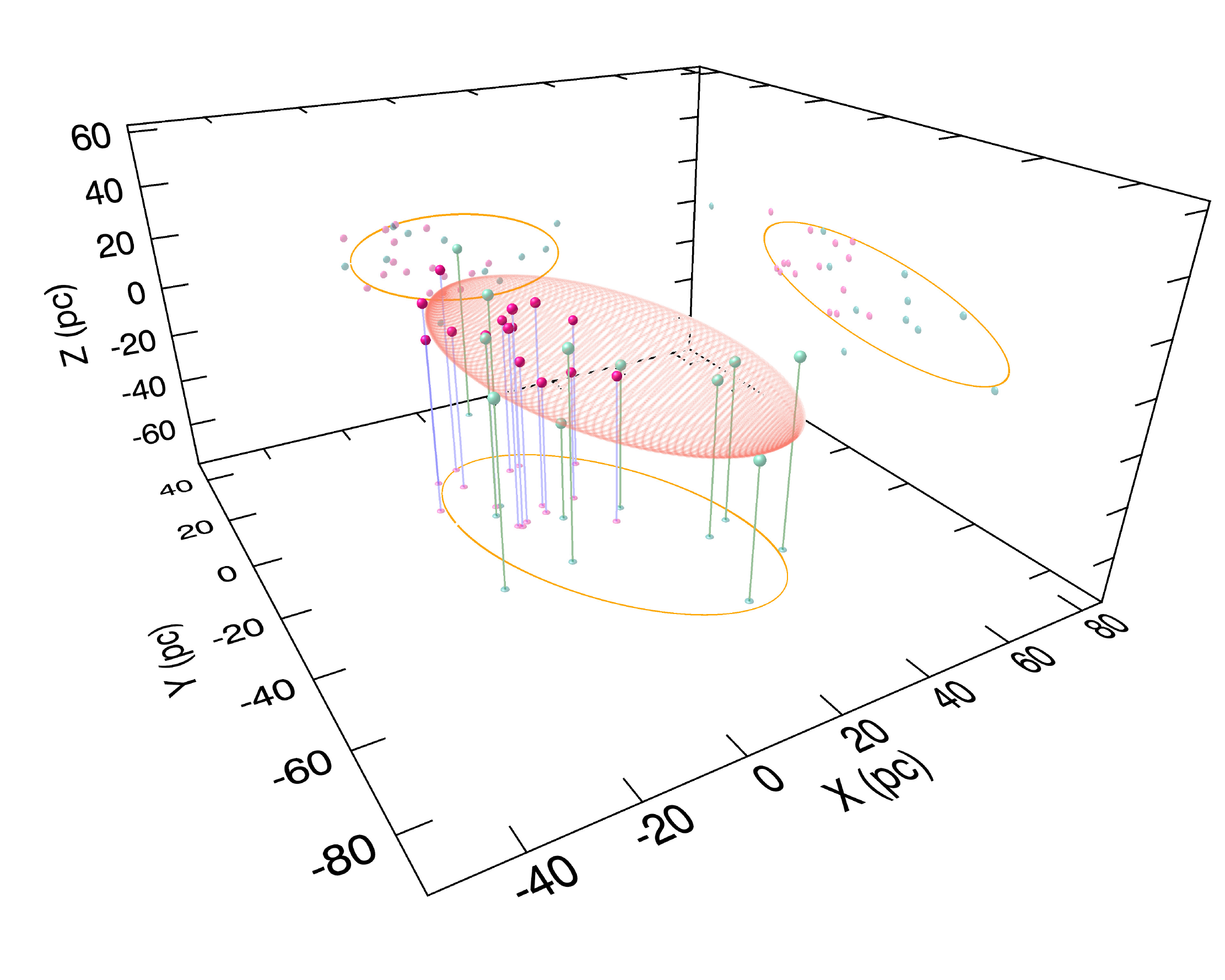}}
	\subfigure[\emph{UVW}]{\includegraphics[width=0.495\textwidth]{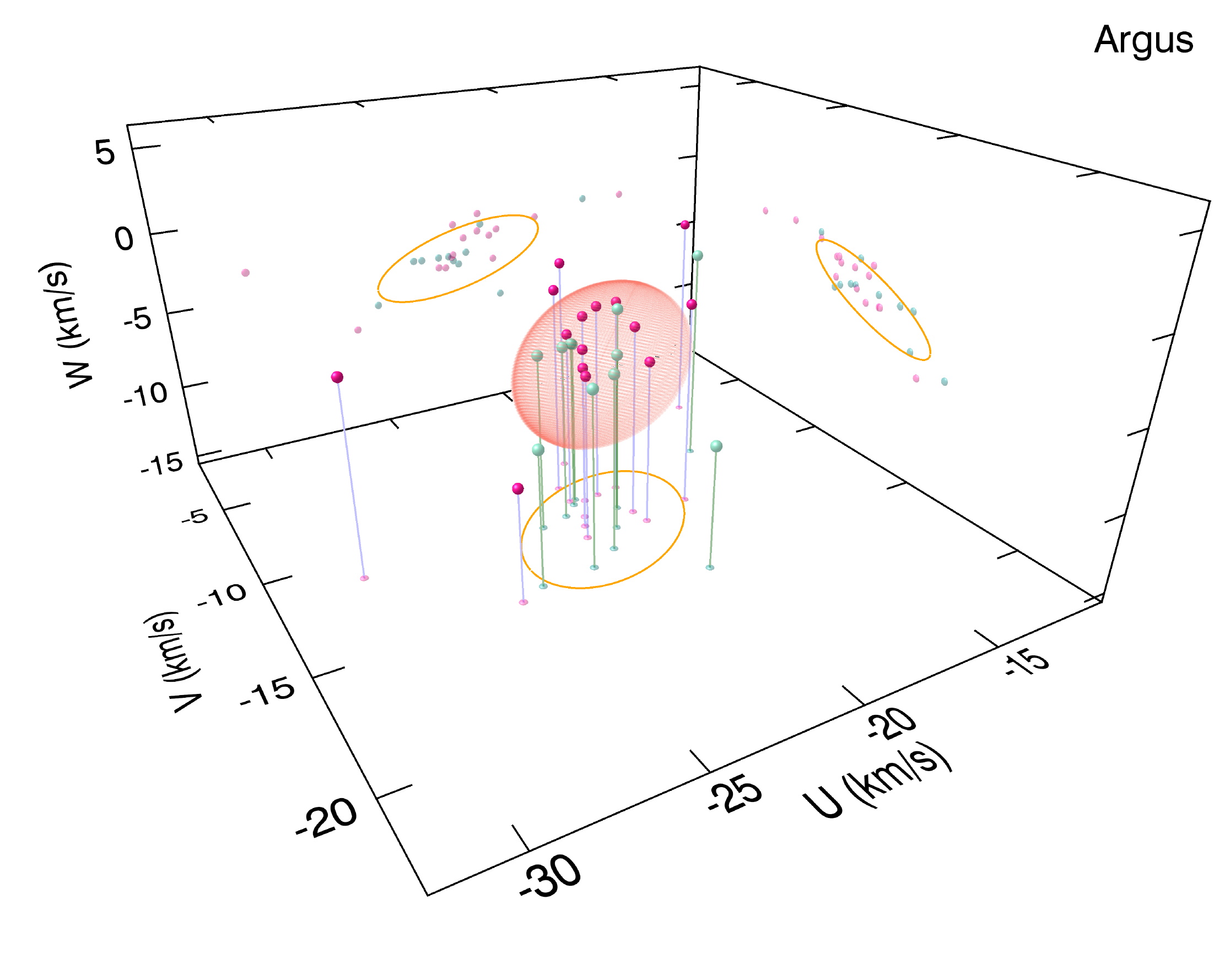}}
	\caption{Most probable galactic positions \emph{XYZ} and space velocities \emph{UVW} based on \href{http://www.astro.umontreal.ca/\textasciitilde gagne/banyanII.php}{BANYAN~II} statistical distances and RVs for all \href{http://www.astro.umontreal.ca/\textasciitilde gagne/BASS.php}{\emph{BASS}} candidate members in ARG compared with bona fide members. Colors and symbols are defined in the same way as in \hyperref[fig:xyzuvw_abdmg]{Figure~\ref*{fig:xyzuvw_abdmg}}.}
	\label{fig:xyzuvw_arg}
\end{figure*}

In Figures~\ref{fig:pm_maps_abdmg}--\ref{fig:pm_maps_arg}, we compare proper motions and sky positions of the \href{http://www.astro.umontreal.ca/\textasciitilde gagne/BASS.php}{\emph{BASS}} sample with currently known bona fide members of YMGs; it can be seen that, as expected, trajectories of candidates in the \href{http://www.astro.umontreal.ca/\textasciitilde gagne/BASS.php}{\emph{BASS}} sample projected on the celestial sphere are consistent with known bona fide members. In \hyperref[fig:CMD]{Figure~\ref*{fig:CMD}}, we use the statistical distances from \href{http://www.astro.umontreal.ca/\textasciitilde gagne/banyanII.php}{BANYAN~II} to display the position of candidates of the \href{http://www.astro.umontreal.ca/\textasciitilde gagne/BASS.php}{\emph{BASS}} sample in two color-magnitude diagrams: absolute $W1$ as a function of $H-W2$, and absolute $W1$ as a function of $J-K_S$. These two CMDs are used as observable in the \href{http://www.astro.umontreal.ca/\textasciitilde gagne/banyanII.php}{BANYAN~II} tool as they are useful to distinguish young $>$ M5 dwarfs from their field counterparts. In Figures~\ref{fig:xyzuvw_abdmg}--\ref{fig:xyzuvw_arg}, we compare the statistical predictions for galactic positions (\emph{XYZ}) and space velocities (\emph{UVW}) of all \href{http://www.astro.umontreal.ca/\textasciitilde gagne/BASS.php}{\emph{BASS}} candidates with those of currently known bona fide members of YMGs, as well as the $1.557\sigma$ contours of the SKM ellipsoids used in \href{http://www.astro.umontreal.ca/\textasciitilde gagne/banyanII.php}{BANYAN~II}. We use $1.557\sigma$ as the 3-dimensional analog to 1$\sigma$ in one dimension in the sense that it encompasses 68\% of objects drawn from a gaussian random PDF.

\newpage\subsection{Mass Estimates}\label{sec:mass}

We used the YMG age and statistical distance associated to the most probable hypothesis from \href{http://www.astro.umontreal.ca/\textasciitilde gagne/banyanII.php}{BANYAN~II} and the \href{http://phoenix.ens-lyon.fr/Grids/BT-Settl/CIFIST2011/ISOCHRONES/}{AMES-Cond} isochrones \citep{2003AA...402..701B} in combination with the CIFIST2011 BT-SETTL atmosphere models (\citealp{2013MSAIS..24..128A}; \citealp{2013AA...556A..15R}) to estimate the mass of all candidates presented here. A uniform distribution spanning the age range of each YMG was used to compare their absolute $J$, $H$, $K_S$, \emph{W1} and \emph{W2} magnitudes with model isochrones in a maximum likelihood analysis. Mass estimates are listed in \hyperref[tab:allbass]{Table~\ref*{tab:allbass}}. The \href{http://www.astro.umontreal.ca/\textasciitilde gagne/BASS.php}{\emph{BASS}} sample comprises 79 new candidate young BDs and 22 candidate planetary-mass objects.

\capstartfalse
\input{bass_table_5_short.tex}
\capstarttrue

\section{A SEARCH FOR NEW COMMON PROPER MOTION PAIRS}\label{sec:common_pm}

Since the \href{http://www.ipac.caltech.edu/2mass/}{\emph{2MASS}} and \href{http://wise2.ipac.caltech.edu/docs/release/allwise/}{\emph{AllWISE}} catalogs provide a fast way to determine proper motions for a large number of targets, we performed a search for common proper motion objects around all candidates in the \href{http://www.astro.umontreal.ca/\textasciitilde gagne/BASS.php}{\emph{BASS}} sample. We used the \href{http://www.astro.umontreal.ca/\textasciitilde gagne/banyanII.php}{BANYAN~II} statistical distance of each candidate to define a projected separation radius of 10,000~AU within which we have searched for any other object with a proper motion respecting the criteria of \cite{2007AJ....133..889L}, albeit with a more conservative filter on allowed proper motion difference. This requires that the separation $\Delta\theta$ (measured in arc seconds) and the proper motion difference $\Delta\mu$ (measured in \masyr) obey the following equations :
\begin{align*}
	\Delta\theta\ \Delta\mu &< 1,000\cdot (\mu/150)^{3.8},\\
	\Delta\mu &< 50.
\end{align*}

These criteria should ensure that the majority of genuine proper motion pairs are recovered, with a minimal amount of contamination from chance alignments. This search allowed us to find 5 new common proper motion pairs and recover 10 which were already known in the literature. Those already known are :
\begin{itemize}
	\item \href{http://simbad.u-strasbg.fr/simbad/sim-id?submit=display&bibdisplay=refsum&bibyear1=1850&bibyear2=$currentYear&Ident=2MASS%20J00451358%2B0015509}{2MASS~J00451358+0015509}$^\ast$ (M3.8) and \\\href{http://simbad.u-strasbg.fr/simbad/sim-id?submit=display&bibdisplay=refsum&bibyear1=1850&bibyear2=$currentYear&Ident=HD%204271}{2MASS~J00451098+0015117} (\href{http://simbad.u-strasbg.fr/simbad/sim-id?submit=display&bibdisplay=refsum&bibyear1=1850&bibyear2=$currentYear&Ident=HD%204271}{HD~4271}; F8; \citealp{2014AJ....147...20N})
	\item \href{http://simbad.u-strasbg.fr/simbad/sim-id?submit=display&bibdisplay=refsum&bibyear1=1850&bibyear2=$currentYear&Ident=2MASS%20J01243060-3355014}{2MASS~J01243060--3355014}$^\ast$ (\href{http://simbad.u-strasbg.fr/simbad/sim-id?submit=display&bibdisplay=refsum&bibyear1=1850&bibyear2=$currentYear&Ident=2MASS%20J01243060-3355014}{GJ~2022~B}; M4.5) and \\\href{http://simbad.u-strasbg.fr/simbad/sim-id?submit=display&bibdisplay=refsum&bibyear1=1850&bibyear2=$currentYear&Ident=2MASS%20J01242767-3355086}{2MASS~J01242767--3355086} (\href{http://simbad.u-strasbg.fr/simbad/sim-id?submit=display&bibdisplay=refsum&bibyear1=1850&bibyear2=$currentYear&Ident=2MASS%20J01242767-3355086}{GJ~2022~AC}; M5+M5; \citealp{1974AA....36..155T})
	\item \href{http://simbad.u-strasbg.fr/simbad/sim-id?submit=display&bibdisplay=refsum&bibyear1=1850&bibyear2=$currentYear&Ident=1RXS%20J020326.8%2B064835}{2MASS~J02033222+0648588}$^\ast$ (estimated M4.5) and \\\href{http://simbad.u-strasbg.fr/simbad/sim-id?submit=display&bibdisplay=refsum&bibyear1=1850&bibyear2=$currentYear&Ident=1RXS%20J020326.8%2B064835}{2MASS~J02032589+0648008} (estimated early-M; \citealp{2012yCat.1322....0Z})
	\item \href{http://simbad.u-strasbg.fr/simbad/sim-id?Ident=%408374929&Name=WISE%20J024202.14-535914.7}{2MASS~J02420204--5359147}$^\ast$ (M4.6) and \\\href{http://simbad.u-strasbg.fr/simbad/sim-id?Ident=%408374930&Name=WISE%20J024204.15-535900.0}{2MASS~J02420404--5359000} (estimated early-M; \citep{2012yCat.1322....0Z})
	\item \href{http://simbad.u-strasbg.fr/simbad/sim-id?submit=display&bibdisplay=refsum&bibyear1=1850&bibyear2=$currentYear&Ident=2MASS%20J03114240-1537183}{2MASS~J03114240--1537183}$^\ast$ (\href{http://simbad.u-strasbg.fr/simbad/sim-id?submit=display&bibdisplay=refsum&bibyear1=1850&bibyear2=$currentYear&Ident=2MASS%20J03114240-1537183}{LP~722--14}; estimated M5.0) and \\\href{http://simbad.u-strasbg.fr/simbad/sim-id?Ident=%40635240&Name=LP%20%20772-15&submit=submit}{2MASS~J03114269--1537327} (\href{http://simbad.u-strasbg.fr/simbad/sim-id?Ident=%40635240&Name=LP%20%20772-15&submit=submit}{LP~722--15}; estimated M2.2; \citealp{1977PMMin..51....1L})
	\item \href{http://simbad.u-strasbg.fr/simbad/sim-id?submit=display&bibdisplay=refsum&bibyear1=1850&bibyear2=$currentYear&Ident=2MASS%20J03283911-1537333}{2MASS~J03283911--1537333}$^\ast$ (\href{http://simbad.u-strasbg.fr/simbad/sim-id?submit=display&bibdisplay=refsum&bibyear1=1850&bibyear2=$currentYear&Ident=2MASS%20J03283911-1537333}{GJ~3229~B}; M3.5) and \\\href{http://simbad.u-strasbg.fr/simbad/sim-id?submit=display&bibdisplay=refsum&bibyear1=1850&bibyear2=$currentYear&Ident=2MASS%20J03283893-1537171}{2MASS~J03283893--1537171} (\href{http://simbad.u-strasbg.fr/simbad/sim-id?submit=display&bibdisplay=refsum&bibyear1=1850&bibyear2=$currentYear&Ident=2MASS%20J03283893-1537171}{GJ~3228~A}; M3.5; \citealp{1991adc..rept.....G})
	\item \href{http://simbad.u-strasbg.fr/simbad/sim-id?submit=display&bibdisplay=refsum&bibyear1=1850&bibyear2=$currentYear&Ident=1RXS%20J035101.5%2B141404}{2MASS~J03505949+1414017}$^\ast$ (M5) and \\\href{http://simbad.u-strasbg.fr/simbad/sim-id?submit=display&bibdisplay=refsum&bibyear1=1850&bibyear2=$currentYear&Ident=1RXS%20J035101.5%2B141404}{2MASS~J03510078+1413398} (M4; \citealp{2001AJ....122.3466M})
	\item \href{http://simbad.u-strasbg.fr/simbad/sim-id?submit=display&bibdisplay=refsum&bibyear1=1850&bibyear2=$currentYear&Ident=2MASS%20J21440795%2B1704372}{2MASS~J21440795+1704372}$^\ast$ (\href{http://simbad.u-strasbg.fr/simbad/sim-id?submit=display&bibdisplay=refsum&bibyear1=1850&bibyear2=$currentYear&Ident=2MASS%20J21440795%2B1704372}{G~126--30}; M4.5) and \\\href{http://simbad.u-strasbg.fr/simbad/sim-id?submit=display&bibdisplay=refsum&bibyear1=1850&bibyear2=$currentYear&Ident=2MASS%20J21440900%2B1703348}{2MASS~J21440900+1703348} (\href{http://simbad.u-strasbg.fr/simbad/sim-id?submit=display&bibdisplay=refsum&bibyear1=1850&bibyear2=$currentYear&Ident=2MASS%20J21440900%2B1703348}{G~126--31}; M4; \citealp{2001AJ....122.3466M})
	\item \href{http://simbad.u-strasbg.fr/simbad/sim-id?submit=display&bibdisplay=refsum&bibyear1=1850&bibyear2=$currentYear&Ident=2MASS%20J23225240-6151114}{2MASS~J23225240--6151114}$^\ast$ (M5) and \\\href{http://simbad.u-strasbg.fr/simbad/sim-id?submit=display&bibdisplay=refsum&bibyear1=1850&bibyear2=$currentYear&Ident=2MASS%20J23225299-6151275}{2MASS~J23225299--6151275}$^\ast$ (L2$\gamma$; \citealp{2014ApJ...783..121G})
	\item \href{http://vizier.u-strasbg.fr/viz-bin/VizieR?-source=&-out.add=_r&-out.add=_RAJ,_DEJ&-sort=_r&-to=&-out.max=20&-meta.ucd=2&-meta.foot=1&-c=23%2010%2021.96%20-07%2048%2053.1&-c.rs=10}{2MASS~J23102196--0748531}$^\ast$ (M5) and \\\href{http://simbad.u-strasbg.fr/simbad/sim-id?submit=display&bibdisplay=refsum&bibyear1=1850&bibyear2=$currentYear&Ident=LTT%209391}{2MASS~J23102471--0748432} (\href{http://simbad.u-strasbg.fr/simbad/sim-id?submit=display&bibdisplay=refsum&bibyear1=1850&bibyear2=$currentYear&Ident=LTT%209391}{HIP~114424}; K0; \citealp{2014AJ....147..160M})	
\end{itemize}

We identified components present in the \href{http://www.astro.umontreal.ca/\textasciitilde gagne/BASS.php}{\emph{BASS}} or \href{http://www.astro.umontreal.ca/\textasciitilde gagne/LP-BASS.php}{\emph{LP-BASS}} catalogs with an asterisk. Any potentially useful information from these matches were already taken into account in \hyperref[sec:lit]{Section~\ref*{sec:lit}}. We discuss the new potential common proper motion pairs below :

\href{http://vizier.u-strasbg.fr/viz-bin/VizieR?-source=&-out.add=_r&-out.add=_RAJ,_DEJ&-sort=_r&-to=&-out.max=20&-meta.ucd=2&-meta.foot=1&-c=04%2035%2030.42%20-64%2049%2057.0&-c.rs=10}{2MASS~J04353042--6449570} from \href{http://www.astro.umontreal.ca/\textasciitilde gagne/BASS.php}{\emph{BASS}} (estimated M8.4 with $J=15.27$) seems to be co-moving with \href{http://vizier.u-strasbg.fr/viz-bin/VizieR?-source=&-out.add=_r&-out.add=_RAJ,_DEJ&-sort=_r&-to=&-out.max=20&-meta.ucd=2&-meta.foot=1&-c=04%2035%2027.09%20-64%2050%2004.2&-c.rs=10}{2MASS~J04352709--6450042} ($J=15.16$) at an angular separation of 22\farcs4 and a proper motion difference of 0.4 \masyr\ (0.05$\sigma$) with respect to a total proper motion of 53.2 \masyr. However, we note that \href{http://vizier.u-strasbg.fr/viz-bin/VizieR?-source=&-out.add=_r&-out.add=_RAJ,_DEJ&-sort=_r&-to=&-out.max=20&-meta.ucd=2&-meta.foot=1&-c=04%2035%2027.09%20-64%2050%2004.2&-c.rs=10}{2MASS~J04352709--6450042} is only $0.11$ magnitudes brighter in the $J$ band, and yet its NIR colors are significantly bluer : it has $J-K_S = 0.42$ and $H-W2 = 0.02$, versus $J-K_S = 1.34$ and $H-W2 = 1.30$ for the \href{http://www.astro.umontreal.ca/\textasciitilde gagne/BASS.php}{\emph{BASS}} candidate. These very blue colors would be indicative of a spectral type earlier than M, which is not consistent with it being at the same distance from the primary, even if the latter was a multiple system. For this reason, \href{http://www.astro.umontreal.ca/\textasciitilde gagne/banyanII.php}{BANYAN~II} rejects it as a probable candidate member of CAR, but if we do not include photometry, then its Bayesian probability for CAR is 31.4\%, with a contamination probability of 21.9\%. We conclude nonetheless that the secondary is most probably not a member of CAR and that this system is possibly a chance alignment, since otherwise it would be hard to reconcile the very different colors and the similar apparent $J$ magnitudes of its components. We note that \cite{2007AJ....133..889L} used their common proper motion criteria only on stars with $\mu > 150$ \masyr\, hence it is possible that it does not perform as well on this system which has only $\mu = 53.2$ \masyr.

\href{http://simbad.u-strasbg.fr/simbad/sim-id?submit=display&bibdisplay=refsum&bibyear1=1850&bibyear2=$currentYear&Ident=2MASS%20J05121347%2B0131539}{2MASS~J05121347+0131539} (\href{http://simbad.u-strasbg.fr/simbad/sim-id?submit=display&bibdisplay=refsum&bibyear1=1850&bibyear2=$currentYear&Ident=2MASS%20J05121347%2B0131539}{NLTT~14667}) from \href{http://www.astro.umontreal.ca/\textasciitilde gagne/LP-BASS.php}{\emph{LP-BASS}} (estimated M4.9 with $J=10.36$) seems to be co-moving with \href{http://vizier.u-strasbg.fr/viz-bin/VizieR?-source=&-out.add=_r&-out.add=_RAJ,_DEJ&-sort=_r&-to=&-out.max=20&-meta.ucd=2&-meta.foot=1&-c=05%2012%2011.70%20%2B01%2031%2015.4&-c.rs=10}{2MASS~J05121170+0131154} ($J=16.39$) at an angular separation of 46\farcs8 and a proper motion difference of 28.6 \masyr\ (0.9$\sigma$) with respect to a total proper motion of 212.4 \masyr. The contrast is significant with $\Delta J = 6.03$, which would point to a late-T spectral type for the secondary if it is at the same distance than the primary. However, we note that the secondary is most probably a contaminating object, since an extended PSF is visible within 10\textquotedbl\ of its \href{http://www.ipac.caltech.edu/2mass/}{\emph{2MASS}} position in the red \href{http://archive.stsci.edu/cgi-bin/dss_form}{\emph{DSS}} filter.

\href{http://simbad.u-strasbg.fr/simbad/sim-id?Ident=%402296995&Name=2MASS%20J14415883-1649008}{2MASS~J14415883--1649008} (\href{http://simbad.u-strasbg.fr/simbad/sim-id?Ident=%402296995&Name=2MASS%20J14415883-1649008}{WT~2090}) from \href{http://www.astro.umontreal.ca/\textasciitilde gagne/LP-BASS.php}{\emph{LP-BASS}} (M4.5 with $J=10.23$) is co-moving with \href{http://simbad.u-strasbg.fr/simbad/sim-id?submit=display&bibdisplay=refsum&bibyear1=1850&bibyear2=$currentYear&Ident=NLTT%2038107}{2MASS~J14415908--1653133} (\href{http://simbad.u-strasbg.fr/simbad/sim-id?submit=display&bibdisplay=refsum&bibyear1=1850&bibyear2=$currentYear&Ident=NLTT%2038107}{Wolf~1501}; M3 with $J=9.35$) at an angular separation of 252\farcs5 and a proper motion difference of 3.8 \masyr\ (0.3$\sigma$) with respect to a total proper motion of 290.3 \masyr. \cite{2010ApJS..190..100K} obtained a NIR spectral type of M3 for \href{http://simbad.u-strasbg.fr/simbad/sim-id?submit=display&bibdisplay=refsum&bibyear1=1850&bibyear2=$currentYear&Ident=NLTT%2038107}{Wolf~1501}. We note that the contrast ratio $\Delta J = 0.88$ is large for their respective spectral types of M3 and M4.5. Both objects are weak candidate members of ABDMG, with respective Bayesian probabilities of 5.4\% and 3.8\% and contamination probabilities of 23.4\% and 26.9\%.

\href{http://vizier.u-strasbg.fr/viz-bin/VizieR?-source=&-out.add=_r&-out.add=_RAJ,_DEJ&-sort=_r&-to=&-out.max=20&-meta.ucd=2&-meta.foot=1&-c=21%2050%2009.33%20%2B05%2058%2010.2&-c.rs=10}{2MASS~J21500933+0558102} from \href{http://www.astro.umontreal.ca/\textasciitilde gagne/LP-BASS.php}{\emph{LP-BASS}} (estimated M4.9 with $J=10.66$) is co-moving with \href{http://vizier.u-strasbg.fr/viz-bin/VizieR?-source=&-out.add=_r&-out.add=_RAJ,_DEJ&-sort=_r&-to=&-out.max=20&-meta.ucd=2&-meta.foot=1&-c=21%2050%2010.11%20%2B05%2058%2013.7&-c.rs=10}{2MASS~J21501011+0558137} from \href{http://www.astro.umontreal.ca/\textasciitilde gagne/LP-BASS.php}{\emph{LP-BASS}} (estimated M4.9 with $J=10.74$) at an angular separation of 12\textquotedbl\ and a proper motion difference of 21.9 \masyr\ (0.8$\sigma$) with respect to a total proper motion of 146.9 \masyr. Their contrast ratio is relatively small with $\Delta J = 0.08$, which is consistent with their similar estimated spectral types. The direction of their \href{http://www.ipac.caltech.edu/2mass/}{\emph{2MASS}}--\href{http://wise2.ipac.caltech.edu/docs/release/allwise/}{\emph{AllWISE}} proper motions is slightly different, which favors ARG for the primary and $\beta$PMG for the secondary. However, both have a somewhat ambiguous membership between ARG and $\beta$PMG; the primary has respective Bayesian probabilities of 8.0\% and 15.6\%, whereas the secondary has 16.6\% and 3.0\%. We thus regard this system as an ambiguous, low-probability candidate member of $\beta$PMG and ARG.

\href{http://vizier.u-strasbg.fr/viz-bin/VizieR?-source=&-out.add=_r&-out.add=_RAJ,_DEJ&-sort=_r&-to=&-out.max=20&-meta.ucd=2&-meta.foot=1&-c=23%2013%2030.55%20-53%2052%2007.9&-c.rs=10}{2MASS~J23133055--5352079} from \href{http://www.astro.umontreal.ca/\textasciitilde gagne/LP-BASS.php}{\emph{LP-BASS}} (estimated M5.7 with $J=12.08$) is co-moving with \href{http://simbad.u-strasbg.fr/simbad/sim-id?Ident=%403322635&Name=HD%20219046}{2MASS~J23133024--5351389} (\href{http://simbad.u-strasbg.fr/simbad/sim-id?Ident=%403322635&Name=HD%20219046}{HD~219046}; $J=8.59$) at an angular separation of 29\farcs1 and a proper motion difference of 17.0 \masyr (1.5$\sigma$). The contrast ratio is consistent with the latter component being a K-type star. We find no additional information in the literature for this system.

\section{A PRELIMINARY INVESTIGATION ON MASS SEGREGATION}\label{sec:mseg}

According to the virial theorem, it is expected that all components of a gravitationally bound astrophysical system will end up with the same average kinetic energy after relaxing to the equilibrium state. Hence, lower-mass members of associations of stars are expected to have a larger velocity than their higher-mass siblings; this effect is called mass segregation. It has already been demonstrated for globular clusters (\citealp{2011MNRAS.413.2345H}; \citealp{2011AA...532A.119O}; \citealp{2013ApJ...764...73P}), however no signs of mass segregation have yet been identified for YMGs. The \href{http://www.astro.umontreal.ca/\textasciitilde gagne/BASS.php}{\emph{BASS}} catalog provides a unique sample on which to test for this effect, since it potentially contains the latest-type, lowest-mass members known to all YMGs.

Instead of relying on mass estimates which are dependent on physical hypotheses inherent to evolutionary models, we use statistical distance predictions from \href{http://www.astro.umontreal.ca/\textasciitilde gagne/banyanII.php}{BANYAN~II} to obtain absolute $W1$ magnitudes for all high probability candidates in the \href{http://www.astro.umontreal.ca/\textasciitilde gagne/BASS.php}{\emph{BASS}} sample. Since members of YMGs are expected to be coeval, their absolute $W1$ magnitude should depend on their mass in a monotonic way, thus providing a more direct way to bring out mass segregation. The \href{http://wise2.ipac.caltech.edu/docs/release/allwise/}{\emph{AllWISE}} $W1$-band is preferred to \href{http://www.ipac.caltech.edu/2mass/}{\emph{2MASS}} bands since it is less affected by clouds in the atmospheres of BDs, which could introduce errors in the absolute magnitude--mass relation. Since the $UVW$ separation to the center of mass of a given YMG is directly related to the kinetic energy of a member with respect to the YMG, it is expected that mass segregation would cause fainter (less massive) objects to be more scattered in the $UVW$ space (i.e. dynamical mass segregation). As a consequence of this, one would also expect that they be more scattered spatially at a given moment in the $XYZ$ space (i.e. spatial mass segregation). 

\begin{figure*}[p]
	\centering
	\subfigure{\includegraphics[width=0.495\textwidth]{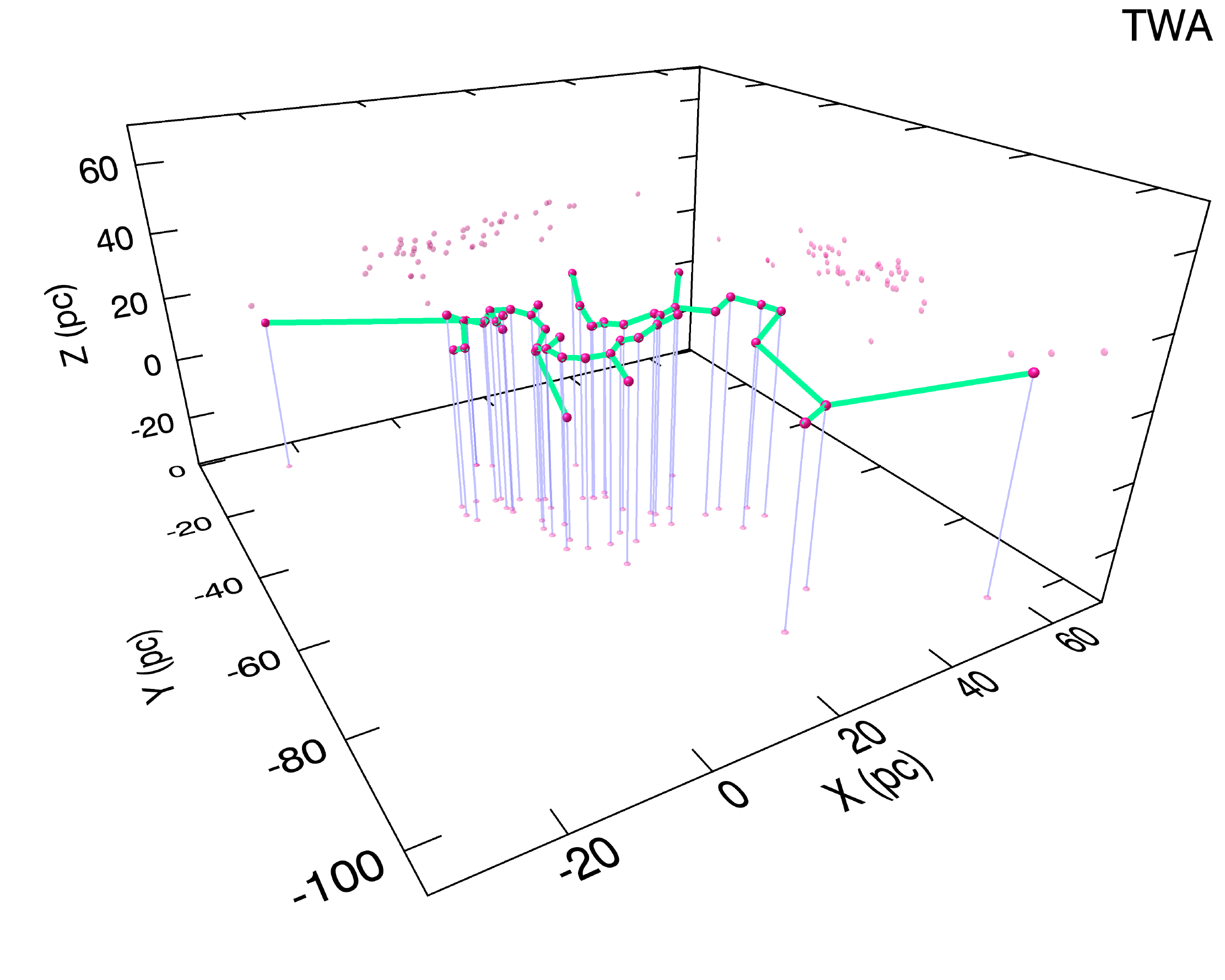}}
	\subfigure{\includegraphics[width=0.495\textwidth]{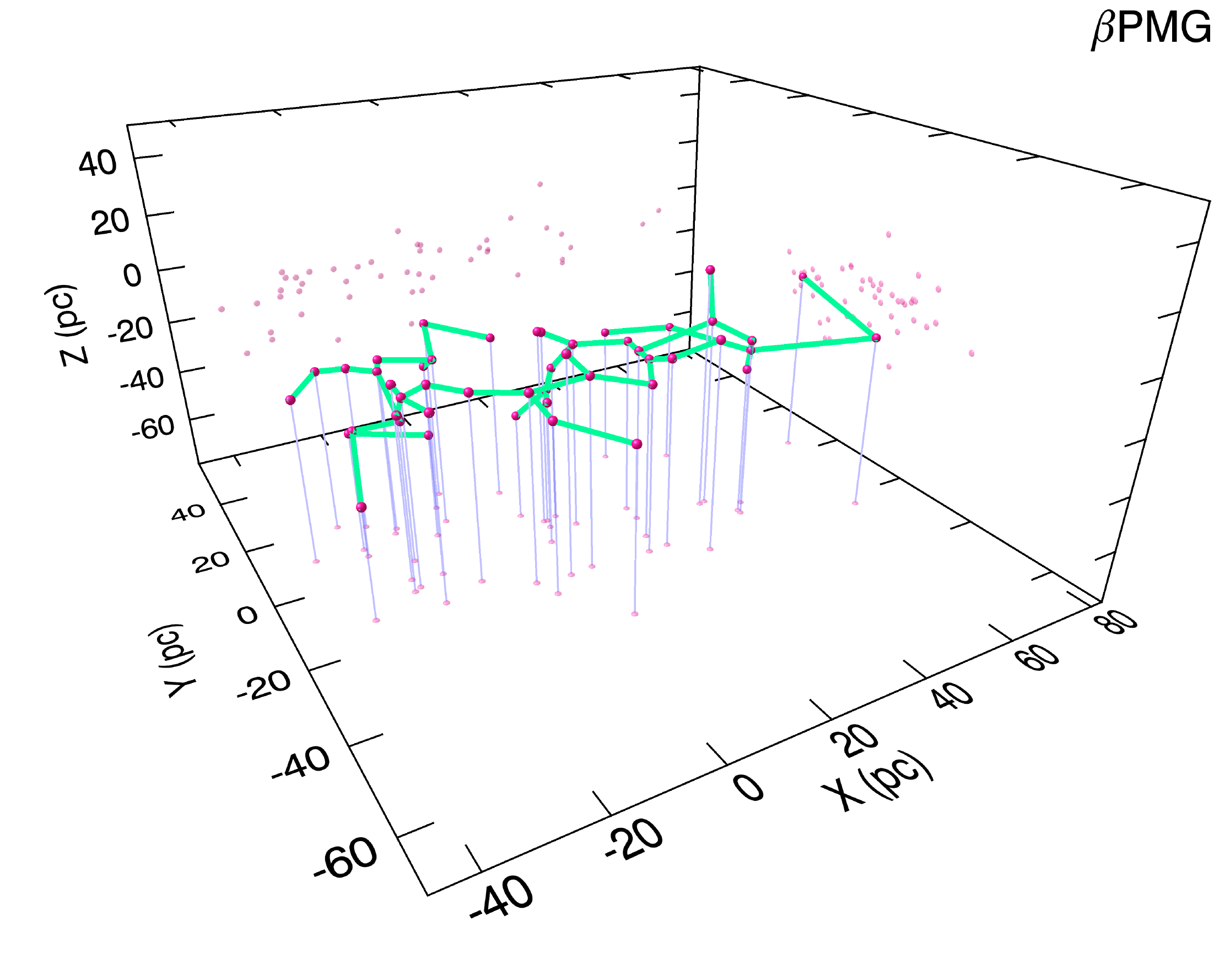}}
	\subfigure{\includegraphics[width=0.495\textwidth]{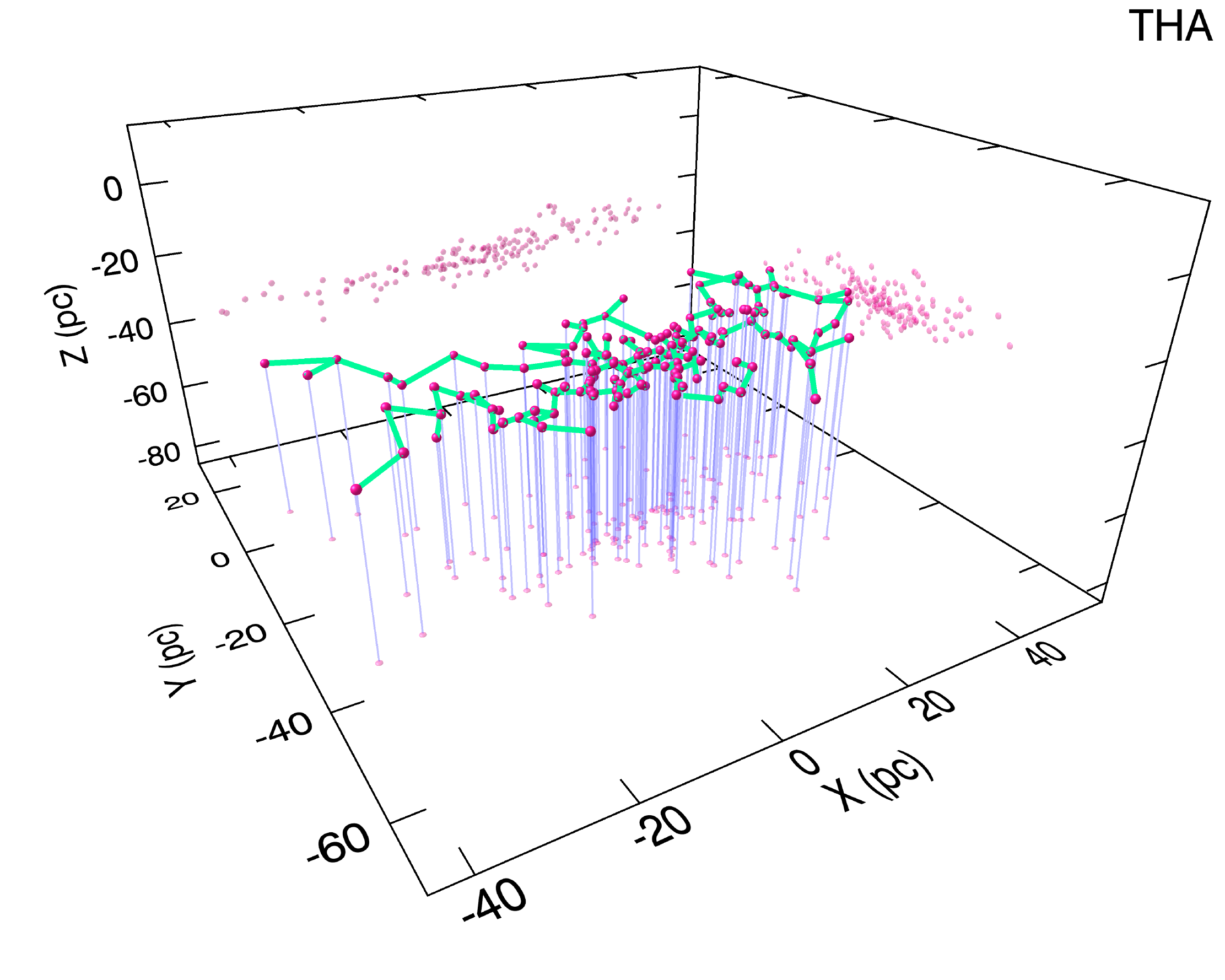}}
	\subfigure{\includegraphics[width=0.495\textwidth]{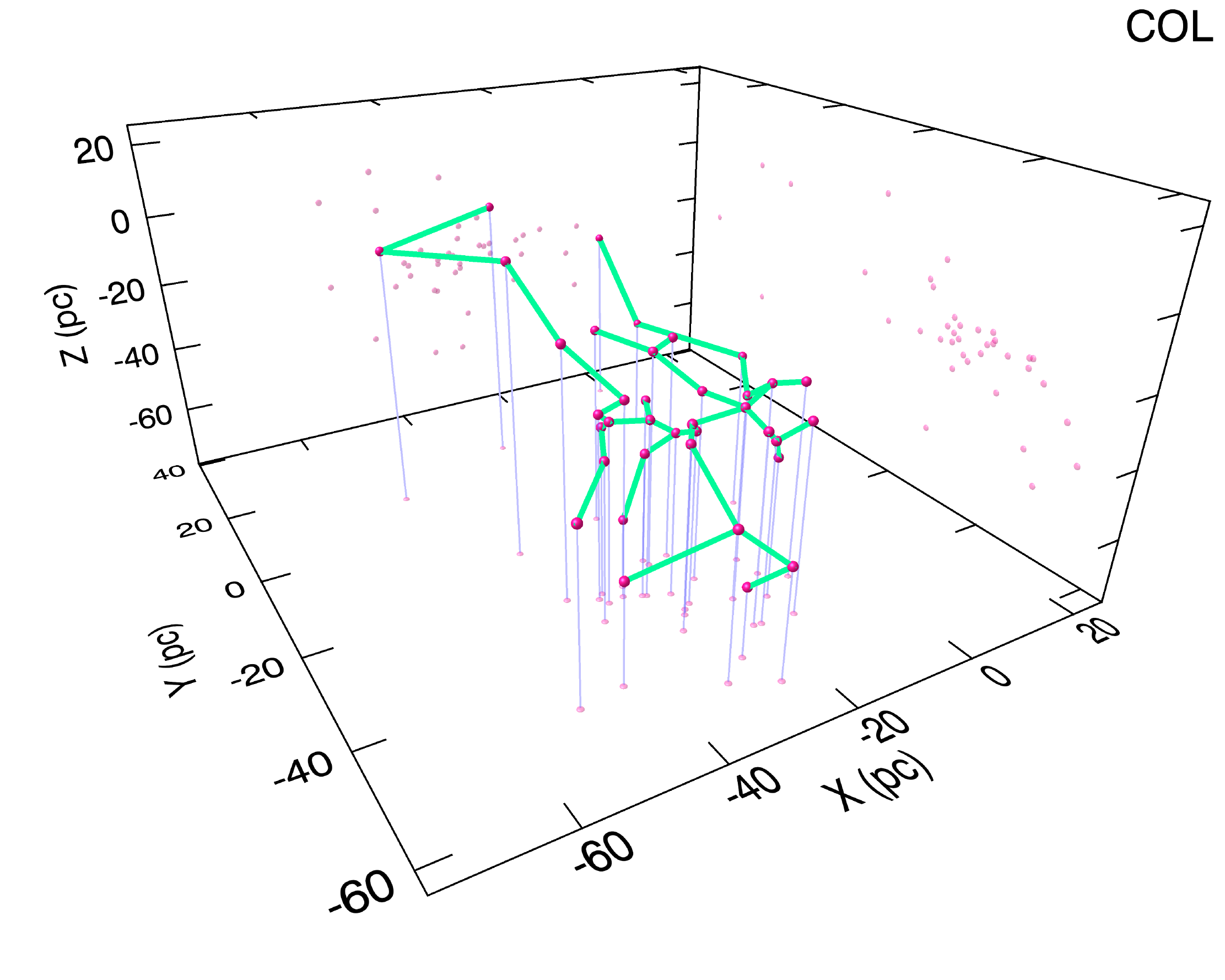}}
	\subfigure{\includegraphics[width=0.495\textwidth]{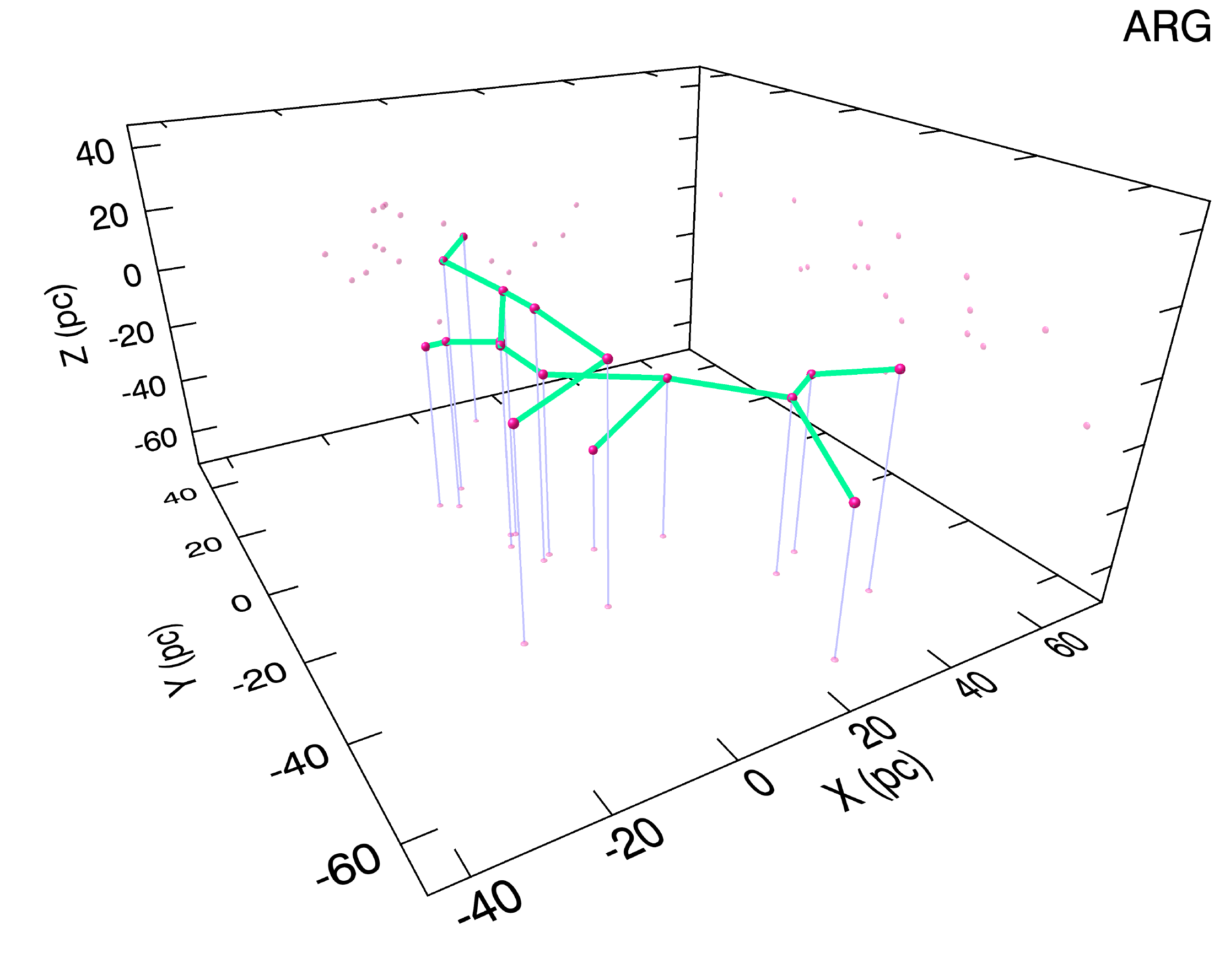}}
	\subfigure{\includegraphics[width=0.495\textwidth]{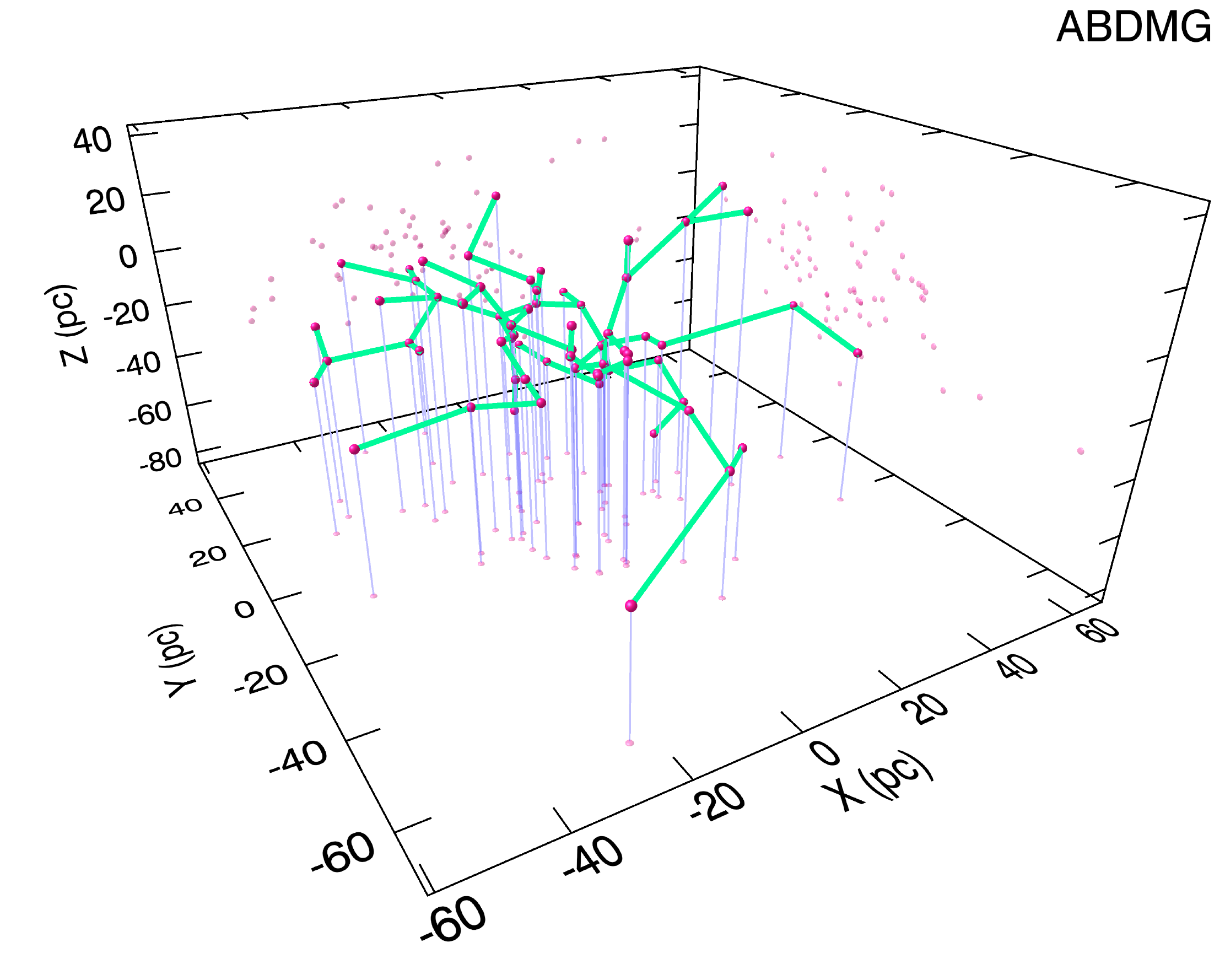}}
	\caption{Minimum spanning trees (MSTs; green lines) in \emph{XYZ} space for bona fide members and high probability \href{http://www.astro.umontreal.ca/\textasciitilde gagne/BASS.php}{\emph{BASS}} candidates (red points and their projections). Blue lines link each data point to its projection on the $XZ$ plane for clarity. The total length of the MSTs for the brightest subsets of objects, compared with a random subset, is a useful diagnosis to determine the presence of mass segregation.}
	\label{fig:mst_1}
\end{figure*}

\begin{figure*}[p]
	\centering
	\subfigure{\includegraphics[width=0.495\textwidth]{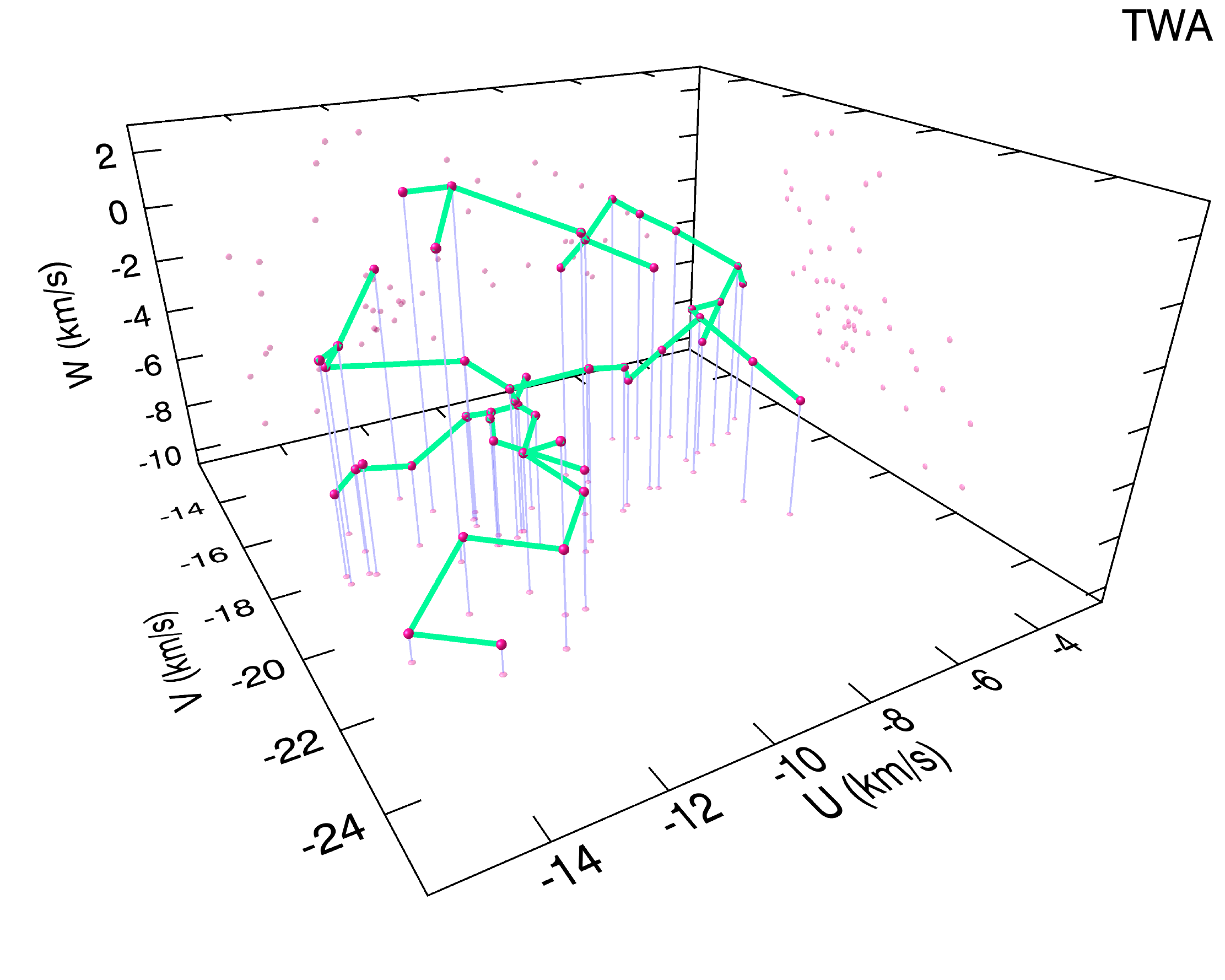}}
	\subfigure{\includegraphics[width=0.495\textwidth]{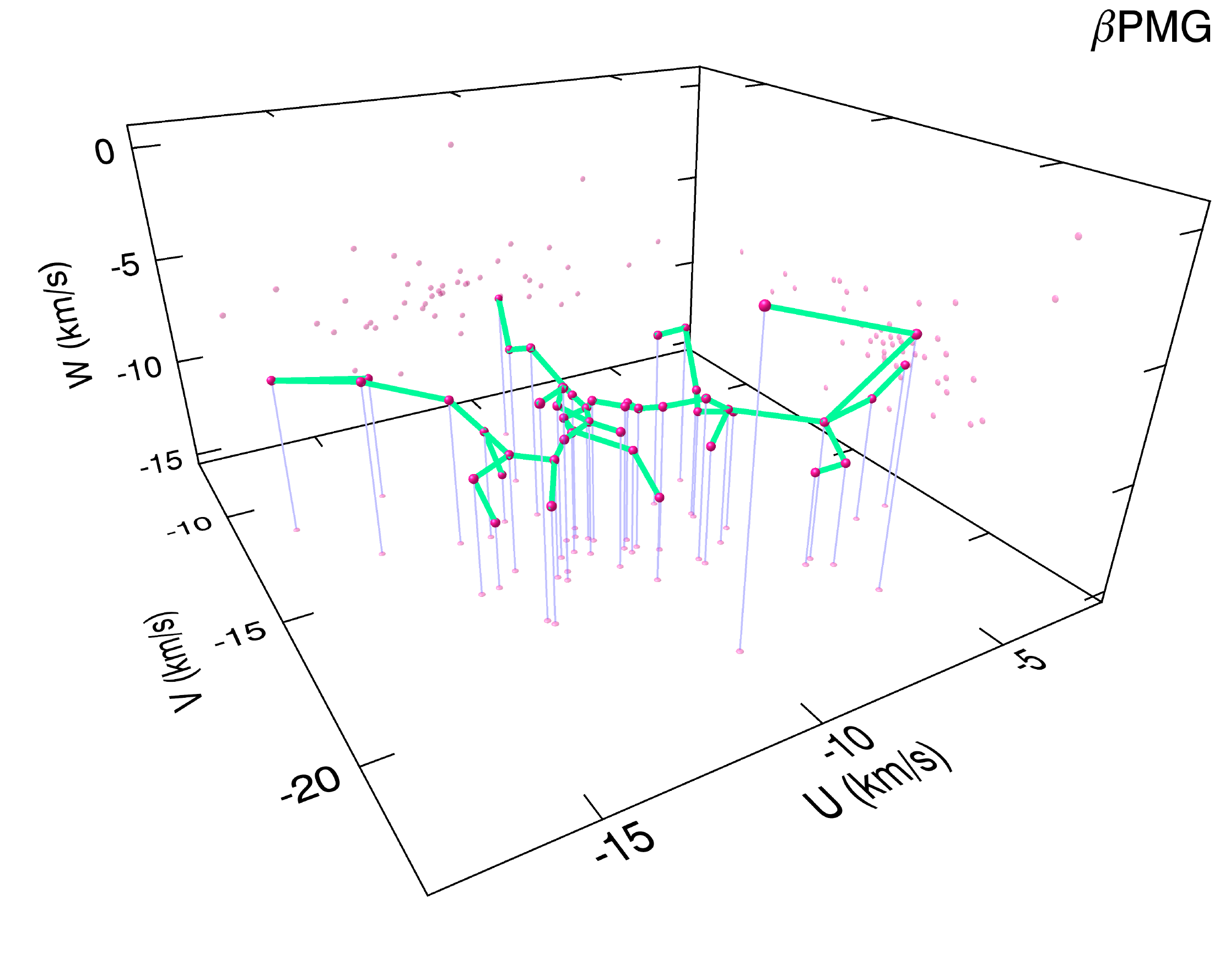}}
	\subfigure{\includegraphics[width=0.495\textwidth]{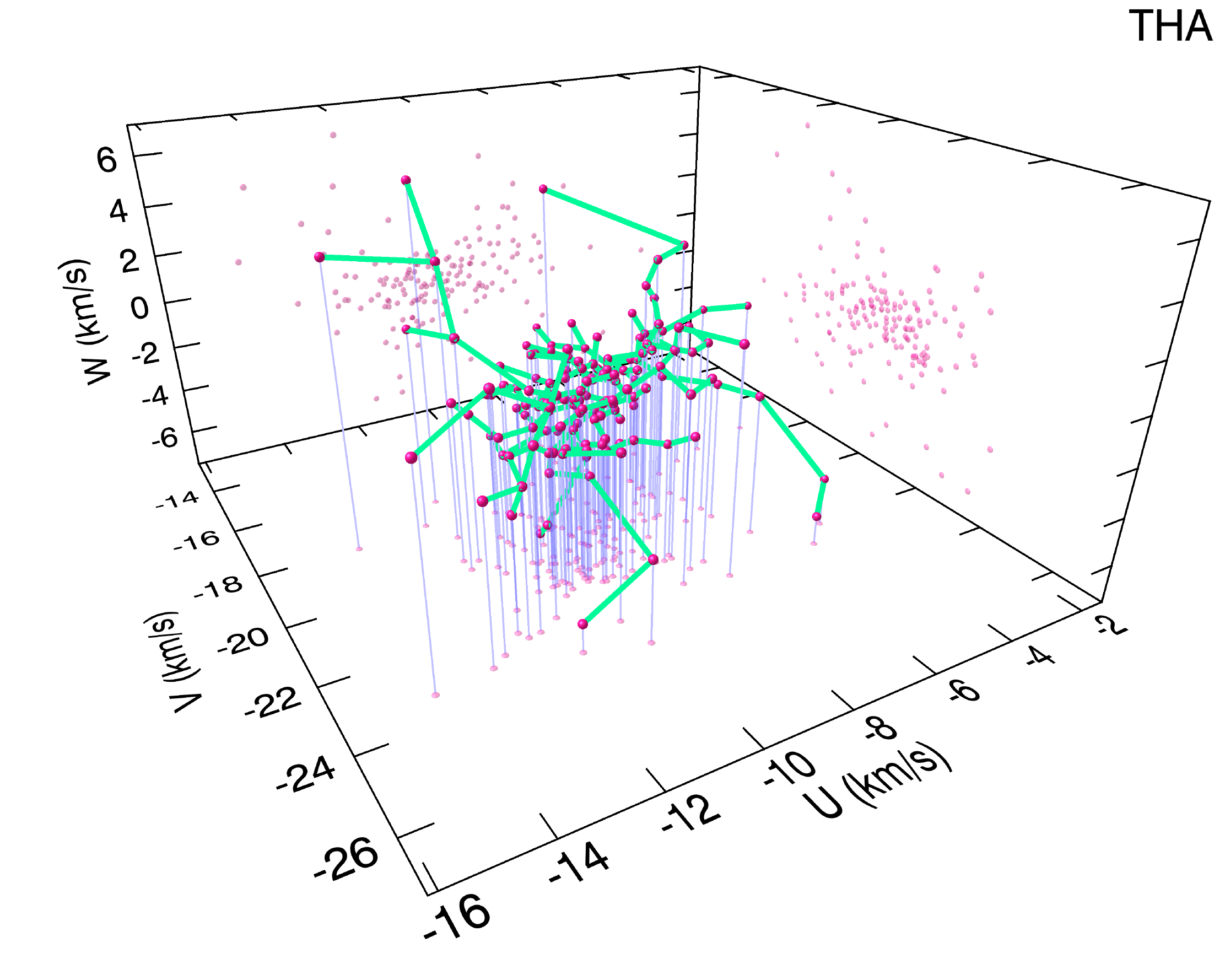}}
	\subfigure{\includegraphics[width=0.495\textwidth]{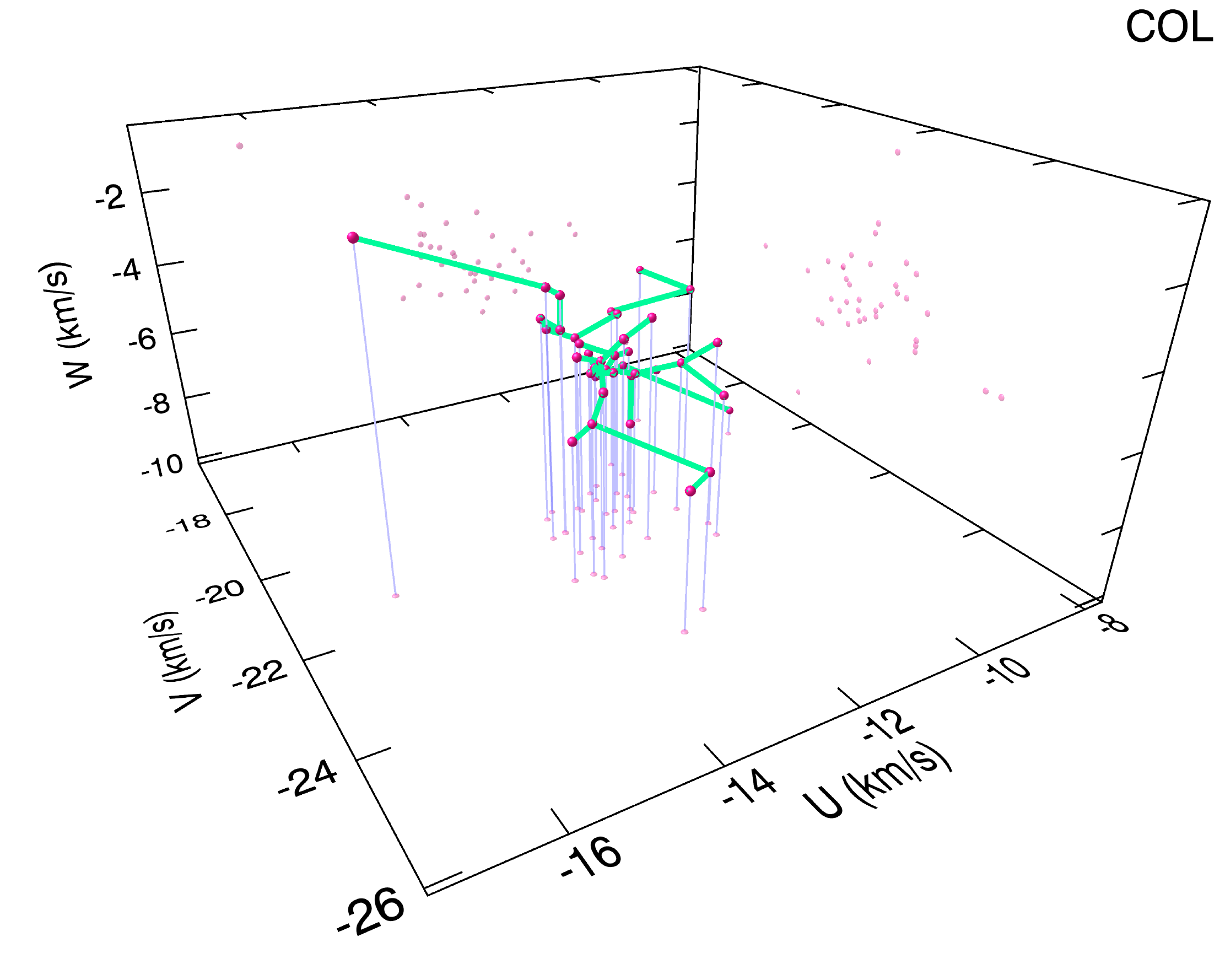}}
	\subfigure{\includegraphics[width=0.495\textwidth]{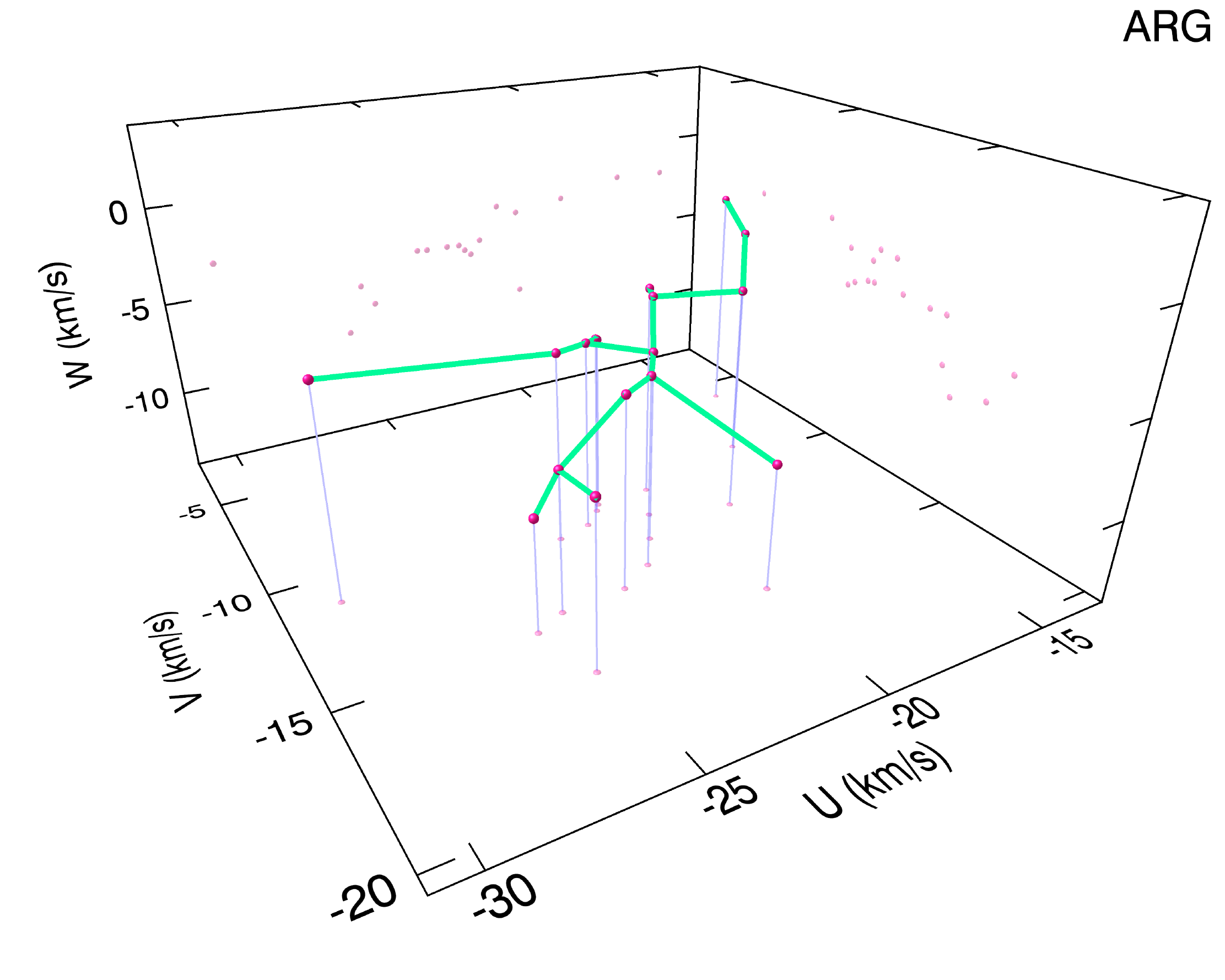}}
	\subfigure{\includegraphics[width=0.495\textwidth]{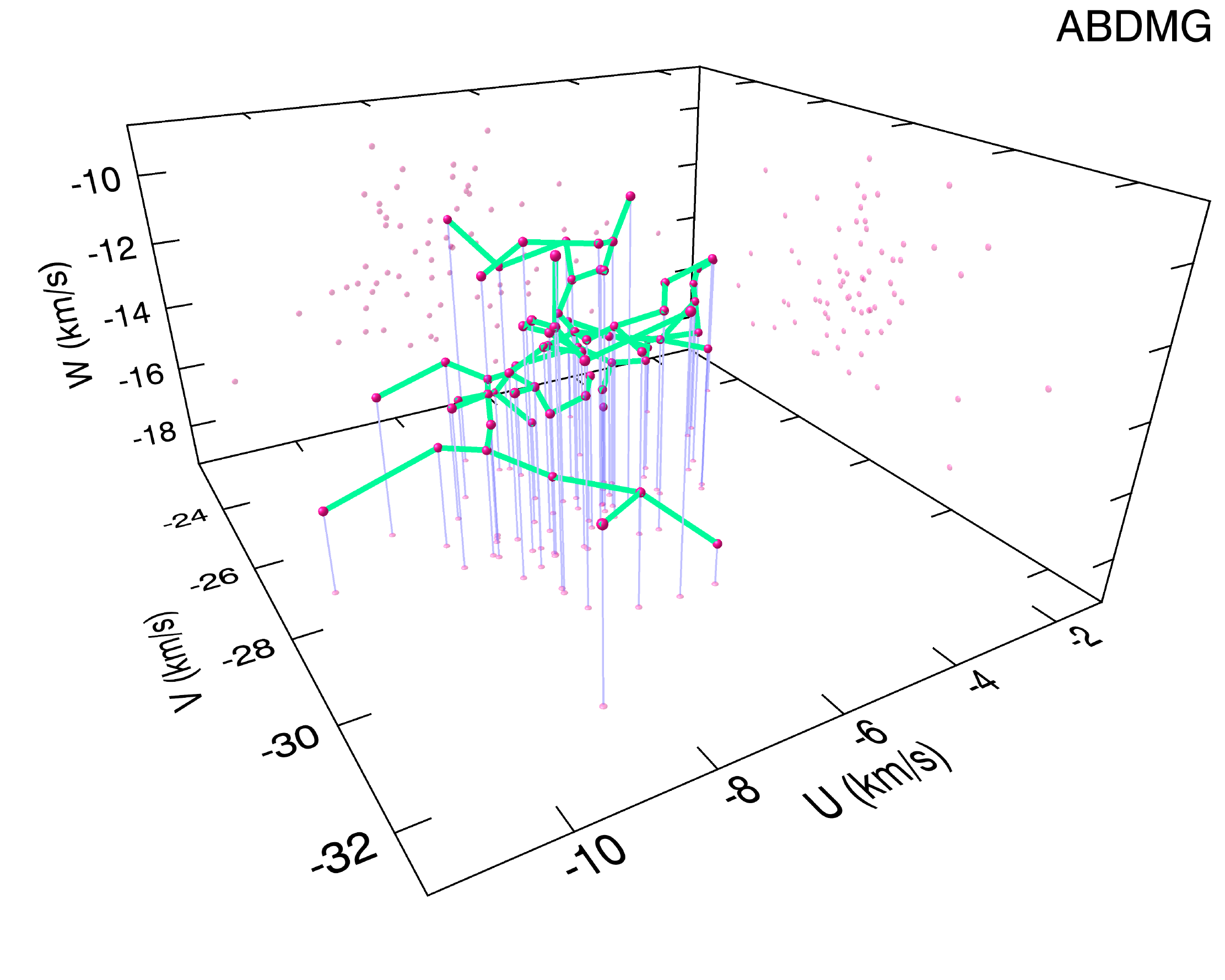}}
	\caption{Minimum spanning trees (MSTs; green lines) in \emph{UVW} space for bona fide members and high probability \href{http://www.astro.umontreal.ca/\textasciitilde gagne/BASS.php}{\emph{BASS}} candidates (red points and their projections). Blue lines link each data point to its projection on the $UV$ plane for clarity. The total length of the MSTs for the brightest subsets of objects, compared with a random subset, is a useful diagnosis to determine the presence of mass segregation.}
	\label{fig:mst_2}
\end{figure*}

\cite{2009MNRAS.395.1449A} devised a quantitative way to assess mass segregation in associations of stars, which is more sensitive than a simple visual characterization, and more importantly does not dependent on the geometry of the members' distribution. They base this characterization on the principle of \emph{Minimum Spanning Trees} (MSTs). For a given distribution of coordinates (e.g. RA and DEC in a bi-dimensional space which is most often used in the case of open clusters), a MST is the shortest network of straight lines that connects all individual points without creating any loop. A mass segregation ratio (MSR) is then defined as :

\begin{align*}
	\Lambda_{\mathrm{MSR}} &= \frac{<l_{\mathrm{norm}}>}{l_{\mathrm{massive}}} \pm \frac{\sigma_{\mathrm{norm}}}{l_{\mathrm{massive}}},\\
\end{align*}

where $l_{\mathrm{massive}}$ is the total length of the MST of the $N$ most massive stars in an association, and $<l_{\mathrm{norm}}>$ and $\sigma_{\mathrm{norm}}$ are respectively the average and standard deviation of a set of Monte Carlo simulations in which the MST network length is determined for a set of $N$ stars randomly selected from the sample. If mass segregation is present, it is expected that $\Lambda_{\mathrm{MSR}}$ will have a value above unity. On the other hand, a value below unity would indicate that massive stars are more scattered than other members. We performed this analysis in both the the \emph{XYZ} and \emph{UVW} 3-dimensional spaces, using the algorithm described by \cite{2004MNRAS.348..589C} to build MSTs. We determined the MSR for values of $N$ spanning 3 to the total number of stars in each YMG, using 100 random subsets in each Monte Carlo simulation. We show resulting MSTs for the full set of $N$ high bona fide members and high probability \href{http://www.astro.umontreal.ca/\textasciitilde gagne/BASS.php}{\emph{BASS}} candidates of each YMG in Figures~\ref{fig:mst_1}--\ref{fig:mst_2}. We sorted stars according to their increasing absolute $W1$ magnitudes instead of decreasing mass when we determined $\Lambda_{\mathrm{MSR}}$, for the reasons mentioned above. This was done for only bona fide members in a first step, and then for bona fide members and all high probability candidates of the \href{http://www.astro.umontreal.ca/\textasciitilde gagne/BASS.php}{\emph{BASS}} catalog taken together.

\begin{figure*}[p]
	\centering
	\subfigure{\includegraphics[width=0.495\textwidth]{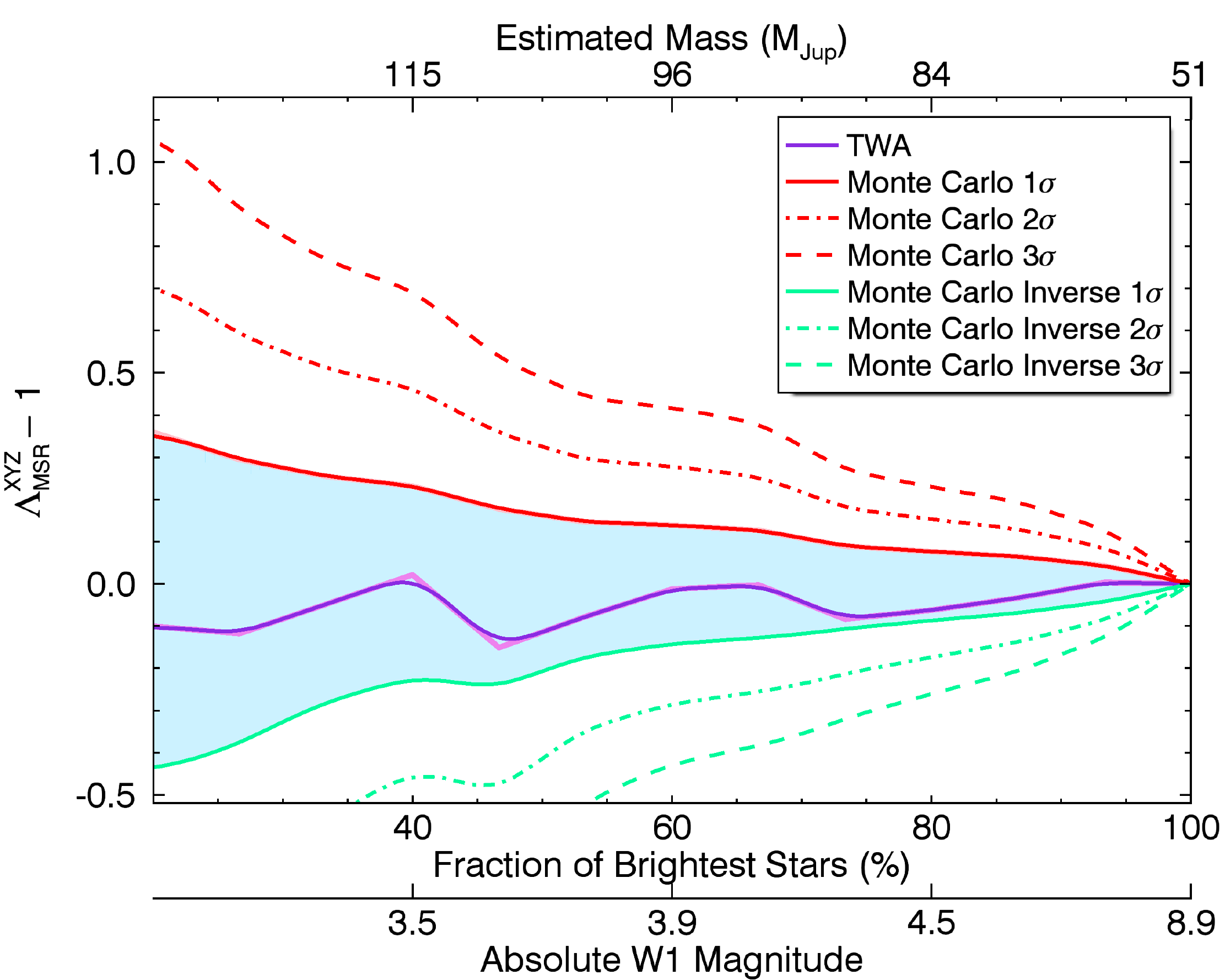}}
	\subfigure{\includegraphics[width=0.495\textwidth]{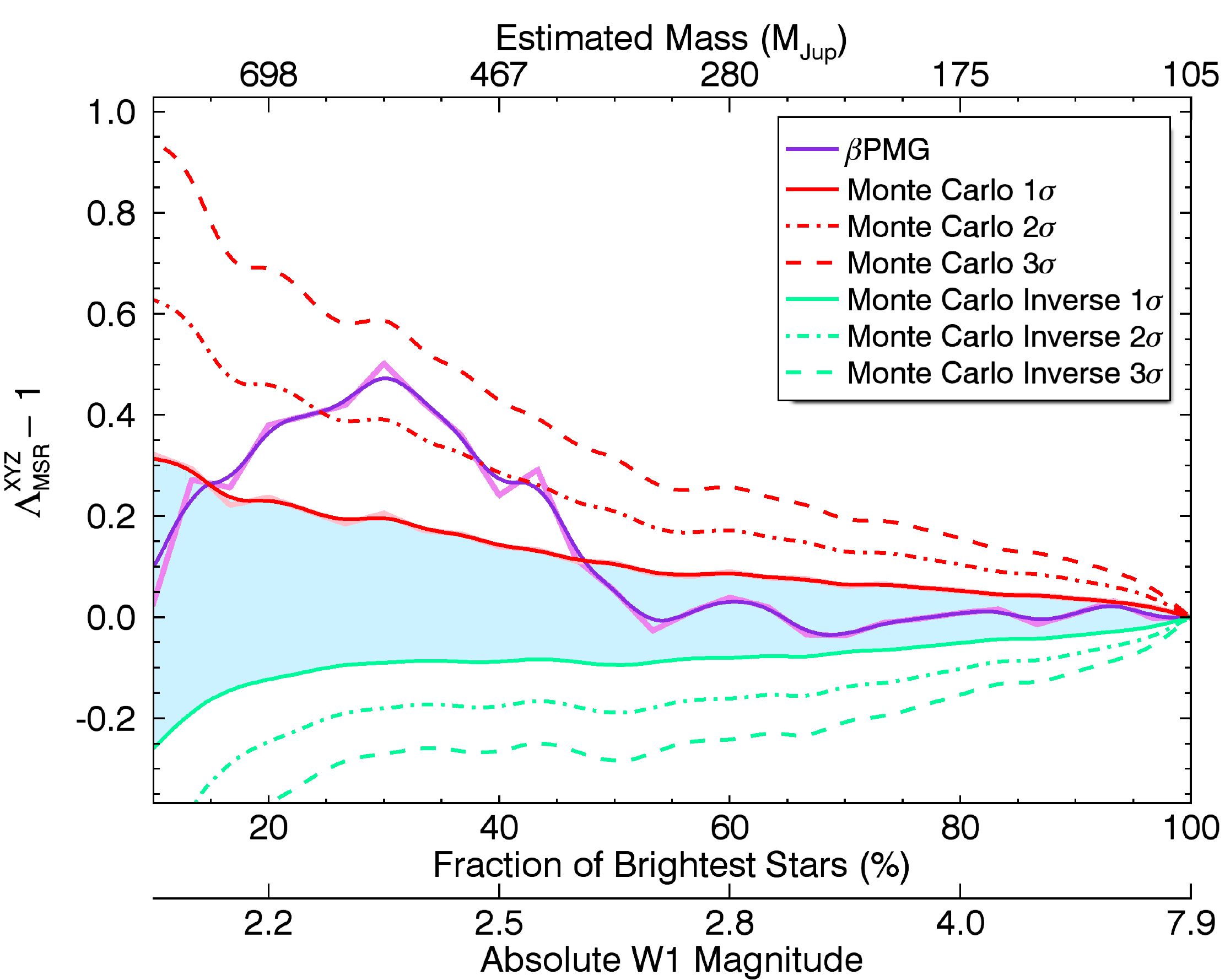}}
	\subfigure{\includegraphics[width=0.495\textwidth]{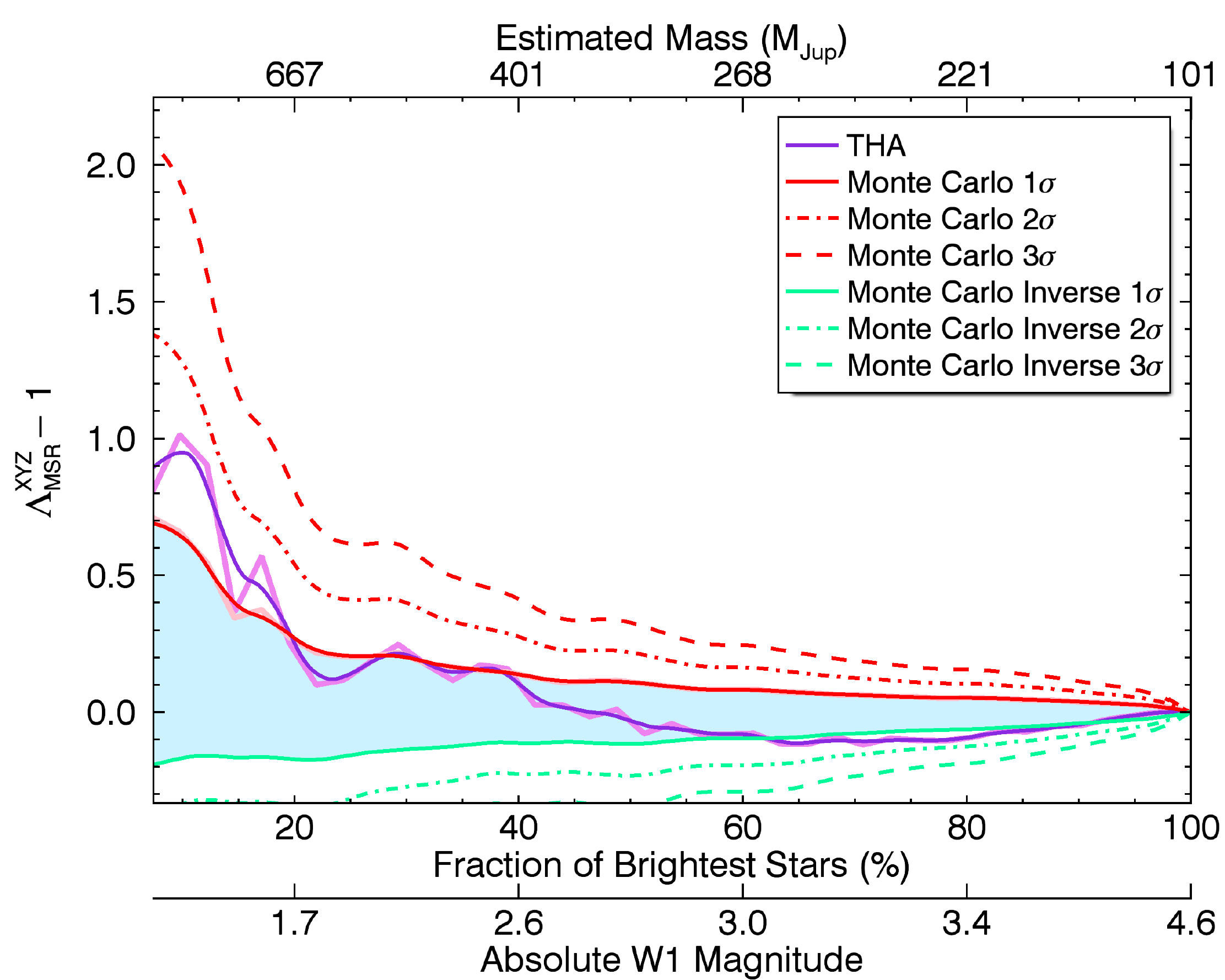}}
	\subfigure{\includegraphics[width=0.495\textwidth]{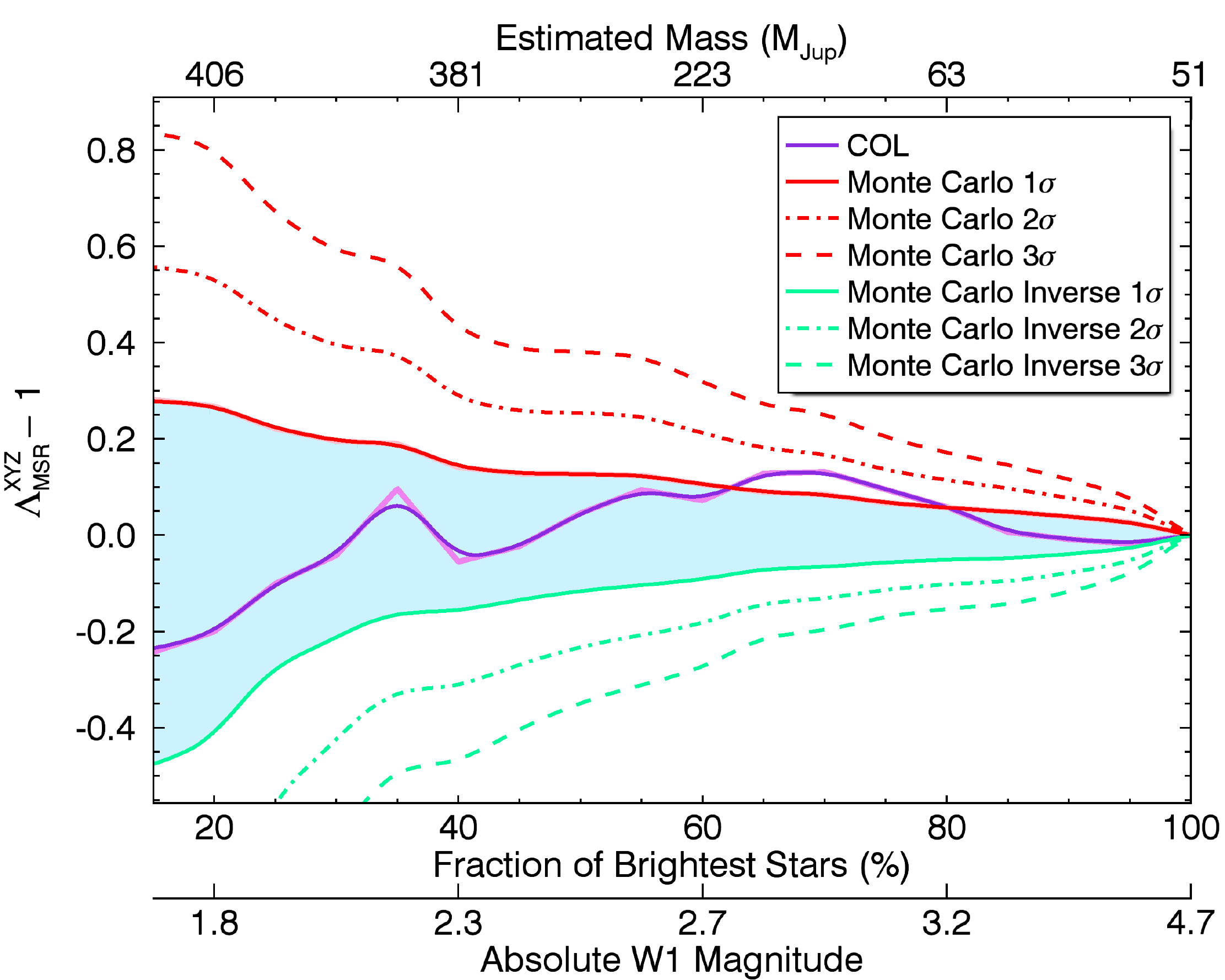}}
	\subfigure{\includegraphics[width=0.495\textwidth]{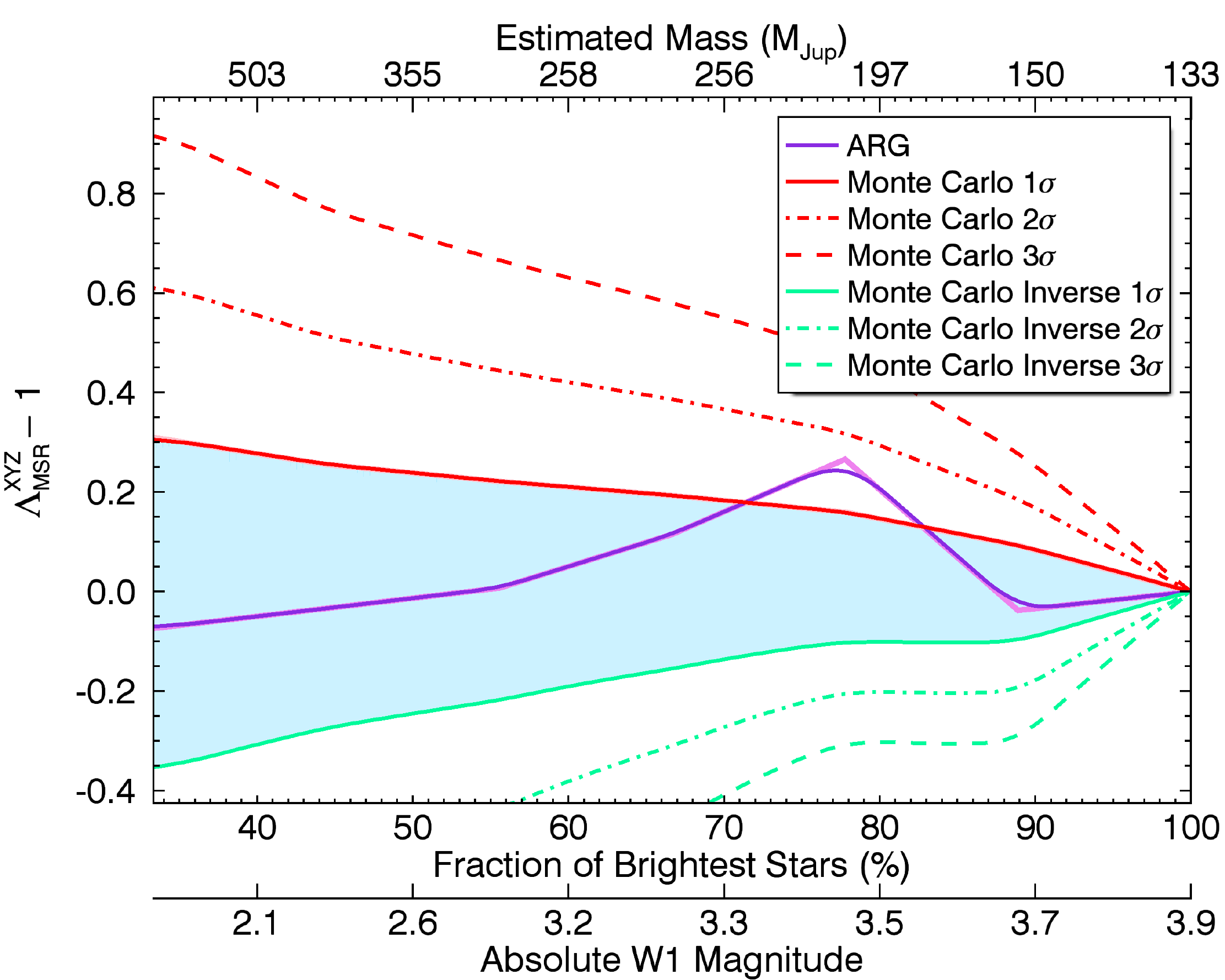}}
	\subfigure{\includegraphics[width=0.495\textwidth]{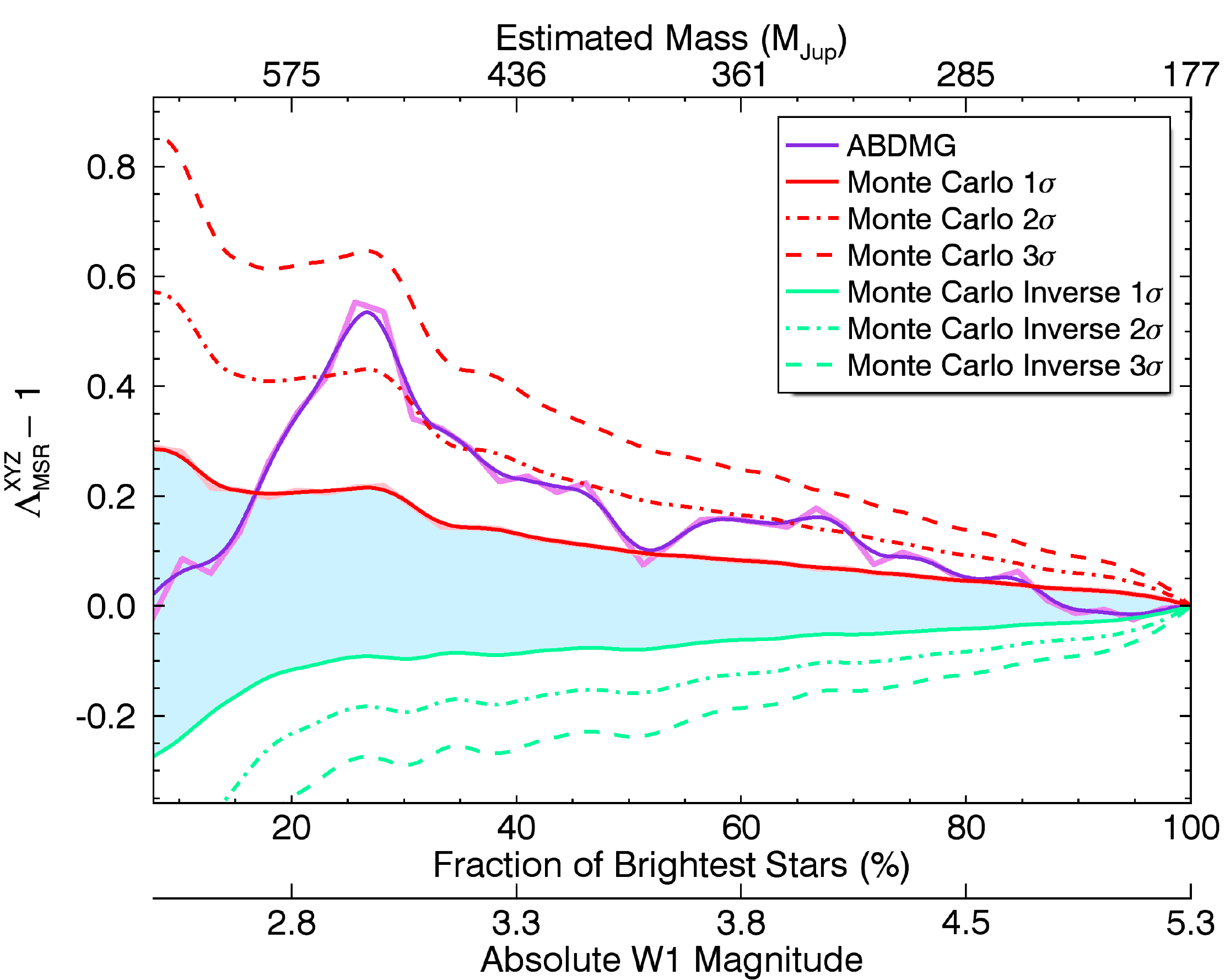}}
	\caption{Spatial mass segregation ratios (MSRs) for bona fide members of YMGs considered here except CAR, as a function of the population fraction of brightest stars that were used in the calculation. Purple curves represent the departure of the MSR from unity, whereas red curves represent results of the Monte Carlo simulation where random stars were chosen instead of the brightest ones. Green curves delimit the region below which the MSR would be smaller than unity with statistical significance (i.e. least massive stars more concentrated towards the center). A MSR (purple curve) located inside the pale blue region indicates no significant difference between the scatter of the brightest or faintest objects. Darker, thick lines represent smoothed versions of the light-colored lines. The segregation mass ratio of CAR does not significantly depart from unity for any value of $N$.}
	\label{fig:segre2_bfide_s}
\end{figure*}

\begin{figure*}[p]
	\centering
	\subfigure{\includegraphics[width=0.495\textwidth]{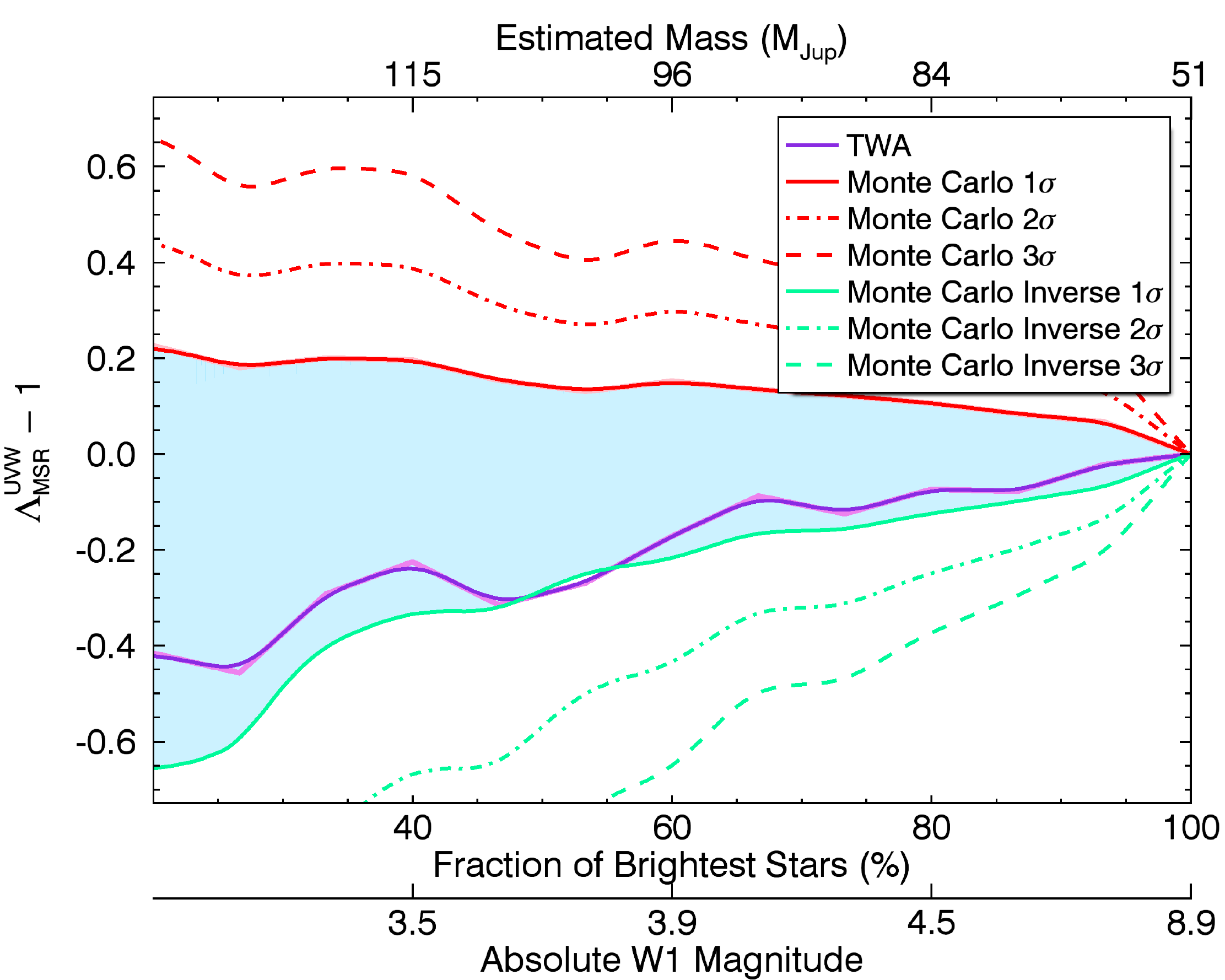}}
	\subfigure{\includegraphics[width=0.495\textwidth]{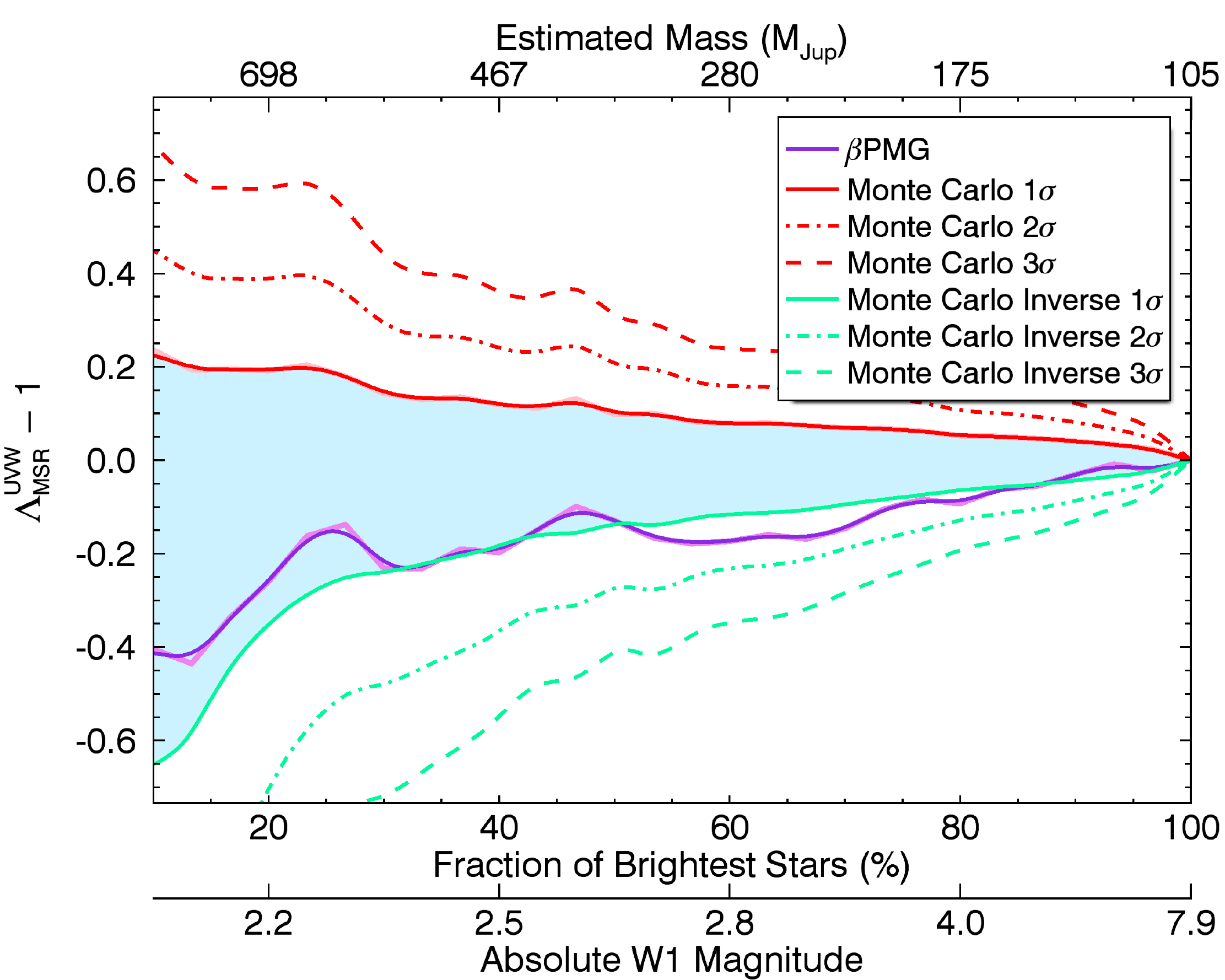}}
	\subfigure{\includegraphics[width=0.495\textwidth]{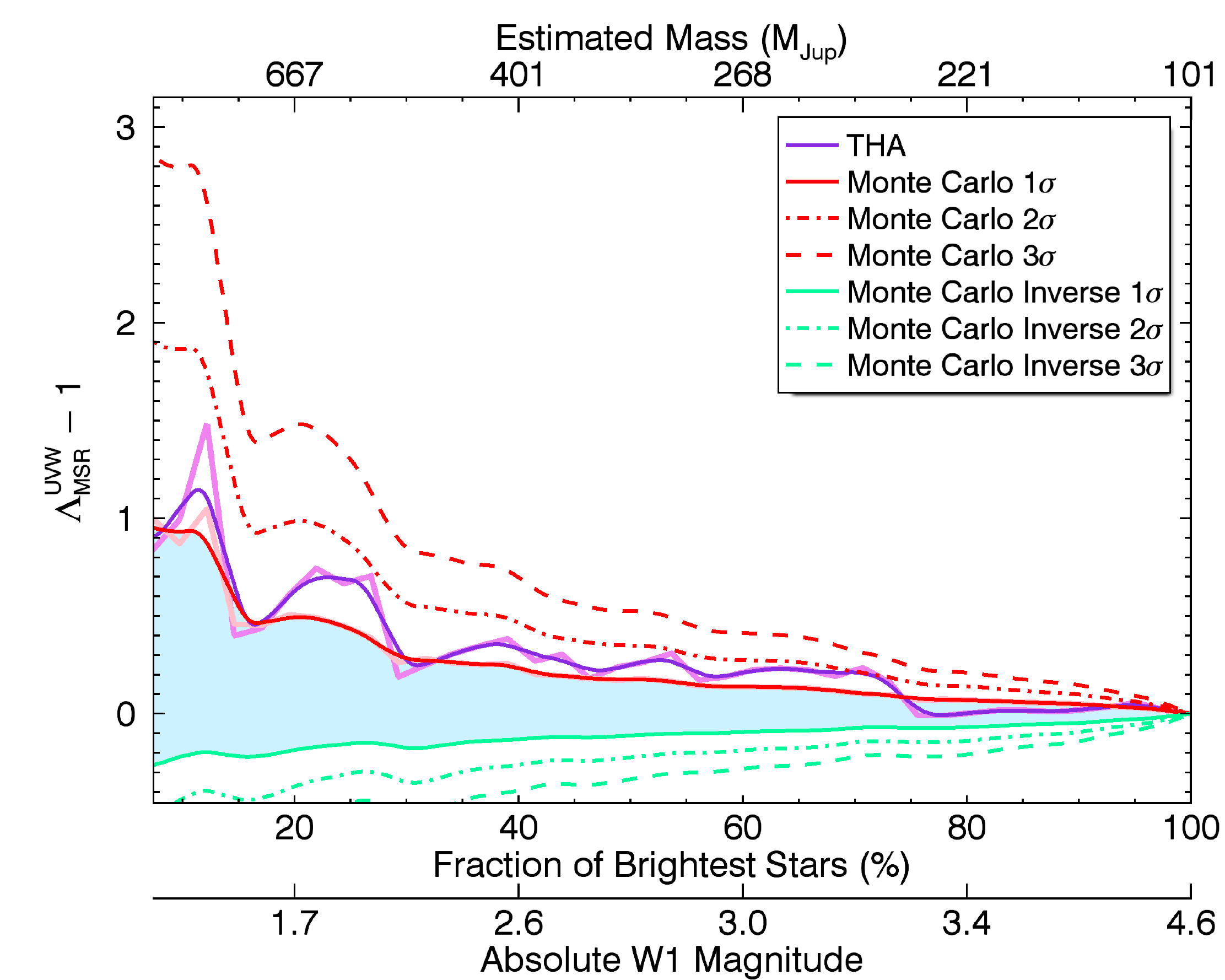}}
	\subfigure{\includegraphics[width=0.495\textwidth]{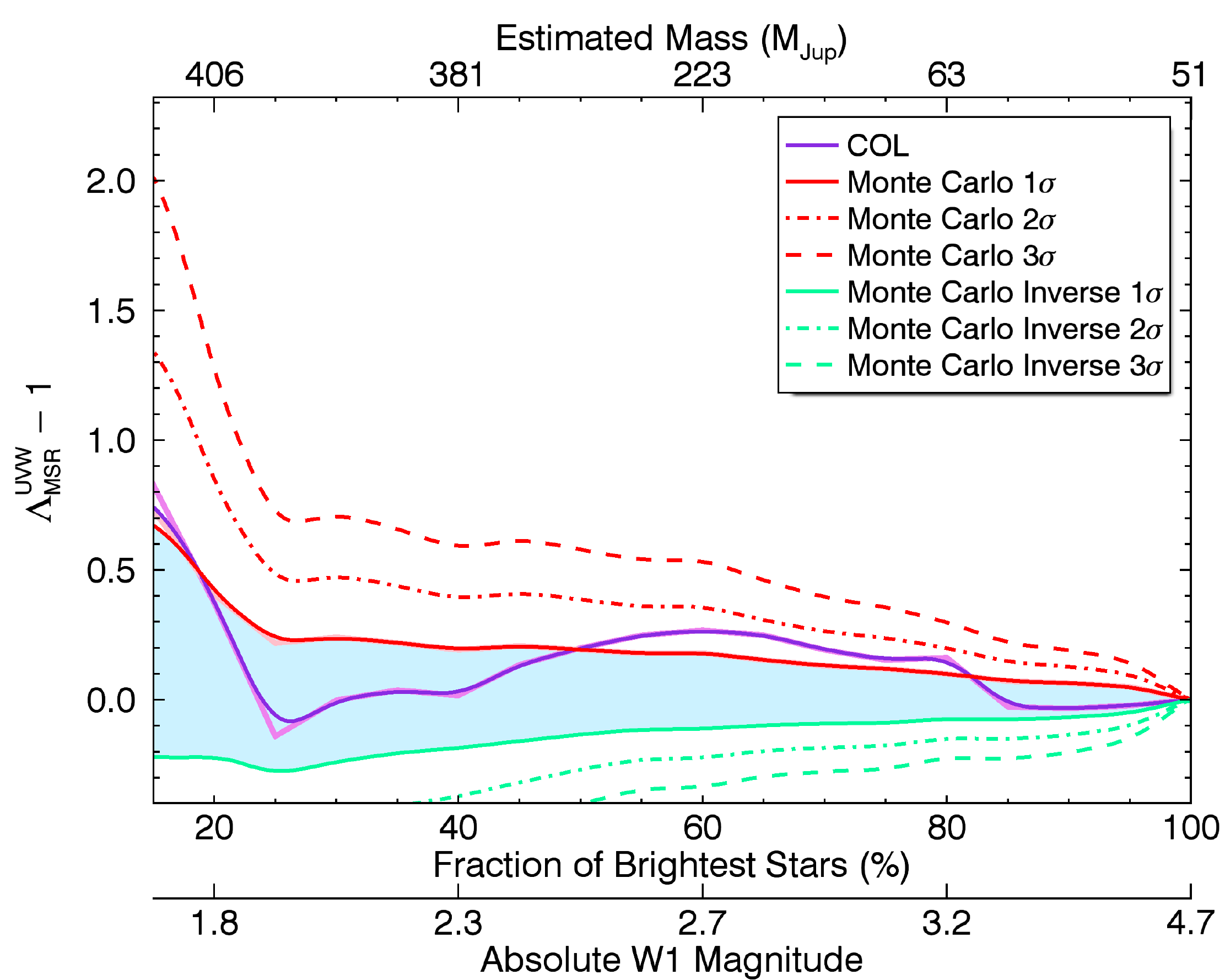}}
	\subfigure{\includegraphics[width=0.495\textwidth]{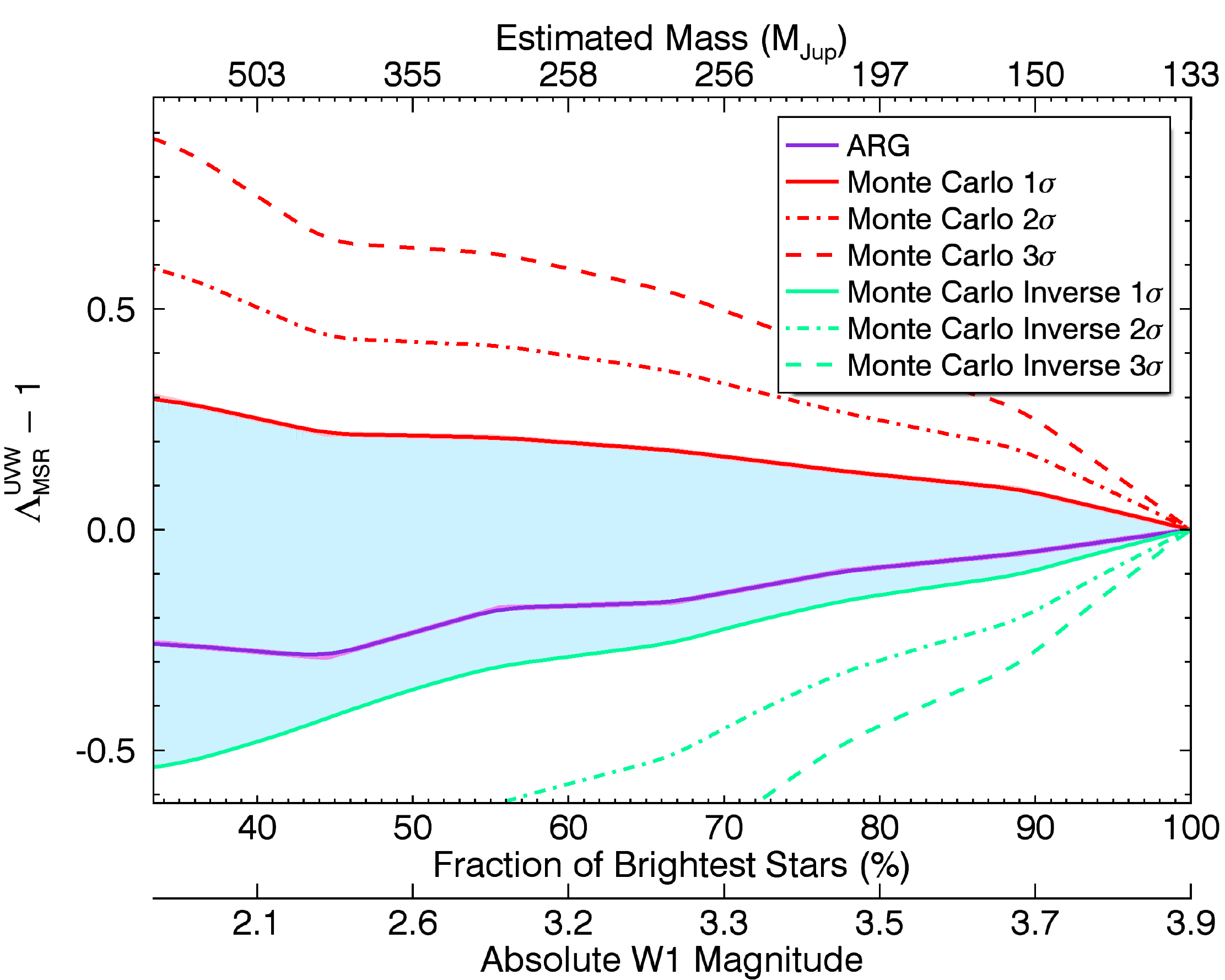}}
	\subfigure{\includegraphics[width=0.495\textwidth]{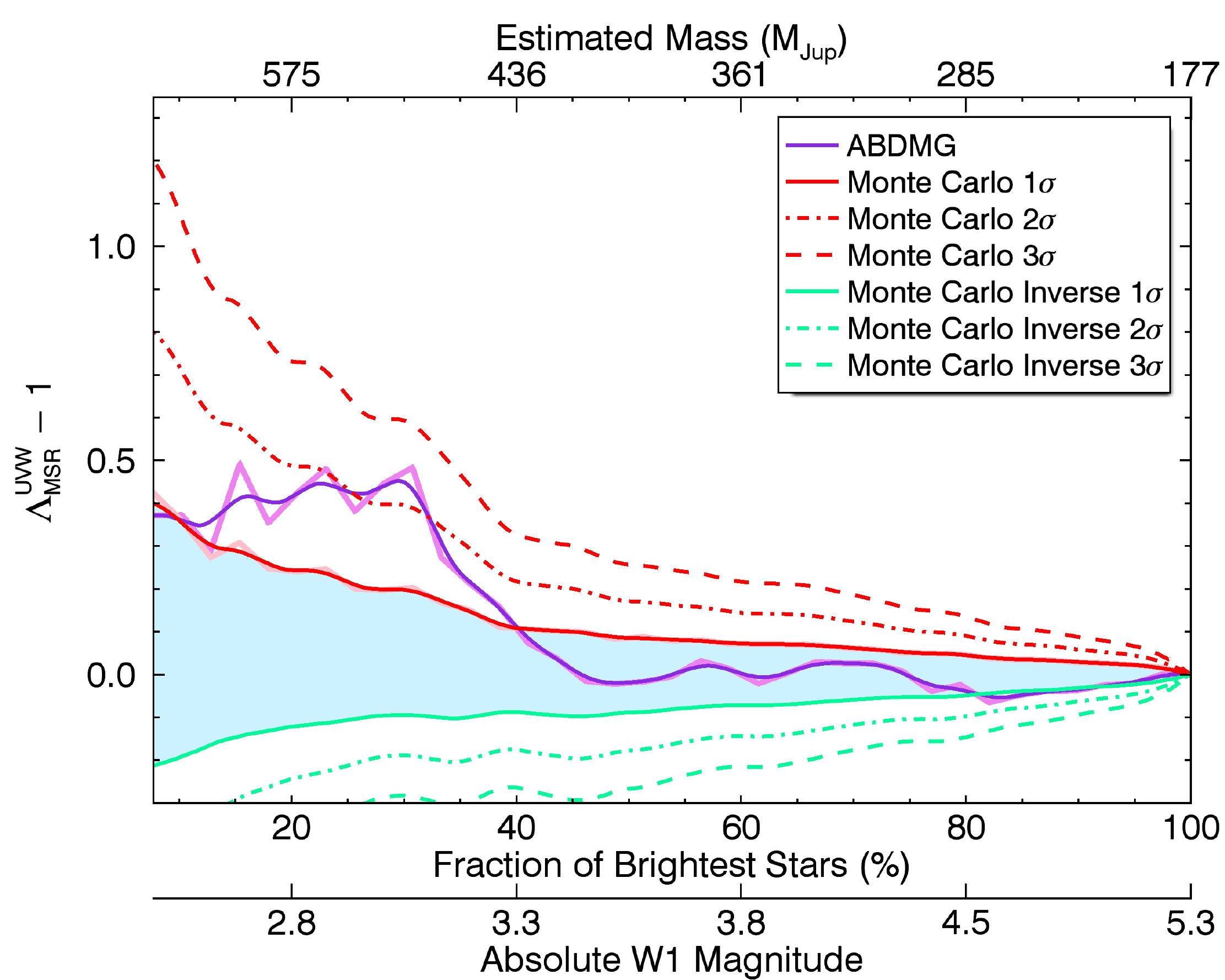}}
	\caption{Same as \hyperref[fig:segre2_bfide_s]{Figure~\ref*{fig:segre2_bfide_s}} for dynamical mass segregation.}
	\label{fig:segre2_bfide_d}
\end{figure*}

\begin{figure*}[p]
	\centering
	\subfigure{\includegraphics[width=0.495\textwidth]{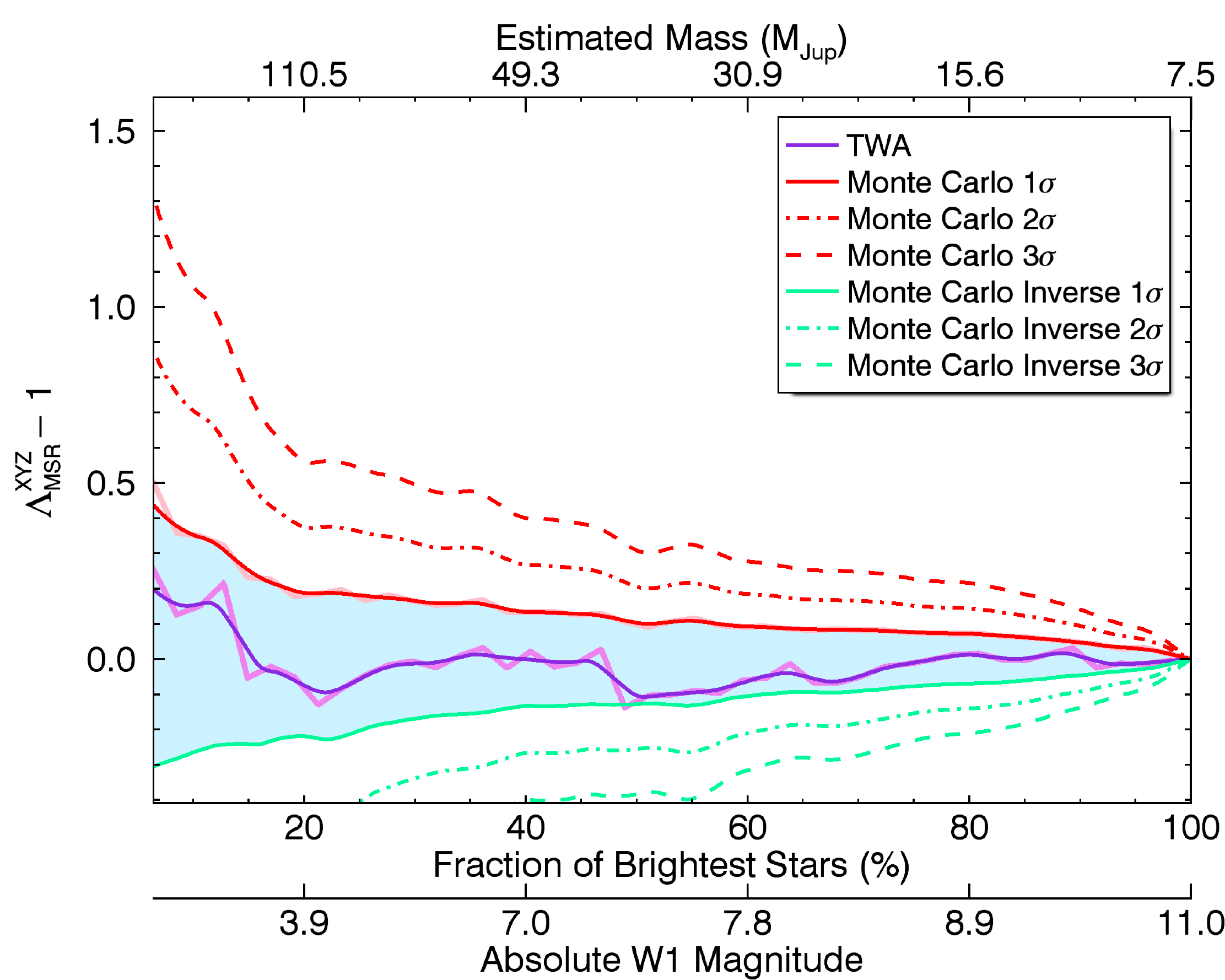}}
	\subfigure{\includegraphics[width=0.495\textwidth]{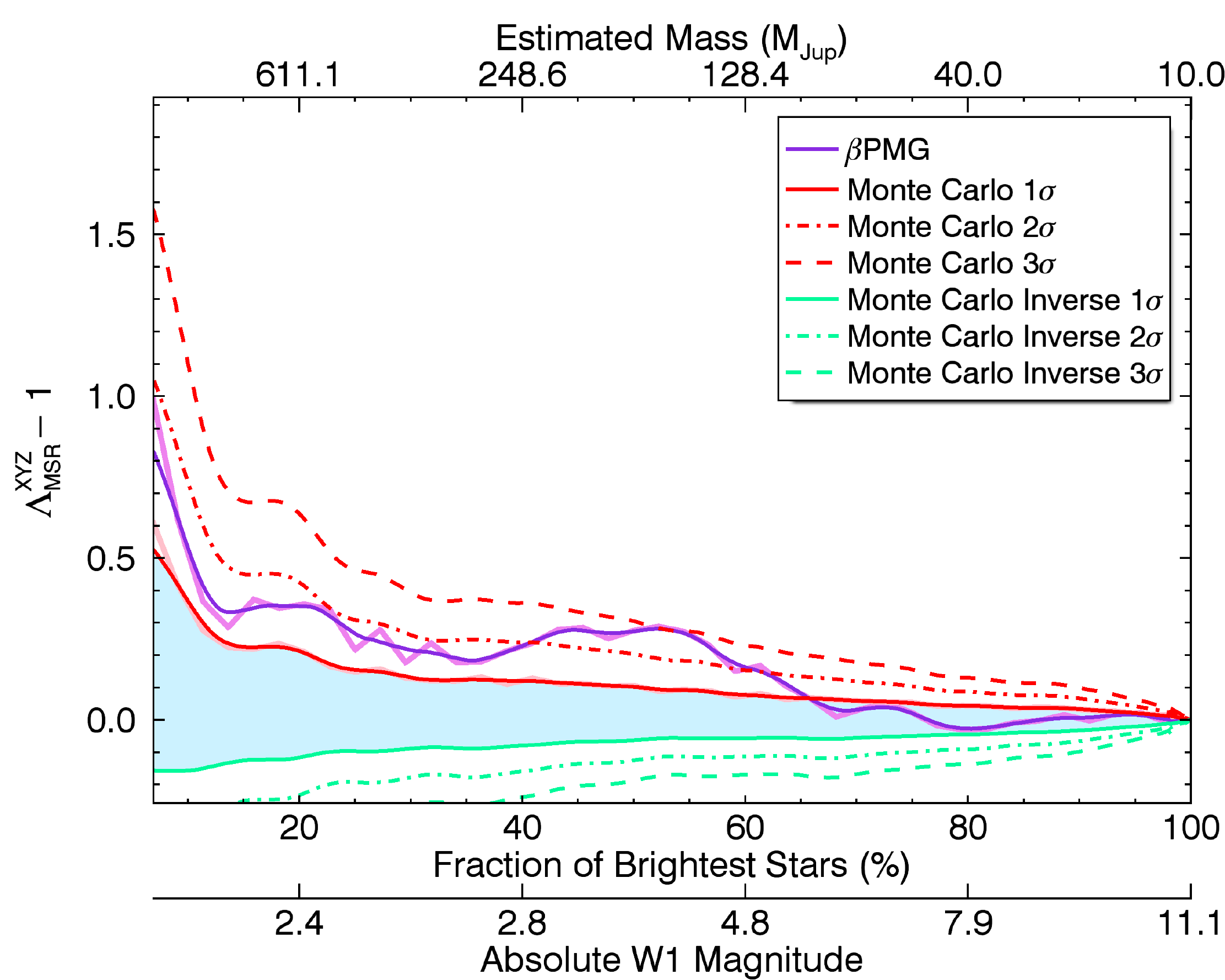}}
	\subfigure{\includegraphics[width=0.495\textwidth]{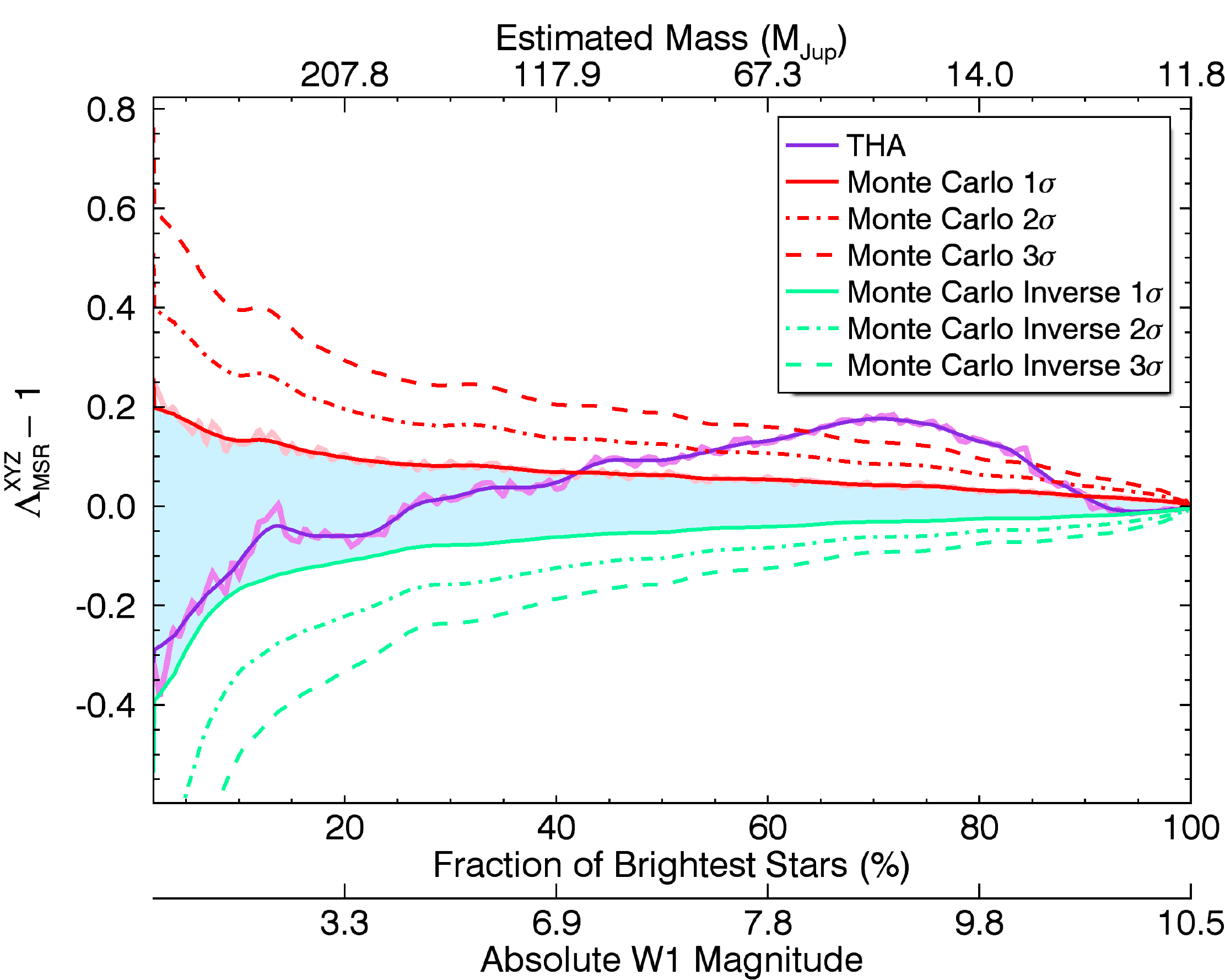}}
	\subfigure{\includegraphics[width=0.495\textwidth]{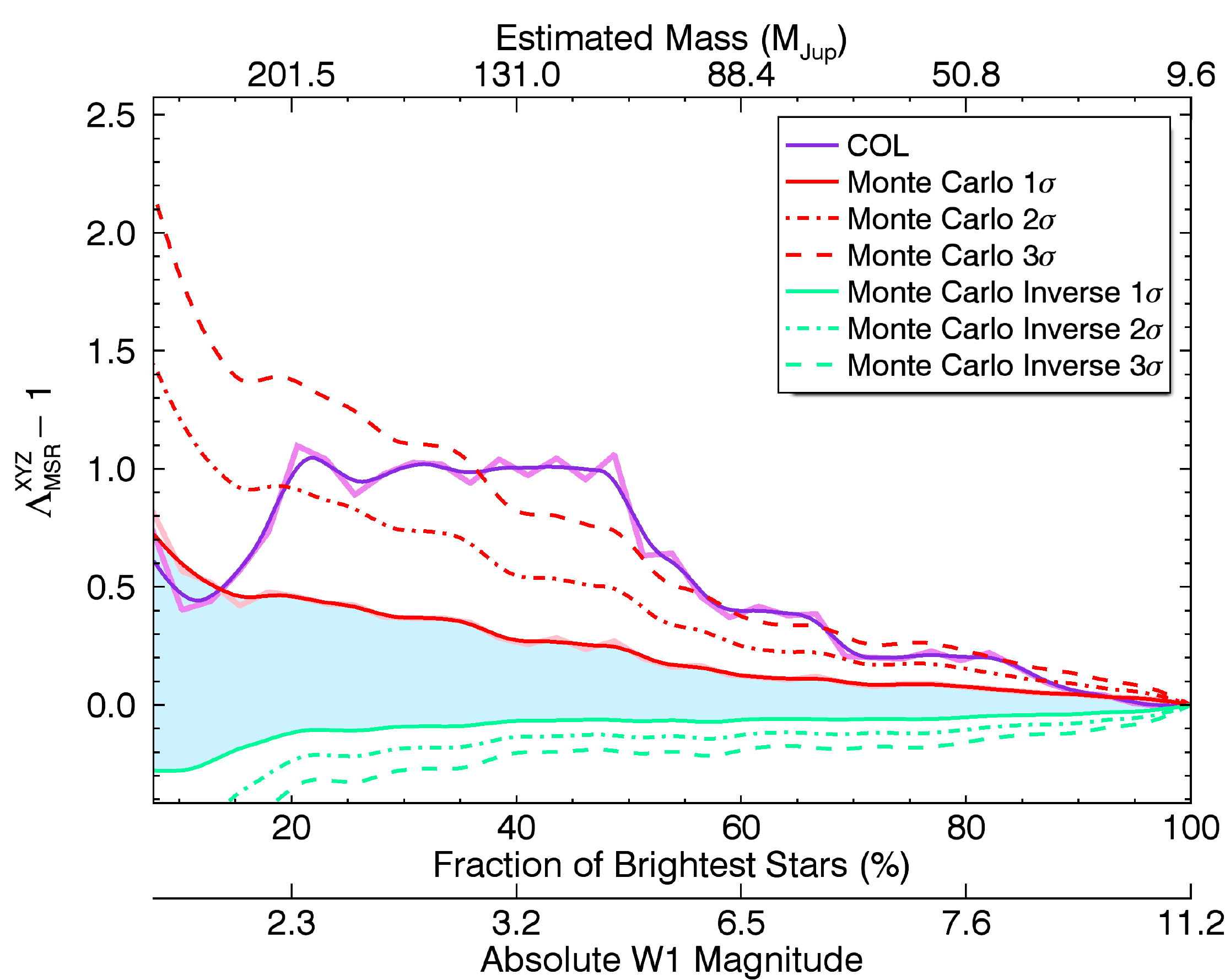}}
	\subfigure{\includegraphics[width=0.495\textwidth]{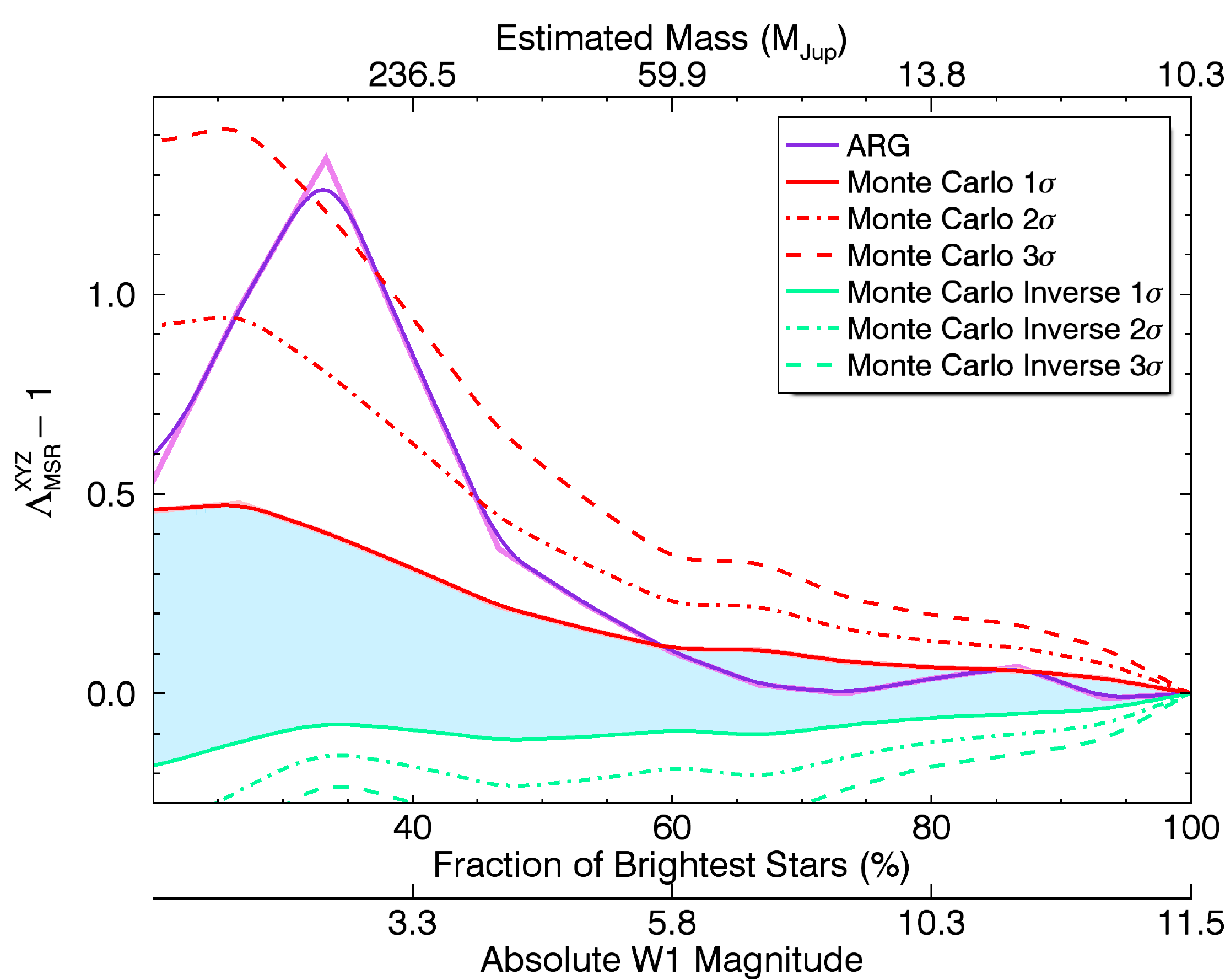}}
	\subfigure{\includegraphics[width=0.495\textwidth]{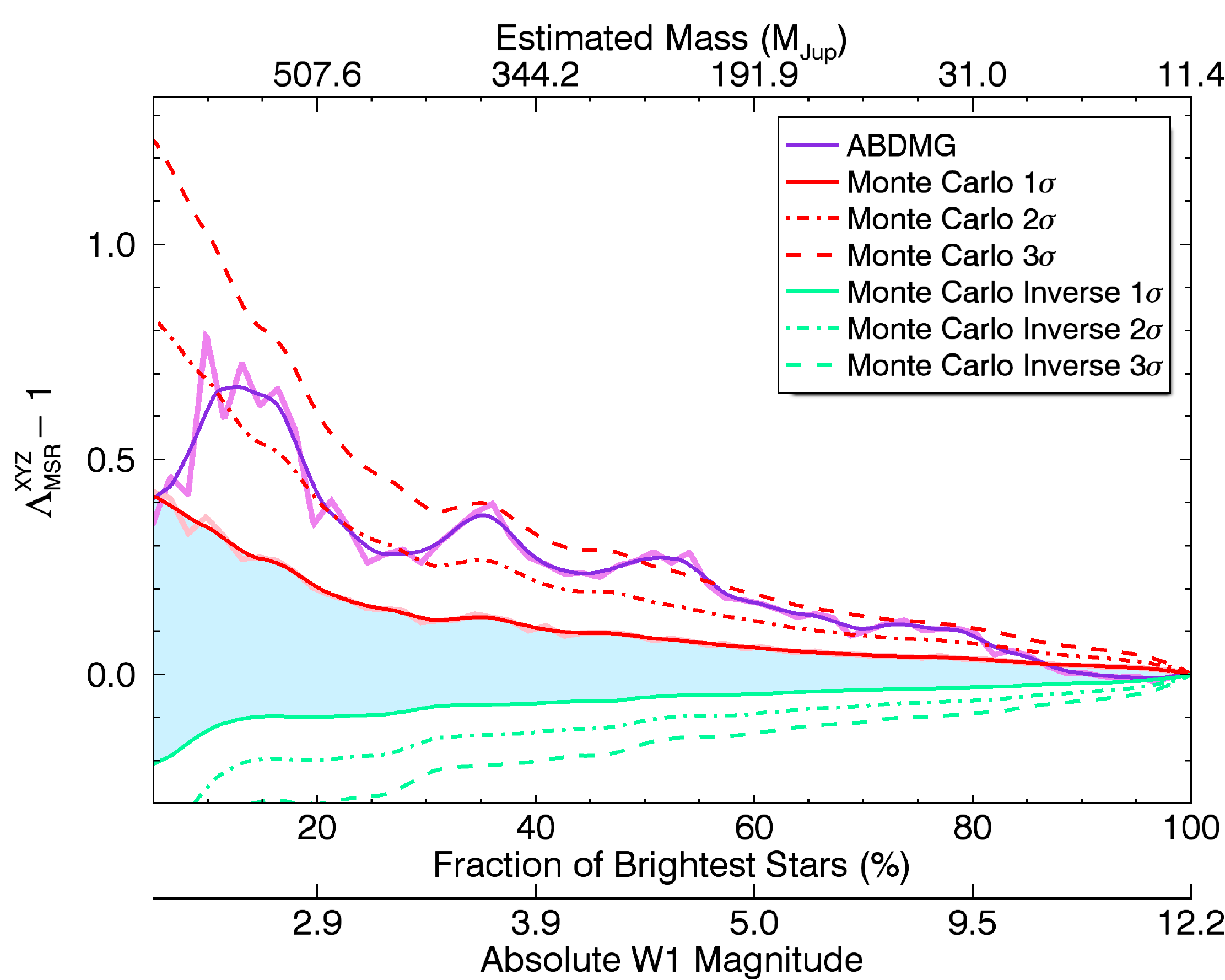}}
	\caption{Same as \hyperref[fig:segre2_bfide_s]{Figure~\ref*{fig:segre2_bfide_s}} with high probability \href{http://www.astro.umontreal.ca/\textasciitilde gagne/BASS.php}{\emph{BASS}} candidates added to the set of bona fide members.}
	\label{fig:segre2_bass_s}
\end{figure*}

\begin{figure*}[p]
	\centering
	\subfigure{\includegraphics[width=0.495\textwidth]{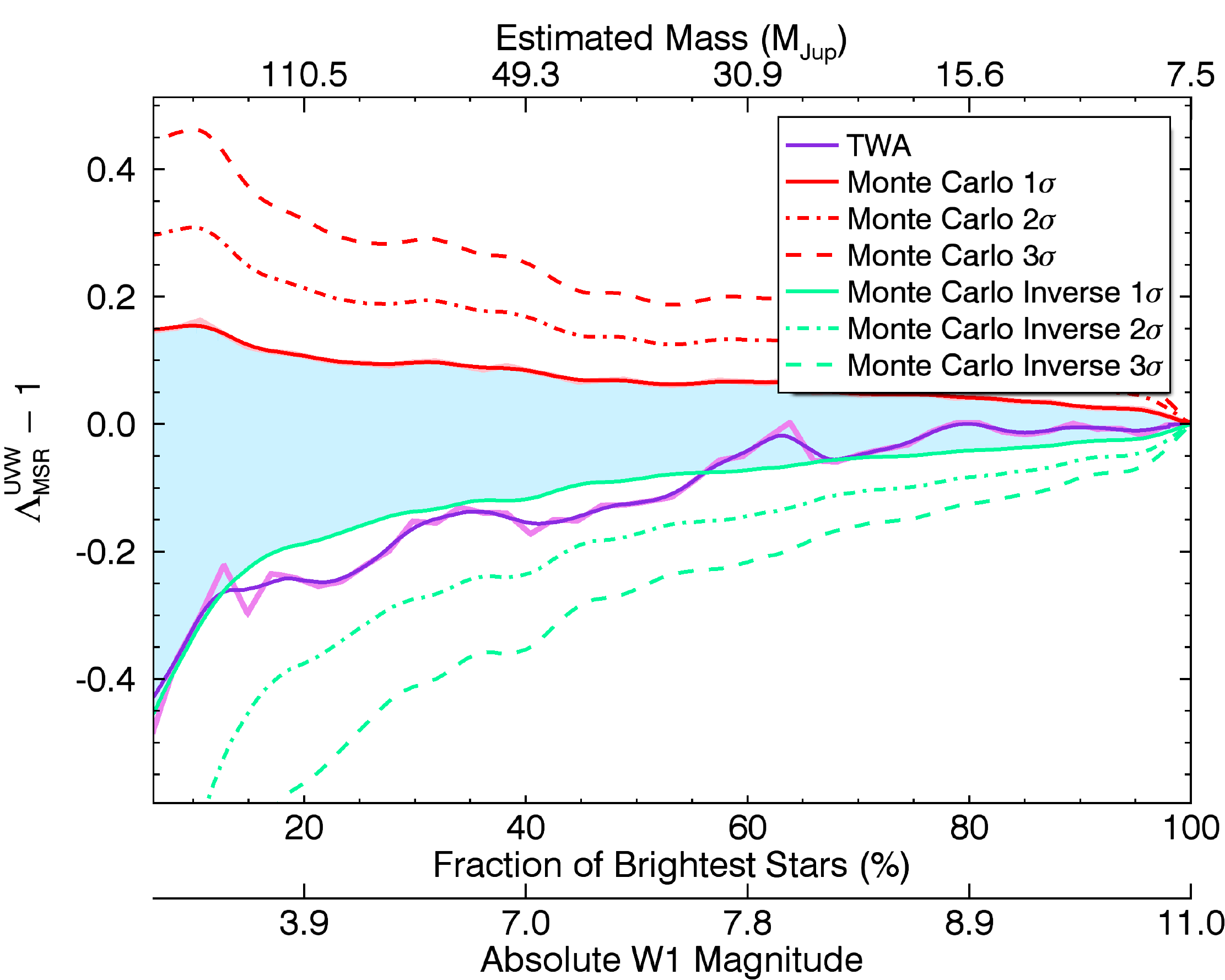}}
	\subfigure{\includegraphics[width=0.495\textwidth]{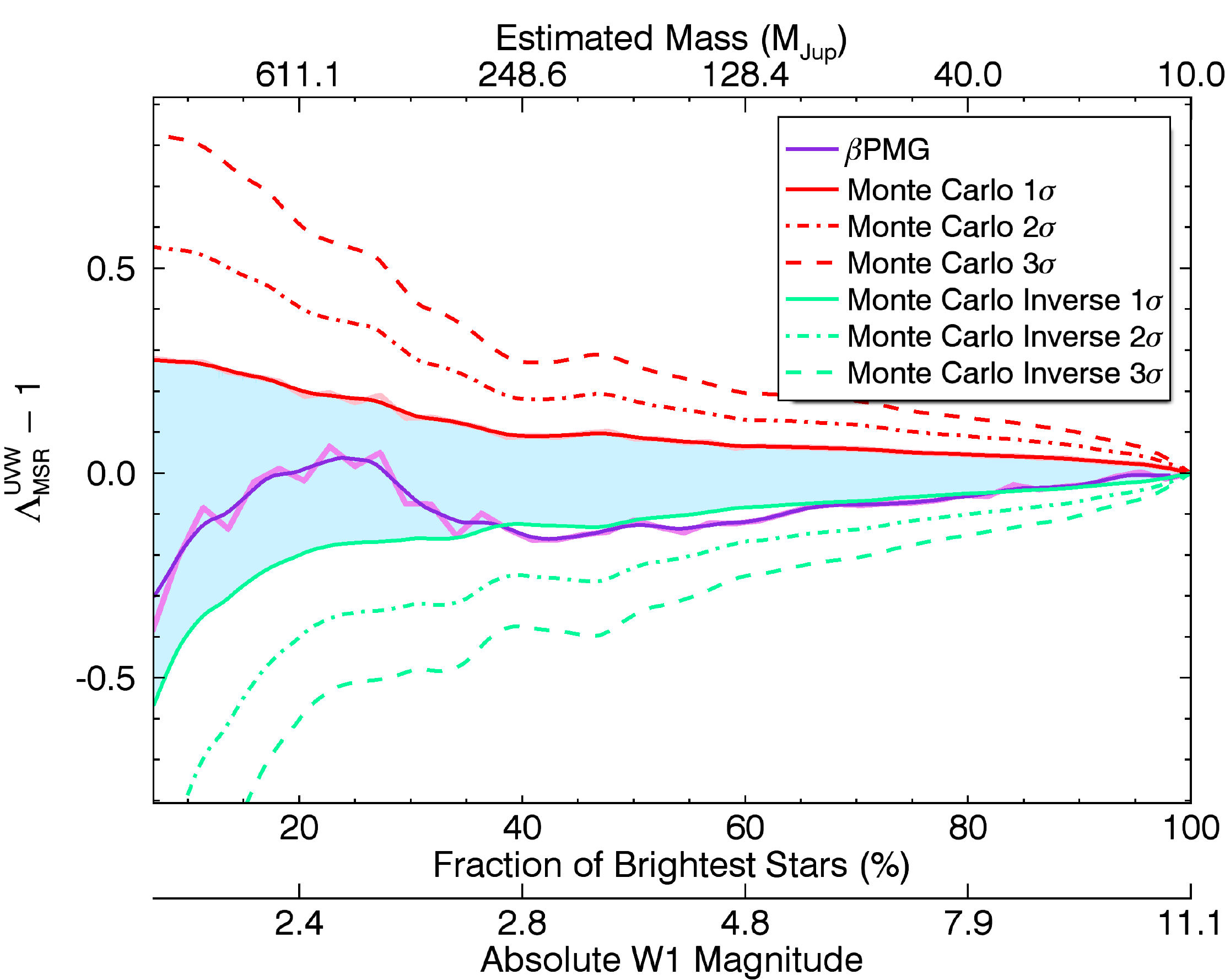}}
	\subfigure{\includegraphics[width=0.495\textwidth]{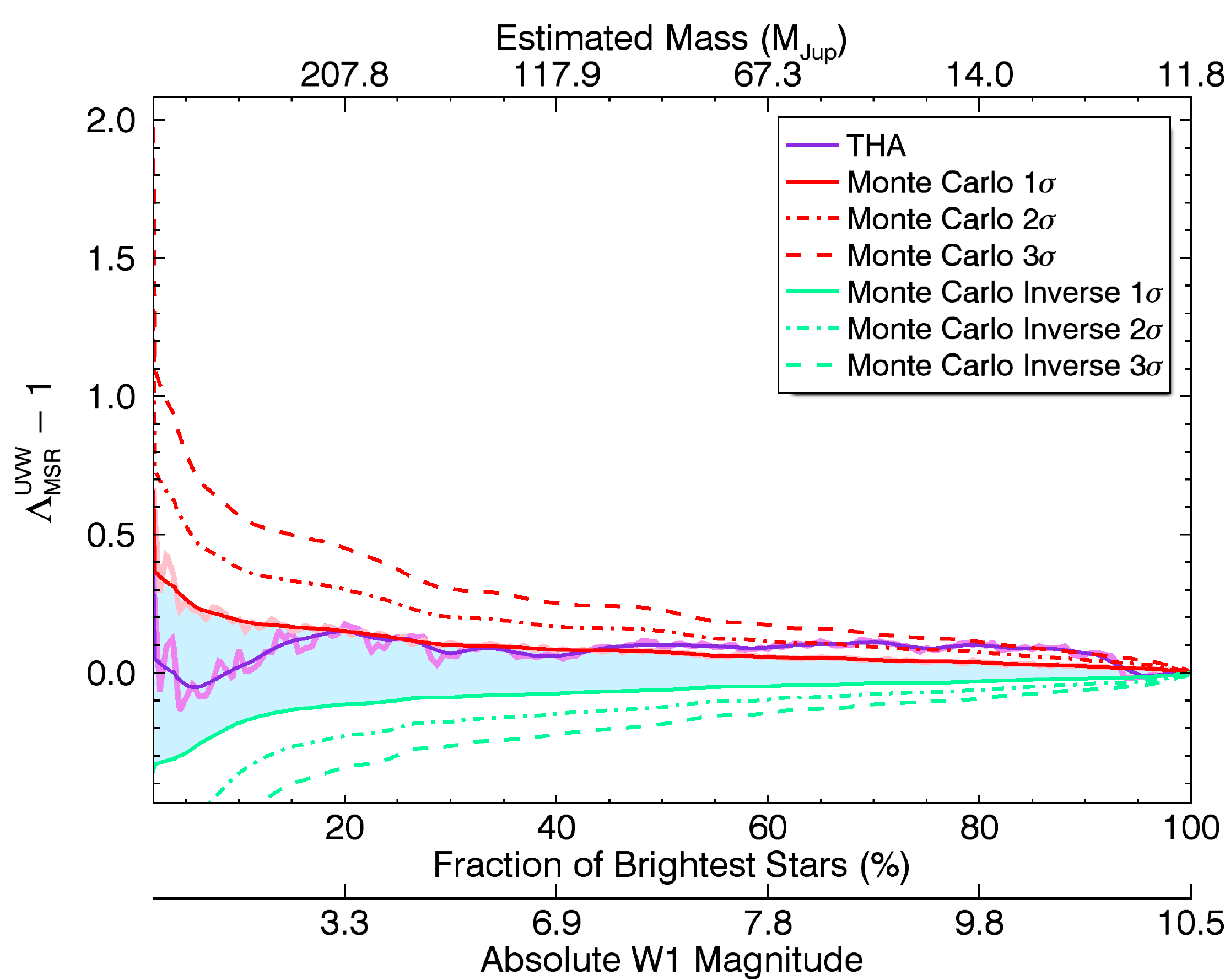}}
	\subfigure{\includegraphics[width=0.495\textwidth]{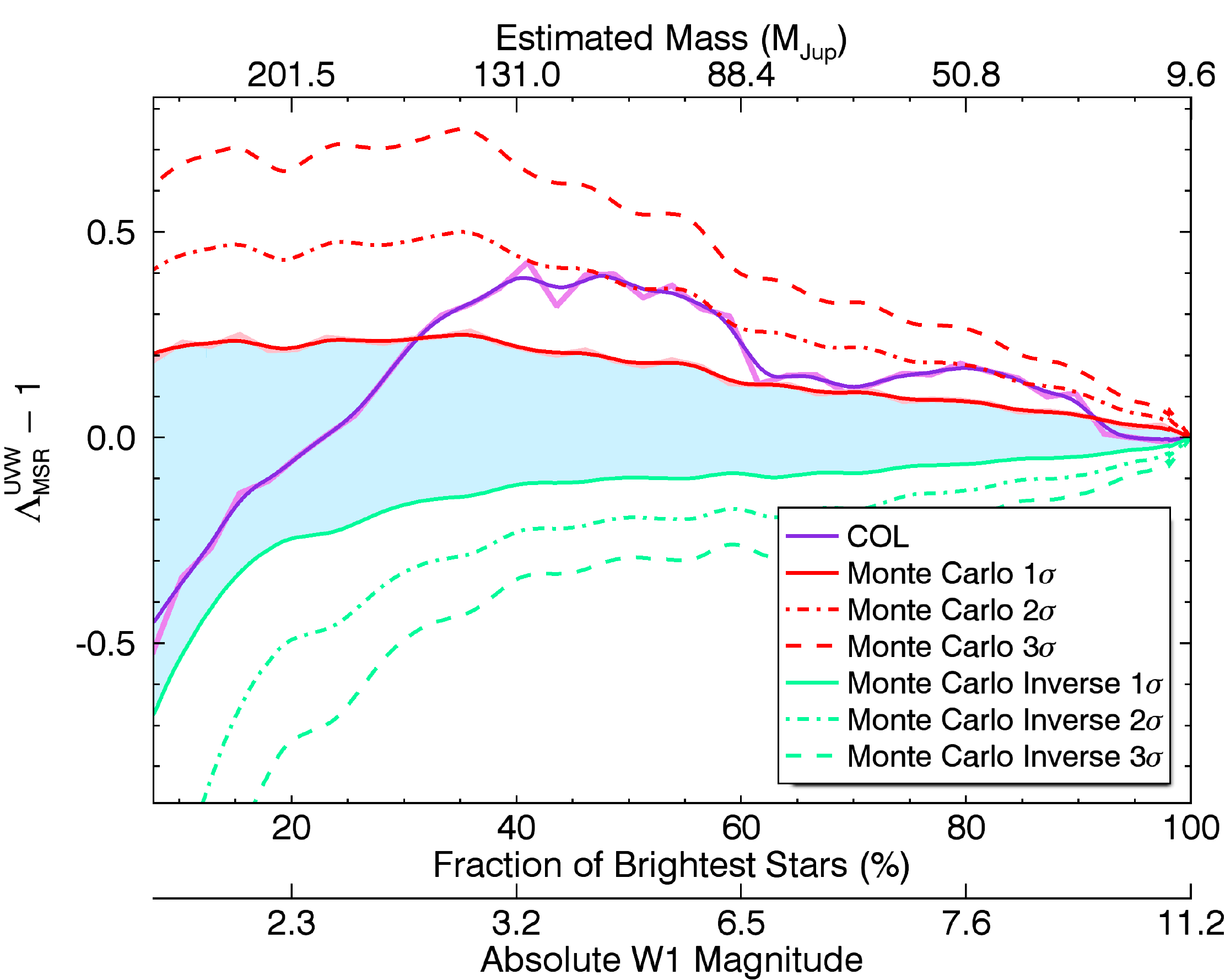}}
	\subfigure{\includegraphics[width=0.495\textwidth]{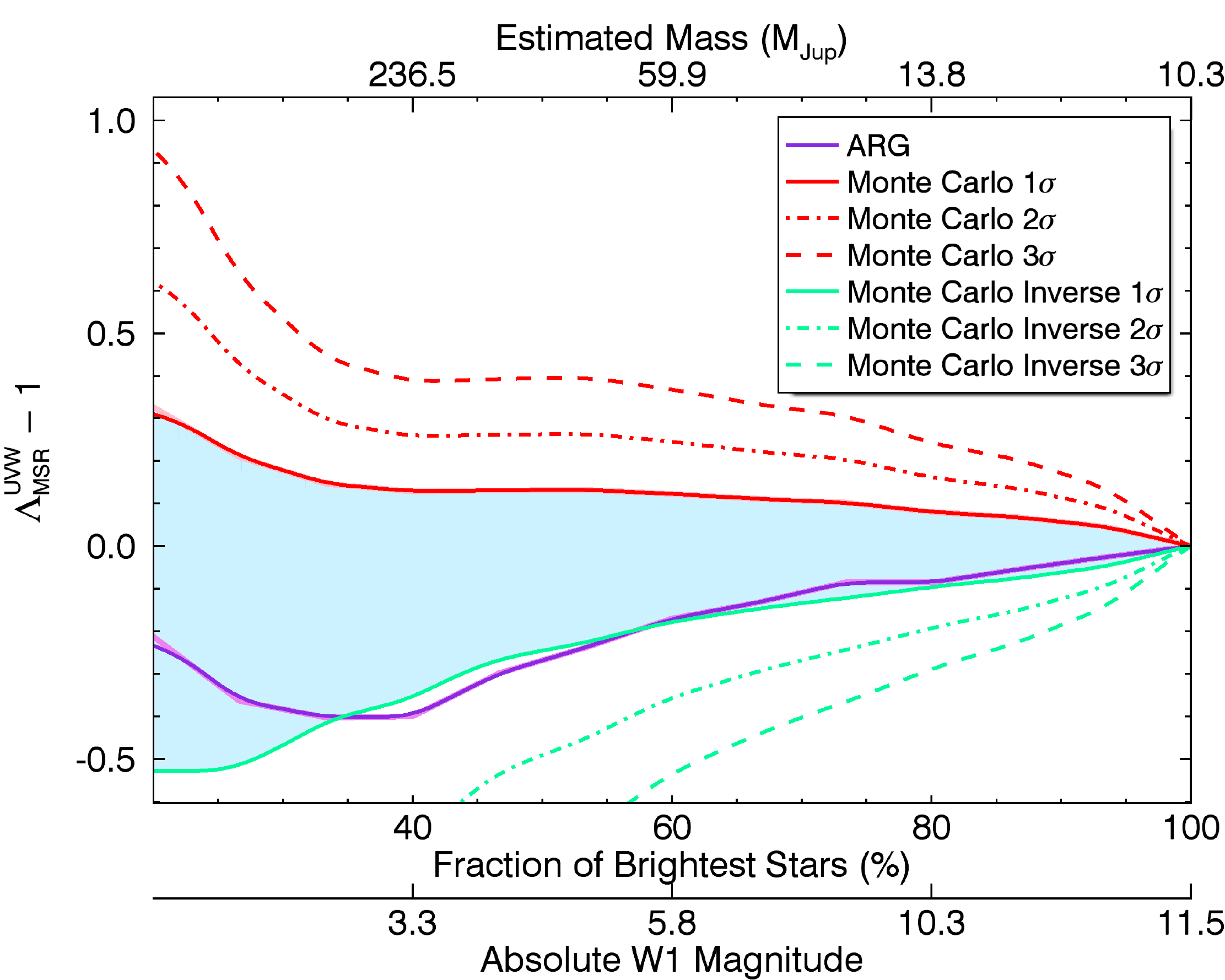}}
	\subfigure{\includegraphics[width=0.495\textwidth]{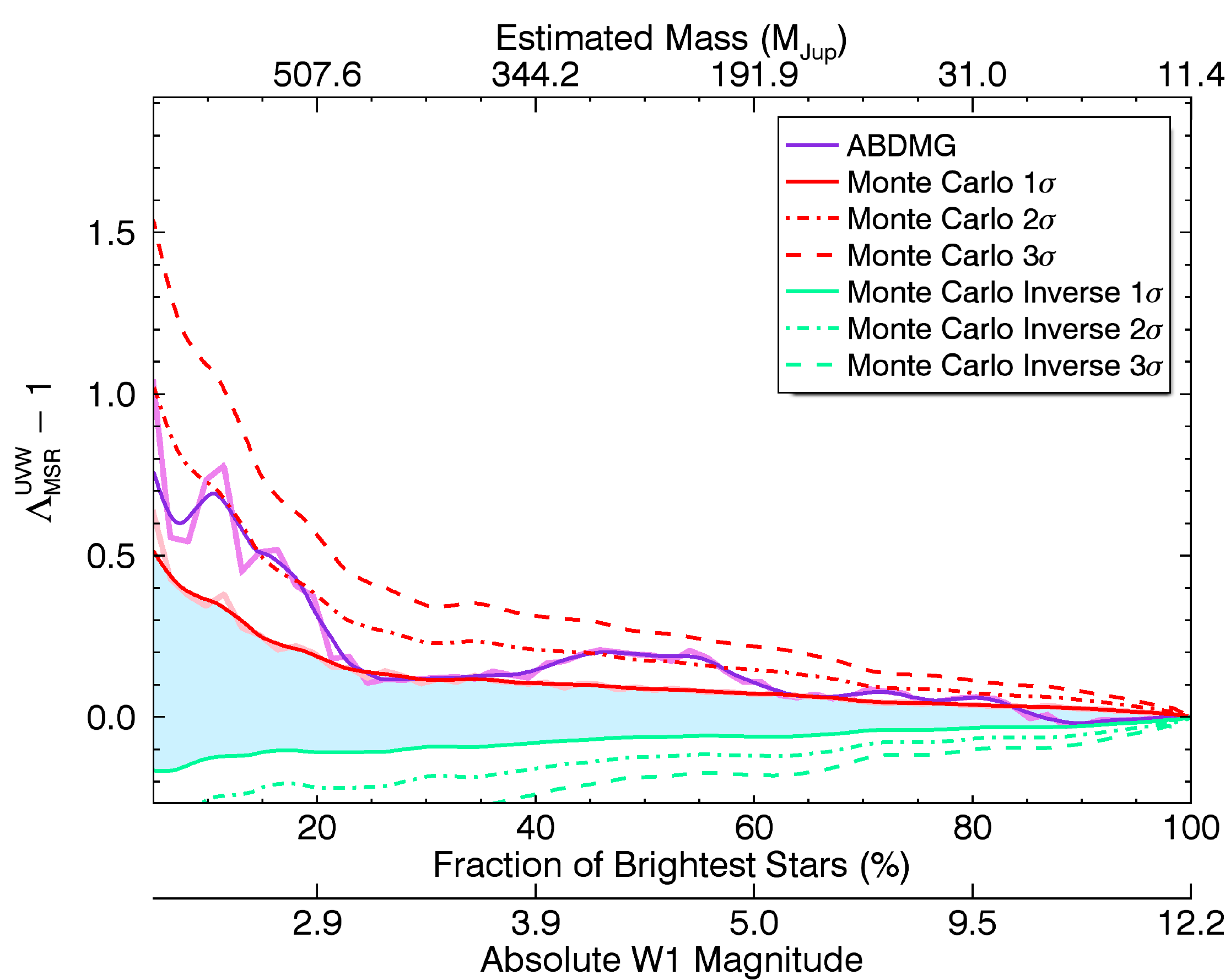}}
	\caption{Same as \hyperref[fig:segre2_bfide_d]{Figure~\ref*{fig:segre2_bfide_d}} with high probability \href{http://www.astro.umontreal.ca/\textasciitilde gagne/BASS.php}{\emph{BASS}} candidates added to the set of bona fide members.}
		\label{fig:segre2_bass_d}
\end{figure*}

We show in Figures~\ref{fig:segre2_bfide_s}--\ref{fig:segre2_bfide_d} the resulting MSRs as a function of $N$ for only bona fide members of each YMG. A MSR larger than one indicates that massive stars are more concentrated towards the center of the distribution, whereas a MSR smaller than one indicates the inverse situation. In most cases with a large statistical significance, the MSR ratio is above unity, which is expected from the physical considerations mentioned above. ABDMG is the only case where both the maximal spatial and dynamical mass segregation are present at $> 2\sigma$, with $2.5\sigma$ and $2.4\sigma$, respectively. $\beta$PMG displays a spatial mass segregation at $2.4\sigma$ and COL displays a dynamical mass segregation at $2.9\sigma$. In some cases ($\beta$PMG, TWA and THA), an inverse spatial or dynamical mass segregation is apparent between 1$\sigma$ and 2$\sigma$, but never at a larger statistical significance. The inclusion of high priority \href{http://www.astro.umontreal.ca/\textasciitilde gagne/BASS.php}{\emph{BASS}} candidates in this analysis (see Figures~\ref{fig:segre2_bass_s}--\ref{fig:segre2_bass_d}) generally increases the significance of the previous results, the only exception being COL. As a consequence, ABDMG, THA and COL display both a maximal dynamical and spatial mass segregation at 2--4$\sigma$ in this situation. Spatial segregation is also apparent for ARG and BPMG at 3.2$\sigma$ and 3.4$\sigma$, respectively. We note that in most cases which are statistically significant, mass segregation only starts appearing at masses lower than 0.3--0.5~$M_\odot$. However, we stress that a follow-up of the \href{http://www.astro.umontreal.ca/\textasciitilde gagne/BASS.php}{\emph{BASS}} sample must be completed before cases other than ABDMG can be considered as significant. We add that even in the case of ABDMG, securing more members will be necessary to increase the statistical significance of this tentative result.

Our analysis does not take account of two effects that could bias our results; (1) the selection criteria imposed to the \href{http://www.astro.umontreal.ca/\textasciitilde gagne/BASS.php}{\emph{BASS}} survey; and (2) the effect of unresolved binaries. To investigate the former effect, we performed a Monte Carlo simulation in which we have drawn a million synthetic objects from each SKM, and rebuilt 500 times the MST corresponding to a random subset of 100 synthetic objects. We repeated this with and without applying the selection filters described in \hyperref[sec:crossmatch]{Section~\ref*{sec:crossmatch}} to assess whether they have any systematic effect on the length of the MST. Any such systematic bias will only affect \href{http://www.astro.umontreal.ca/\textasciitilde gagne/BASS.php}{\emph{BASS}} candidates, which all have masses lower than currently known bona fide members. Hence, if this bias systematically shrinks the MST length, we will have underestimated mass segregation in the analysis described above, and vice versa. We found that our selection bias did not significantly affect the dynamical mass segregation: in all cases, they decreased the length of the dynamical MST with a statistical significance between 0 and 0.1$\sigma$. However, the spatial mass segregation was affected by our selection filters: in all cases, the average length of the MST has also decreased, with statistical significances of $\sim$~1.5$\sigma$ (ABDMG), $\sim$~1.8$\sigma$ (ARG and TWA), $\sim$~2.2$\sigma$ (CAR), $\sim$~2.8$\sigma$ ($\beta$PMG) and $\sim$~3$\sigma$ (COL; THA was unaffected). We have thus likely underestimated any positive spatial mass segregation in our analysis, as well as overestimated any negative spatial mass segregation. Since all of the statistically significant spatial mass segregation ratios obtained here are positive (less massive objects are more spread out), this does not change the conclusions of our analysis, except that we might generally underestimate the statistical significance of these conclusions.

Since we did not account of known and unknown unresolved binaries in our analysis and because the $W1$ flux of an object always falls rapidly when decreasing its mass, we will have systematically overestimated the total mass of unresolved systems. However, there is no apparent reason that would cause the fraction of multiple systems in a given YMG to correlate with \emph{XYZUVW}. Hence, the effect of ignoring unresolved systems will be the same as overestimating the mass and luminosity of a random subset of members that we considered isolated. This addition of noise will thus tend to draw the MSR closer to unity, as well as increase the measurement error on the MSR. As a consequence, this simplification will has made us less sensitive to the detection of any mass segregation, whether it be positive or negative.



\section{SUMMARY AND CONCLUSIONS}\label{sec:conclusion}

We used the \href{http://www.ipac.caltech.edu/2mass/}{\emph{2MASS}} and \href{http://wise2.ipac.caltech.edu/docs/release/allwise/}{\emph{AllWISE}} surveys to perform the first systematic all-sky survey for $\geq$~M5 candidate members of YMGs. We identified a total of 275 M4--L7 candidate members, from which 153 are new strong candidates with an expected overall contamination of 13\% from field stars, from which 79 are expected to be brown dwarfs, and 22 are expected to be planetary-mass objects. We searched for all additional information available in the literature for the \href{http://www.astro.umontreal.ca/\textasciitilde gagne/BASS.php}{\emph{BASS}} sample to update membership probability, and show that we recover 60\% of known $\geq$~M5 candidates to YMGs, whereas most of the remaining 40\% were missed due to the quality filters used to minimize false-positives. Three new common proper motion pairs were discovered among low-probability candidates. We finally used this unique sample to tentatively identify signs of mass segregation in YMGs. We find marginal evidence for mass segregation in ABDMG even when considering only bona fide members, and this result extends to THA and COL when high probability \href{http://www.astro.umontreal.ca/\textasciitilde gagne/BASS.php}{\emph{BASS}} candidates are taken into account. The \href{http://www.astro.umontreal.ca/\textasciitilde gagne/BASS.php}{\emph{BASS}} sample will open the door to the identification of BD members of YMGs, and has already proved extremely fruitful from a number of discoveries previously published. Extensive NIR and optical spectroscopic follow-ups are ongoing and have already enabled the discovery of several new young BDs which will be presented in upcoming papers. Complementary data can be found at our group's website \url{http://www.astro.umontreal.ca/mbderg} and \url{http://www.astro.umontreal.ca/\textasciitilde gagne}, and the \href{http://www.astro.umontreal.ca/\textasciitilde gagne/banyanII.php}{BANYAN~II} web tool is publicly available at \url{http://www.astro.umontreal.ca/\textasciitilde gagne/banyanII.php}.

\acknowledgments
We thank the anonymous referee of this Paper, which provided us with valuable comments that significantly increased the quality of this work. The authors would also like to thank Kelle Cruz, Jacqueline K. Faherty, Philippe Delorme, Adric Riedel, Lo\"ic Albert, Rebecca Oppenheimer, Eric Mamajek, Brendan Bowler, David Blank, Am\'elie Simon and Jonathan Foster for useful comments and discussions and Adric Riedel for sharing data. This work was supported in part through grants from the the Fond de Recherche Qu\'eb\'ecois - Nature et Technologie and the Natural Science and Engineering Research Council of Canada. This research has benefitted from the SpeX Prism Spectral Libraries, maintained by Adam Burgasser at \url{http://pono.ucsd.edu/\textasciitilde adam/browndwarfs/spexprism}, and the \emph{Database of Ultracool Parallaxes} at \url{http://www.cfa.harvard.edu/\textasciitilde tdupuy/plx/Database\_of\_\\ Ultracool\_Parallaxes.html}. This research made use of; the \href{http://simbad.u-strasbg.fr/}{SIMBAD} database and \href{http://vizier.u-strasbg.fr/viz-bin/VizieR}{VizieR} catalog access tools, operated at Centre de Donn\'ees astronomiques de Strasbourg, France \citep{2000AAS..143...23O}; data products from the Two Micron All Sky Survey, which is a joint project of the University of Massachusetts and the Infrared Processing and Analysis Center (IPAC)/California Institute of Technology (Caltech), funded by the National Aeronautics and Space Administration (NASA) and the National Science Foundation \citep{2006AJ....131.1163S}; data products from the Wide-field Infrared Survey Explorer, which is a joint project of the University of California, Los Angeles, and the Jet Propulsion Laboratory (JPL)/Caltech, funded by NASA \citep{2010AJ....140.1868W}; the NASA/IPAC Infrared Science Archive, which is operated by the JPL, Caltech, under contract with NASA; the M, L, and T dwarf compendium housed at \url{http://DwarfArchives.org} and maintained by Chris Gelino, Davy Kirkpatrick, and Adam Burgasser.

\clearpage
\appendix

\section{APPENDIX~A: THE INPUT SAMPLE OF NEARBY POTENTIAL $>$ M5 DWARFS} \label{App:input}

We present in \hyperref[tab:input]{Table~\ref*{tab:input}} the complete sample of 98\,970 potential $>$ M5, nearby objects in which we searched for candidate members to YMGs, which will might prove useful to study the kinematics of such red objects. This table includes all observables that were fed to \href{http://www.astro.umontreal.ca/\textasciitilde gagne/banyanII.php}{BANYAN~II} to determine the Bayesian probability: \href{http://www.ipac.caltech.edu/2mass/}{\emph{2MASS}} and \href{http://wise2.ipac.caltech.edu/docs/release/allwise/}{\emph{AllWISE}} magnitudes, sky position and proper motion determined from the \href{http://www.ipac.caltech.edu/2mass/}{\emph{2MASS}}--\href{http://wise2.ipac.caltech.edu/docs/release/allwise/}{\emph{AllWISE}} cross-match. This list was built from the selection criteria described in 
\hyperref[sec:crossmatch]{Section~\ref*{sec:crossmatch}}, which produced the two following SQL statements that we used to query the \href{http://www.ipac.caltech.edu/2mass/}{\emph{2MASS}} and \href{http://wise2.ipac.caltech.edu/docs/release/allwise/}{\emph{AllWISE}} all-sky catalogs, respectively, on the \href{http://irsa.ipac.caltech.edu/}{IRSA} service :
\\
\noindent\textbullet\ \href{http://www.ipac.caltech.edu/2mass/}{\emph{2MASS}} :
\begin{lstlisting}[frame=single,language=SQL,breaklines=true,linewidth=0.96\textwidth,basicstyle=\ttfamily\scriptsize]
(GLAT > 15 OR GLAT < -15) AND (J_M-H_M) >= 0.506 AND (J_M-H_M) < 2  AND (H_M-K_M) >= 0.269 AND (H_M-K_M) < 1.6 AND (NOT rd_flg LIKE '%0%') AND (NOT rd_flg LIKE '%6%') AND (NOT rd_flg LIKE '%9%') AND bl_flg = '111' AND cc_flg = '000' AND gal_contam = '0' AND J_M > 2 AND H_M > 2 AND K_M > 2 AND (NOT ph_qual LIKE '%D%') AND (NOT ph_qual LIKE '%E%') AND (NOT ph_qual LIKE '%F%') AND (NOT ph_qual LIKE '%X%') AND (NOT ph_qual LIKE '%U%') AND (NOT ph_qual LIKE '%CC%') AND (NOT ph_qual='CAC') AND (NOT ph_qual='CBC') AND PROX > 6.4 AND mp_flg = '0' AND (b_m_opt is null OR (b_m_opt - J_M) >= 4.048) AND (vr_m_opt is null OR (vr_m_opt - J_M) >= 2.63) AND (b_m_opt is null OR vr_m_opt is null OR (b_m_opt - vr_m_opt) >= 1.3)
\end{lstlisting}
\textbullet\ \href{http://wise2.ipac.caltech.edu/docs/release/allwise/}{\emph{AllWISE}} :
\begin{lstlisting}[frame=single,language=SQL,breaklines=true,linewidth=0.96\textwidth,basicstyle=\ttfamily\scriptsize]
(GLAT > 15 OR GLAT < -15) AND (W1MPRO - W2MPRO) >= 0.168 AND (W1MPRO - W2MPRO) < 2.5 AND ( W3SNR < 5 OR (NOT W3SAT = 0) OR ( (W1MPRO - W2MPRO) > (0.96*(W2MPRO - W3MPRO)-0.96) ) ) AND (cc_flags NOT LIKE '_D__' AND cc_flags NOT LIKE 'D___' AND cc_flags NOT LIKE '_O__' AND cc_flags NOT LIKE 'O___' AND cc_flags NOT LIKE '_P__' AND cc_flags NOT LIKE 'P___' AND cc_flags NOT LIKE '_H__' AND cc_flags NOT LIKE 'H___') AND (EXT_FLG = '0' OR EXT_FLG = '1') AND W1SNR > 5 AND W2SNR > 5 AND W1RCHI2 < 5 AND W2RCHI2 < 5 AND W1MPRO > 2 AND W2MPRO > 2 AND W1SAT < 0.002 AND W2SAT < 0.002 AND (PH_QUAL LIKE 'AA%' OR PH_QUAL LIKE 'AB%' OR PH_QUAL LIKE 'BA%' OR PH_QUAL LIKE 'BB%') AND (tmass_key is null OR (R_2MASS >= 0.3 AND (j_m_2MASS - h_m_2MASS) >= 0.506 AND (j_m_2MASS - h_m_2MASS) < 2 AND (h_m_2MASS - k_m_2MASS) >= 0.269 AND (h_m_2MASS - k_m_2MASS) < 1.6 AND (k_m_2MASS - w1mpro) >= 0.153 AND (k_m_2MASS - w1mpro) < 2))
\end{lstlisting}

\capstartfalse
\input{bass_table_A1_short.tex}
\capstarttrue

\section{APPENDIX~B: MARGINALLY RED CANDIDATES} \label{App:marg}

We present here the \emph{Low-Priority} \emph{BASS} (\href{http://www.astro.umontreal.ca/\textasciitilde gagne/LP-BASS.php}{\emph{LP-BASS}}) sample, consisting of all candidates which were rejected from the \href{http://www.astro.umontreal.ca/\textasciitilde gagne/BASS.php}{\emph{BASS}} sample because they were less than 1$\sigma$ redder than the field in the $M_{W1}$ versus $J-K_S$ and $M_{W1}$ versus $H-W2$ CMD diagrams as indicated by the statistical distance of their most probable \href{http://www.astro.umontreal.ca/\textasciitilde gagne/banyanII.php}{BANYAN~II} membership. However, we still only include candidates which are redder than the field sequence. Using the same method as described in the Paper, we estimate contamination fractions of $\sim$ 26\% and $\sim$ 80\% in the high and modest-probability \href{http://www.astro.umontreal.ca/\textasciitilde gagne/LP-BASS.php}{\emph{LP-BASS}} samples. We thus discourage the use of this for statistical studies or time-consuming follow-ups. However, since the spread in the NIR colors of young objects in the two CMD mentioned above are large, we expect that a fraction of young objects will be rejected by our conservative filter which requires candidates to be $> 1\sigma$ redder than the field. It is thus likely that this sample will contain a considerable fraction of true members of YMGs. Candidate members in the \href{http://www.astro.umontreal.ca/\textasciitilde gagne/LP-BASS.php}{\emph{LP-BASS}} are also being following spectroscopically to identify signs of youth, albeit with a lower priority. Results will be presented in subsequent papers.

In \hyperref[tab:lit_annex]{Table~\ref*{tab:lit_annex}}, we show all measurements in the literature which are useful in constraining the membership of the \href{http://www.astro.umontreal.ca/\textasciitilde gagne/LP-BASS.php}{\emph{LP-BASS}} candidate members. We use these measurements to refine results from \href{http://www.astro.umontreal.ca/\textasciitilde gagne/banyanII.php}{BANYAN~II}, and report the final probability and most probable YMG for all \href{http://www.astro.umontreal.ca/\textasciitilde gagne/LP-BASS.php}{\emph{LP-BASS}} objects in \hyperref[tab:allbass_ann]{Table~\ref*{tab:allbass_ann}}.

We note that \href{http://simbad.u-strasbg.fr/simbad/sim-id?submit=display&bibdisplay=refsum&bibyear1=1850&bibyear2=$currentYear&Ident=2MASS%20J00455663%2B3347109}{2MASS~J00455663+3347109} (\href{http://simbad.u-strasbg.fr/simbad/sim-id?submit=display&bibdisplay=refsum&bibyear1=1850&bibyear2=$currentYear&Ident=2MASS%20J00455663%2B3347109}{G~132--25}) had three distinct trigonometric distance measurements in the literature with one being very discrepant : \cite{2002AJ....123.2806R} report $68.0~\pm~18.5$ pc from the \href{http://vizier.u-strasbg.fr/viz-bin/VizieR?-meta.foot&-source=I/238A}{Yale catalog} \citep{1995gcts.book.....V}, \cite{2013MNRAS.435.1083K} measure $20.1~\pm~2.1$ pc, and \cite{2014ApJ...784..156D} measure $17.4~\pm~1.3$ pc. We thus consulted the \href{http://vizier.u-strasbg.fr/viz-bin/VizieR?-meta.foot&-source=I/238A}{Yale catalog} directly to verify the measurement. Sky coordinates are reported as of 1900 in the catalog; we thus used the \href{http://idlastro.gsfc.nasa.gov/ftp/pro/astro/precess.pro}{\emph{precess}} IDL routine from the \href{http://idlastro.gsfc.nasa.gov/}{IDL Astronomy Users Library}\footnote{Available at \url{http://idlastro.gsfc.nasa.gov/}} to precess the coordinates of \href{http://simbad.u-strasbg.fr/simbad/sim-id?submit=display&bibdisplay=refsum&bibyear1=1850&bibyear2=$currentYear&Ident=2MASS%20J00455663%2B3347109}{G~132--25} back to this epoch. We find RA=00h40m32.625s, DEC=33\textdegree14$'$21\farcs78. The closest entry in the \href{http://vizier.u-strasbg.fr/viz-bin/VizieR?-meta.foot&-source=I/238A}{Yale catalog} is that of \href{http://simbad.u-strasbg.fr/simbad/sim-id?submit=display&bibdisplay=refsum&bibyear1=1850&bibyear2=$currentYear&Ident=LP%20294-2}{LP~294--2}, at a distance of 4$'$. Since \href{http://simbad.u-strasbg.fr/simbad/sim-id?submit=display&bibdisplay=refsum&bibyear1=1850&bibyear2=$currentYear&Ident=LP%20294-2}{LP~294--2} has a distinct \href{http://www.ipac.caltech.edu/2mass/}{\href{http://www.ipac.caltech.edu/2mass/}{\emph{2MASS}}} counterpart (\href{http://simbad.u-strasbg.fr/simbad/sim-id?submit=display&bibdisplay=refsum&bibyear1=1850&bibyear2=$currentYear&Ident=LP%20294-2}{2MASS~J00461297+3350108}), we conclude the most probable explanation is that the trigonometric distance of \href{http://simbad.u-strasbg.fr/simbad/sim-id?submit=display&bibdisplay=refsum&bibyear1=1850&bibyear2=$currentYear&Ident=LP%20294-2}{LP~294--2} has been misattributed to \href{http://simbad.u-strasbg.fr/simbad/sim-id?submit=display&bibdisplay=refsum&bibyear1=1850&bibyear2=$currentYear&Ident=2MASS%20J00455663%2B3347109}{G~132--25} in \cite{2002AJ....123.2806R}. We thus rejected this measurement and combined the two others to obtain $18.1~\pm~1.3$ pc in \hyperref[tab:lit_annex]{Table~\ref*{tab:lit_annex}}.

\capstartfalse
\input{bass_table_B1.tex}
\capstarttrue

\capstartfalse
\input{bass_table_B2_short.tex}
\capstarttrue

\bibliographystyle{apj}

\input{BASS_Arxiv.bbl}
\end{document}

%% file: bass_table_1.tex
\onecolumngrid
\LongTables
\begin{deluxetable*}{lcccccc}
\tablecolumns{7}
\tablecaption{Expected Completeness of the \href{http://www.astro.umontreal.ca/\textasciitilde gagne/BASS.php}{\emph{BASS}} Survey.\label{tab:incompleteness}}
\tablehead{\colhead{YMG Name} & \colhead{$|b| \leq 15$\textdegree} & \colhead{$\mu \leq 30$~\masyr} & \colhead{SFRs\tablenotemark{a}} & \colhead{SFRs or $|b| \leq 15$\textdegree\ or $\mu \leq 30$~\masyr\tablenotemark{b}} & \colhead{Contamination $\geq 50$\%} & \colhead{Expected Completeness}}
\startdata
ARG        & 42.1\%    & 0.5\%     &  0.6\% & 42.6\%    & 89.6\% & 6.0\% \\
COL        & 15.7\%    & 23.4\%    &  1.8\% & 36.4\%    & 59.7\% & 25.6\% \\
$\beta$PMG & 25.2\%    & 0.8\%     &  3.4\% & 28.3\%    & 60.0\% & 28.7\% \\
ABDMG      & 20.7\%    & 1.1\%     &  1.6\% & 22.8\%    & 59.6\% & 31.2\% \\
CAR        & 41.2\%    & 2.7\%     &  0.1\% & 42.9\%    & 9.9\%  & 51.4\% \\
TWA        & 19.7\%    & 0.4\%     &    0\% & 20.0\%    & 10.3\% & 71.8\% \\
THA        & $<$ 0.1\% & $<$ 0.1\% &    0\% & $<$ 0.1\% & 10.0\% & 90.0\% \\
\enddata
\tablenotetext{a}{Expected fraction of members aligned with Orion, Taurus, Chamaeleon and Upper Scorpius (see 
\hyperref[sec:crossmatch]{Section~\ref*{sec:crossmatch}}).}
\tablenotetext{b}{Filters on position and proper motion are not independent.}
\end{deluxetable*}
\twocolumngrid

%% file: bass_table_2_short.tex
\onecolumngrid
\LongTables
\begin{deluxetable*}{llcccccccccclcc}
\tablecolumns{15}
\tablecaption{All-Sky Search for > M5 Candidates in Young Moving Groups.\label{tab:bass1}}
\tablehead{\colhead{\href{http://www.ipac.caltech.edu/2mass/}{\emph{2MASS}}} & \colhead{Estim.} & \colhead{} & \multicolumn{3}{c}{\href{http://www.ipac.caltech.edu/2mass/}{\emph{2MASS}}} & \colhead{} & \multicolumn{2}{c}{\href{http://wise2.ipac.caltech.edu/docs/release/allwise/}{\emph{AllWISE}}} & \colhead{} & \colhead{$\mu_\alpha\cos\delta$} & \colhead{$\mu_\delta$}  & \colhead{Member-} & \colhead{Bayesian} & \colhead{Contamination} \\
\cline{4-6}
\cline{8-9}
\colhead{Designation} & \colhead{SpT} & \colhead{} & \colhead{$J$} & \colhead{$H$} & \colhead{$K_S$} & \colhead{} & \colhead{$W1$} & \colhead{$W2$} & \colhead{} & \colhead{(\masyr)} & \colhead{(\masyr)} & \colhead{ship} & \colhead{Prob. (\%)} & \colhead{Prob. (\%)}}
\startdata
\cutinhead{Candidates with a High Probability}
00011217+1535355 & L3.2 & & $15.52$ & $14.51$ & $13.71$ & & $12.97$ & $12.54$ & & $139.6 \pm 7.8$ & $-183.5 \pm 11.8$ & ABDMG  & $79.1$ & $1.6$\\
00040288-6410358 & L2.5 & & $15.79$ & $14.83$ & $14.01$ & & $13.41$ & $12.96$ & & $77.7 \pm 3.0$ & $-56.1 \pm 8.4$ & THA\tablenotemark{a} & $99.9$ & $< 0.1$\\
00041589-8747254 & M5.7 & & $12.90$ & $12.20$ & $11.86$ & & $11.65$ & $11.41$ & & $77.3 \pm 2.0$ & $-29.9 \pm 9.2$ & THA  & $55.4$ & $< 0.1$\\
00065794-6436542 & M6.9 & & $13.39$ & $12.66$ & $12.17$ & & $11.74$ & $11.42$ & & $92.7 \pm 3.1$ & $-71.0 \pm 7.3$ & THA\tablenotemark{a} & $99.9$ & $< 0.1$\\
00111532-3756553 & M5.7 & & $12.15$ & $11.60$ & $11.22$ & & $11.02$ & $10.79$ & & $105.7 \pm 5.0$ & $-77.4 \pm 7.4$ & THA  & $80.2$ & $< 0.1$\\
00182834-6703130 & M9.6 & & $15.46$ & $14.48$ & $13.71$ & & $13.19$ & $12.80$ & & $83.6 \pm 2.9$ & $-65.0 \pm 9.3$ & THA\tablenotemark{a} & $99.8$ & $< 0.1$\\
00191296-6226005 & M9.7 & & $15.64$ & $14.62$ & $13.96$ & & $13.38$ & $12.96$ & & $66.1 \pm 2.9$ & $-50.6 \pm 8.4$ & THA  & $99.5$ & $< 0.1$\\
00212774-6351081 & M4.0 & & $11.02$ & $10.48$ & $10.11$ & & $9.91$ & $9.66$ & & $83.0 \pm 2.9$ & $-57.6 \pm 7.2$ & THA  & $99.8$ & $< 0.1$\\
00235732-5531435 & M4.5 & & $11.11$ & $10.55$ & $10.24$ & & $10.07$ & $9.87$ & & $92.3 \pm 3.4$ & $-67.7 \pm 7.4$ & THA\tablenotemark{a} & $99.5$ & $< 0.1$\\
00305785-6550058\tablenotemark{b} & M2.1 & & $9.82$ & $9.24$ & $8.95$ & & $8.79$ & $8.61$ & & $70.3 \pm 2.9$ & $-51.9 \pm 8.7$ & THA & $99.1$ & $< 0.1$\\
\cutinhead{Candidates with a Modest Probability}
00160844-0043021 & L4.0 & & $16.33$ & $15.23$ & $14.54$ & & $13.84$ & $13.39$ & & $138.3 \pm 9.9$ & $-33.7 \pm 14.2$ & BPMG & $18.8$ & $36.4$\\
00192626+4614078 & M5.9 & & $12.60$ & $11.94$ & $11.50$ & & $11.28$ & $11.02$ & & $119.6 \pm 6.1$ & $-82.5 \pm 6.9$ & ABDMG  & $53.3$ & $17.5$\\
00274534-0806046 & M5.3 & & $11.57$ & $10.97$ & $10.61$ & & $10.41$ & $10.18$ & & $111.5 \pm 7.0$ & $-59.9 \pm 6.7$ & BPMG & $45.6$ & $35.1$\\
00390342+1330170 & M5.1 & & $10.94$ & $10.37$ & $10.06$ & & $9.84$ & $9.65$ & & $109.8 \pm 6.8$ & $-96.5 \pm 7.0$ & BPMG & $57.9$ & $15.3$\\
00464841+0715177 & M8.2 & & $13.89$ & $13.18$ & $12.55$ & & $12.09$ & $11.64$ & & $97.0 \pm 9.2$ & $-60.3 \pm 7.3$ & BPMG\tablenotemark{a} & $78.5$ & $28.4$\\
00581143-5653326 & L6.1 & & $16.78$ & $15.55$ & $14.55$ & & $13.76$ & $13.24$ & & $197.4 \pm 6.2$ & $46.0 \pm 12.2$ & ARG  & $80.4$ & $32.9$\\
01033203+1935361 & L6.2 & & $16.29$ & $14.90$ & $14.15$ & & $13.18$ & $12.70$ & & $303.0 \pm 13.4$ & $16.6 \pm 7.2$ & ARG  & $31.7$ & $16.9$\\
01525534-6329301 & M4.7 & & $10.17$ & $9.60$ & $9.26$ & & $9.06$ & $8.84$ & & $130.0 \pm 3.5$ & $7.0 \pm 6.4$ & BPMG  & $71.4$ & $22.1$\\
02534448-7959133 & M5.4 & & $11.34$ & $10.74$ & $10.38$ & & $10.18$ & $9.97$ & & $81.7 \pm 2.2$ & $90.3 \pm 9.3$ & BPMG  & $71.8$ & $24.9$\\
03390160-2434059 & M3.7 & & $10.90$ & $10.34$ & $9.97$ & & $9.72$ & $9.52$ & & $56.3 \pm 5.7$ & $-12.7 \pm 6.0$ & COL & $60.5$ & $32.9$\\
\enddata
\tablenotetext{a}{The binary hypothesis is more probable than the single hypothesis (see \hyperref[sec:filters]{Section~\ref*{sec:filters}}).}
\tablenotetext{b}{Object from the \href{http://irsa.ipac.caltech.edu/Missions/wise.html}{\emph{WISE}} catalog rather than \href{http://wise2.ipac.caltech.edu/docs/release/allwise/}{\emph{AllWISE}}.}
\tablenotetext{~}{This table is available in its entirety in the online journal. The complete table has 263 rows (239 high probability candidates and 24 modest probability candidates).}
\end{deluxetable*}
\twocolumngrid

%% file: bass_table_3.tex
\onecolumngrid
\LongTables
\begin{deluxetable*}{lllcccllc}
\tablecolumns{9}
\tablecaption{Candidates With Additional Information in the Literature.\label{tab:lit}}
\tablehead{\colhead{\href{http://www.ipac.caltech.edu/2mass/}{\emph{2MASS}}} & \colhead{Measured} & \colhead{Signs of} & \colhead{RV} & \colhead{Trig.}  & \colhead{Multipli-} & \colhead{Known} & \colhead{Updated} & \colhead{Updated} \\
\colhead{Name} & \colhead{SpT\tablenotemark{a}} & \colhead{Youth\tablenotemark{b}} & \colhead{(\kms)} & \colhead{Dist. (pc)} & \colhead{city\tablenotemark{c}} & \colhead{Membership} & \colhead{Membership} & \colhead{Prob. (\%)}}
\startdata
00011217+1535355 & L4:$^{51}$ & $\cdots$ & $\cdots$ & $\cdots$ & $\cdots$ & $\cdots$ & ABDMG & $77.8$\\
00040288-6410358 & L1$\gamma^{49}$ & OR$^{49}$ & $\cdots$ & $\cdots$ & $\cdots$ & THA$^{49,31}$ & THA & $> 99.9$\\
00065794-6436542 & M9:$^{82}$ & OH$^{72}$ & $\cdots$ & $\cdots$ & $\cdots$ & THA$^{31}$ & THA & $> 99.9$\\
00160844-0043021 & L5.5$^{51}$ & $\cdots$ & $\cdots$ & $\cdots$ & $\cdots$ & $\cdots$ & BPMG & $19.1$\\
00192626+4614078 & M8$^{94}$ & LH$^{83,94}$ & $-19.5 \pm 3.0^{83}$ & $\cdots$ & $\cdots$ & ABDMG$^{92,31}$ & ABDMG & $92.1$\\
00212774-6351081 & M5.5$^{50}$ & $\cdots$ & $\cdots$ & $\cdots$ & $\cdots$ & $\cdots$ & THA & $99.8$\\
00235732-5531435 & M4.1$^{53}$ & $\cdots$ & $5.3 \pm 0.7^{53}$ & $\cdots$ & $\cdots$ & THA$^{53}$ & THA & $99.8$\\
00325584-4405058 & L0$\gamma^{16,71}$ & OITRH$^{16,71}$ & $\cdots$ & $26.4 \pm 3.3^{28}$ & $\cdots$ & BPMG$^{31}$ & BPMG & $97.7$\\
00354313+0233137 & M5+M6$^{55}$ & $\cdots$ & $\cdots$ & $\cdots$ & AB$^{55}$ & $\cdots$ & ABDMG & $88.4$\\
00374306-5846229 & L0$\gamma^{82,16}$ & OR$^{16}$ & $\cdots$ & $\cdots$ & $\cdots$ & THA$^{31}$ & THA & $99.9$\\
00390342+1330170 & $\cdots$ & XN$^{92}$ & $\cdots$ & $\cdots$ & $\cdots$ & ABDMG$^{92}$ & BPMG & $91.9$\\
00413538-5621127 & M6.5+M9$^{94}$ & VHLA$^{90}$ & $2.8 \pm 1.9^{83,34,31}$ & $\cdots$ & AB$^{94}$ & THA$^{31}$ & THA & $> 99.9$\\
00452143+1634446 & L2$\beta^{82,16}$ & OITRH$^{16}$ & $3.3 \pm 0.2^{4}$ & $17.5 \pm 0.6^{114}$ & $\cdots$ & ARG$^{31}$ & ARG & $98.0$\\
00464841+0715177 & M9$^{82,118}$ & $\cdots$ & $\cdots$ & $\cdots$ & $\cdots$ & $\cdots$ & BPMG & $77.0$\\
00514081-5913320 & M4.4$^{53}$ & $\cdots$ & $6.3 \pm 1.3^{53}$ & $\cdots$ & $\cdots$ & THA$^{53}$ & THA & $99.9$\\
01033203+1935361 & L6$\beta^{28,119}$ & OITR$^{27,28}$ & $\cdots$ & $21.3 \pm 3.4^{28}$ & $\cdots$ & ARG$^{31}$ & ARG & $78.2$\\
01033563-5515561 & M5.5$^{19,53}$ & OHU$^{19,89}$ & $5.2 \pm 1.6^{68,53}$ & $47.2 \pm 3.1^{89}$ & AB$^{19}$ & THA;CAR$^{19,89}$ & THA & $99.9$\\
01134031-5939346 & M5.0$^{53}$ & $\cdots$ & $11.9 \pm 6.7^{53}$ & $\cdots$ & $\cdots$ & THA$^{53}$ & THA & $99.7$\\
01174748-3403258 & L1$\beta^{14,2}$ & TRM$^{7,112,2}$ & $\cdots$ & $\cdots$ & $\cdots$ & THA$^{31}$ & THA & $99.6$\\
01180670-6258591 & M5.1$^{53}$ & L$^{53}$ & $9.3 \pm 1.3^{53}$ & $\cdots$ & $\cdots$ & THA$^{53}$ & THA & $> 99.9$\\
01231125-6921379 & M8$^{94}$ & UL$^{83}$ & $10.9 \pm 3.0^{83}$ & $42.2 \pm 4.8^{87}$ & $\cdots$ & THA$^{31}$\tablenotemark{d} & THA & $> 99.9$\\
01243060-3355014 & M4.5$^{89}$ & OU$^{89}$ & $18.3 \pm 0.5^{100}$ & $25.3 \pm 0.8^{100,89}$ & C$^{106}$ & ABDMG$^{88,100}$\tablenotemark{d} & ABDMG & $> 99.9$\\
01294256-0823580 & M5$^{81}$ & $\cdots$ & $\cdots$ & $\cdots$ & $\cdots$ & $\cdots$ & BPMG & $66.2$\\
01344601-5707564 & M4.9$^{53}$ & L$^{53}$ & $11.1 \pm 6.3^{53}$ & $\cdots$ & $\cdots$ & THA$^{53}$ & THA & $99.8$\\
01372781-4558261 & M5.0$^{53}$ & L$^{53}$ & $13.5 \pm 1.4^{53}$ & $\cdots$ & $\cdots$ & THA$^{53}$ & THA & $97.8$\\
01415823-4633574 & L0$\gamma^{120,16}$ & OITRHM$^{120,16}$ & $12.0 \pm 15.0^{53}$ & $\cdots$ & $\cdots$ & THA$^{31}$ & THA & $99.5$\\
01443191-4604318 & M5.5$^{76}$ & $\cdots$ & $\cdots$ & $\cdots$ & $\cdots$ & $\cdots$ & THA & $99.1$\\
01504543-5716488 & M5.5$^{53}$ & L$^{53}$ & $9.3 \pm 1.7^{53}$ & $\cdots$ & $\cdots$ & THA$^{53}$ & THA & $> 99.9$\\
01531463-6744181 & L2:$^{82}$ & $\cdots$ & $\cdots$ & $\cdots$ & $\cdots$ & $\cdots$ & THA & $99.9$\\
01532494-6833226 & M5.1$^{90,53}$ & N$^{90}$ & $9.8 \pm 1.4^{53}$ & $\cdots$ & $\cdots$ & THA$^{90,53}$ & THA & $> 99.9$\\
02153328-5627175 & M5.4$^{90,53}$ & LN$^{53}$ & $11.3 \pm 5.7^{53}$ & $\cdots$ & $\cdots$ & THA$^{90,53}$ & THA & $99.8$\\
02180960-6657524 & M4.5$^{53}$ & L$^{53}$ & $11.0 \pm 1.2^{53}$ & $\cdots$ & $\cdots$ & THA$^{53}$ & THA & $> 99.9$\\
02192210-3925225 & M4.9$^{53}$ & L$^{53}$ & $10.6 \pm 0.7^{53}$ & $\cdots$ & $\cdots$ & THA$^{53}$ & THA & $> 99.9$\\
02212859-6831400 & M8:$^{82}$ & OR$^{27}$ & $\cdots$ & $39.4 \pm 5.6^{28}$ & $\cdots$ & $\cdots$ & ABDMG & $40.8$\\
02215494-5412054 & M8$\beta^{82,27}$ & OR$^{16}$ & $\cdots$ & $\cdots$ & $\cdots$ & THA$^{31}$ & THA & $99.8$\\
02235464-5815067 & L0$\gamma^{82,27}$ & OR$^{82}$ & $\cdots$ & $\cdots$ & $\cdots$ & THA$^{31}$ & THA & $> 99.9$\\
02251947-5837295 & M9$^{82,27}$ & O$^{82}$ & $\cdots$ & $\cdots$ & $\cdots$ & THA$^{31}$ & THA & $99.9$\\
02294869-6906044 & M4.6$^{53}$ & L$^{53}$ & $13.0 \pm 1.2^{53}$ & $\cdots$ & $\cdots$ & THA$^{53}$ & THA & $> 99.9$\\
02321934-5746117 & M4.4$^{90,53}$ & $\cdots$ & $11.2 \pm 0.7^{53}$ & $\cdots$ & $\cdots$ & THA$^{53}$ & THA & $> 99.9$\\
02340093-6442068 & L0$\gamma^{29}$ & OR$^{29}$ & $\cdots$ & $\cdots$ & $\cdots$ & THA$^{49,31}$ & THA & $99.8$\\
02401209-5305527 & M9.5$^{72}$ & $\cdots$ & $\cdots$ & $\cdots$ & $\cdots$ & $\cdots$ & THA & $99.9$\\
02411151-0326587 & L0$\gamma^{64,15,16,48}$ & OTR$^{15,16,2}$ & $\cdots$ & $46.7 \pm 5.7^{114}$ & $\cdots$ & THA$^{31}$ & THA & $98.3$\\
02435103-5432194 & M9$^{82}$ & $\cdots$ & $\cdots$ & $\cdots$ & $\cdots$ & $\cdots$ & THA & $99.9$\\
02501167-0151295 & $\cdots$ & $\cdots$ & $\cdots$ & $33.1 \pm 4.9^{107}$ & $\cdots$ & $\cdots$ & BPMG & $88.3$\\
02523550-7831183 & M4.4$^{53}$ & $\cdots$ & $12.8 \pm 1.3^{53}$ & $\cdots$ & $\cdots$ & THA$^{53}$ & THA & $98.6$\\
02534448-7959133 & M5.5$^{76}$ & H$^{56}$ & $\cdots$ & $\cdots$ & $\cdots$ & $\cdots$ & BPMG & $50.1$\\
03014892-5903021 & M9$^{82,72}$ & $\cdots$ & $\cdots$ & $\cdots$ & $\cdots$ & $\cdots$ & THA & $99.9$\\
03032042-7312300 & L2$\gamma^{49}$ & OR$^{49}$ & $\cdots$ & $\cdots$ & $\cdots$ & THA$^{49,31}$ & THA & $78.2$\\
03050556-5317182 & M5.4$^{90,53}$ & N$^{90}$ & $12.1 \pm 2.2^{53}$ & $\cdots$ & $\cdots$ & THA$^{90,53}$ & THA & $99.9$\\
03093877-3014352 & M4.7$^{53}$ & L$^{53}$ & $12.5 \pm 2.3^{53}$ & $\cdots$ & $\cdots$ & THA$^{53}$ & THA & $99.9$\\
03114544-4719501 & M4.3$^{90,53}$ & N$^{90}$ & $11.3 \pm 0.5^{53}$ & $\cdots$ & $\cdots$ & THA$^{90,53}$ & THA & $> 99.9$\\
03152363-5342539 & M5.2$^{90}$ & N$^{90}$ & $\cdots$ & $\cdots$ & $\cdots$ & THA$^{90}$ & THA & $99.9$\\
03164512-2848521 & L0:$^{14}$ & $\cdots$ & $\cdots$ & $\cdots$ & $\cdots$ & $\cdots$ & ABDMG & $77.2$\\
03231002-4631237 & L0$\gamma^{82,27}$ & ORL$^{16}$ & $\cdots$ & $\cdots$ & $\cdots$ & THA$^{31}$ & THA & $99.7$\\
03252938-4312299 & M9$^{82,72}$ & $\cdots$ & $\cdots$ & $\cdots$ & $\cdots$ & $\cdots$ & THA & $78.9$\\
03264225-2102057 & L4$^{15}$ & ORL$^{15}$ & $\cdots$ & $\cdots$ & $\cdots$ & ABDMG$^{31}$ & ABDMG & $98.9$\\
03363144-2619578 & M5.7$^{90}$ & N$^{90}$ & $\cdots$ & $43.5 \pm 3.8^{89}$ & $\cdots$ & THA$^{90}$ & THA & $99.9$\\
03390160-2434059 & M5.9$^{90}$ & N & $\cdots$ & $\cdots$ & $\cdots$ & COL$^{90}$ & COL & $77.8$\\
03393521-3525440 & L0$\beta^{2,44,27}$ & TLM$^{27,112,2}$ & $9.3 \pm 1.7^{83,77}$ & $6.41 \pm 0.04^{22}$ & $\cdots$ & CAS$^{86,31}$ & ARG & $87.6$\\
03421621-6817321 & L2:$^{15}$ & R$^{27}$ & $\cdots$ & $\cdots$ & $\cdots$ & THA$^{31}$ & THA & $99.7$\\
03550477-1032415 & M8.5$^{14,27}$ & $\cdots$ & $\cdots$ & $\cdots$ & $\cdots$ & $\cdots$ & BPMG & $39.5$\\
03552337+1133437 & L5$\gamma^{82,27}$ & OITRL$^{16}$ & $11.9 \pm 0.2^{4}$ & $9.1 \pm 0.1^{61,28}$ & AB$^{3}$ & ABDMG$^{61,31}\tablenotemark{d}$ & ABDMG & $99.7$\\
03572695-4417305 & M9$\beta$+L1.5$\beta^{60}$ & OR$^{16}$ & $\cdots$ & $\cdots$ & AB$^{60}$ & THA$^{31}$ & THA & $99.9$\\
03582255-4116060 & L5$^{82,27}$ & $\cdots$ & $\cdots$ & $\cdots$ & $\cdots$ & $\cdots$ & BPMG & $36.8$\\
04174743-2129191 & M8$^{15,27}$ & $\cdots$ & $\cdots$ & $\cdots$ & $\cdots$ & $\cdots$ & THA & $57.7$\\
04210718-6306022 & L5$\gamma^{82,27}$ & OIRL$^{15}$ & $\cdots$ & $\cdots$ & $\cdots$ & ARG;BPMG$^{31}$ & ARG & $97.7$\\
04362788-4114465 & M8$\gamma^{15,2}$ & OITR$^{15.2}$ & $\cdots$ & $\cdots$ & $\cdots$ & COL$^{31}$ & COL & $97.6$\\
04433761+0002051 & M9$\gamma^{15,2}$ & OITVRHL$^{27,72,31,2}$ & $17.1 \pm 3.0^{83}$ & $\cdots$ & $\cdots$ & BPMG$^{31,92}$ & BPMG & $99.8$\\
04532647-1751543 & L3:$^{14}$ & $\cdots$ & $\cdots$ & $\cdots$ & $\cdots$ & $\cdots$ & COL & $95.8$\\
04533604-2835349 & $\cdots$ & $\cdots$ & $22.5 \pm 6.7^{52}$ & $\cdots$ & $\cdots$ & $\cdots$ & COL & $87.6$\\
05002100+0330501 & L3$\gamma^{82,27}$ & $\cdots$ & $15.9 \pm 0.2^{4}$ & $\cdots$ & $\cdots$ & $\cdots$ & ABDMG & $62.8$\\
05012406-0010452 & L3$\gamma^{82,16,2}$ & OTRL$^{27}$ & $\cdots$ & $14.7 \pm 2.8^{28,114}$ & $\cdots$ & FIELD$^{31}$ & CAR & $97.7$\\
05120636-2949540 & L4:$^{14}$ & R$^{48}$ & $\cdots$ & $\cdots$ & $\cdots$ & BPMG$^{31}$ & BPMG & $33.8$\\
05181131-3101529 & M6.5$^{12}$ & $\cdots$ & $\cdots$ & $\cdots$ & $\cdots$ & $\cdots$ & COL & $93.7$\\
05361998-1920396 & L2$\gamma^{29}$ & OITR$^{29,2}$ & $\cdots$ & $39.0 \pm 14.0^{28}$ & $\cdots$ & COL$^{31}$ & COL & $96.6$\\
06022216+6336391 & L1:$^{82}$ & $\cdots$ & $\cdots$ & $\cdots$ & $\cdots$ & $\cdots$ & ABDMG & $26.1$\\
06420559+4101599 & L/Tp$^{65}$ & R$^{65}$ & $\cdots$ & $\cdots$ & $\cdots$ & ABDMG$^{31}$ & ABDMG & $38.4$\\
06524851-5741376 & M8$\beta^{82,27}$ & OR$^{82,27}$ & $\cdots$ & $32.0 \pm 3.3^{28}$ & AB$^{10}$ & ABDMG$^{31}$ & CAR & $87.9$\\
08095903+4434216 & L6$^{51,116}$ & $\cdots$ & $\cdots$ & $\cdots$ & $\cdots$ & $\cdots$ & ARG & $30.7$\\
09455843-3253299 & M4.5$^{85}$ & X$^{85}$ & $\cdots$ & $\cdots$ & $\cdots$ & $\cdots$ & ARG & $89.2$\\
09532126-1014205 & L0$^{15}$ & $\cdots$ & $\cdots$ & $\cdots$ & $\cdots$ & $\cdots$ & CAR & $63.7$\\
10284580-2830374 & M5$^{96}$ & $\cdots$ & $\cdots$ & $\cdots$ & $\cdots$ & TWA$^{96}$ & TWA & $96.3$\\
10582800-1046304 & M4$^{91}$ & $\cdots$ & $\cdots$ & $\cdots$ & $\cdots$ & $\cdots$ & TWA & $4.3$\\
10584787-1548172 & L3$^{36}$ & $\cdots$ & $\cdots$ & $17.3 \pm 0.3^{18}$ & $\cdots$ & $\cdots$ & ARG & $93.1$\\
11020983-3430355 & M8.5$\gamma^{28,116}$ & $\cdots$ & $\cdots$ & $56.4 \pm 1.6^{104}$ & $\cdots$ & TWA$^{116}$ & TWA & $99.8$\\
11393382-3040002 & M4.7$^{96}$ & $\cdots$ & $\cdots$ & $\cdots$ & $\cdots$ & TWA$^{96}$ & TWA & $99.0$\\
11395113-3159214 & M8$\gamma^{38,82,2}$ & OITRM$^{38,112}$ & $11.2 \pm 2.0^{69}$ & $28.5 \pm 3.5^{28}$ & $\cdots$ & TWA$^{38,69}$\tablenotemark{d} & TWA & $99.8$\\
12073346-3932539 & M8$^{48}$ & ORL$^{48,17,2}$ & $\cdots$ & $52.3 \pm 1.1^{24}$ & Ab$^{10}$ & TWA$^{38,69}$\tablenotemark{d} & TWA & $99.6$\\
12074836-3900043 & L1$\gamma^{32}$ & OITR$^{32}$ & $\cdots$ & $\cdots$ & $\cdots$ & TWA$^{32}$ & TWA & $99.7$\\
12474428-3816464 & M9$\gamma^{32}$ & ITR$^{32}$ & $\cdots$ & $\cdots$ & $\cdots$ & TWA$^{32}$ & TWA & $47.1$\\
13262009-2729370 & L5$^{38,82}$ & $\cdots$ & $\cdots$ & $\cdots$ & $\cdots$ & $\cdots$ & ARG & $23.3$\\
14252798-3650229 & L5$^{51,82}$ & $\cdots$ & $5.4 \pm 0.3^{4}$ & $11.6 \pm 0.1^{22}$ & $\cdots$ & $\cdots$ & ABDMG & $99.6$\\
17571539+7042011 & M7.5$^{37}$ & U$^{57}$ & $-12.4 \pm 0.6^{103,20}$ & $19.1 \pm 0.4^{57}$ & $\cdots$ & $\cdots$ & ARG & $91.0$\\
19564700-7542270 & L0$\gamma^{15}$ & OR$^{90}$ & $\cdots$ & $\cdots$ & $\cdots$ & THA$^{31}$ & THA & $85.2$\\
20004841-7523070 & M9$^{94}$ & OR$^{90}$ & $11.8 \pm 1.0^{34}$ & $\cdots$ & $\cdots$ & CAS;BPMG$^{34,31}$ & BPMG & $98.2$\\
20111744-2917584 & M5.5$^{81}$ & $\cdots$ & $\cdots$ & $\cdots$ & $\cdots$ & $\cdots$ & ARG & $49.3$\\
20224803-5645567 & M5.5$^{12}$ & $\cdots$ & $\cdots$ & $\cdots$ & $\cdots$ & $\cdots$ & THA & $59.2$\\
20291446-5456116 & M4.3$^{53}$ & $\cdots$ & $-1.4 \pm 1.2^{53}$ & $\cdots$ & $\cdots$ & THA$^{53}$ & THA & $71.4$\\
20330186-4903105 & $\cdots$ & $\cdots$ & $\cdots$ & $16.3 \pm 5.0^{89}$ & $\cdots$ & $\cdots$ & BPMG & $99.1$\\
20334670-3733443 & M5$^{81}$ & $\cdots$ & $\cdots$ & $\cdots$ & $\cdots$ & $\cdots$ & BPMG & $80.0$\\
20414283-3506442 & L2:$^{15}$ & $\cdots$ & $\cdots$ & $\cdots$ & $\cdots$ & $\cdots$ & ABDMG & $14.4$\\
20423672-5425263 & M4.0$^{53}$ & $\cdots$ & $-1.4 \pm 1.7^{53}$ & $\cdots$ & $\cdots$ & THA$^{53}$ & THA & $94.8$\\
21083826-4244540 & M4.4$^{53}$ & $\cdots$ & $-4.9 \pm 1.9^{53}$ & $\cdots$ & $\cdots$ & THA$^{53}$ & THA & $84.4$\\
21265040-8140293 & L3$\gamma^{82}$ & OR$^{90}$ & $\cdots$ & $\cdots$ & $\cdots$ & THA$^{31}$ & THA & $85.1$\\
21420580-3101162 & L2$^{58,27,8}$ & $\cdots$ & $\cdots$ & $\cdots$ & $\cdots$ & $\cdots$ & ABDMG & $12.6$\\
21490499-6413039 & M4.5$^{85,53}$ & X$^{85}$ & $0.4 \pm 5.1^{53}$ & $\cdots$ & $\cdots$ & THA$^{53}$ & THA & $99.7$\\
21543454-1055308 & L4$\beta^{33}$ & ITR$^{33}$ & $\cdots$ & $\cdots$ & $\cdots$ & ARG$^{33}$ & ARG & $58.6$\\
22060961-0723353 & M5.5$^{78}$ & $\cdots$ & $\cdots$ & $\cdots$ & $\cdots$ & $\cdots$ & ABDMG & $82.1$\\
22064498-4217208 & L2$^{14}$ & R$^{14}$ & $\cdots$ & $\cdots$ & $\cdots$ & ABDMG$^{31}$ & ABDMG & $95.2$\\
22244102-7724036 & M4.2$^{53}$ & $\cdots$ & $8.5 \pm 1.4^{53}$ & $\cdots$ & $\cdots$ & THA$^{53}$ & THA & $99.2$\\
22400144+0532162 & $\cdots$ & $\cdots$ & $\cdots$ & $23.6 \pm 2.7^{23}$ & $\cdots$ & $\cdots$ & BPMG & $79.0$\\
22443167+2043433 & L6$\gamma^{82,2}$ & ITRLM$^{90}$ & $\cdots$ & $\cdots$ & $\cdots$ & ABDMG$^{31}$ & ABDMG & $99.8$\\
22444835-6650032 & M4.8$^{53}$ & L$^{53}$ & $0.7 \pm 1.7^{53}$ & $\cdots$ & $\cdots$ & THA$^{53}$ & THA & $99.7$\\
22583200+1014589 & $\cdots$ & $\cdots$ & $\cdots$ & $23.1 \pm 1.4^{23}$ & $\cdots$ & $\cdots$ & ABDMG & $98.3$\\
23130558-6127077 & M4.5$^{53}$ & L$^{53}$ & $2.9 \pm 2.3^{53}$ & $\cdots$ & $\cdots$ & THA$^{53}$ & THA & $99.8$\\
23225240-6151114 & M5$^{31}$ & $\cdots$ & $\cdots$ & $\cdots$ & A$^{31}$ & THA$^{31}$ & THA & $98.7$\\
23225299-6151275 & L2$\gamma^{82}$ & OR$^{16}$ & $\cdots$ & $\cdots$ & B$^{31}$ & THA$^{31}$ & THA & $> 99.9$\\
23225384+7847386 & M5$^{66}$ & UC$^{66}$ & $-17.0 \pm 1.3^{46}$ & $19.1 \pm 5.5^{66,23}$ & B$^{66}$ & Pleiades$^{74}$ & CAR & $29.7$\\
23255604-0259508 & L3$^{8}$ & $\cdots$ & $\cdots$ & $\cdots$ & $\cdots$ & $\cdots$ & ABDMG & $29.8$\\
23392527+3507165 & L3.5$^{82,8}$ & $\cdots$ & $\cdots$ & $\cdots$ & $\cdots$ & $\cdots$ & BPMG & $10.6$\\
23424333-6224564 & M4.3$^{53}$ & $\cdots$ & $5.1 \pm 4.6^{53}$ & $\cdots$ & $\cdots$ & THA$^{53}$ & THA & $99.6$\\
23520507-1100435 & M7$^{14,58,15}$ & $\cdots$ & $\cdots$ & $\cdots$ & $\cdots$ & $\cdots$ & ABDMG & $42.0$\\
\enddata
\tablenotetext{a}{The $\beta$ and $\gamma$ symbols stand for low-gravity and very low-gravity, \emph{p} stands for peculiar, and a semi-colon indicates an uncertain spectral type.}
\tablenotetext{b}{A capital letter means the object displays the associated sign of youth. O: lower-than normal equivalent width of atomic species in the optical spectrum, I: same but in the NIR spectrum, T: a triangular-shaped $H$-band continuum, V: high rotational velocity, X: X-ray emission, R: redder-than-normal colors for given spectral type, U: over luminous, H: H$\alpha$ emission, L: Li absorption, A: signs of accretion, M: signs of low gravity from atmospheric models fitting, N: bright NUV emission and C: Companion to a young star. A question mark following a flag indicates that the result is uncertain.}
\tablenotetext{c}{AB: Unresolved binary, B or C: Resolved companion.}
\tablenotetext{d}{Bona fide member.}
\tablenotetext{~}{References to the table : \\
(1)~\citealp{2007AJ....133..971A}; (2)~\citealp{2013ApJ...772...79A}; (3)~\citealp{2010ApJ...715..724B}; (4)~\citealp{2010ApJ...723..684B}; (5)~\citealp{2005AJ....130.1871B}; (6)~\citealp{2011AJ....142..103B}; (7)~\citealp{2008ApJ...681..579B}; (8)~\citealp{2010ApJ...710.1142B}; (9)~\citealp{2007ApJ...667..520C}; (10)~\citealp{2012A&A...548A..33C}; (11)~\citealp{2005AJ....130..337C}; (12)~\citealp{2005A&A...441..653C}; (13)~\citealp{2002AJ....123.2828C}; (14)~\citealp{2003AJ....126.2421C}; (15)~\citealp{2007AJ....133..439C}; (16)~\citealp{2009AJ....137.3345C}; (17)~\citealp{2009A&A...508..833D}; (18)~\citealp{2002AJ....124.1170D}; (19)~\citealp{2013A&A...553L...5D}; (20)~\citealp{2012ApJ...747L..38T}; (21)~\citealp{2011AJ....141....7D}; (22)~\citealp{2014AJ....147...94D}; (23)~\citealp{2014ApJ...784..156D}; (24)~\citealp{2008A&A...477L...1D}; (25)~\citealp{2012ApJS..201...19D}; (26)~Finder Charts; (27)~\citealp{2009AJ....137....1F}; (28)~\citealp{2012ApJ...752...56F}; (29)~\citealp{2013AJ....145....2F}; (30)~\citealp{2005A&A...435L...5F}; (31)~\citealp{2014ApJ...783..121G}; (32)~\citealp{2014ApJ...785L..14G}; (33)~\citealp{2014ApJ...792L..17G}; (34)~\citealp{2010MNRAS.409..552G}; (35)~\citealp{2009AJ....137..402G}; (36)~\citealp{2002ApJ...564..466G}; (37)~\citealp{2000AJ....120.1085G}; (38)~\citealp{2002ApJ...575..484G}; (39)~\citealp{2004ApJS..150..455G}; (40)~\citealp{2003A&A...401..677G}; (41)~\citealp{1997AJ....113.1458H}; (42)~\citealp{1999A&A...341..163H}; (43)~\citealp{2012ApJ...754...44J}; (44)~\citealp{2009ApJ...704..975J}; (45)~\citealp{2007MNRAS.374..445K}; (46)~\citealp{2007yCat.3254....0K}; (47)~\citealp{2013MNRAS.435.1083K}; (48)~\citealp{2008ApJ...689.1295K}; (49)~\citealp{2010ApJS..190..100K}; (50)~\citealp{2011ApJS..197...19K}; (51)~\citealp{2004AJ....127.3553K}; (52)~\citealp{2013AJ....146..134K}; (53)~\citealp{2014AJ....147..146K}; (54)~\citealp{2011A&A...530A..31L}; (55)~\citealp{2008MNRAS.384..150L}; (56)~\citealp{2010ApJ...708.1482L}; (57)~\citealp{2009AJ....137.3632L}; (58)~\citealp{2006PASP..118..659L}; (59)~\citealp{2008ApJ...689..436L}; (60)~\citealp{2010ApJ...722..311L}; (61)~\citealp{2013AN....334...85L}; (62)~\citealp{2013ApJ...777L..20L}; (63)~\citealp{2007ApJ...669L..97L}; (64)~\citealp{2009ApJ...703..399L}; (65)~\citealp{2013ApJS..205....6M}; (66)~\citealp{2007ApJ...658..480M}; (67)~\citealp{2013ApJ...762...88M}; (68)~\citealp{2014ApJ...788...81M}; (69)~\citealp{2005ApJ...634.1385M}; (70)~\citealp{2013AJ....145...52M}; (71)~\citealp{2013AJ....146..161M}; (72)~\citealp{2010A&A...517A..53M}; (73)~\citealp{2001AJ....122.3466M}; (74)~\citealp{2001MNRAS.328...45M}; (75)~\citealp{2014AJ....147...20N}; (76)~\citealp{2006A&A...446..515P}; (77)~\citealp{2002AJ....124..519R}; (78)~\citealp{2003AJ....126.3007R}; (79)~\citealp{2004AJ....128..463R}; (80)~\citealp{2006AJ....132..891R}; (81)~\citealp{2007AJ....133.2825R}; (82)~\citealp{2008AJ....135..580R}; (83)~\citealp{2009ApJ...705.1416R}; (84)~\citealp{2006MNRAS.373..705R}; (85)~\citealp{2006AJ....132..866R}; (86)~\citealp{2003A&A...400..297R}; (87)~\href{http://www.astro.umontreal.ca/\textasciitilde gagne/ARRiedel2014inprep.php}{A.~R.~Riedel et al., in preparation}; (88)~\citealp{2012PhDT.......100R}; (89)~\citealp{2014AJ....147...85R}; (90)~\citealp{2013ApJ...774..101R}; (91)~\citealp{2012ApJ...748...93R}; (92)~\citealp{2012AJ....143...80S}; (93)~\citealp{2012AJ....144..109S}; (94)~\citealp{2007AJ....133.2258S}; (95)~\citealp{2010AJ....139.1808S}; (96)~\citealp{2012ApJ...757..163S}; (97)~\citealp{2014AJ....147...34S}; (98)~\citealp{2010A&A...512A..37S}; (99)~\citealp{2009ApJ...699..649S}; (100)~\citealp{2012ApJ...758...56S}; (101)~\citealp{2008AJ....136..421Z}; (102)~\citealp{2005AJ....129..413S}; (103)~\citealp{2010PASP..122.1195T}; (104)~\citealp{2008A&A...489..825T}; (105)~\citealp{2009A&A...506..799B}; (106)~\citealp{1974A&A....36..155T}; (107)~\citealp{1996MNRAS.281..644T}; (108)~\citealp{1996MNRAS.281..644T}; (109)~\citealp{2007ASSL..350.....V}; (110)~\citealp{2004AJ....127.2948V}; (111)~\citealp{2008AJ....135..785W}; (112)~\citealp{2011A&A...529A..44W}; (113)~\citealp{2012yCat.1322....0Z}; (114)~\citealp{2014A&A...568A...6Z}; (115)~\citealp{2002ApJS..141..503N}; (116)~\citealp{2005A&A...430L..49S}; (117)~\citealp{2006AJ....131.2722C}; (118)~\citealp{2003IAUS..211..197W}; (119)~\citealp{2000AJ....120..447K}; (120)~\citealp{2006ApJ...639.1120K}.
}
\end{deluxetable*}
\twocolumngrid

%% file: bass_table_4.tex
\LongTables
\begin{deluxetable*}{llcl}
\tablecolumns{4}
\tablecaption{Known YMG Candidate Members not Recovered in \href{http://www.astro.umontreal.ca/\textasciitilde gagne/BASS.php}{\emph{BASS}}.\label{tab:miss}}
\tablehead{\href{http://www.ipac.caltech.edu/2mass/}{\emph{2MASS}} & \colhead{Measured} & \colhead{Known} & \colhead{Reason for} \\
\colhead{Designation} & \colhead{SpT\tablenotemark{a}} & \colhead{Candidacy} & \colhead{Rejection\tablenotemark{b}}}
\startdata
00332386-1521309 & L4$\beta$ & ARG$^{31}$ & $HW2_{CMD}$\\
00470038+6803543 & L7p & ABDMG$^{31}$ & $b$, $2M_\#$\\
01112542+1526214 & M5+M6 & $\beta$PMG$^{67}$ & $W1_{SAT}$\\
01291221+3517580 & L4 & ARG$^{31}$ & $HW2_{CMD}$\\
01424687-5126469 & M6.5 & COL$^{90}$ & $J-H$, $2M_{PH}$, $\sigma\mu$, $\mu$, $P$, $C$\\
02535980+3206373 & M7p & $\beta$PMG$^{31}$ & $HW2_{CMD}$, $P$\\
03214475-3309494 & M5.8 & COL$^{90}$ & $V-J$, $2M_{PH}$, $P$, $C$\\
03244305-2733230 & M5.5 & COL$^{90}$ & $K_S-W1$, $\mu$, $P$, $C$\\
03350208+2342356 & M8.5 & $\beta$PMG$^{100}$ & $W1_{SAT}$, $W2_{SAT}$, $C$\\
04062677-3812102 & L0$\gamma$ & COL$^{31}$ & $P$, $C$\\
05184616-2756457 & L1$\gamma$ & COL$^{31}$ & $\mu$\\
06195260-2903592 & M6 & COL$^{31}$ & $\mu$\\
06322402-5010349 & L3 & ABDMG$^{31}$ & $HW2_{CMD}$, $\sigma\mu$, $C$\\
07285117-3015527 & M5 & ABDMG$^{100}$ & $b$, $W1_{SAT}$, $2M_\#$\\
09445422-1220544 & M5 & ARG$^{67}$ & $W1-W2$, $W1_{SAT}$\\
10042066+5022596 & L3$\beta$ & ABDMG$^{31}$ & $W1_{SAT}$, $P$, $C$\\
10172689-5354265 & M5 & $\beta$PMG$^{105}$ & $b$, $J-H$, $W1_{SAT}$, $2M_\#$\\
11321831-3019518 & M5 & TWA$^{67}$ & $H-K_S$, $K_S-W1$\\
11324116-2652090 & M5 & TWA$^{69}$ & $H-K_S$, $K_S-W1$, $2M_{CC}$, $W1_{SAT}$, $W_{CC}$\\
12242443-5339088 & M5 & $\beta$PMG$^{67}$ & $b$, $K_S-W1$, $HW2_{CMD}$, $2M_\#$\\
12451416-4429077 & M9.5p & TWA$^{31}$ & $2M_\#$\\
13142039+1320011 & M7 & ABDMG$^{93}$ & $P$\\
16002647-2456424 & M7.5p & ABDMG$^{31}$ & $JK_{CMD}$, $HW2_{CMD}$, USco, $2M_\#$, $P$, $C$\\
16471580+5632057 & L9p & ARG$^{31}$ & $P$, $C$\\
17410280-4642218 & L7p & $\beta$PMG;ABDMG$^{97}$ & $b$, $2M_{CC}$, $2M_\#$\\
18450097-1409053 & M5 & ARG$^{67}$ & $b$, $W1-W2$, $2M_{CC}$, $2M_{PROX}$, $W1_{SAT}$, $W2_{SAT}$, $2M_\#$, $P$\\
21011544+1756586 & L7.5 & ABDMG$^{31}$ & $2M_{PH}$, $2M_{CC}$, $2M_\#$\\
21103096-2710513 & M5 & $\beta$PMG$^{67}$ & $WISE$\\
21140802-2251358 & L7 & $\beta$PMG$^{62}$ & $2M_{PH}$\\
21354554-4218343 & M5.2 & THA$^{53}$ & $B-V$, $P$, $C$\\
21374019+0137137 & M5 & $\beta$PMG$^{93}$ & $H-K_S$, $W1_{SAT}$\\
21481633+4003594 & L6 & ARG$^{31}$ & $b$, $2M_\#$\\
22081363+2921215 & L3$\gamma$ & $\beta$PMG$^{31}$ & $P$, $C$\\
23204705-6723209 & M5 & THA$^{67}$ & $V-J$, $2M_{PH}$, $2M_{CC}$, $\sigma\mu$\\
23512200+3010540 & L5.5 & ARG$^{31}$ & $B-J$, $\chi^2_{W1}$, $2M_{BL}$, $HW2_{CMD}$\\
\enddata
\tablenotetext{a}{Measured in the NIR unless symbol otherwise specified.}
\tablenotetext{b}{This column contains codes corresponding to the filters that rejected an object from the \href{http://www.astro.umontreal.ca/\textasciitilde gagne/BASS.php}{\emph{BASS}} catalog; (1)~$WISE$ -- No entry in the \href{http://irsa.ipac.caltech.edu/Missions/wise.html}{\emph{WISE}} and \href{http://wise2.ipac.caltech.edu/docs/release/allwise/}{\emph{AllWISE}} catalogs, (2)~$b$ -- Absolute Galactic latitude is too low, (3)~$B-V$ color is too blue, (4)~$B-J$ color is too blue, (5)~$V-J$ color is too blue, (6)~$J-H$ color is too blue, (7)~$H-K_S$ color is too blue, (8)~$K_S-W1$ color is too blue, (9)~$W1-W2$ color is too blue, (10)~$\chi^2_{W1}$ -- the reduced $\chi^2$ from the adjusted profile in the $W1$ band is too large, (11) $2M_{PH}$ -- \href{http://www.ipac.caltech.edu/2mass/}{\emph{2MASS}} photometric quality is too low, (12) $2M_{BL}$ -- A blend flag is suspicious in \href{http://www.ipac.caltech.edu/2mass/}{\emph{2MASS}}, (13) $2M_{CC}$ -- A contamination flag is suspicious in \href{http://www.ipac.caltech.edu/2mass/}{\emph{2MASS}}, (14) $2M_{PROX}$ -- A close-by \href{http://www.ipac.caltech.edu/2mass/}{\emph{2MASS}} source is unresolved in \href{http://wise2.ipac.caltech.edu/docs/release/allwise/}{\emph{AllWISE}}, (15) $W1_{SAT}$ -- $W1$ magnitude is saturated, (16) $W2_{SAT}$ -- $W2$ magnitude is saturated, $W_{CC}$ -- A contamination flag is suspicious in \href{http://wise2.ipac.caltech.edu/docs/release/allwise/}{\emph{AllWISE}}, (17) $JK_{CMD}$ -- The object falls to the left of the $M_{W1}$ versus $J-K_S$ field sequence using its statistical distance, (18) $HW2_{CMD}$ -- The object falls to the left of the $M_{W1}$ versus $H-W2$ field sequence using its statistical distance, (19) Usco -- The object is too close to Upper Scorpius, (20) $2M_\#$ -- The object has too many immediate neighbours in \href{http://www.ipac.caltech.edu/2mass/}{\emph{2MASS}}, (21) $\sigma\mu$ -- the \href{http://www.ipac.caltech.edu/2mass/}{\emph{2MASS}}--\href{http://wise2.ipac.caltech.edu/docs/release/allwise/}{\emph{AllWISE}} proper motion is not precise enough, (22) $\mu$ -- The proper motion is too low, (23) $P$ -- The Bayesian probability is too low, (24) $C$ -- The contamination probability is too high. See Sections~\ref{sec:crossmatch}--\ref{sec:filters} for detailed descriptions of these respective filters.}
\tablenotetext{~}{References to this table are identical to those of \hyperref[tab:lit]{Table~\ref*{tab:lit}}.}
\end{deluxetable*}

%% file: bass_table_5_short.tex
\onecolumngrid
\LongTables
\begin{deluxetable*}{lllcccccc}
\tablecolumns{9}
\tablecaption{The Complete \href{http://www.astro.umontreal.ca/\textasciitilde gagne/BASS.php}{\emph{BASS}} Catalog.\label{tab:allbass}}
\tablehead{\colhead{\href{http://www.ipac.caltech.edu/2mass/}{\emph{2MASS}}} & \colhead{Spectral} & \colhead{Probable} & \colhead{Bayesian} & \colhead{Contamination} & \colhead{Estimated Mass} & \colhead{Statistical} & \colhead{Statistical}\\
\colhead{Designation} & \colhead{Type\tablenotemark{a}} & \colhead{Membership} & \colhead{Prob. (\%)} & \colhead{Prob. (\%)} & \colhead{Range (\Mjup)} & \colhead{Distance (pc)} & \colhead{RV (\kms)}}
\startdata
\cutinhead{Candidates with a High Probability}
00011217+1535355 & L4: & ABDMG & $77.8$ & $1.8$ & $17.5^{+0.8}_{-1.1}$ & $27.3 \pm 1.6$ & $-6.5 \pm 2.0$\\
00040288-6410358 & L1$\gamma$ & THA & $> 99.9$ & $< 0.1$ & $12.8 \pm 0.3$ & $45.0 \pm 2.4$ & $6.5 \pm 2.5$\\
00041589-8747254 & (M5.7) & THA & $55.4$ & $< 0.1$ & $60.9^{+8.8}_{-7.1}$ & $51.8 \pm 3.6$ & $11.3 \pm 2.2$\\
00065794-6436542 & M9: & THA & $> 99.9$ & $< 0.1$ & $20.5^{+1.1}_{-13.9}$ & $41.4 \pm 2.4$ & $6.2 \pm 2.4$\\
00111532-3756553 & (M5.7) & THA & $80.2$ & $< 0.1$ & $60.6^{+8.6}_{-6.7}$ & $38.2^{+2.0}_{-2.4}$ & $1.5 \pm 2.2$\\
00182834-6703130 & (M9.6) & THA & $99.8$ & $< 0.1$ & $13.3 \pm 0.3$ & $43.8^{+2.8}_{-2.4}$ & $6.9 \pm 2.5$\\
00191296-6226005 & (M9.7) & THA & $99.5$ & $< 0.1$ & $13.3^{+0.3}_{-0.4}$ & $46.6^{+2.4}_{-2.8}$ & $6.7 \pm 2.5$\\
00192626+4614078 & M8 & ABDMG & $92.1$ & $4.1$ & $87.1^{+8.5}_{-8.6}$ & $37.8 \pm 3.2$ & $-19.5 \pm 3.0$\\
00212774-6351081 & M5.5 & THA & $99.8$ & $< 0.1$ & $158.3^{+19.9}_{-18.5}$ & $44.2^{+2.8}_{-2.4}$ & $6.8 \pm 2.4$\\
00235732-5531435 & M4.1 & THA & $99.8$ & $< 0.1$ & $133.1^{+17.4}_{-14.6}$ & $41.4 \pm 2.4$ & $5.3 \pm 0.7$\\
\cutinhead{Candidates with a Modest Probability}
00160844-0043021 & L5.5 & BPMG & $19.1$ & $36.1$ & $9.6 \pm 0.3$ & $30.9^{+2.8}_{-3.2}$ & $3.3 \pm 1.8$\\
00274534-0806046 & (M5.3) & BPMG & $45.6$ & $35.1$ & $66.9 \pm 4.2$ & $32.1 \pm 2.8$ & $4.4 \pm 1.5$\\
00464841+0715177 & M9 & BPMG & $77.0$ & $26.9$ & $15.0^{+0.1}_{-0.3}$ & $33.8^{+2.8}_{-3.2}$ & $3.2 \pm 1.7$\\
00581143-5653326 & (L6.1) & ARG & $80.4$ & $32.9$ & $10.3^{+0.7}_{-0.3}$ & $25.3^{+2.8}_{-2.4}$ & $2.6 \pm 2.0$\\
01525534-6329301 & (M4.7) & BPMG & $71.4$ & $22.1$ & $107.6^{+6.8}_{-7.8}$ & $23.7 \pm 2.4$ & $14.7 \pm 1.7$\\
02534448-7959133 & M5.5 & BPMG & $50.1$ & $30.9$ & $66.9 \pm 4.9$ & $28.9^{+2.8}_{-3.2}$ & $12.0 \pm 2.1$\\
03390160-2434059 & M5.9 & COL & $77.8$ & $31.9$ & $204.7^{+6.6}_{-3.5}$ & $59.4^{+5.6}_{-6.0}$ & $18.6 \pm 1.8$\\
03473987-4114014 & (M5.3) & COL & $38.0$ & $45.4$ & $77.2^{+11.0}_{-10.5}$ & $71.0^{+8.8}_{-8.0}$ & $19.7 \pm 1.7$\\
03510460-5701469 & (M5.1) & COL & $17.6$ & $47.4$ & $88.3^{+12.0}_{-11.7}$ & $68.6^{+8.8}_{-8.0}$ & $19.1 \pm 1.7$\\
03550477-1032415 & M8.5 & BPMG & $39.5$ & $38.5$ & $26.4^{+3.5}_{-4.2}$ & $35.0^{+4.4}_{-4.8}$ & $17.7 \pm 1.8$\\
\enddata
\tablenotetext{a}{Spectral types in parentheses were estimated from \href{http://www.ipac.caltech.edu/2mass/}{\emph{2MASS}}--\href{http://wise2.ipac.caltech.edu/docs/release/allwise/}{\emph{AllWISE}} colors (see \hyperref[sec:filters]{Section~\ref*{sec:spt}}).}
\tablenotetext{b}{The binary hypothesis is more probable than the single hypothesis (see \hyperref[sec:filters]{Section~\ref*{sec:filters}}).}
\tablenotetext{~}{This table is available in its entirety in the online journal. The complete table has 252 rows.}
\end{deluxetable*}
\twocolumngrid

%% file: bass_table_A1_short.tex
\onecolumngrid
\LongTables
\begin{deluxetable*}{lcccclccccc}
\tablecolumns{11}
\tablecaption{All-Sky input sample of nearby, potential $>$ M5 objects.\label{tab:input}}
\tablehead{\multicolumn{4}{c}{\href{http://www.ipac.caltech.edu/2mass/}{\emph{2MASS}}} & \colhead{} & \multicolumn{3}{c}{\href{http://wise2.ipac.caltech.edu/docs/release/allwise/}{\emph{AllWISE}}} & \colhead{} & \colhead{$\mu_\alpha$} & \colhead{$\mu_\delta$}\\
\cline{1-4}
\cline{6-8}
\colhead{Designation}  & \colhead{$J$} & \colhead{$H$} & \colhead{$K_S$} & \colhead{} & \colhead{Designation} & \colhead{$W1$} & \colhead{$W2$} & \colhead{} & \colhead{(\masyr)} & \colhead{(\masyr)}}
\startdata
00000027-1534494 & $10.47 \pm 0.02$ & $9.90 \pm 0.02$ & $9.63 \pm 0.02$ & & 000000.46-153448.4 & $9.40 \pm 0.02$ & $9.22 \pm 0.02$ & & $240.5 \pm 9.3$ & $87.4 \pm 7.4$\\
00000058-2621542 & $12.83 \pm 0.02$ & $12.27 \pm 0.02$ & $11.98 \pm 0.02$ & & 000000.60-262154.9 & $11.78 \pm 0.02$ & $11.59 \pm 0.02$ & & $27.2 \pm 5.2$ & $-63.1 \pm 6.2$\\
00000160-7721530 & $15.66 \pm 0.08$ & $15.09 \pm 0.09$ & $14.76 \pm 0.13$ & & 000002.10-772152.6 & $14.33 \pm 0.03$ & $14.06 \pm 0.04$ & & $151.6 \pm 3.8$ & $32.9 \pm 10.0$\\
00000296+2541349 & $13.34 \pm 0.02$ & $12.82 \pm 0.02$ & $12.51 \pm 0.03$ & & 000002.98+254134.4 & $12.29 \pm 0.02$ & $12.09 \pm 0.02$ & & $29.0 \pm 5.2$ & $-40.4 \pm 6.1$\\
00000497+3740328 & $15.66 \pm 0.05$ & $15.15 \pm 0.08$ & $14.82 \pm 0.10$ & & 000004.96+374033.4 & $14.57 \pm 0.03$ & $14.32 \pm 0.05$ & & $-16.0 \pm 6.7$ & $56.3 \pm 9.6$\\
00000540-5418547 & $14.23 \pm 0.03$ & $13.70 \pm 0.04$ & $13.39 \pm 0.04$ & & 000005.29-541855.4 & $13.20 \pm 0.02$ & $13.03 \pm 0.03$ & & $-85.2 \pm 3.7$ & $-64.8 \pm 8.5$\\
\enddata
\tablenotetext{~}{This table is available in its entirety in the online journal. The complete table has 98\,970 rows.}
\end{deluxetable*}
\twocolumngrid

%% file: bass_table_B1.tex
\onecolumngrid
\LongTables
\begin{deluxetable*}{lllcccllc}
\tablecolumns{9}
\tablecaption{\href{http://www.astro.umontreal.ca/\textasciitilde gagne/LP-BASS.php}{\emph{LP-BASS}} Candidates With Additional Information in the Literature.\label{tab:lit_annex}}
\tablehead{\colhead{\href{http://www.ipac.caltech.edu/2mass/}{\emph{2MASS}}} & \colhead{Measured} & \colhead{Signs of} & \colhead{RV} & \colhead{Trig.}  & \colhead{Multipli-} & \colhead{Known} & \colhead{Updated} & \colhead{Updated} \\
\colhead{Designation} & \colhead{SpT\tablenotemark{a}} & \colhead{Youth\tablenotemark{b}} & \colhead{(\kms)} & \colhead{Dist. (pc)} & \colhead{city\tablenotemark{c}} & \colhead{Membership} & \colhead{Membership} & \colhead{Prob. (\%)}}
\startdata
00165057-7122387 & $\cdots$ & $\cdots$ & $-3.4 \pm 3.0^{54}$ & $\cdots$ & $\cdots$ & $\cdots$ & THA & $36.6$\\
00192753-3620153 & M5.5$^{12}$ & $\cdots$ & $\cdots$ & $\cdots$ & $\cdots$ & $\cdots$ & THA & $11.5$\\
00281434-3227556 & M5$^{85}$ & $\cdots$ & $\cdots$ & $\cdots$ & $\cdots$ & $\cdots$ & BPMG & $30.4$\\
00303013-1450333 & L7$^{14,47}$ & $\cdots$ & $\cdots$ & $26.7 \pm 3.2^{110}$ & $\cdots$ & $\cdots$ & ARG & $24.1$\\
00425349-6117384 & M4.2$^{53}$ & $\cdots$ & $6.9 \pm 1.0^{53}$ & $\cdots$ & $\cdots$ & THA$^{53}$ & THA & $99.9$\\
00455663+3347109 & M4.5+M5.5$^{55}$ & $\cdots$ & $\cdots$ & $18.1 \pm 1.3^{47,23}$ & AB$^{55}$ & $\cdots$ & ARG & $89.7$\\
00551459+4511019 & $\cdots$ & $\cdots$ & $\cdots$ & $35.8 \pm 3.3^{23}$ & $\cdots$ & $\cdots$ & ABDMG & $19.3$\\
00584253-0651239 & L0$^{14,48,27}$ & $\cdots$ & $\cdots$ & $29.6 \pm 3.5^{71}$ & $\cdots$ & $\cdots$ & BPMG & $88.6$\\
01000219-6156270 & M6$^{12}$ & $\cdots$ & $\cdots$ & $\cdots$ & $\cdots$ & $\cdots$ & THA & $99.1$\\
01044008+1129485 & $\cdots$ & N$^{92}$ & $\cdots$ & $\cdots$ & $\cdots$ & ABDMG$^{92}$ & BPMG & $76.5$\\
01234181-3833496 & M4.5$^{78}$ & $\cdots$ & $18.4 \pm 6.3^{101}$ & $\cdots$ & $\cdots$ & $\cdots$ & BPMG & $0.6$\\
01253196-6646023 & M4.2$^{53}$ & $\cdots$ & $7.1 \pm 5.1^{53}$ & $\cdots$ & $\cdots$ & THA$^{53}$ & THA & $99.7$\\
01275875-6032243 & M4.2$^{90,53}$ & N & $9.1 \pm 2.5^{53}$ & $\cdots$ & $\cdots$ & THA$^{90,53}$ & THA & $> 99.9$\\
01283025-4921094 & M4.1$^{53}$ & $\cdots$ & $6.5 \pm 5.7^{53}$ & $\cdots$ & $\cdots$ & THA$^{53}$ & THA & $99.3$\\
01375879-5645447 & M3.9$^{53}$ & $\cdots$ & $8.5 \pm 0.6^{53}$ & $\cdots$ & $\cdots$ & THA$^{53}$ & THA & $99.9$\\
01534955+4427284 & $\cdots$ & $\cdots$ & $\cdots$ & $20.2 \pm 1.2^{23}$ & $\cdots$ & $\cdots$ & ARG & $98.5$\\
02001992-6614017 & M4.3$^{90,53}$ & N & $11.8 \pm 1.1^{53}$ & $\cdots$ & $\cdots$ & THA$^{90,53}$ & THA & $> 99.9$\\
02025788-3136262 & M4.0$^{90}$ & N & $\cdots$ & $\cdots$ & $\cdots$ & FIELD$^{90}$ & COL & $40.7$\\
02030658-5545420 & M4.5$^{90}$ & N & $\cdots$ & $\cdots$ & $\cdots$ & ABDMG$^{90}$ & THA & $99.9$\\
02033222+0648588 & $\cdots$ & $\cdots$ & $\cdots$ & $\cdots$ & C$^{113}$ & ABDMG$^{92}$ & BPMG & $64.5$\\
02123372-6049185 & M6.5$^{34}$ & $\cdots$ & $13.1 \pm 0.2^{34}$ & $\cdots$ & $\cdots$ & $\cdots$ & THA & $94.8$\\
02190228+2352550 & M3.6$^{99}$ & X$^{99}$ & $15.7 \pm 0.7^{100}$ & $20.6 \pm 0.8^{23}$ & $\cdots$ & $\cdots$ & ARG & $72.2$\\
02294569-5541496 & M4.8$^{53}$ & L$^{53}$ & $11.5 \pm 1.0^{53}$ & $\cdots$ & $\cdots$ & THA$^{53}$ & THA & $> 99.9$\\
02341866-5128462 & M4.3$^{53}$ & $\cdots$ & $10.9 \pm 0.9^{53}$ & $\cdots$ & $\cdots$ & THA$^{53}$ & THA & $> 99.9$\\
02351494+0247534 & $\cdots$ & $\cdots$ & $\cdots$ & $17.8 \pm 1.0^{23}$ & $\cdots$ & $\cdots$ & BPMG & $77.6$\\
02383255-7528065 & M4.1$^{53}$ & $\cdots$ & $12.3 \pm 0.6^{53}$ & $\cdots$ & $\cdots$ & THA$^{53}$ & THA & $98.9$\\
02412721-3049149 & M4.7$^{90,53}$ & ON$^{90}$ & $18.2 \pm 1.1^{53}$ & $\cdots$ & $\cdots$ & THA$^{90,53}$ & BPMG & $88.3$\\
02420204-5359147 & M4.6$^{90,53}$ & N & $11.5 \pm 2.3^{53}$ & $\cdots$ & $\cdots$ & THA$^{90,53}$ & THA & $> 99.9$\\
02591904-5122341 & M5.4$^{53}$ & L$^{53}$ & $11.0 \pm 2.3^{53}$ & $\cdots$ & $\cdots$ & THA$^{53}$ & THA & $> 99.9$\\
03090022-4924513 & M4.5$^{84}$ & $\cdots$ & $\cdots$ & $\cdots$ & $\cdots$ & $\cdots$ & ARG & $18.8$\\
03104941-3616471 & M4.3$^{90,53}$ & N & $13.8 \pm 1.6^{53}$ & $\cdots$ & $\cdots$ & THA$^{90,53}$ & THA & $> 99.9$\\
03341065-2130343 & M6$^{14}$ & $\cdots$ & $19.0 \pm 0.8^{34}$ & $\cdots$ & $\cdots$ & IC 2391?$^{34}$ & BPMG & $22.9$\\
03370359-1758079 & L4.5$^{1}$ & $\cdots$ & $\cdots$ & $\cdots$ & $\cdots$ & $\cdots$ & ARG & $11.6$\\
03561624-3915219 & M5.0$^{90,53}$ & N & $16.7 \pm 0.7^{53}$ & $\cdots$ & $\cdots$ & THA$^{90,53}$ & THA & $99.9$\\
04032484+0824508 & $\cdots$ & X$^{92}$ & $\cdots$ & $\cdots$ & $\cdots$ & ABDMG$^{92}$ & BPMG & $96.7$\\
04054799-1515399 & M8$^{45}$ & $\cdots$ & $\cdots$ & $\cdots$ & $\cdots$ & $\cdots$ & THA & $70.1$\\
04111790-0556489 & M9$^{111}$ & $\cdots$ & $20.1 \pm 5.0^{111}$ & $\cdots$ & $\cdots$ & $\cdots$ & COL & $20.6$\\
04133609-4413325 & M3.9$^{90,53}$ & N & $16.4 \pm 1.4^{53}$ & $\cdots$ & $\cdots$ & THA$^{90,53}$ & THA & $99.6$\\
04231498-1533245 & $\cdots$ & $\cdots$ & $\cdots$ & $22.4 \pm 1.0^{23}$ & AB$^{26}$ & $\cdots$ & BPMG & $93.3$\\
04390494-0959012 & M6$^{13}$ & $\cdots$ & $\cdots$ & $\cdots$ & $\cdots$ & $\cdots$ & ABDMG & $15.4$\\
04475779-5035200 & M4.0$^{53}$ & $\cdots$ & $18.6 \pm 0.9^{53}$ & $\cdots$ & $\cdots$ & THA$^{53}$ & COL & $72.0$\\
05195412-0723359 & M4+M4.5$^{85,43}$ & X$^{85}$ & $\cdots$ & $\cdots$ & AB$^{43}$ & $\cdots$ & COL & $89.7$\\
06142994-6318559 & $\cdots$ & $\cdots$ & $\cdots$ & $\cdots$ & Ab$^{73}$ & $\cdots$ & ARG & $89.0$\\
06313103-8811365 & M5$^{102}$ & $\cdots$ & $\cdots$ & $\cdots$ & $\cdots$ & $\cdots$ & ARG & $28.1$\\
07135309-6545115 & $\cdots$ & $\cdots$ & $\cdots$ & $\cdots$ & AB$^{26}$ & $\cdots$ & CAR & $91.5$\\
07140394+3702459 & M8$^{94,82,27}$ & $\cdots$ & $\cdots$ & $12.5 \pm 0.7^{23}$ & $\cdots$ & $\cdots$ & ARG & $74.9$\\
07355465+3333459 & M4.5$^{78}$ & $\cdots$ & $\cdots$ & $32.4 \pm 2.5^{23}$ & $\cdots$ & $\cdots$ & ABDMG & $26.2$\\
10023100-2814280 & M4+M6$^{80}$ & $\cdots$ & $\cdots$ & $\cdots$ & AB$^{43}$ & $\cdots$ & CAR & $93.7$\\
10134260-2759586 & M5$^{38}$ & $\cdots$ & $\cdots$ & $\cdots$ & $\cdots$ & TWA$^{38}$ & CAR & $43.5$\\
10451718-2607249 & M8$^{38,82,27}$ & $\cdots$ & $\cdots$ & $\cdots$ & $\cdots$ & $\cdots$ & ABDMG & $18.2$\\
15031325-2840134 & M5$^{78}$ & $\cdots$ & $\cdots$ & $\cdots$ & $\cdots$ & $\cdots$ & ABDMG & $4.3$\\
20042845-3356105 & M4.5$^{85}$ & X$^{85}$ & $\cdots$ & $\cdots$ & $\cdots$ & $\cdots$ & BPMG & $93.8$\\
21144103-4339531 & $\cdots$ & $\cdots$ & $2.7 \pm 0.3^{34}$ & $\cdots$ & $\cdots$ & CAS$^{34}$ & ABDMG & $74.5$\\
21272613-4215183 & M8$^{82}$ & $\cdots$ & $-7.6 \pm 0.3^{34}$ & $34.6 \pm 7.5^{108}$ & $\cdots$ & Pleiades$^{34}$ & BPMG & $82.5$\\
21380269-5744583 & M3.7$^{53}$ & $\cdots$ & $-0.5 \pm 1.3^{53}$ & $\cdots$ & $\cdots$ & THA$^{53}$ & THA & $98.7$\\
21414678-2704542 & M4.5$^{78}$ & $\cdots$ & $\cdots$ & $\cdots$ & $\cdots$ & $\cdots$ & ABDMG & $45.7$\\
22021125-1109461 & M6.5$^{77}$ & $\cdots$ & $-9.4 \pm 1.0^{40}$ & $\cdots$ & $\cdots$ & $\cdots$ & ABDMG & $84.9$\\
22043859-1832204 & M4.5$^{5}$ & $\cdots$ & $-7.2 \pm 3.8^{52}$ & $\cdots$ & $\cdots$ & $\cdots$ & BPMG & $26.3$\\
22294830-4858285 & M4.5$^{80}$ & $\cdots$ & $\cdots$ & $\cdots$ & $\cdots$ & $\cdots$ & BPMG & $21.1$\\
22302626-0142063 & M4$^{5}$ & $\cdots$ & $\cdots$ & $\cdots$ & $\cdots$ & $\cdots$ & ABDMG & $14.6$\\
22541103+1606546 & M4$^{42}$ & $\cdots$ & $\cdots$ & $30.2 \pm 1.3^{23}$ & $\cdots$ & $\cdots$ & ARG & $68.8$\\
23261182+1700082 & M4.5+M6$^{43}$ & $\cdots$ & $\cdots$ & $\cdots$ & AB$^{43}$ & $\cdots$ & BPMG & $66.8$\\
23301129-0237227 & M6$^{81}$ & $\cdots$ & $\cdots$ & $\cdots$ & $\cdots$ & $\cdots$ & BPMG & $42.8$\\
23310161-0406193 & M8+L3$^{9}$ & $\cdots$ & $-12.86 \pm 0.09^{9,115}$ & $26.1 \pm 0.4^{109}$ & AB$^{9}$ & $\cdots$ & ABDMG & $0.5$\\
23524562-5229593 & M4.6$^{53}$ & L$^{53}$ & $3.1 \pm 0.7^{53}$ & $\cdots$ & $\cdots$ & THA$^{53}$ & THA & $99.9$\\
\enddata
\tablenotetext{a}{The $\beta$ and $\gamma$ symbols stand for low-gravity and very low-gravity, \emph{p} stands for peculiar, and a semi-colon indicates an uncertain spectral type.}
\tablenotetext{b}{A capital letter means the object displays the associated sign of youth. O: lower-than normal equivalent width of atomic species in the optical spectrum, I: same but in the NIR spectrum, T: a triangular-shaped $H$-band continuum, V: high rotational velocity, X: X-ray emission, R: redder-than-normal colors for given spectral type, U: over luminous, H: H$\alpha$ emission, L: Li absorption, A: signs of accretion, M: signs of low gravity from atmospheric models fitting, N: bright NUV emission and C: Companion to a young star. A question mark following a flag indicates that the result is uncertain.}
\tablenotetext{c}{AB: Unresolved binary, B or C: Resolved companion.}
\tablenotetext{~}{References to this table are identical to those of Table~\ref{tab:lit}.}
\end{deluxetable*}
\twocolumngrid

%% file: bass_table_B2_short.tex
\onecolumngrid
\LongTables
\begin{deluxetable*}{lllcccccc}
\tablecolumns{9}
\tablecaption{The Complete \href{http://www.astro.umontreal.ca/\textasciitilde gagne/LP-BASS.php}{\emph{LP-BASS}} Catalog.\label{tab:allbass_ann}}
\tablehead{\colhead{\href{http://www.ipac.caltech.edu/2mass/}{\emph{2MASS}}} & \colhead{Spectral} & \colhead{Probable} & \colhead{Bayesian} & \colhead{Contamination} & \colhead{Estimated Mass} & \colhead{Statistical} & \colhead{Statistical}\\
\colhead{Designation} & \colhead{Type\tablenotemark{a}} & \colhead{Membership} & \colhead{Prob. (\%)} & \colhead{Prob. (\%)} & \colhead{Range (\Mjup)} & \colhead{Distance (pc)} & \colhead{RV (\kms)}}
\startdata
\cutinhead{Candidates with a High Probability}
00081980-2559449 & (M5.8) & ABDMG & $60.1$ & $5.3$ & $87.1^{+8.2}_{-7.8}$ & $36.2^{+2.4}_{-2.0}$ & $10.0 \pm 2.0$\\
00091768+0603461 & (M5.2) & ABDMG & $36.6$ & $2.4$ & $156.3^{+14.4}_{-12.8}$ & $25.3 \pm 1.6$ & $-2.0 \pm 2.0$\\
00165057-7122387 & (M5.7) & THA & $36.6$ & $< 0.1$ & $57.7^{+8.1}_{-6.3}$ & $47.4 \pm 3.2$ & $-3.4 \pm 3.0$\\
00165242-7640540 & (M5.3) & THA & $31.8$ & $< 0.1$ & $85.0^{+11.1}_{-8.9}$ & $45.4^{+3.2}_{-2.8}$ & $6.4 \pm 2.4$\\
00200551-5359372 & (M6.2) & THA & $98.9$ & $< 0.1$ & $36.2^{+9.3}_{-6.7}$ & $39.8^{+2.4}_{-2.0}$ & $5.3 \pm 2.4$\\
00303013-1450333 & L7 & ARG & $24.1$ & $2.6$ & $10.4^{+0.6}_{-0.4}$ & $26.7 \pm 3.2$ & $4.3 \pm 2.0$\\
00381489-6403529 & (M8.6) & THA & $99.7$ & $< 0.1$ & $15.3^{+0.7}_{-6.1}$ & $44.2 \pm 2.4$ & $7.5 \pm 2.4$\\
00425349-6117384 & M4.2 & THA & $99.9$ & $< 0.1$ & $123.0^{+15.6}_{-13.1}$ & $42.6 \pm 2.4$ & $6.9 \pm 1.0$\\
00455663+3347109 & M4.5+M5.5 & ARG & $89.7$ & $0.1$ & $86.4^{+8.4}_{-7.7}$ & $18.1 \pm 1.3$ & $4.3 \pm 1.4$\\
00474453+4159428 & (M3.7) & BPMG & $49.0$ & $14.4$ & $169.0^{+10.5}_{-11.0}$ & $30.5 \pm 2.8$ & $-3.2 \pm 2.2$\\
\cutinhead{Candidates with a Modest Probability}
00085614-2813211 & (L8.9) & BPMG & $21.5$ & $21.8$ & $6.1 \pm 0.1$ & $16.1 \pm 1.2$ & $5.8 \pm 1.5$\\
00102936-0746487 & (M6.2) & ABDMG & $18.2$ & $19.9$ & $74.3^{+7.0}_{-6.6}$ & $43.8^{+3.2}_{-2.8}$ & $3.3 \pm 2.1$\\
00192753-3620153 & M5.5 & THA & $11.5$ & $43.0$ & $60.3^{+8.6}_{-6.8}$ & $37.8^{+2.0}_{-2.4}$ & $0.9 \pm 2.2$\\
00193193-0554404 & (M5.0) & BPMG & $30.8$ & $44.2$ & $89.9^{+6.4}_{-6.1}$ & $33.8^{+3.6}_{-3.2}$ & $3.2 \pm 1.7$\\
00281434-3227556 & M5 & BPMG & $30.4$ & $45.7$ & $168.7^{+10.6}_{-11.1}$ & $32.1^{+2.8}_{-3.2}$ & $8.1 \pm 1.5$\\
00324451+2744454 & (M5.0) & BPMG & $17.4$ & $36.0$ & $93.9 \pm 5.6$ & $35.8 \pm 3.2$ & $-3.6 \pm 2.0$\\
00465095+3822416 & (M5.5) & ARG & $15.8$ & $28.2$ & $77.3^{+7.9}_{-8.2}$ & $33.8^{+3.2}_{-3.6}$ & $2.3 \pm 1.7$\\
00473149-1424425 & (M4.8) & BPMG & $54.3$ & $34.2$ & $100.0^{+5.7}_{-6.3}$ & $30.9 \pm 2.8$ & $6.9 \pm 1.5$\\
00584590+2430511 & (M5.8) & BPMG & $24.8$ & $27.3$ & $46.8 \pm 2.6$ & $31.3 \pm 2.8$ & $2.2 \pm 2.1$\\
01012488-2412472 & (M6.0) & BPMG & $12.1$ & $30.2$ & $41.5^{+2.8}_{-2.5}$ & $23.3 \pm 2.0$ & $9.3 \pm 1.5$\\
\enddata
\tablenotetext{a}{Spectral types in parentheses were estimated from \href{http://www.ipac.caltech.edu/2mass/}{\emph{2MASS}}--\href{http://wise2.ipac.caltech.edu/docs/release/allwise/}{\emph{AllWISE}} colors (see \hyperref[sec:filters]{Section~\ref*{sec:spt}}).}
\tablenotetext{b}{The binary hypothesis is more probable than the single hypothesis (see \hyperref[sec:filters]{Section~\ref*{sec:filters}}).}
\tablenotetext{~}{This table is available in its entirety in the online journal. The complete table has 249 rows.}
\end{deluxetable*}
\twocolumngrid